\documentclass[twocolumn,rmp,aps,amssymb,showkeys,floatfix]{revtex4}

\usepackage{graphicx}
\usepackage{dcolumn}
\usepackage{bm}
\usepackage{bbm}
\usepackage{physics}
\usepackage[utf8]{inputenc}
\usepackage[T1]{fontenc}
\usepackage{hhline}
\usepackage{enumitem}
\usepackage{mathtools}
\usepackage{tabularx}
\usepackage{color}
\usepackage{booktabs}
\usepackage{multirow}

\setlist[enumerate]{label*=\arabic*.}
\UseRawInputEncoding

\begin{document}

\title{Continuous-variable quantum communication}%

\author{Vladyslav C. Usenko}%
 \email{usenko@optics.upol.cz}
\affiliation{%
Department of Optics, Palacky University, 17. listopadu 12, 77146 Olomouc, Czech Republic
}%

\author{Antonio Ac\'in}%
\affiliation{%
ICFO-Institut de Ciencies Fotoniques, Barcelona Institute of Science and Technology, 08860 Castelldefels (Barcelona), Spain
}%
\affiliation{%
ICREA - Instituci\'o Catalana de Recerca i Estudis Avan\c{c}ats, 08010 Barcelona, Spain
}%

\author{Romain All\'eaume}%
\affiliation{%
T\'el\'ecom Paris-LTCI, Institut Polytechnique de Paris, 19 Place Marguerite Perey, 91120 Palaiseau, France
}%

\author{Ulrik L. Andersen}%
\affiliation{%
Center for Macroscopic Quantum States (bigQ), Department of Physics, Technical University of Denmark, 2800 Kongens Lyngby, Denmark
}%

\author{Eleni Diamanti}%
\affiliation{%
Sorbonne Universit\'e, CNRS, LIP6, 4 Place Jussieu, F-75005 Paris, France
}%

\author{Tobias Gehring}%
\affiliation{\relax%
Center for Macroscopic Quantum States (bigQ), Department of Physics, Technical University of Denmark, 2800 Kongens Lyngby, Denmark
}%

\author{Adnan A.E. Hajomer}%
\affiliation{\relax\relax%
Center for Macroscopic Quantum States (bigQ), Department of Physics, Technical University of Denmark, 2800 Kongens Lyngby, Denmark
}%

\author{Florian Kanitschar}
 \affiliation{Vienna Center for Quantum Science and Technology (VCQ), Atominstitut, Technische Universit\"at Wien, Stadionallee 2, 1020 Vienna, Austria}
\affiliation{AIT  Austrian  Institute  of  Technology,  Center  for  Digital  Safety and Security,  Giefinggasse  4,  1210  Vienna, Austria}

\author{Christoph Pacher}%
\affiliation{AIT  Austrian  Institute  of  Technology,  Center  for  Digital  Safety and Security,  Giefinggasse  4,  1210  Vienna, Austria}
\affiliation{fragmentiX Storage Solutions GmbH, Wohllebengasse 10/7, 1040 Vienna, Austria}

\author{Stefano Pirandola}%
\affiliation{%
Department of Computer Science, University of York, York YO10 5GH, United Kingdom
}%

\author{Valerio Pruneri}%
\affiliation{\relax%
ICFO-Institut de Ciencies Fotoniques, Barcelona Institute of Science and Technology, 08860 Castelldefels (Barcelona), Spain
}%
\affiliation{%
ICREA - Instituci\'o Catalana de Recerca i Estudis Avan\c{c}ats, 08010 Barcelona, Spain
}%

\date{\today}

\begin{abstract}
Tremendous progress in experimental quantum optics in recent decades has enabled the advent of
quantum technologies, one of which is quantum communication. Aimed at novel methods for more
secure or more efficient information transfer, quantum communication has developed into an active
field of research and proceeds toward full-scale implementations and industrialization. 
Continuous-variable methods of multiphoton quantum state preparation, manipulation, and coherent detection, as
well as the respective theoretical tools of phase-space quantum optics, offer the possibility of making
quantum communication efficient, applicable, and accessible, thus boosting the development of the
field. We review the methodology, techniques, and protocols of continuous-variable quantum communication
from the first theoretical ideas through milestone implementations to recent developments. The review covers quantum key distribution as well as other quantum communication
schemes that are suggested on the basis of continuous-variable states and measurements. 
\end{abstract}

\maketitle

\tableofcontents

\section{Introduction}
Quantum communication~\cite{Gisin2007} is the research field at the intersection of quantum physics and information science that studies and develops methods for the distribution of quantum states and for information exchange using those states. While  associated mostly with quantum-based information security, enabled by quantum cryptography~\cite{Pirandola2020}, quantum communication also provides the basis for other communication tasks, that rely on the principles of quantum physics. Quantum states of light are
well suited to be used for quantum communication because
of their relatively low coupling to the environment and long
coherence times, so immense advances in quantum optics in
recent decades have also boosted quantum communication. As an integral part of the second quantum revolution~\cite{Dowling2003}, quantum communication became a quantum technology that revolutionized the communication itself by enabling essentially new principles and properties of information exchange. 

In this review we present the field of quantum communication based on continuous variables of light, that can be effectively manipulated using practical quantum-optical techniques and measured by means of highly efficient coherent detection. Unlike discrete-variable (DV) quantum communication, which typically relies on direct photodetection and which is based largely on the corpuscular treatment of light, ideally using single-photon states, continuous-variable (CV) quantum communication can be seen as utilizing wave-type properties of light through manipulation and coherent detection of generally multiphoton quantum states. Since it is complementary to the DV approach, CV quantum communication can offer certain advantages, both quantitative and qualitative, particularly concerning efficiency, cost, or scalability. During the more than two decades of its existence, CV quantum communication has developed from basic ideas to practical realizations and even industrial implementations. Yet, it remains an active field of research aimed at making CV quantum communication more efficient, mature, useful and cost-effective. In this review we therefore discuss principles, applications, practical aspects, limitations and perspectives of CV quantum communication, providing an overview of this modern and dynamic field of research.
%

The review is structured as follows: In Sec. \ref{sec:preliminaries} we recapitulate the methods of quantum communication, which have historically been developed on the basis of DV states, starting with quantum key distribution (QKD) protocols then describing protocols beyond QKD, and  also provide an overview of CV systems covering CV states, including the broadly used Gaussian ones, respective Gaussian operations, and coherent detection methods. In Sec. \ref{sec:cvqkd} we review CV QKD, which is the most advanced and developed branch of CV quantum communication, covering its protocols, theoretical security proofs, tools for its practical implementation and security aspects of the implementation, postprocessing algorithms used in CV QKD, and the recent advancements in the field. In Sec. \ref{sec:cvbeyond} we discuss the methods of CV quantum communication beyond QKD, which are less advanced but are promising for further development. We conclude the review with a summary and future outlook.
Our review is the first, to our knowledge, to cover all major methods of CV quantum communication, presenting principles, modern trends and recent developments in CV QKD, as well as CV protocols beyond QKD. 
Several reviews related to CV quantum communication were published earlier. Quantum cryptography using DV states and measurements was discussed by~\cite{Gisin2002} who introduced basic principles and concepts that are also used in CV QKD. The practical security of QKD was reviewed by~\cite{Scarani2009} who focused largely on DV QKD, but also briefly discussed CV protocols. A CV quantum information review by~\cite{Braunstein2005} covered the underlining principles and methods behind CV quantum communication, such as CV quantum entanglement, and described several quantum communication methods, including QKD, dense coding, and quantum teleportation, as well as other applications of CV information, such as quantum computation. The Gaussian quantum information review by~\cite{Weedbrook2012a} described  tools of quantum information processing using Gaussian states, such as covariance-matrix formalism, which largely remain relevant for the field of CV quantum communication, and covered many of the ongoing developments in the field, such as finite-size security analysis or discretely modulated protocols, while also describing quantum computation with Gaussian states. The review on CV quantum information by~\cite{Adesso2014} presented mathematical structures behind Gaussian quantum information and the quantification of CV correlations. Security of CV QKD and its state-of-the-art experimental implementations were covered in the review by~\cite{Diamanti2015}. The review on QKD with practical devices by~\cite{Xu2020} reported relevant topics for QKD security, for example, quantum side channels and quantum hacking, while briefly discussing how they apply to CV QKD. The review on advances in quantum cryptography by~\cite{Pirandola2020} described modern trends in the field of QKD with much attention to CV-QKD protocols, their security analysis and their practical implementation. The recent review on CV QKD by~\cite{zhang2024continuous} contains practical details of CV-QKD implementations. 
For convenience, in Table \ref{tab:abbr} we present a List of Abbreviations, used in the review.

\begin{table}[!]
    \centering
    \begin{tabular}{|p{2.5cm}|p{6cm}|}
    \hline
     \leavevmode\centering\textbf{Abbreviation} & \leavevmode\centering\arraybackslash\textbf{Meaning}    \\
    \hhline{|=|=|}
1sDI & one-sided device independent \\
AWGN & additive white Gaussian noise \\
AEP & asymptotic equipartition property\\
BB84 & Bennett-Brassard 1984 \\
CV & continuous variable  \\
cw & continuous wave  \\
DI & device independent \\
DM & discrete modulation \\
DR & direct reconciliation \\
DSP & digital signal processing \\
DV & discrete variable \\
E91 & Ekert 1991 \\
EAL & evaluation assurance level \\
EAT & entropy accumulation theorem \\
EB & entanglement based \\
EOM & electro-optical modulator \\
EPR & Einstein-Podolsky-Rosen \\
FER & frame error rate \\
FFT & fast Fourier transform \\
GG02 & Grosshans-Grangier 2002 \\
GKP & Gottesman-Kitaev-Preskill \\
GM & Gaussian modulation \\
i.i.d. & independent and identically distributed \\
InP & indium phosphide \\
IQ & in-phase and quadrature \\
KTP & potassium titanyl phosphate \\
LDPC & low-density parity check \\
LHL & leftover hashing lemma \\
LLO & local local oscillator \\
LO & local oscillator \\  
MD & multidimensional \\
MDI & measurement-device independent \\
NLA & noiseless linear amplifier \\
OPA & optical parametric amplifier \\
OPO & optical parametric oscillator \\
P\&M & prepare and measure \\
PDC & parametric down-conversion \\
PIA & phase-insensitive amplifier \\
PLOB & Pirandola-Laurenza-Ottaviani-Banchi \\
PP & protection profile \\
PQC & postquantum cryptographic \\
PSA & phase-sensitive amplifier \\
QA & quantum authentication \\
QAM & quadrature amplitude modulation\\
QBER & quantum bit error rate \\
QDC & quantum dense coding \\
QDS & quantum digital signatures \\
QKD & quantum key distribution \\
QOT & quantum oblivious transfer \\
QPSK & quadrature phase-shift keying\\
QSDC & quantum secure direct communication \\
QT & quantum teleportation \\
RCI & reverse coherent information \\
rf & radio frequency \\
RR & reverse reconciliation \\
SDP & semidefinite programming \\
SNR & signal-to-noise ratio \\
SNU & shot-noise unit \\
TMSV & two-mode squeezed vacuum \\
UD & unidimensional \\
WDM & wavelength division multiplexing \\
     \hline
    \end{tabular}
        \caption{List of abbreviations used in the review.}
    \label{tab:abbr}
\end{table}

\section{Preliminaries}
\label{sec:preliminaries}

As, historically, most of the quantum communication methods were first developed on the basis of DV techniques, we start our review with an introduction to DV quantum communication.

\subsection{Quantum key distribution}
\label{sec:qkd}
QKD~\cite{Gisin2002, Scarani2009, Pirandola2020} is one of the most mature and important directions within quantum communication, having as its goal the development of protocols for provably secure distribution of secret cryptographic keys. Together with classical encryption algorithms, QKD provides legitimate users with the methods of secure communication. Such methods are then based on the principles of quantum physics and generally do not rely on mathematical complexity assumptions, which can be overturned either by the development of quantum computing or progress in classical computations. 

QKD was first suggested as the BB84 protocol by~\cite{Bennett1984} who considered polarization states of single photons as \textit{qubits} (quantum bit states, generally being in superpositions of the two basis states). By using nonorthogonal states and relying on the no-cloning theorem, which forbids perfect copying of unknown quantum states~\cite{Wootters1982}, the protocol enables the trusted (honest) parties to ensure the security of the key, obtained from polarization encoding and measurements and processed over an authenticated classical channel. By randomly switching the preparation and measurement bases, the trusted parties accumulate the so-called raw key, which is then sifted during the bases reconciliation stage and is further classically processed to obtain the secret key. 

The \textit{entanglement-based} (EB) E91 QKD protocol suggested by~\cite{Ekert1991} almost a decade later uses entangled-photon pairs (and hence, entangled qubits) such that each of the photons is measured by one of the trusted parties. By randomly switching between three bases settings, the trusted parties can either verify the violation of Bell inequalities (in the incompatible bases)~\cite{Aspect1982} or already obtain the sifted key once the bases coincide. The security of the scheme relies on the fact that an eavesdropper attempting to intercept the key would break the nonlocal correlations and stop the Bell inequality violation. This reasoning was later extended to the notion of device-independent security, which is discussed further in this subsection.~\cite{Bennett1992c} later showed that entangled-photon pairs can be used equivalently to implement the BB84 protocol without the need for Bell inequality violation, hence randomly switching between two detection bases. This equivalence between the prepare-and-measure (P\&M) version of the protocol, when Alice prepares the states and Bob performs their measurement, and the EB version became an important feature of QKD protocols, enabling their formal information-theoretical security analysis.

The first experimental test of QKD using the BB84 protocol was reported by~\cite{Bennett1992d} who used weak coherent pulses to emulate single-photon states. However, it was pointed out that such a realization is vulnerable to photon-number-splitting attack (also known as beam-splitting attack), allowing an eavesdropper to exploit multiphoton pulses for error-free interception of key bits.~\cite{Brassard2000} later showed that such an attack in combination with limited detection efficiency and dark counts (when a photodetector, working in Geiger mode, produces a false click in the absence of an incoming photon) strongly limits the practical applicability of QKD. To overcome the limitation, modified QKD protocols were proposed, such as SARG~\cite{Scarani2004}, in which every key bit is encoded into a pair of nonorthogonal states, hence reducing the efficiency of photon-number-splitting attack for an eavesdropper, or the decoy-state protocol~\cite{Hwang2003}, using additional multiphoton states to detect the splitting. These developments enabled numerous experimental realizations of QKD in long-distance fibers~\cite{Korzh2015} and in free-space~\cite{Ursin2007} or intercontinental satellite-based channels~\cite{Liao2018}. 

The security analysis of QKD protocols is largely defined by the set of assumptions on the protocol implementation and on the capabilities of a malicious eavesdropper. 
One major assumption and necessity to achieve a secure QKD protocol is that the classical communication of the QKD protocol is transmitted through an authenticated channel to prevent an eavesdropper from performing a man-in-the-middle attack either on the classical channel or jointly on the classical and quantum channels. An authenticated classical channel can be established from a nonauthenticated one by using message authentication codes that in turn consume short keys. 

The analysis of attacks on the quantum channel alone is typically based on evaluating the Devetak-Winter rate~\cite{Devetak2005} for the achievable secret key, which characterizes the information advantage of the trusted parties over an eavesdropper, capable of storing probe quantum states in a quantum memory. While not generally relying on particular attack models, this approach has to be extended toward the finite-size regime, when statistical aspects of limited datasets have to be taken into account~\cite{Tomamichel2012}. The security of a QKD protocol can then be quantified in the context of composable security~\cite{Portmann2022}, allowing us to evaluate the security of an entire cryptographic system, which contains the given protocol. 

However, gaps between theoretical security and practical implementations are always possible and can be exploited by an eavesdropper as, for example, in the blinding attack on the single-photon detectors~\cite{Lydersen2010}. To rule out such \textit{quantum hacking} attacks, the assumptions on the devices used for QKD can be waived in the device-independent approach.

An important feature of the E91 protocol was the reliance of its security on the violation of Bell inequalities. While this effect is not required for the EB implementation of BB84 and similar QKD protocols, it can be used to waive the assumption on the trusted nature of the devices, namely, the source and detectors. The resulting \textit{device-independent} (DI) protocol~\cite{Acin2007,Vazirani2014} can in principle be secure without making assumptions on the internal working of the devices. In practice this means that the trusted parties do not need to fully characterize and calibrate the devices, relying solely on the loophole-free violation of Bell inequality [typically its Clauser-Horne-Shimony-Holt version~\cite{Clauser1969}] from the measurement statistics, which can be seen as part of the more general device self-checking approach~\cite{Mayers1998}. However, implementation of DI QKD protocols remains demanding in terms of highly pure entanglement, which is required for their security, and high sensitivity to practical imperfections, which particularly limits the distances at which DI QKD can be implemented.

It is also possible to remove only the trust assumption on the measurement devices, which can rule out quantum hacking attacks on the detectors. Such attacks are more likely to happen since the detectors are susceptible to incoming light that is potentially manipulated by an attacker. In this intermediate approach, called \textit{measurement-device-independent} (MDI) QKD, the trusted parties send their modulated states only to a middle station, which then performs a Bell-type measurement and announces the outcomes, allowing the trusted parties to establish correlations on their data, sufficient for distilling a secret key~\cite{Lo2012,Braunstein2012}. The detection can then be assumed to be under the full control of an eavesdropper because tampering with the detection outcomes can be revealed by the trusted parties. While managing to rule out the attacks on the measurement devices, the MDI protocols also generally offer higher robustness and better performance than DI QKD. Another intermediate scenario is possible in which only one of the trusted parties trusts their measurement apparatus, which is the case of \textit{one-sided device-independent} protocols (1sDI)~\cite{Branciard2012}

\subsection{Quantum communication beyond QKD}

While QKD is the most commonly developed application of quantum communication, the field offers several other directions in which quantum effects bring enhancements (particularly physics-based security instead of  computational security) to communication methods (mainly cryptographic primitives). In addition, we give an overview on the major ones that are also being developed within CV quantum communication.

\textit{Quantum secure direct communication} (QSDC) is the family of protocols aimed at ensuring the security of the directly transmitted information so that its acceptably small part is gained by an eavesdropper while not relying on QKD. This allows trusted users to directly and deterministically share their secret messages while making the protocols more technically demanding than QKD (see~\cite{Pan2024} for a recent review on QSDC). The first QSDC protocol was proposed by~\cite{Long2002} as the deterministic QKD of a pregenerated key with no information leakage. Deterministic secure quantum communication was proposed in ~\cite{Beige2002} using single-photon states to simultaneously encode two qubits in different bases that are accompanied by a classically shared secret key. Another deterministic secure direct communication protocol was proposed on the basis of entangled pairs and a quantum memory on the sender side~\cite{Bostroem2002}. A protocol for directly sharing a secret key encoded in the sequences of single-photon states, sent back and forth between the trusted parties and stored in quantum memories, was suggested by~\cite{Deng2004}. It was later extended to EB realization~\cite{Deng2003} with reportedly higher performance. QSDC was then experimentally tested using single-photon~\cite{Hu2016} and entangled states~\cite{Zhang2017} with atomic quantum memories. 

\textit{Quantum dense coding} (QDC; also referred to as quantum super-dense coding) allows communicating parties to share more classical bits of information with fewer shared qubits. QDC was first theoretically proposed by~\cite{Bennett1992b} and was based on an entangled pair distributed between two parties so that, after an operation on one particle, its transfer, and joint measurement on the other side of the channel, two classical bits (encoded into one of four possible qubit operations) become known to both parties. QDC was further combined with QSDC by~\cite{Wang2005} and experimentally tested using atomic qubit states~\cite{Schaetz2004} or photonic qubits distributed over optical fibers~\cite{Williams2017}.

\textit{Quantum digital signature} (QDS) protocols are the quantum counterpart of the classical digital signature methods, which are used to verify the authenticity of communicated messages. Like the classical protocols, the quantum digital signature supposes the use of private and public secret keys. The keys are related through a quantum analog of the classical one-way functions. The QDS principles were first described by~\cite{Gottesman2001b} who showed that multiple remote receivers can validate the signature of a message sender. This is achieved by each of the receivers having a copy of the sender's public quantum key (also referred to as a quantum signature), created, using a quantum one-way function, from a classical private key. The QDS protocol with a sender and two receivers was experimentally demonstrated by~\cite{Clarke2012} who used phase-encoded coherent states of light. The signed messages were shown to be secure against forging and repudiation when typical attack scenarios were considered. The receiver, however, requires a long-term quantum memory for storing the public quantum key. The signed message is then verified by interfering the stored key with a set of states created according to the classical description of the sender's private key that is sent along with the message. This requirement was waived by~\cite{Dunjko2014} who allowed the remote receiver to measure the public quantum key instead of storing it in the quantum memory. The classical measurement outcomes are then used to verify the messages, which is however probabilistic. The protocol was realized experimentally~\cite{Collins2014} using only linear-optical components and photodetectors, enabling a more efficient measurement technique with higher probabilities of success.

\textit{ Quantum authentication} (QA) is a family of quantum protocols for verification of the identity of a message sender and/or the integrity of the message. In this regard, it is closely related to QDS, which can be seen as multiuser authentication, and to QKD, which requires authentication in the key postprocessing stage but can also provide authentication once the secret key is distributed. Furthermore, QA schemes can vary depending on whether a message is classical or quantum (i.e., whether it is a sequence of classical data or quantum states). The possibility of using quantum resources for message authentication was first discussed by~\cite{Crepeau1995} who showed that two trusted parties were able to verify their mutual knowledge of a classical data string without revealing the data themselves (this way, the QA scheme is related to the oblivious transfer protocols discussed in the next subsection). Quantum protocols for the identification of classical messages combining classical identification and QKD were proposed and experimentally tested by~\cite{Dusek1999}, depending on whether or not the public channel between the trusted parties is fully controlled by an eavesdropper. Quantum enhancement of classical message authentication was shown by~\cite{Curty2001} with a one-qubit key reportedly being a sufficient resource for proving the integrity of a classical bit. Furthermore, the efficiency of the QA of classical messages was improved to 2 bits by one qubit~\cite{Hong2017}. Authentication of quantum messages by their encryption was addressed by~\cite{Barnum2002}, who presented an asymptotically optimal protocol for authenticating one qubit with two classical bits along with the proof of impossibility to digitally sign quantum states. 

\textit{Quantum oblivious transfer} (QOT) is the quantum version of a communication scheme in which a sending party transfers one or more of many possible messages to a receiver while not knowing which messages were sent (and if there were any). This essential cryptographic primitive was shown to provide security of multipartite computation~\cite{Kilian1988}. Typically the so-called one-out-of-two scheme is studied, in which one message is communicated out of two, with possible generalizations to the schemes when $k$ out of $n$ messages are communicated. The one-out-of-$n$ case is commonly referred to as the private database query. Development of QOT protocols was first based largely on a reduction to the quantum bit commitment~\cite{Bennett1992e} until the latter was shown to be insecure against quantum attacks~\cite{Mayers1997,Lo1997a}. Furthermore, it was shown that unconditionally secure QOT is also impossible~\cite{Lo1997}. The QOT protocols are therefore developed by either imposing limitations on the eavesdropping, for example, the noisy storage model, in which qubits, stored by a dishonest receiver, undergo  decoherence~\cite{Wehner2008}, or allowing partial information leakage to an eavesdropper~\cite{Chailloux2010}. More recently, a so-called simulation-based security framework was introduced for QOT, showing that it is possible to construct a practical protocol for QOT only with minimal assumptions, in particular, using one-way functions~\cite{Diamanti2024}. 

In the previously described schemes, classical information is communicated using quantum states. In contrast, \textit{quantum teleportation} (QT) offers the possibility of communicating a quantum state (and hence quantum information, for example, a qubit) between remote parties~\cite{Bennett1993}, assisted by classical communication. This is achieved by presharing a quantum resource (an entangled state), performing joint \textit{Bell measurement} on the input state and the local part of the entangled state (projective measurement in the Bell basis) at the side of one of the parties, and classically communicating the outcomes to another party, which performs respective operation on the other part of the entangled state to obtain ideally an exact copy of the teleported quantum state. The quantum state is then transferred without directly sending the physical system that realizes it, while the original quantum sate is destroyed by the measurement, hence not violating the no-cloning theorem. Since the sending and receiving parties  generally do not know the teleported state, QT is closely related to QKD and can be seen as a quantum encryption~\cite{Gisin2002}. In practice, when entangled resource and device performance are not perfect, the receiver obtains an imperfect copy of the teleported state and the quality of the QT is characterized by fidelity (overlap) $F$ between the states. Successful (true) QT then should demonstrate fidelity higher than what can be achieved via the best classical strategy, fox example, through quantum state measurement and reconstruction, in the qubit case limited by $F=2/3$. QT was first tested using polarization photon states~\cite{Bouwmeester1997,Boschi1998} and later extensively developed and advanced (see~\cite{Pirandola2015b,Hu2023} for recent reviews), also toward CV realizations, which can offer unconditional teleportation, as we discuss in Sec. \ref{sec:cvbeyond}.

There are many more protocols and cryptographic primitives that are reportedly enabled or enhanced by quantum effects compared to their classical counterparts. The interplay between some of the protocols is yet to be fully clarified in the ongoing development of quantum communication. The previous introduction nevertheless provides a basis for discussion of the major CV quantum communication schemes in the next sections. Before proceeding, we first recapitulate the basic notions from CV quantum information along with the major properties of the CV quantum states that are useful for quantum communication protocols.

\subsection{Continuous-variable quantum systems}
CV quantum systems are defined on infinite-dimensional Hilbert spaces, and described by observables with continuous eigenspectra (see~\cite{Braunstein2005} and~\cite{Weedbrook2012a} for reviews). A radiation mode $k$ of the quantized electromagnetic field is represented by a harmonic oscillator with frequency $\omega_k$ and the Hamiltonian $\hat{H}_k=\hbar\omega_k(\hat{a}_k^{\dag}\hat{a}_k+1/2)$, where $\hat{a}_k^{\dag}$ and $\hat{a}_k$ are the creation and annihilation operators of the mode $k$. Setting $\hbar=2$, we define the dimensionless \textit{quadrature operators} $\hat{x}_k$ and $\hat{p}_k$ of the field in the mode $k$ as the real and imaginary parts of the creation and annihilation operators, $\hat{a}_k=(\hat{x}+i\hat{p})/2$ and $\hat{a}^{\dag}_k=(\hat{x}-i\hat{p})/2$, from which follow $\hat{x}_k=\hat{a}^{\dag}_k+\hat{a}_k$ and $\hat{p}_k=i(\hat{a}^{\dag}_k-\hat{a}_k)$. Following the commutation relations for the bosonic operators, $[\hat{a}^{\dag}_l,\hat{a}_m]=\delta_\text{lm}$, we obtain $[\hat{x}_l,\hat{p}_m]=2i\delta_\text{lm}$. The conjugate quadrature operators $\hat{x}_k,\hat{p}_k$ of mode $k$ then have the meaning of position and momentum of a harmonic oscillator associated with this mode. 

By introducing the variance of an operator $\hat{A}$ as $V(\hat{A})=\langle(\Delta\hat{A})^2\rangle=\langle\hat{A}^2\rangle-\langle\hat{A}\rangle^2$, where $\Delta\hat{A}=\hat{A}-\langle\hat{A}\rangle$ is the deviation from the mean, we obtain the Heisenberg uncertainty principle for quadrature operators in mode $k$ as $V(\hat{x}_k)V(\hat{p}_k)\geq 1$. In our notation the quadrature variance of a vacuum state is equal to 1, which defines the shot-noise unit (SNU).

For an $N$-mode state we can arrange the quadrature operators into a column vector $\hat{\mathbf{r}}=\{\hat{x}_1,\hat{p}_1,\ldots,\hat{x}_N,\hat{p}_N\}^T$. The commutation relations can then be expressed as $[\hat{r}_i,\hat{r}_j]=2i\Omega_{ij}$ through the elements of the symplectic form
\begin{equation}
    \Omega=\oplus_{k=1}^N
    \begin{pmatrix}
        0 & 1 \\
        -1 & 0
    \end{pmatrix}.
\end{equation}
A CV quantum state with a density matrix $\hat{\rho}$ can be represented in phase space by means of the Wigner function, which is the (quasi-)probability distribution for the dimensionless quadrature observables introduced earlier. It can be written through the eigenstates of the quadrature operators as
\begin{equation}
W(x,p)=\frac{1}{(2\pi)^N}\int_{\mathbb{R}^N}\langle x-y|\hat{\rho}|x+y\rangle e^{ipy}d^Ny
\label{eq:wigner}
\end{equation}
for $x,p \in \mathbb{R}^N$. The first statistical moments (mean values) of quadratures define the displacement vector $\mathbf{\bar{r}}=\langle \hat{\mathbf{r}}\rangle = Tr[\hat{\rho}\hat{\mathbf{r}}]$. The second moments form the \textit{covariance matrix} $\gamma$ with elements
\begin{equation}
    \gamma_{i,j}=\frac{1}{2}\langle\{\Delta\hat{r}_i\Delta\hat{r}_j\}\rangle,
\end{equation}
where $\{\cdot\}$ is the anticommutator. The diagonal elements of the covariance matrix then give the quadrature variances of a particular mode $i$ as $\gamma_{i,i}=\langle \hat{r}_i^2\rangle - \langle \hat{r}_i\rangle^2$, while the off-diagonal elements represent the quadrature correlations between the modes. The uncertainty principle can be generalized in terms of the covariance matrix as $\gamma+i\Omega\geq 0$, which can be seen as a condition on the physicality of a state represented by a covariance matrix $\gamma$.

Alternatively to the phase-space representation, the CV states can be represented in the Fock (number) basis, composed by the eigenstates of the photon-number operator, $\hat{n}=\hat{a}^{\dag}\hat{a}$, such that $\hat{n}|n\rangle=n|n\rangle$. The number basis $\{|n\rangle\}_0^\infty$ then spans the infinite-dimensional Hilbert space of a given mode.

\subsubsection{Gaussian quantum states}

An important class of CV states is the states for which the Wigner function is Gaussian (see~\cite{Weedbrook2012a} for a review on Gaussian quantum information) and can thus be represented through the displacement vector $\mathbf{\bar{r}}$ and the covariance matrix $\gamma$ as
\begin{equation}
W(\mathbf{r})=\frac{1}{(2\pi)^{N}\sqrt{\text{det}\gamma}}
e^{-\frac{1}{2}(\mathbf{r}-\mathbf{\bar{r}})^T\gamma^{-1}(\mathbf{r}-\mathbf{\bar{r}})}
\end{equation}
for $\mathbf{r} \in \mathbb{R}^{2N}$ and where $det$ stands for the determinant of a matrix. Importantly, Gaussian states are then explicitly described by the first (displacement vector) and second (covariance matrix) moments of the quadratures, which largely simplifies the analysis of those infinite-dimensional states.

The typical example of a Gaussian state commonly used in CV quantum information is the \textit{coherent state}, which is an eigenstate of the annihilation operator, $\hat{a}|\alpha\rangle=\alpha|\alpha\rangle$, with a generally complex eigenvalue $\alpha$. In the Fock basis, a coherent state is then represented as
\begin{equation}
    |\alpha\rangle=e^{-\frac{|\alpha|^2}{2}}\sum_{n=0}^\infty{\frac{\alpha^n}{\sqrt{n}}|n\rangle}.
\end{equation}
Coherent states are characterized by the quadrature mean values $\mathbf{\bar{r}}_\text{coh}=\{2\Re(\alpha),2\Im(\alpha)\}$ and the $2\times 2$ diagonal covariance matrix  $\gamma_\text{coh}=\text{diag}(1,1)$. The vacuum state can then be seen as the zero-mean coherent state with $\mathbf{\bar{r}}_\text{coh}=\{0,0\}$ but the same covariance matrix $\gamma_\text{vac}=\text{diag}(1,1)$, containing variances of one SNU for either of the quadratures. Note that a coherent state is a minimum-uncertainty state that saturates equality in the Heisenberg principle, $V(\hat{x}_k)V(\hat{p}_k)=1$. The mean photon number of a coherent state, obtained as the mean of the photon-number operator $\hat{n}\equiv \hat{a}^{\dag}\hat{a}$, is $\langle \hat{n} \rangle_\text{coh} = |\alpha|^2$. 

A Gaussian (quadrature-)\textit{squeezed state} is also a minimum-uncertainty state, but with different quadrature variances, which is reflected in the covariance matrix $\gamma_\text{sq}=\text{diag}(e^{-2r},e^{2r})$. Assuming that $r>0$, so that the $\hat{x}$ quadrature is squeezed, and denoting the squeezed quadrature variance as $V_S \equiv e^{-2r}$, so that $V_S<1$, we obtain the covariance matrix $\gamma_\text{sq}=\text{diag}(V_S,1/V_S)$, where the $\hat{p}$ quadrature with variance $V_S>1$ is antisqueezed (the opposite situation with $\hat{p}$ quadrature squeezing is obtained at $r<0$). The displacement of squeezed states can be zero (squeezed vacuum states) or nonzero (squeezed coherent states). 

A \textit{thermal state} is a Gaussian state with a diagonal covariance matrix $\gamma_\text{th}=\text{diag}(V,V)$ such that the variances of both quadratures are $V>1$ and the state does not have the minimum uncertainty. The quadrature variance $V$ is related to the mean photon number of a thermal state as $V = 2\langle \hat{n} \rangle_\text{th}+1$. A thermal state with zero mean photons reduces to the vacuum state with $V=1$.

Introducing the Gaussian purity, which is expressed through the determinant of the covariance matrix as  $\mu=1/\sqrt{\det\gamma}$~\cite{Paris2003}, we note that coherent and squeezed states are pure with $\mu=1$, while a thermal state is not.  

The two-mode squeezed vacuum (TMSV) state is an entangled two-mode Gaussian state that, in the Fock representation, is given by
\begin{equation}
    |\lambda\rangle=\sqrt{1-\lambda^2}\sum_{n=0}^\infty(-\lambda)^n|n,n\rangle,
\end{equation}
where the state parameter $\lambda$ is expressed through the squeezing as $\lambda=\tanh{r}$ and $|n,n\rangle$ represents the number state with the same number of photons $n$ in each of the two squeezed vacuum modes. TMSV states have zero displacement, and their two-mode covariance matrix reads
\begin{equation}
    \gamma_\text{TMSV}=
            \begin{pmatrix}
            V\mathbb{I} & \sqrt{V^2-1}\mathbb{Z} \\
            \sqrt{V^2-1}\mathbb{Z} & V\mathbb{I}
            \end{pmatrix},
\end{equation}
where the quadrature variance $V=\cosh 2r$. Each of the two modes of a TMSV state is in a thermal state (after tracing out another mode), that is mixed, yet the overall two-mode state is pure with $\text{det}(\gamma_\text{TMSV})=1$. It can therefore be seen as a purification of a single-mode thermal state. TMSV states are quadrature-entangled and provide maximum entanglement for the given mean photon number. In the limit of infinite squeezing $r \to \infty$ with perfect correlation (anticorrelation) between $\hat{x}$ ($\hat{p}$) quadratures in each of the modes, TMSV states become equivalent to an EPR pair in quantum mechanics (named after the well-known Einstein-Podolsky-Rosen paradox). They are also sometimes therefore referred to as CV EPR states.

\subsubsection{Gaussian operations}
In the description and analysis of CV quantum communication, we rely largely on the physical operations on CV Gaussian states that preserve their Gaussianity (hence, the Gaussian character of the Wigner function). Those operations, referred to as Gaussian ones~\cite{Weedbrook2012a}, are described in the Heisenberg picture by affine maps of the form $(S,\mathbf{d}):\hat{\mathbf{r}}\to S\hat{\mathbf{r}}+\mathbf{d}$ with $\mathbf{d}\in \mathbb{R}^{2N}$ and $S$ a $2N \times 2N$ real matrix. Preservation of the commutation relations implies that the transformation is symplectic; hence, $S\Omega S^T=\Omega$. Then the action on the quadrature moments is given by $\mathbf{\bar{r}}\to S\mathbf{\bar{r}}+\mathbf{d}$ and $\gamma \to S\gamma S^T$.

A \textit{displacement operation} is a single-mode Gaussian unitary defined as the Bogolyubov transformation $\hat{a}\to\hat{a}+\alpha$. It transforms the quadrature operators accordingly as $\hat{\mathbf{r}} \to \hat{\mathbf{r}}+\mathbf{d}_{\alpha}$ through the displacement vector $\mathbf{d}_{\alpha}=\{2\Re(\alpha),2\Im(\alpha)\}$, therefore shifting the quadrature mean values by the elements of the displacement vector and leaving the covariance matrix unchanged [hence, the transformation matrix $S_{\alpha}=\mathbb{I}$ is the $2\times 2$ identity matrix, $\mathbb{I}= \text{diag}(1,1)$]. The coherent state can be obtained via a displacement operation acting on a vacuum state.

\textit{Phase rotation}, another single-mode Gaussian unitary, is the Bogolyubov transformation $\hat{a} \to e^{i\theta}\hat{a}$, that transforms the quadratures as $\mathbf{\hat{r}}\to S_{\theta}\mathbf{\hat{r}}$ and is characterized by zero displacement and the transformation matrix
\begin{equation}
    S_{\theta}=
        \begin{pmatrix}
        \cos\theta & \sin\theta \\
        -\sin\theta & \cos\theta
        \end{pmatrix}.
\end{equation}
\textit{Single-mode squeezing} is represented by the transformation $\hat{a}\to \hat{a}\cosh r-\hat{a}^{\dag}\sinh r$ with the squeezing parameter $r \in \mathbb{R}$, which is equivalent to mapping 
$\mathbf{\hat{r}}\to S_{r}\mathbf{\hat{r}}$ with
\begin{equation}
        S_{r}=
        \begin{pmatrix}
        e^{-r} & 0 \\
        0 & e^r
        \end{pmatrix}.
\end{equation}
Squeezed states can then be obtained via the action of the single-mode squeezing operation on a vacuum or a coherent state.

\textit{Beam splitter transformation} is a two-mode Gaussian unitary, characterized for the modes 1 and 2 with operators $\hat{a}_1$ and $\hat{a}_2$ via the Bogolyubov transformation
\begin{equation}
        \begin{pmatrix}
       \hat{a_1} \\
       \hat{a_2}
        \end{pmatrix}\to
        \begin{pmatrix}
        \sqrt{T} & \sqrt{1-T} \\
        -\sqrt{1-T} & \sqrt{T}
        \end{pmatrix}
        \begin{pmatrix}
       \hat{a_1} \\
       \hat{a_2}
        \end{pmatrix},
        \label{eq:bsa}
\end{equation}
where the coupling ratio $\sqrt{T}$ defines the transmittance $T$ and reflectance $1-T$ of the beam splitter. Equivalently, the two-mode quadrature vector $\hat{\mathbf{r}}_{12}=\{\hat{x}_1,\hat{p}_1,\hat{x}_2,\hat{p}_2\}^T$ is transformed as $\hat{\mathbf{r}}_{12}\to S_{T}\hat{\mathbf{r}}_{12}$ with the transformation matrix
\begin{equation}
        S_{T}=
        \begin{pmatrix}
        \sqrt{T}\mathbb{I} & \sqrt{1-T}\mathbb{I} \\
        -\sqrt{1-T}\mathbb{I} & \sqrt{T}\mathbb{I}
        \end{pmatrix}.
\end{equation}
Finally, the \textit{two-mode squeezing} operation on the modes 1 and 2 is defined by $\hat{a}_1\to\hat{a}_1\cosh r+\hat{a}_2^{\dag}\sinh r$ and $\hat{a}_2\to\hat{a}_2\cosh r+\hat{a}_1^{\dag}\sinh r$ through the two-mode squeezing parameter $r \in \mathbb{R}$. The transformation for the quadrature vector $\hat{\mathbf{r}}_{12}$ then reads $\hat{\mathbf{r}}_{12}\to S_{2r}\hat{\mathbf{r}}_{12}$ with the transformation matrix for the two-mode squeezing
\begin{equation}
        S_{2r}=
        \begin{pmatrix}
        \cosh r \mathbb{I} &  \sinh r \mathbb{Z} \\
        \sinh r \mathbb{Z} &  \cosh r \mathbb{I}
        \end{pmatrix},
\end{equation}
where the Pauli matrix $\mathbb{Z}=\text{diag}(1,-1)$. TMSV states can then be obtained via the two-mode squeezing operation acting on a two-mode vacuum state.

\subsubsection{Gaussian channels\label{sec:Gaussian_Channels}}
In CV quantum communication the most typical medium is an optical fiber, which is well modeled by a bosonic Gaussian channel. A Gaussian channel is a completely-positive trace-preserving map $\mathcal{E}$ transforming Gaussian states into Gaussian states. In particular, a single-mode Gaussian channel can be represented by the following transformations on the first two statistical moments, $\bar{\mathbf{r}}$ and $\gamma$, of an input Gaussian state
\begin{equation}
    \bar{\mathbf{r}} \rightarrow\mathbf{T}\bar{\mathbf{r}}+\mathbf{d},~~\gamma\rightarrow \mathbf{T} \gamma \mathbf{T}^{T}+\mathbf{N},
\end{equation}
where $\mathbf{d}$ is a two-dimensional displacement vector, while the $2 \times2$ transmittance matrix $\mathbf{T}$ and noise matrix $\mathbf{N}$ satisfy the properties $\mathbf{N}^{T}=\mathbf{N}>0$ and $\det \mathbf{N} \ge (\det \mathbf{T}-1)^2$.

A single-mode Gaussian channel can be reduced to a \textit{canonical form} for which $\mathbf{d}=\mathbf{0}$ and the aforementioned matrices take diagonal expressions~\cite{Weedbrook2012a}. The most important form is certainly the thermal-loss channel, which is characterized by $\mathbf{T}=\sqrt{T} \mathbb{I}$ and $\mathbf{N}=(1-T)(2\bar{n}_N+1)\mathbb{I}$. Here the transmittance $T$ represents the fraction of photons surviving at the output of the channel, while the thermal noise $\bar{n}_N$ describes extra thermal photons that are injected into the channel from the external environment (this noise is often rewritten in the form of excess noise, as we later discuss). 

The dilation of a thermal-loss channel is a beam splitter transformation with the same transmittance $T$, where the environmental port is used to inject $\bar{n}_N$ thermal photons. When the environmental thermal state is purified, it is further dilated into an entangling cloner where one mode of a TMSV state is injected into the beam splitter. This is what forms the entangling cloner attack, which is discussed in Sec. \ref{subsec:ind}. 

Other relevant canonical forms of a single-mode Gaussian channel are the quantum amplifier and the additive-noise channel. However, these forms play a minor role in QKD communications, where loss is always present. In general, any canonical form can be dilated into a collective Gaussian attack~\cite{Pirandola2008}.

\subsubsection{Coherent detection}

\begin{figure}
    \centering
    \includegraphics[width=0.9\linewidth]{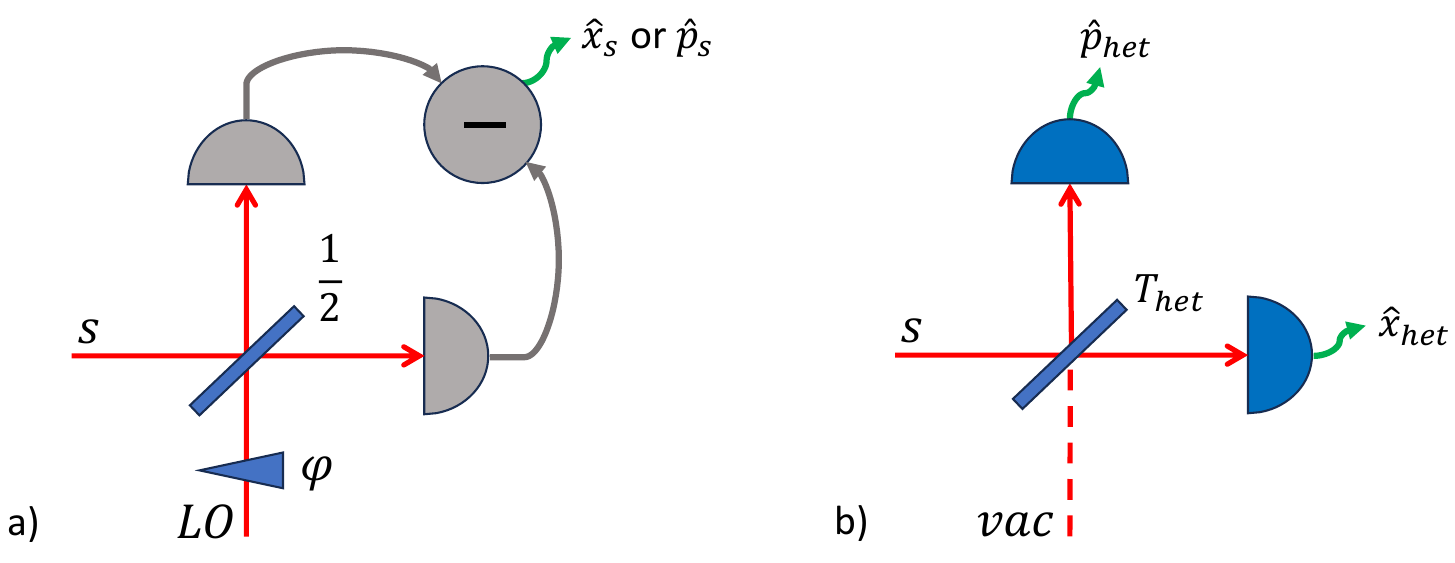}
    \caption{(a) Homodyne detection of an incoming signal in mode $s$ based on a balanced coupling with the local oscillator $LO$ and the subsequent subtraction and scaling of the signals from the photodetectors (given in gray), resulting in the measured value of an $\hat{x}_s$ or $\hat{p}_s$ quadrature observable of the signal. (b) Heterodyne (double homodyne) detection of an incoming signal in mode $s$, based on a generally unbalanced splitting coupling to a vacuum mode $vac$ with transmittance $T_\text{het}$ and subsequent measurement of the $\hat{x}_\text{het}$ and $\hat{p}_\text{het}$ quadrature observables, which are the linear combinations of the quadratures in modes $s$ and $vac$, using the homodyne detectors on the outputs of the beam splitter (given in blue). The LO modes and the structure of each of the two homodyne detectors (balanced coupling to LO and the photocurrent difference scheme) are omitted for simplicity.}
    \label{fig:hom_het}
\end{figure}

The quadrature observables are measured by means of coherent detection, which, for a single-quadrature measurement, is called \textit{homodyne detection} and is the major detection type used in CV quantum information. This is in contrast to the DV protocols, which are based largely on direct photodetection either using avalanche photodiodes, working in the Geiger mode, and registering the presence of an incoming signal on a single-photon level or using photon-number-resolving detectors. Homodyne detection results in the value $x$ or $p$ of the $\hat{x}$ or $\hat{p}$ quadrature respectively and is therefore a projective measurement on the infinitely-squeezed eigenstates of the quadrature operators, $|x\rangle$ or $|p\rangle$. The quadrature probability distributions are the marginal integrals of the Wigner function over the complementary quadrature,
\begin{equation}
    P(x)=\int_{\mathbb{R}}W(x,p)dp, P(p)=\int_{\mathbb{R}} W(x,p)dx\ .
    \label{eq:quadprob}
\end{equation}
In practice, homodyne detection is realized by coupling an incoming signal mode $s$ with another mode that is in a strongly displaced coherent state (referred to as the local oscillator, LO) on a balanced beam splitter (hence, $T=1/2$), see Fig.\,\ref{fig:hom_het}(a). The two output modes of the beam splitter are individually measured with two photodetectors, that produce photocurrents proportional to the number of incoming photons. Those photon numbers are then subtracted and the resulting quantity is proportional to the value of a quadrature, defined by the phase of the LO. Indeed, by transforming the field operators of the signal mode $\hat{a}_s$ and the LO $\hat{a}_\text{LO}$ after the balanced beam splitter [Eq. (\ref{eq:bsa})] as $\hat{a'}=(\hat{a}_s+\hat{a}_\text{LO})/\sqrt{2}$ and $\hat{a'}_\text{LO}=(-\hat{a}_s+\hat{a}_\text{LO})/\sqrt{2}$ and parametrizing the strong coherent LO as $\hat{a}_\text{LO}=\alpha_\text{LO}e^{i\phi}$, where $\alpha_\text{LO}\in\mathbb{R}$ and $\phi$ is the LO phase, we find that the photon-number difference between the beam splitter outputs is $\hat{a'}_s^{\dag}\hat{a'}_s-\hat{a'}_\text{LO}^{\dag}\hat{a'}_\text{LO}=\hat{a}^{\dag}\alpha_\text{LO}e^{i\phi}+\hat{a}\alpha_\text{LO}e^{-i\phi}$. Hence, by setting $\phi$ equal to either $\phi=0$ or $\phi=\pi/2$, one can detect either $\hat{x}$ or $\hat{p}$-quadrature observables of the single mode, thereby compensating for the photon-number fluctuations of the signal and the LO. Moreover, by continuously changing the phase of the LO one can perform state tomography and reconstruct the Wigner function [Eq. (\ref{eq:wigner})].
In the case of an $N$-mode system the probability distribution of a measured quadrature [Eq. (\ref{eq:quadprob})] is obtained by integrating the Wigner function over the complementary quadrature in the same mode and the quadratures of all other $N-1$ modes. The covariance matrix of the conditional state of the residual $N-1$ modes of system $A$ after knowing the outcome $q_B\in\{x_B,p_B\}$ of a $\hat{q}_B\in\{\hat{x}_B,\hat{p}_B\}$ quadrature measurement on the single-mode system $B$ is given by~\cite{Eisert2002,Fiurasek2002a}
\begin{equation}
\gamma_{A|q_B}=\gamma_{A}-\sigma_{A,B}(Q\gamma_B Q)^{MP}\sigma_{A,B}^T,
\label{eq:condmathom}
\end{equation}
where $\gamma_A$ and $\gamma_B$ are the covariance matrices of the subsystems $A$ and $B$ and $\sigma_{A,B}$ is the correlation matrix between the subsystems $A$ and $B$ before the measurement; thus, the state of the overall $N$-mode system AB before the measurement is characterized by the covariance matrix
\begin{equation}\label{eq:CovarianceMatrix}
\gamma_{AB}=
    \begin{pmatrix}
        \gamma_A & \sigma_{A,B} \\
        \sigma_{A,B} & \gamma_B
    \end{pmatrix}
\end{equation}
(note that the covariance matrix can be always rearranged such that a single-mode system $B$ resides in mode $N$), $MP$ stands for the Moore-Penrose (pseudo-)inverse of a generally singular matrix, and matrix $Q=\text{diag}(1,0)$ for the $\hat{x}_B$ quadrature measurement and $Q=\text{diag}(0,1)$ for the measurement of the quadrature $\hat{p}_B$. 

A homodyne measurement yields the value of the measured quadrature while being fully uninformative about the complementary one. As an alternative, the \textit{heterodyne detection} (also referred to as the double homodyne detection), which projects the measured state on a coherent one, yields information on both of the complementary quadratures in the measured mode while being limited by the uncertainty principle. In practice, the heterodyne measurement is realized by a balanced splitting of a signal mode on a beam splitter with $T=1/2$ and then $\hat{x}$ and $\hat{p}$ quadrature homodyne measurements on the outputs of the beam splitter (set by the phases of the respective LO beams). The conditional state of a subsystem $A$ of a generally $N$-mode system $AB$ after the heterodyne measurement on the single-mode subsystem $B$ is then given by~\cite{Weedbrook2012a}
\begin{equation}
    \gamma_{A|x_B,p_B}=\gamma_{A}-\sigma_{A,B}(\gamma_B+\mathbb{I})^{MP}\sigma_{A,B}^T.
\end{equation}
Heterodyne measurement can generally be unbalanced, that is set by the beam splitter transmittance $T_\text{het}$, as shown in Fig. \ref{fig:hom_het}(b). The quadratures measured on the beam splitter outputs are then the linear combinations of the input quadratures according to Eq. (\ref{eq:bsa}) and read $\hat{x}_\text{het}=\sqrt{T_\text{het}}\hat{x}_{s}+\sqrt{1-T_\text{het}}\hat{x}_\text{vac}$ and $\hat{p}_\text{het}=-\sqrt{1-T_\text{het}}\hat{p}_{s}+\sqrt{T_\text{het}}\hat{p}_\text{vac}$, where $\hat{x}_\text{vac}$ and $\hat{p}_\text{vac}$ are the quadratures of the vacuum mode. The variances of the measured quadratures are then equal to $T_\text{het}V^x_s+1-T_\text{het}$ and $(1-T_\text{het})V^p_s+T_\text{het}$, where $V^{\{x,p\}}_s$ are $\hat{x}$ and $\hat{p}$ quadrature variances of the signal, combined with the fractions of vacuum noise, which can be seen as a penalty for the simultaneous measurements of the complementary quadratures. The covariance matrix of a generally $N$-mode state conditioned on the generally unbalanced heterodyne measurement can be obtained by introducing the vacuum state in mode $N+1$, splitting the measured mode $N$ with this vacuum mode with the ratio $T_\text{het}$ and double application of the conditioning [Eq. (\ref{eq:condmathom})] on $\hat{x}$- and $\hat{p}$-quadrature measurements in modes $N,N+1$. 
\subsection{Continuous-variable quantum information theory}
The outcomes of the measurements on continuous-variable states are typically discretized and represented by discretely distributed classical random variables. Information contained in such variable $X$, which takes values from the support (encoding alphabet) $\mathbb{X}$ with the probability distribution $p(X)$ is well known to be defined by the \textit{Shannon entropy} $H(X)=-\sum_{X\in\mathbb{X}}{p(X)\log{p(X)}}$. We omit the logarithm base here, but further, with no loss of generality, we take logarithms for the entropic measures base 2 and quantify the information in Shannon units or bits. Similarly, joint $H(X,Y)=-\sum\sum_{X\in\mathbb{X},Y\in\mathbb{Y}}{p(X,Y)\log{p(X,Y)}}$ and conditional $H(X|Y)=-\sum\sum_{X\in\mathbb{X},Y\in\mathbb{Y}}{p(X,Y)\log{[p(X,Y)/p(Y)]}}$ entropies can be defined by adding the random variable $Y$, with support $\mathbb{Y}$ and with individual $p(Y)$ and joint $p(X,Y)$ probability distributions. The mutual entropy $H(X:Y)$, which we refer to here as the \textit{mutual information} $I_{XY}=-\sum\sum_{X\in\mathbb{X},Y\in\mathbb{Y}}{p(X,Y)\log{(p(X,Y)/[p(X)p(Y)])}}=H(X)-H(X|Y)=H(Y)-H(Y|X)=H(X)+H(Y)-H(X,Y)$ quantifies the capacity of a classical channel with discrete input and output $X$ and $Y$. 

Theoretical evaluation of properties of continuous-variable states can be efficiently performed by considering the observables as continuously distributed random variables. For such a variable $X$  with the probability density function $f(X)$ on support $\mathbb{X}$, the information contained in this variable is characterized by the differential entropy $h(X)=-\int_\mathbb{X}f(x)\log{f(x)}dx$. Similarly, the conditional $h(X|Y)=-\int_{\mathbb{X,Y}}f(x,y)\log{f(x|y)}dxdy$ entropy can be defined with the continuous variable $Y$ on $\mathbb{Y}$ with $f(y)$ and conditional density function $f(x|y)=f(x,y)/f(y)$, expressed through the joint density function $f(x,y)$. The mutual information between the continuous variables $X,Y$ is then $I_{XY}=-\int_\mathbb{X}f(x,y)\log{[f(x,y)/[f(x)f(y)]]}dxdy$. 

For a Gaussian-distributed random variable $X$ with variance $V_X$ the differential entropy simplifies to $h(X) = (1/2)\log{V_X}+C$, where $C:= (1/2)\log{(2\pi e)}$. Similarly, the conditional entropy on variable $Y$ with variance $V_Y$ becomes $H(X|Y)=(1/2)\log{V_{X|Y}}+C$, where the conditional variance $V_{X|Y}=V_X-C_{XY}^2/V_Y$ is expressed through the correlation (covariance) $C_{XY}=\langle xy \rangle$ between the values $x$ and $y$ of the variables $X$ and $Y$. Then the mutual information for two Gaussian random variables $I_{XY}=(1/2)\log{[(V_XV_Y)/(V_XV_Y-C_{XY}^2)]}$ is independent of the scaling and can be equivalently expressed as 
\begin{equation}
    I_{XY}=\frac{1}{2}\log{\frac{V_X}{V_{X|Y}}}=
    \frac{1}{2}\log{\frac{V_Y}{V_{Y|X}}}.
    \label{eq:MI}
\end{equation}

The amount of information contained in a quantum state with density matrix $\hat{\rho}$ is given by the quantum \textit{von Neumann entropy} $S(\hat{\rho})=-Tr\hat{\rho}\log{\hat{\rho}}$. The von Neumann entropy of a pure state is then 0, while a thermal state maximizes the von Neuman entropy at the given mean photon number. The quantum entropy can be extended to definitions of conditional, joint and mutual entropies, similarly to the classical case. 

If a classical discrete random variable $X$ with support $\mathbb{X}$ and probability distribution $p(X)$ is encoded by the alphabet of quantum states $\hat{\rho}(X)$, then the information on $X$, which can be extracted from the measurements of the resulting quantum ensemble with outcomes $Y$ is upper limited as $I_{XY} \leq \chi(p(X),\hat{\rho}(X))$ by the \textit{Holevo bound} (also referred to as the Holevo information or Holevo quantity):
\begin{equation}
    \chi(p(X),\hat{\rho}(X))=S\Big(\sum_{\mathbb{X}}{p(X)\hat{\rho}(X)}\Big)-\sum_{\mathbb{X}}{p(X)S(\hat{\rho}(X)}).
    \label{eq:holevo}
\end{equation}
For continuously distributed classical random variables the sums in Eq. (\ref{eq:holevo}) turn into integrals over the continuous support of the random variable $X$. 

If we consider a classical, generally noisy channel $\mathcal{N}$ that maps the input $X$ to the output $Y$, $X \xrightarrow{\mathcal{N}} Y$, then the channel capacity is given by the mutual information $I_{XY}$ maximized over all possible input sets $X$. Similarly, the classical capacity of a memoryless quantum channel $\mathcal{M}$, independently mapping quantum input states to the output ones, $\hat{\rho}(X) \xrightarrow{\mathcal{M}} \mathcal{M}[\hat{\rho}(X)]$, is given by the Holevo bound $\chi(p(X),\mathcal{M}[\hat{\rho}(X)])$, maximized over all possible sources $p(X),\hat{\rho}(X)$.

By choosing the orthogonal basis $|i\rangle$, in which the density matrix of a state is diagonal, $\hat{\rho}=\sum_i{\lambda_i|i\rangle\langle i|}$, the von Neumann entropy can be expressed as the Shannon entropy of the eigenvalues $\lambda_i$, $S(\hat{\rho})=-\sum_i{\lambda_i \log{\lambda_i}}$. For Gaussian states such diagonalization can be performed in terms of the covariance matrix $\gamma$ by applying a symplectic transformation $S$ such that
\begin{equation}
    S\gamma S^T=\bigoplus_{k=1}^N
    \begin{pmatrix}
        \lambda_k & 0 \\
        0 & \lambda_k
    \end{pmatrix}:=\gamma^{\oplus},
\end{equation}
where $\{\lambda_k\}$ are the \textit{symplectic eigenvalues} of the covariance matrix $\gamma$ and the diagonal matrix $\gamma^{\oplus}$ is referred to as the \textit{Williamson form} of the covariance matrix (following the Williamson theorem~\cite{Williamson1936}, proving that the diagonalization is always possible for any $2N\times 2N$ real positive-definite matrix). The symplectic eigenvalues can be found as the eigenvalues of the matrix $|i\Omega\gamma|$ and define the von Neumann entropy of a state with the covariance matrix $\gamma$ as~\cite{Serafini2004}
\begin{equation}
    S(\gamma)=\sum_i^N{G\Bigg(\frac{\lambda_i-1}{2}\Bigg)},
\label{eq:sofgamma}
\end{equation}
where $G(x)=(x+1)\log{(x+1)}-x\log{x}$ is the bosonic entropy function~\cite{Serafini2005}. A symplectic transformation therefore represents a Gaussian state as a direct product of thermal states with variances $\lambda_i$ whose entropies are then expressed through the mean photon numbers $\bar{n}_i$ and sum up to the von Neumann entropy of the entire state. 

Having defined the CV quantum states, their observables, typical measurements, operations, and information quantities, we can proceed to the description of CV quantum communication. We start with the major application of quantum key distribution.

\section{Continuous-variable quantum key distribution}
\label{sec:cvqkd}
The idea to use quadrature observables measured by homodyne detection for QKD was first stated by~\cite{Ralph1999} using simple binary encoding and analyzed in terms of the quantum bit error rate (QBER) introduced by an eavesdropper during the interception of the key bits. While the use coherent signal states was shown to be inefficient within this framework, the EPR correlations obtained by coupling two displaced squeezed states, subsequently sent to the remote trusted party after a random time delay on one of the beams, can be advantageous in protecting the key bits against practical attacks, resulting in higher QBER for each intercepted bit. The use of modulated single-mode squeezed vacuum states for CV QKD was suggested by~\cite{Hillery2000} as a means of overcoming the vulnerability of the binary-encoded coherent-state protocol to practical attacks such as the beam-splitting or intercept-resend attack, when an eavesdropper measures the incoming signal and then resends the modulated signal according to the measured results. The security of the scheme was generalized by~\cite{Gottesman2001} using quantum error-correcting codes for encoding qubits in the infinite-dimensional Hilbert space of the CV quantum systems. The possibility of using entangled CV states prepared using modulated coherent states entering a nondegenerate parametric amplifier and measured separately via the trusted parties for binary CV QKD was discussed by~\cite{Reid2000} where practical eavesdropping attacks where considered. 

The distribution of continuous Gaussian keys was first proposed using squeezed states by~\cite{Cerf2001}. This all-continuous CV-QKD protocol allowed evaluation of information shared between the trusted parties and leaking to an eavesdropper upon Gaussian individual attacks (which we discuss in Sec. \ref{sec:theosec}) using the previously described formalism of Gaussian quantum information. The Gaussian CV-QKD protocol using experimentally more feasible coherent states was subsequently proposed by~\cite{Grosshans2002} and is known as the GG02 protocol. It was shown to be secure against individual attacks at channel transmittance above $50\%$. To overcome this limitation, the coherent-state CV-QKD protocol using postselection was proposed by~\cite{Silberhorn2002a}. Alternatively, the use of reverse reconciliation was shown to provide theoretical security of the coherent-state protocol at any level of channel attenuation by~\cite{Grosshans2003}. This result indicated the practical feasibility of CV QKD and led to its rapid development and implementation, while defining the basic principles of Gaussian CV QKD. However, the applicability of the protocols was still largely limited by the postprocessing (error-correction) efficiency for Gaussian-distributed data~\cite{Lodewyck2007}, which we discuss in Sec. \ref{Sec:Post_Processing}. To overcome this limitation, discrete-modulated (DM) CV QKD was proposed by~\cite{Leverrier2009a}. The subsequent development of CV QKD then continued along the lines of Gaussian or DM protocols toward improving their applicability and security, as we discuss in the forthcoming sections.

\subsection{CV-QKD protocols}
\label{sec:protocols}
\begin{figure}
    \centering
    \includegraphics[width=0.9\linewidth]{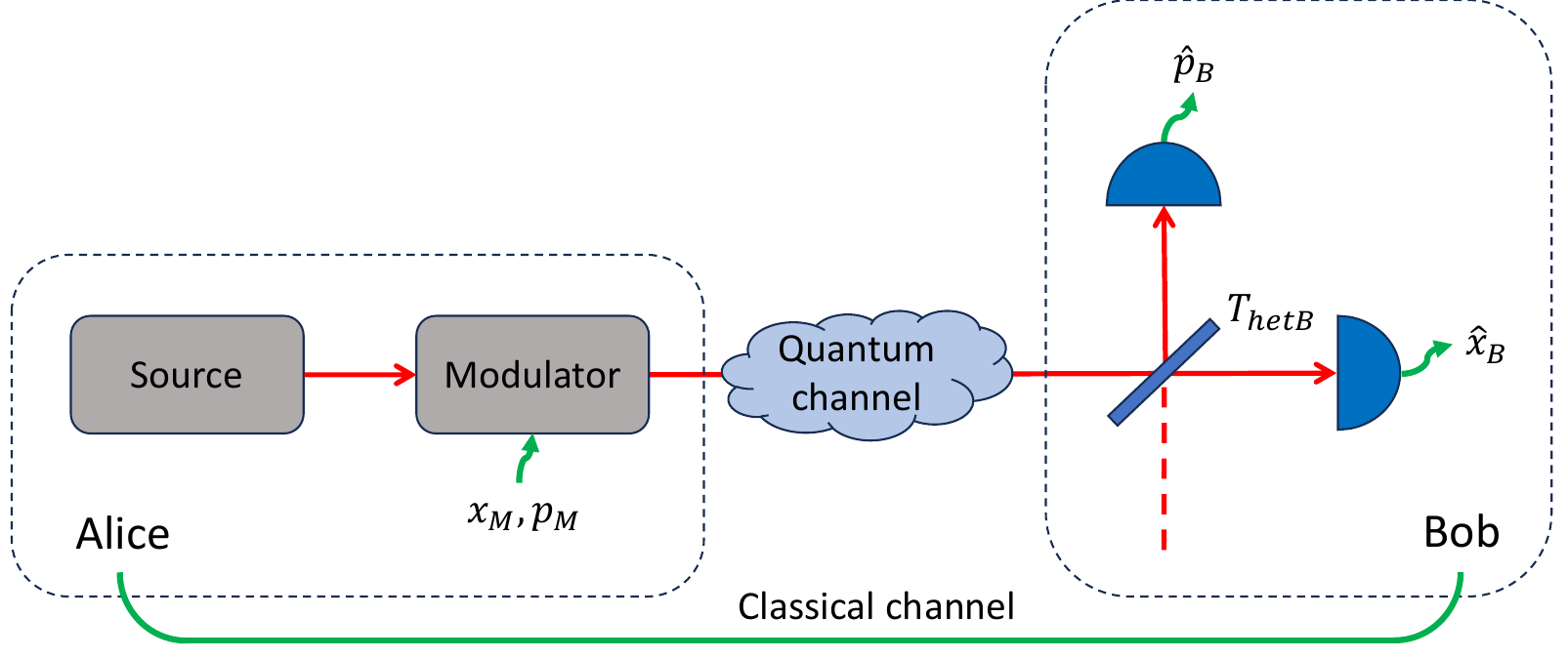}
    \caption{Generic P\&M CV-QKD scheme. Alice generates a signal state using the source, applies the displacement according to the pregenerated data $x_M$ and $p_M$ on the modulator, and sends the modulated states through the quantum channel to the remote party Bob, who performs the heterodyne detection with a balancing $T_\text{hetB}$, obtaining his data from the measurement of $\hat{x}_B$ and $\hat{p}_B$. The trusted parties then use the authenticated classical channel to perform error correction and privacy amplification.}
    \label{fig:P_and_M}
\end{figure}

\begin{figure}
    \centering
    \includegraphics[width=0.9\linewidth]{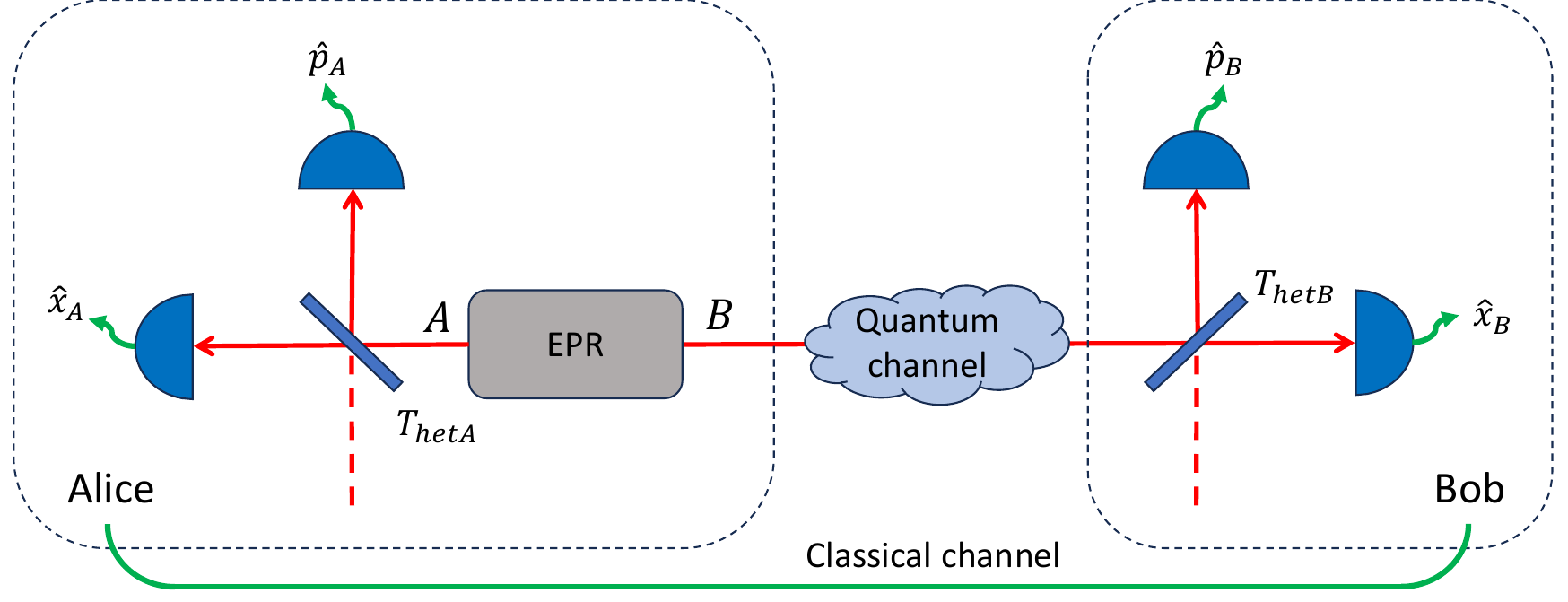}
    \caption{Generic EB CV-QKD scheme. Alice generates a two-mode entangled state using the source EPR (a TMSV state in practice), performs the heterodyne detection on her mode $A$ with balancing $T_\text{hetA}$, obtaining her data from the measurement of $\hat{x}_A$ and $\hat{p}_A$ and sends the mode $B$ through the quantum channel to the remote party Bob. He performs the heterodyne detection with balancing $T_\text{hetB}$, obtaining his data from the measurement of $\hat{x}_B$ and $\hat{p}_B$. The trusted parties then use the authenticated classical channel to perform error correction and privacy amplification.}
    \label{fig:EPR}
\end{figure}

A generic CV-QKD protocol between two trusted parties, that are traditionally denoted as Alice (\textit{A}) and Bob (\textit{B}), consists of the following major steps, as illustrated in Fig. \ref{fig:P_and_M}:

\begin{enumerate}
    \item \textbf{State preparation.} Alice generates random data sequences (strings) $\mathbf{x_M}$ and $\mathbf{p_M}$, containing key values to be encoded in the modulated quantum states\footnote{We omit a discussion of random number generation for QKD and other quantum communication tasks as it goes beyond the scope of this review. Quantum random number generators (particularly those based on continuous variables) can be used for this purpose, as reviewed by~\cite{HerreroCollantes2017}.}. Those random values either are taken from the zero-centered Gaussian distributions $\mathcal{N}(0,V_x),\mathcal{N}(0,V_p)$ in case of the Gaussian protocols, or follow a discrete modulation profile (binary, quaternary or more complex constellations with different probability distributions). Alice then prepares a signal state (coherent or squeezed) and applies displacement to one or both of the quadratures, depending on a particular protocol. In a standard Gaussian scheme (for example, GG02), Alice modulates the signal up to a symmetrical thermal state, but asymmetrical modulation profiles are also possible as we discuss below. Alice verifies the modulation by monitoring the modulated signal (tapping off a part or switching to fully measuring it with her local detector).
 
    \item \textbf{State distribution.} The modulated states are sent to the receiving party, Bob, through a quantum channel, which is assumed to be fully controlled by an eavesdropper, Eve.

    \item \textbf{State measurement.} The remote (receiving) party, Bob, performs the detection of the incoming signal, using homodyne or heterodyne detection, hence obtaining data sequences $\mathbf{x_B}$ and $\mathbf{p_B}$. In the case of homodyne detection, Bob randomly switches between measuring $\hat{x}$ and $\hat{p}$ quadratures. This case implies the necessity of a key sifting (basis reconciliation) procedure, where Alice and Bob keep the data, encoded in and decoded from the same quadratures. The state measurement concludes the quantum phase of the protocol.

    \item \textbf{Postprocessing.} In the postprocessing stage, the trusted parties perform all necessary procedures with the classical data encoded and obtained from the measurements. This typically includes error correction, when a reference trusted party sends the correcting trusted party the syndromes of their data, which are compared. The respective data are discarded in the case of mismatch and kept otherwise. The case in which Alice, who is the sender of the quantum signals, is also the reference side of the error correction, sending her syndromes to Bob, is called direct reconciliation (DR). The opposite situation in which Bob is the reference side of the error correction, is referred to as reverse reconciliation (RR). Further in the postprocessing stage the trusted parties also typically perform the parameter estimation, by estimating the relevant parameters of their states and the quantum channel. This allows the trusted parties to bound the information leaked to the eavesdropper Eve. Depending on the protocol and implementation, the parameter estimation can be performed either after or before the error correction. The trusted parties can also apply postselection to their data, as we discuss in Sec. \ref{Subsec:postselection}. Finally, they perform privacy amplification in order to minimize the information on the key, which is available to Eve. This procedure results in the secret keys, which are ideally identical and completely unknown to the eavesdropper. We discuss the postprocessing procedures in Sec. \ref{Sec:Post_Processing}.
\end{enumerate}
The previously described scheme in which Alice prepares the modulated signal states according to her pregenerated data and Bob performs the detection to obtain his measured data is the P\&M version of CV QKD. Alternatively and equivalently, Gaussian CV-QKD protocols can be implemented such that both of the trusted parties are measuring respective parts of a bipartite entangled state - typically a TMSV in the case of symmetrical modulation; however, a more advanced entangled state preparation is needed for asymmetrical modulations, as we later discuss and as summarized in Table \ref{tab:protocols}. Such EB realization of CV QKD is shown in Fig. \ref{fig:EPR} for the standard (generic) Gaussian protocols. In this case, the state preparation consists of generating an entangled state while Alice is also performing a measurement (homodyne or heterodyne) on her side to obtain her data sequences. The non-Gaussian DM protocols can be equivalently approximated using the EB schemes. Since Eve ideally cannot distinguish between P\&M and EB versions of the protocols, the latter became not only a technical alternative but also an ansatz for theoretical security analysis, as we discuss in Sec. \ref{sec:theosec}.

While more difficult to implement in practice, EB CV QKD can have several advantages. First, there is no need to pregenerate the random keys in the case of an EB protocol, as the data sequences are obtained from the measurements and their randomness is provided by the stochastic nature of the quantum measurement. Second, the trusted parties directly characterize the bipartite state, shared over the quantum channel, which simplifies the parameter estimation and the evaluation of the information leakage. Otherwise, in the P\&M protocol, the trusted parties generally have to derive an equivalent entangled state in order to perform the security analysis of the protocol, as we discuss in Sec. \ref{sec:theosec}. Finally, strongly nonclassical features of the entangled states can provide better efficiency and robustness of the protocols than the P\&M realizations based on coherent states, as we discuss in Sec. \ref{subsec:sqz_sig}. Nevertheless, coherent-state P\&M protocols remain more technically feasible and are a standard choice for current CV-QKD implementations.

Note that the schemes in Fig. \ref{fig:P_and_M} and \ref{fig:EPR} represent the conceptual design of the CV-QKD protocols. They omit essential technical details, in particular, the signal monitoring on the sender side and the LO beams needed for the homodyne detectors. The details of the practical realization are discussed in Sec. \ref{sec:practimpl}. 

\subsubsection{Coherent-state protocols}
The aforementioned GG02 protocol~\cite{Grosshans2003} is based on symmetrical Gaussian modulation in both quadratures of coherent states and subsequent homodyne detection on the receiving side; hence, Bob is randomly switching between the quadrature measurements by choosing $T_\text{hetB}=\{0,1\}$. This implies key sifting to be performed by Alice and Bob. The no-switching protocol~\cite{Weedbrook2004} is similarly based on symmetrical Gaussian modulation of coherent states and heterodyne detection at Bob's side ($T_\text{hetB}=1/2$) and does not require key sifting, as key bits are obtained simultaneously from both of the quadratures. The phase-space symmetries of the no-switching protocol enable the most general security proofs for this type of protocol (finite-size composable security against general attacks)~\cite{Leverrier2015}, while most of the other CV-QKD protocols rely on  security against collective attacks, taking into account some finite-size effects, as we discuss in Sec. \ref{sec:theosec}. The unidimensional (UD) coherent-state CV-QKD protocol~\cite{Usenko2015} uses Gaussian modulation in one of the quadratures and homodyne detection of the same kind (note that measurement of the unmodulated quadrature is still needed for parameter estimation). The DM coherent-state protocols~\cite{Leverrier2009a} assume a plethora of discrete, non-Gaussian modulation profiles and homodyne or heterodyne modulation. The security and performance of those schemes are discussed in in Sec. \ref{sec:theosec}.

\subsubsection{Squeezed-state protocols}
The first Gaussian CV-QKD protocol~\cite{Cerf2001} is based on the symmetrical Gaussian modulation of the squeezed states, when Alice randomly switches between the squeezing of two orthogonal quadratures and applies Gaussian modulation to the squeezed one. Bob then performs homodyne detection by switching between $T_\text{hetB}=\{0,1\}$, and the trusted parties perform basis reconciliation, keeping those elements of the key that were modulated and measured in the same squeezed quadrature. With certain adaptations and thanks to the intrinsic symmetry, the squeezed-state homodyne-detection protocol also enables composable security against general attacks~\cite{Furrer2012}, as we discuss in Sec. \ref{sec:theosec}. The squeezed-state protocol with heterodyne detection is also possible, but was shown to be theoretically suboptimal~\cite{GarciaPatron2007} in the asymptotic regime, as it does not benefit from measurement of the antisqueezed quadrature, while it contains the vacuum noise from the balanced coupling of the heterodyne detection. However, heterodyne detection may become useful in the finite-size regime, where measurement of the antisqueezed quadrature may improve the parameter estimation and the resulting key rate~\cite{Oruganti2025}. The Gaussian UD protocol can also be extended to squeezed signal states~\cite{Usenko2018}. Signal squeezing was shown to be helpful to improve the robustness of the protocols to channel noise~\cite{Garcia2009}, especially when additional modulation of squeezed states beyond the level of antisqueezing is used~\cite{Madsen2012} (as we discuss in Sec. \ref{subsec:sqz_sig}), and to inefficient postprocessing~\cite{Usenko2011}. Squeezed signal states can also be used to perform DM CV QKD~\cite{Denys2021}, but this avenue, to our knowledge, has not been explored in practice.

\subsubsection{Entanglement-based protocols}
Entanglement-based Gaussian CV-QKD protocols as shown in Fig. \ref{fig:EPR} are generally equivalent to the P\&M ones~\cite{Grosshans2003a}. A conceptual difference occurs when an entangled source is placed in the middle of the channel (the scheme, accordingly referred to as entanglement-in-the-middle)~\cite{Weedbrook2013} . However, the presence of the channel in both arms of the entangled state makes the protocol effectively combine DR and RR scenarios and limits the tolerable channel transmittance, as we discuss in Sec. \ref{subsec:limits}.

\subsubsection{Two-way protocols}
CV QKD can be extended to the two-way protocols when one of the trusted parties (for example, Bob) prepares the modulated coherent or squeezed state, and sends it to another trusted party (respectively, Alice) through an untrusted quantum channel, Alice then performs additional modulation on the state and sends it through the same quantum channel back to Bob, who performs homodyne or heterodyne detection. The main variants of the two-way CV-QKD protocols, both P\&M and EB, were introduced by~\cite{Pirandola2008a}. 
Later extensions involved its realization at different wavelengths~\cite{Weedbrook2014}. These protocols enable higher robustness to channel noise than one-way protocols.

\subsubsection{Measurement-device-independent protocols}
\begin{figure}
    \centering
    \includegraphics[width=0.9\linewidth]{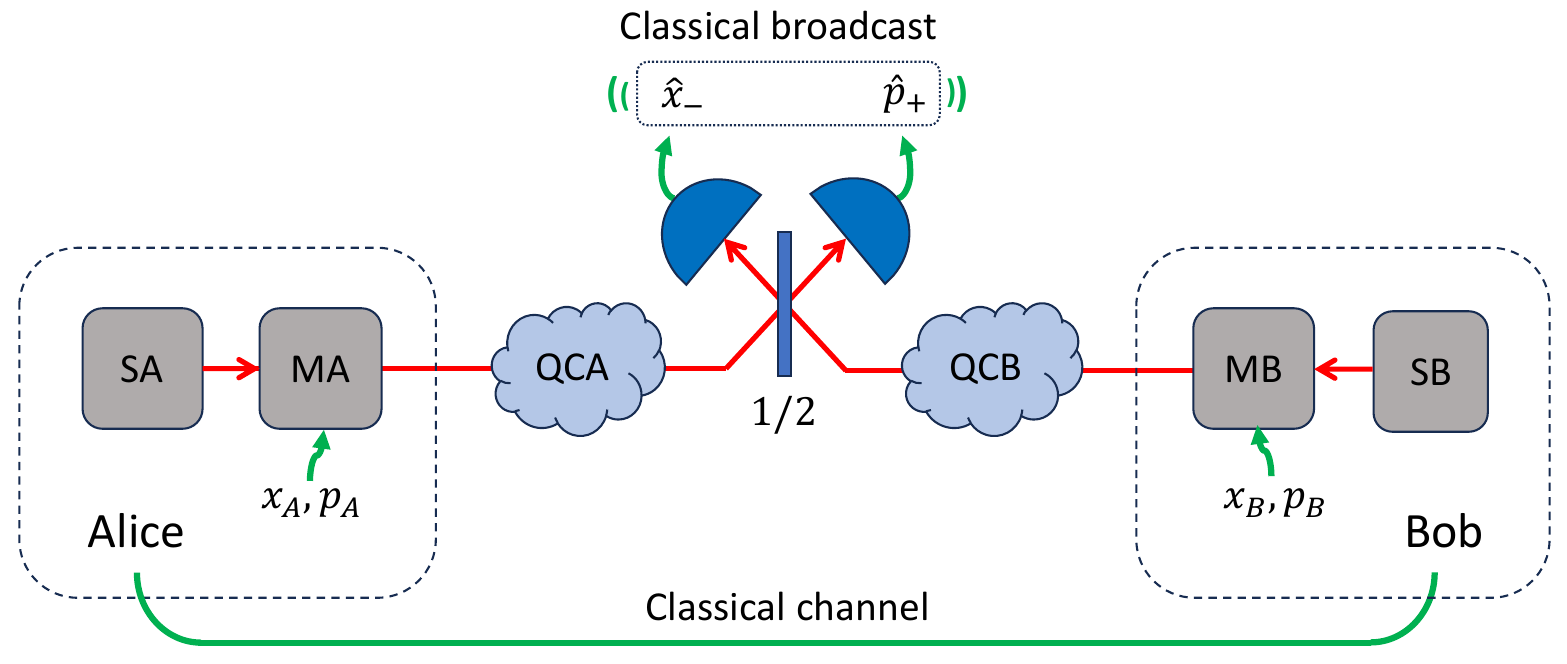}
    \caption{CV-MDI-QKD scheme. Alice and Bob generate their signal states, with each using their own source (SA and SB) and perform modulation to encode their data on a respective modulator (MA and MB). The modulated signal travels through the quantum channels (QCA and QCB) to the middle station, where they are coupled on a balanced beam splitter with transmittance $1/2$. The outputs $\hat{x}_-$ and $\hat{p}_+$ are measured and classically broadcasted.
    The trusted parties then use the authenticated classical channel to perform error correction and privacy amplification.}
    \label{fig:MDI}
\end{figure}

CV quantum states and homodyne detection can be used to implement MDI protocols, as first demonstrated in ~\cite{Pirandola2015}; see Fig.~\ref{fig:MDI}. Alice and Bob independently perform state preparation (signal state generation and modulation) on their sides, encoding their key data. The modulated signal travels through the quantum channels to a middle station (relay) where the signals are interfered and detected in complementary quadratures (this is also known as CV Bell measurement, first proposed in the context of CV quantum teleportation~\cite{Braunstein1998}, as discussed in Sec. \ref{sec:cv-qt}). The detection results are broadcast, which allows Alice and Bob to establish correlations between their data and perform error correction, parameter estimation, and privacy amplification. The measurement in this case can be fully controlled by Eve because its imperfections are attributed to the channel. Simultaneously, this makes the protocol sensitive to device imperfections and the configuration with the relay in the middle is the least optimal as we discuss in Sec. \ref{subsec:limits}. 

\subsubsection{Thermal-state protocols}
In CV QKD, while typically pure coherent states are used, studies have shown that ``noisy'' thermal states are also viable. Initial research~\cite{Filip2008,Usenko2010a} demonstrated their effectiveness with RR and signal purification. Other studies~\cite{Weedbrook2010,Weedbrook2012,Weedbrook2014,Papanastasiou2018, Papanastasiou2021} showed that thermal states could be directly employed in CV QKD (without input purification). This led to CV QKD being extended to longer wavelengths. In this context~\cite{Ottaviani2020} explored the use of the terahertz spectrum for short-range terrestrial communications. Finally, the composable finite-size security of microwave wireless communications was shown by~\cite{Pirandola2021a}.  

\subsection{Theoretical security}

\label{sec:theosec}

The theoretical security of QKD protocols relies on a set of assumptions, made about the protocol implementation and the capabilities of an eavesdropper. The underlying assumption of any security proof is that the trusted parties properly follow the protocol steps. However, deviations from a perfect implementation are unavoidable in practice and should be properly taken into account, as we discuss in Sec. \ref{subsec:implsec}.  Theoretical security then first focuses on eavesdropping attacks in the quantum channel, which is assumed to be fully controlled by an eavesdropper. In addition, implementation-specific attacks can target and exploit practical device imperfections and side channels. The attacks in the quantum channel follow the general scenario of Eve preparing her ancillary states, performing interaction of the ancillae with the quantum signal states, sent through the channel, and measuring the ancilla states to obtain the information on the key. The task of the theoretical security analysis is then to evaluate the upper bound on Eve's information based on the disturbance introduced to the signals. 

The most basic type of eavesdropping attack in the quantum channel is the \textit{individual attack}, when Eve individually (one by one) prepares her ancillae and also individually measures them after the interaction with the signal. In the asymptotic limit of infinitely many signals, exchanged by Alice and Bob, which implies perfect parameter estimation, Eve's information on the key is then bounded by the classical (Shannon) mutual information between her data and the data of the reference side of the postprocessing (Alice in the DR scenario and Bob in the RR one). The Csisz\'ar-K\"orner theorem for classical secure communication~\cite{Csiszar1978} can then be applied to QKD~\cite{Maurer1993} such that the amount of the secret key that can be extracted from the data shared between Alice and Bob is lower-bounded by their information advantage over Eve:
\begin{equation}
    \Delta I^{\to}=I_{AB}-I_{AE}, \Delta I^{\gets}=I_{AB}-I_{BE}
    \label{eq:Kind}
\end{equation}
in either of the reconciliation scenarios, which are denoted as $\to$ for DR and $\gets$ for RR. 

In the case of more advanced \textit{collective attacks}, Eve is able to store her ancilla states in a quantum memory and later apply an optimal collective measurement after the basis reconciliation, reaching the amount of information given by the Holevo bound [Eq. (\ref{eq:holevo})] between Eve's system (set of ancilla states) and the subsystem measured by the trusted party, which is the reference side of the postprocessing. The secret key is then lower-bounded by the \textit{Devetak-Winter rate}~\cite{Devetak2005} in the asymptotic regime for either of the reconciliation scenarios given by
\begin{equation}
    R^{\to}=I_{AB}-\chi_{AE}, R^{\gets}=I_{AB}-\chi_{BE}
    \label{eq:DWrate}
\end{equation}

The most general case of \textit{coherent} (also called general) attacks implies that Eve is able to not only collectively measure her ancilla states but also to collectively optimally prepare them. In the asymptotic limit, coherent attacks on CV-QKD are usually reduced to the collective ones as the signal states are independent and identically distributed (i.i.d.) and collective preparation does not bring any advantage to Eve, but if this is not the case or the data ensembles are limited, a rigorous treatment of coherent attacks is required, as we discuss in the forthcoming sections. 

In the finite-size regime, the efficiency of parameter estimation is limited and should be taken into account. The most general security analysis of QKD then also takes into account the fact that key distribution is part of a more complex cryptographic system and should be described in a composable way, allowing the security of the key to be quantified. 

\subsubsection{Notions of composable security}\label{sec:Composable_Securty}
The primary purpose behind generating cryptographic keys lies in their application within larger cryptographic settings, such as in conjunction with one-time pad encryption. Consequently, the security of the generated key must be guaranteed, regardless of its subsequent use. Composable security, in essence, leverages the security of subprotocols to establish the security of the entire cryptographic scheme. Historically, early QKD security proofs relied on a noncomposable security definition, for example,~\cite{Shor2000} where the information gathered by an adversary was quantified. However, this approach was shown to be insecure when applied within a larger framework~\cite{Koenig2007}. Inspired by the development of composability in classical cryptography~\cite{Canetti2000, Canetti2001, Pfitzmann2001}, this notion was extended to the quantum realm by~\cite{BenOr2004a, BenOr2004b, Unruh2004, Renner2006}. In the subsequent discussion, we loosely follow~\cite{Portmann2022}; see this review for a more detailed exploration of composability.

In contrast to earlier security definitions, for composable security we do not aim to quantify the amount of information an adversary has gathered but rather avoid defining security directly by instead comparing the real protocol to the ideal version of the protocol. This approach is known as the real-world-ideal-world paradigm. It can be illustrated as a virtual game where the adversary is presented with two black boxes, one containing the real protocol and the other containing the ideal protocol, without any indication of which box corresponds to which protocol. The adversary's task is to act as a distinguisher, attempting to tell apart the real protocol from the ideal one. They may choose inputs and receive all outputs that are not private for Alice or Bob according to the protocol, randomly (with probability $1/2$) either from the box containing the real protocol or the ideal one. 
 
Within this paradigm, a real cryptographic system is called perfectly secure if the adversary cannot distinguish the black box containing the real protocol from the one containing the ideal protocol, i.e., if their guessing probability is exactly $1/2$, which is equal to the probability of random guessing. 

To describe this idea in a mathematical framework, 
suppose that we have two protocols $\mathcal{P}_0$ and $\mathcal{P}_1$, and assume that a distinguisher $\mathcal{D}$ is provided with the outputs of both protocols with equal probability. The distinguisher then makes a guess $G \in \{0,1\}$, and $C=0$ or $C=1$ denotes which protocol produced the output. The distinguishability is then defined as
\begin{equation}
    d(\mathcal{P}_0, \mathcal{P}_1) := \sup_{\mathcal{D}} \left|\text{Pr}\left[G=0|C=0\right] - \text{Pr}\left[G=0|C=1\right]  \right|,
\end{equation}
where the supremum is taken over all distinguishers $\mathcal{D}$. It can be shown that for quantum states the right measure to capture the notion of distinguishability is half the trace norm (see~\cite[Chapter 4]{Barnett2009} for an explanation),
\begin{equation}
    d(\mathcal{P}_0, \mathcal{P}_1) = \frac{1}{2}\left|\left| \hat{\rho}_{K_AK_BE}^{\text{ideal}} - \hat{\rho}_{K_AK_BE}^{\text{real}} \right|\right|_1, 
\end{equation}
where the key of an ideal protocol is completely random, uniformly distributed, and completely decoupled from the adversary's knowledge. Hence, it is in tensor-product structure with Eve's system,
\begin{equation}
\begin{aligned}
      &\hat{\rho}_{K_A K_B E}^{\text{ideal}} :=\\
      &p^{\perp} \ketbra{\perp_A, \perp_B} \otimes \hat{\rho}_E^{\perp} + \tau_{UU} \otimes \hat{\rho}_E,  
\end{aligned}
\end{equation}
where $\tau_{UU}:=(1/N) \sum_{x=0}^{N-1} \ketbra{x,x}$, while the density matrix corresponding to the real protocol has the more general form
\begin{equation}
\begin{aligned}
 &\hat{\rho}_{K_AK_BE}^{\text{real}} =\\
 &p^{\perp} \ketbra{\perp_A, \perp_B} \otimes \hat{\rho}_E^{\perp} + \sum_{x,y=0}^{N-1} P_{x,y}\ketbra{x,y}\otimes \hat{\rho}_E^{x,y}.   
\end{aligned}
\end{equation}
Here $\perp_{A,B}$ denotes the event of the protocol aborting on Alice's or Bob's side and $p^{\perp}$ is the probability of that event. 

Generally, one cannot expect to achieve perfect security in (real) cryptographic routines. However, we may allow the adversary a certain small distinguishing advantage $\epsilon \geq 0$. In addition, a QKD protocol might abort sometimes in which case it is trivially secure, as no key is produced. Then we call a QKD protocol $\epsilon$-secure if 
\begin{align}
  \left( 1-p^{\perp}\right)  \frac{1}{2}\left|\left| \hat{\rho}_{K_AK_BE}^{\text{ideal}} - \hat{\rho}_{K_AK_BE}^{\text{real}} \right|\right|_1 \leq \epsilon.
\end{align}
By applying the triangle inequality~\cite{Araki1970} this can be rewritten as
\begin{align*}
    & \left( 1-p^{\perp}\right) \frac{1}{2}\left|\left| \hat{\rho}_{K_AK_BE}^{\text{ideal}} - \hat{\rho}_{K_AK_BE}^{\text{real}} \right|\right|_1 \\
    \leq & \text{Pr}\left[K_A \neq K_B\right] + (1-p^{\perp})\frac{1}{2} \left|\left| \hat{\rho}_{AE}^{\text{real}'} - \tau_U\otimes \hat{\rho}_E' \right|\right|_1.
\end{align*}
The first term,
\begin{equation}
\text{Pr}\left[K_A \neq K_B\right] \leq \epsilon_{\text{cor}}
\end{equation}
is called the correctness condition and $\epsilon_{\text{cor}}$ bounds the probability that the protocol does not abort and Alice and Bob do not share the same key, while the second term, 
\begin{equation}
    (1-p^{\perp})\frac{1}{2} \left|\left| \hat{\rho}_{AE}^{\text{real}'} - \tau_U\otimes \hat{\rho}_E' \right|\right|_1 \leq \epsilon_{\text{sec}},
\end{equation}
is called secrecy condition and $\epsilon_{\text{sec}}$ bounds the joint probability of not aborting and the private information being known to Eve. Thus, if one manages to prove a protocol $\epsilon_{\text{cor}}$-correct and $\epsilon_{\text{sec}}$-secret, this implies that the protocol is $\epsilon:= \epsilon_{\text{cor}} + \epsilon_{\text{sec}}$ secure. 

As previously mentioned, a protocol that aborts all the time is trivially secure, albeit not overly useful. This notion is captured by the completeness or robustness of a protocol. A QKD protocol is called $\epsilon_{\text{comp}}$-complete if
\begin{equation}
    \text{Pr}\left[ \perp | \textsc{Honest} \right] \leq \epsilon_{\text{comp}}.
\end{equation}
By $\textsc{Honest}$ we denote the honest implementation of the protocol, wherein all devices and the channel act as expected. In particular, no eavesdropping is taking place. Thus, if the protocol is $\epsilon_{\text{comp}}$-complete, under honest conditions, it does not abort, except with the probability $\epsilon_{\text{comp}}$.

\subsubsection{Individual attacks and channel models}
\label{subsec:ind}
Security of Gaussian CV QKD against individual attacks was first shown for the coherent-state protocol by~\cite{Grosshans2003} and then extended to the squeezed-state protocol using equivalent EB representation. The security argument is based on the Heisenberg uncertainty principle, which prevents Eve from precisely simultaneously measuring both the $\hat{x}$ and $\hat{p}$ quadratures. We assume that Alice and Bob ideally share a pure two-mode entangled state in modes $A$ and $B$, and mode $E$ is available to Eve for individual measurements after the interaction with mode $B$ in a quantum channel. Then, after the projective quadrature measurements on modes $A$ and $E$ or $B$ and $E$, the uncertainty principle can be generalized in terms of the conditional variances as 
\begin{equation}
    V_{A|B}V_{A|E}\geq 1,  V_{B|A}V_{B|E}\geq 1.
    \label{eq:Heisbounds}
\end{equation}
This allows the asymptotic key rate Eq. (\ref{eq:Kind}) to be bounded using Eq. (\ref{eq:MI}) for the Shannon mutual information of the Gaussian-distributed data as
\begin{equation}
    \Delta I_{\infty}^{\to} \geq \log{\frac{1}{V_{A|B}}}, \Delta I_{\infty}^{\gets} \geq \log{\frac{1}{V_{B|A}}}.
\end{equation}
Here and afterward we evaluate the key rate in bits (taking logarithm base 2) per transmitted symbol, i.e., a signal state, which is also called bits per pulse or bits per channel use, as it is independent of the system clock (repetition) rate. In the practical realizations of QKD the relevant figure of merit is bits per second (bps), which characterizes the actual speed of the secret key distribution and is scaled by the system clock rate.

A typical quantum channel that describes the optical fiber links well is the \textit{thermal-loss} Gaussian channel defined by the channel transmittance (also referred to as channel loss or attenuation) $T$ and the excess noise of variance $\nu$, characterized with respect to the channel input. The quadrature observable measured at the receiver is then $\hat{x}_B=\sqrt{T}(\hat{x}_S+\hat{x}_M+\hat{x}_N)+\sqrt{1-T}\hat{x}_0$, where $\hat{x}_S$ is the quadrature of the signal state prior to modulation with variance $V_S$ (such that $V_S=1$ for the coherent-state protocol), $\hat{x}_M$ is the value of modulation (displacement) applied by the sender, taken from a zero-centered Gaussian distribution with variance $V_M$, $\hat{x}_N$ is the contribution from the channel excess noise with variance $\nu$, and $\hat{x}_0$ is the contribution from the vacuum noise with variance 1, same for the $\hat{p}$ quadrature. The modulated signal of variance $V=V_S+V_M$ in either of the quadratures is then changed to $T(V+\nu)+1-T$ on the channel output. Note that often the channel noise introduced by the channel with respect to the input is joined into a single parameter $\chi=\nu+(1-T)/T$, which simplifies the expression for the variance of the signal on the channel output to $T(V+\chi)$. The channel excess noise can be caused by numerous factors, such as, for example, residual modulation noise, coupling to other optical modes (particularly, in the coexistence with other optical signals), and imperfect noise estimation in the finite-size regime. The parametrization of the channel noise by $\nu$ (or $\chi$), which is then downscaled by the channel transmittance $T$ (often used in the theoretical predictions of CV-QKD performance), can therefore be less relevant for the practical scenarios, where the total noise from different sources is observed on the channel output. An equivalent parametrization of the excess noise on the channel output (and hence the noise beyond the vacuum noise due to losses) as $\nu_\text{out}=T\nu$, such that the variance measured by Bob is $TV+1-T+\nu_\text{out}$, can then be more adequate for predicting the performance of the protocols (and is used in Fig. \ref{fig:KRplot}). 

\begin{figure}
    \centering
    \includegraphics[width=0.9\linewidth]{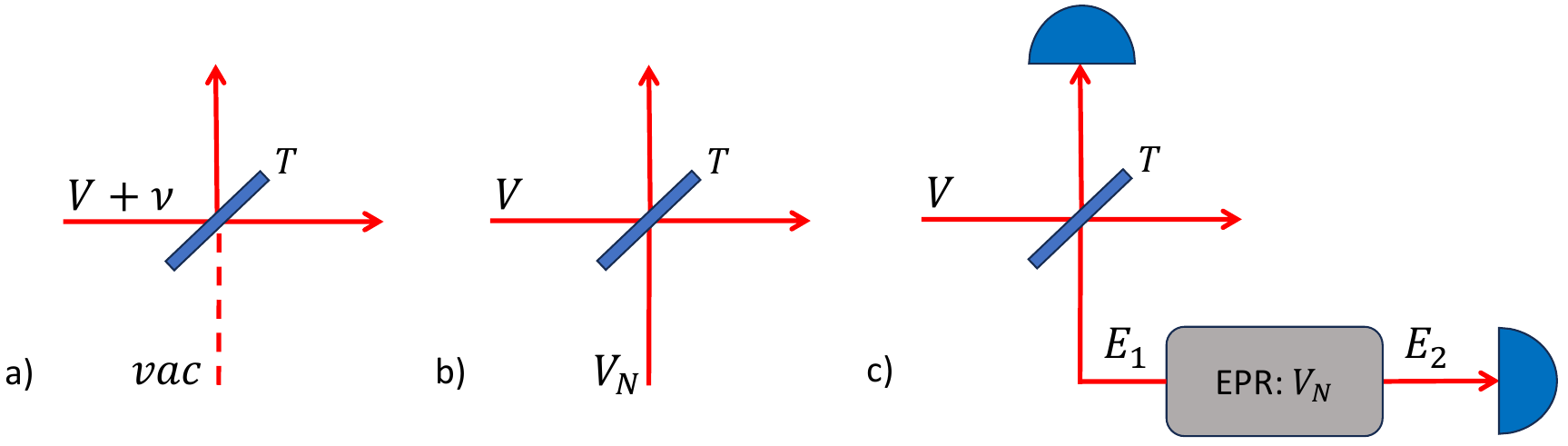}
    \caption{(a) Gaussian thermal-loss channel, characterized by the excess noise $\nu$, added to the modulated signal of variance $V$ at the channel input, followed by the pure loss $T$ as coupling to a vacuum mode (vac). (b) Equivalent representation of the channel as a coupling $T$ of the modulated signal $V$ to a thermal-noise mode with variance $V_N$. (c) Entangling cloner attack as the purification of the thermal noise $V_N$ by a TMSV (EPR) source with variance $V_N$ in modes $E_1$ (coupled to the signal) and $E_2$, both of which are available for Eve for individual or collective measurements.}
    \label{fig:TL}
\end{figure}

For a perfectly implemented P\&M protocol, the correlation between Alice's data, having variance $V_M$, and the data measured by Bob, is also downscaled by the channel transmittance as $\sqrt{T}V_M$. Equivalently, the thermal-loss channel can be represented as a coupling of the signal mode to a thermal-noise mode with variance $V_N$ at the coupling ratio $T$, as shown in Fig. \ref{fig:TL}(b). Equivalence then holds at $V_N=1+T\nu/(1-T)$. 
Eve's optimal individual attack in such a channel, which allows her to saturate the bounds given by Eq. (\ref{eq:Heisbounds}) set by the uncertainty principle, can be realized by means of an entangling cloner~\cite{Grosshans2003a}, as shown in Fig. \ref{fig:TL}(c). An entangling cloner is a TMSV state with quadrature variance $V_N$ in modes $E_{1,2}$, so one mode is coupled to the signal and both modes are available to Eve for individual measurement. Such an attack can be seen as Eve's purification of the noise $V_N$, which allows her to improve the knowledge of the noise added in the channel by conditionally preparing a squeezed state of variance $1/V_N$. 

While individual attacks are the most restrictive for an eavesdropper and hence are optimistic for the trusted parties, they still give important theoretical insights into the limitations of the protocols, particularly as insecurity against individual attacks implies insecurity against more advanced attack types such as collective or coherent attacks. Consequently, the key rate for individual attacks is an upper bound on the key rates for collective and coherent attacks. In the forthcoming subsections we proceed to the security of CV QKD against collective attacks.

\subsubsection{Asymptotic security via Gaussian extremality}
An important tool for showing the security of the Gaussian CV QKD is the Gaussian extremality theorem~\cite{Wolf2006}, which states that for fixed first and second moments, the distillable secret key rate against collective attacks is lower bounded by the secret key rate provided by a Gaussian state with these first and second moments, 
\begin{equation}\label{eq:GaussianExtremality}
    R^{\text{coll}}(\hat{\rho}) \geq R^{\text{coll}}(\hat{\rho}_{\text{Gauss}}).
\end{equation}
This theorem turned into the workhorse of CV security analyses and gave rise to an entire family of security proofs that rely on estimating the covariance matrix of the objective quantum state. Eq. (\ref{eq:GaussianExtremality})--and consequently the obtained secure key rates--is tight if the modulation is Gaussian and essentially tight if the modulation is close to Gaussian. Therefore, arguments of this type have been applied to both Gaussian-modulated and DM protocols, as we now elaborate on.

\paragraph{Gaussian-modulated CV-QKD protocols}
For collective attacks the Devetak-Winter formula
$R_\infty^{\to}$ and $R_\infty^{\gets}$ [Eq. (\ref{eq:DWrate})] provides a bound for the secure key rate in the asymptotic limit. While the mutual information between trusted parties $I_{AB}$ is assessed similarly to the previously described individual attacks (as it does not depend on the type of attack in the given quantum channel), the estimation of the bound on Eve's information is different for collective attacks. Early applications of the Gaussian extremality theorem~\cite{Wolf2006} to Gaussian-modulated protocols proved the optimality of Gaussian attacks~\cite{Navascues2006,Garcia2006}. In the case of pure loss (thermal-loss channel with no excess noise, $\nu=0$), the information available to Eve and described by the Holevo bound can then be directly obtained from the state in the initial vacuum mode after the beam splitter $T$, as shown in Fig. \ref{fig:TL}(a)~\cite{Grosshans2005}. For the noisy channels, $\nu > 0$, the purification argument is used to assess Eve's information: since Eve is able to control all the untrusted noise, added in the quantum channel, the overall state of the three-partite system $ABE$ is pure. The entropic triangle inequality~\cite{Araki1970} turns into $S(E)=S(AB)$. Similarly, after projective homodyne measurements at Alice's or Bob's system $A$ or $B$, the equalities $S(E|A)=S(AB|A)=S(B|A)$ and $S(E|B)=S(AB|B)=S(A|B)$ hold. Therefore, the Holevo bound can be directly evaluated from the state shared between Alice and Bob using Eq. (\ref{eq:sofgamma}), without the need to explicitly model the quantum channel (and allowing for multiple instances of such), so that all the impurity (noise) in the bipartite state of $AB$ is attributed to Eve. Note that it requires purification (by building equivalent EPR states) of all presumably trusted noise in the devices of Alice and Bob. Typically, the trust assumption applies to the sending and receiving stations (given inside the dashed boxes in Fig. \ref{fig:P_and_M}). We discuss the incorporation of this noise into the security analysis in Sec. \ref{subsec:implsec}. Security of CV-MDI QKD is obtained similarly after the covariance matrix, shared between the trusted parties in the EB version of the protocol, is conditioned by the publicly announced  outcomes of the measurements on $\hat{x}_-$ and $\hat{p}_+$. Note that unlike the previously described purification-based approach, the Holevo bound, limiting Eve's accessible information in the case of collective attacks on CV QKD, can be obtained from the two-mode entangling cloner, as shown in Fig. \ref{fig:TL}(c). While for a single-mode quantum channel it yields the same results as the purification-based analysis, it requires additional optimization in the multimode case~\cite{Pirandola2015} to account for possible correlations between the cloners.

Gaussian collective attacks on coherent-state CV QKD were generalized in the asymptotic regime by~\cite{Pirandola2008}, showing the extremality of canonical attacks, which correspond to the thermal-loss channels and can be explicitly described by the entangling cloner model, as shown in Fig. \ref{fig:TL}(c).

Security against collective attacks in the asymptotic regime can be extended directly to security against general (coherent) attacks using the de Finetti representation theorem for infinite-dimensional systems~\cite{Renner2009}. Alternatively, in the case of passive eavesdropping (noiseless channels with $\nu=0$), asymptotic collective attacks reduce to the individual ones. 

While practical security of QKD protocols can be shown only in the finite-size regime, the asymptotic security proofs remain essential for establishing theoretical security bounds of the protocols and can be used to preliminarily assess the possibility of performing QKD under given conditions. Specifically, asymptotic bounds can be used to establish limits on the performance of the protocols, as we later discuss.

\paragraph{DM CV-QKD protocols}
As mentioned at the beginning of this section, the Gaussian extremality argument applies generally and, therefore -- at the cost of introducing looseness for modulation patterns that do not closely resemble Gaussian modulation well -- can also be used to analyze the security of DM CV-QKD protocols. This examination was first carried out by~\cite{Leverrier2009}, who analyzed a four-state protocol with coherent states known as quadrature phase shift keying (QPSK). Although this work showed that in regimes of low signal-to-noise ratio (SNR), DM protocols can outperform Gaussian-modulated ones owing to more efficient error-correction codes. The reported key rates are still pessimistic, as they overestimate the adversary's knowledge about the key.

More recent methods for DM protocols combine the Gaussian extremality argument with semidefinite programming (SDP) theory. Historically first,~\cite{Ghorai2019a} analyzed a QPSK protocol with heterodyne detection. They quantified the quality of the correlations between Alice and Bob via two experimentally accessible parameters, the variance $v$ and the covariance $c$. While Alice's variance $V_A$ is known and Bob's variance $V_B = v$ is determined via measurements, the only unknown quantity in the covariance matrix $\begin{pmatrix}
    V_A \mathbbm{1}_2 & Z \sigma_z \\
    Z \sigma_z & V_B \mathbbm{1}_2
\end{pmatrix}$ [see also Eq. (\ref{eq:CovarianceMatrix})] representing Alice and Bob's joint state is the covariance $\sigma_{AB}$, denoted by the authors as $Z$, which can be calculated via
\begin{equation}
\label{eq:Zcovariance}
    Z = \Tr{\left(\hat{a}\hat{b}+\hat{a}^{\dagger}\hat{b}^{\dagger}\right)\hat{\rho}_{AB}}.
\end{equation}
It is well established that for fixed $V_A$ and $V_B$, the Holevo quantity $\chi(Y:E)$ is a decreasing function in $Z$. Thus, finding the smallest $Z$ yields the maximum $\chi(Y:E)$ and consequently minimizes $R^{\infty}$. Therefore, let $\Pi$ be the orthogonal projector onto the space spanned by the four coherent states prepared by Alice and define $\hat{C}:= \Pi\hat{a}\Pi \otimes \hat{b} + \Pi \hat{a}^{\dagger}\Pi\otimes \hat{b}^{\dagger}$. The minimal $Z$ compatible with Bob's measurements is determined by solving the following semidefinite program
\begin{equation}
    \begin{aligned}
    \min & \Tr{\hat{C} \hat{X}}\\
    \text{subject to }&\\
    & \Tr{\hat{B}_0 \hat{X}} = v,\\
    & \Tr{\hat{B}_1 \hat{X}} = c,\\
    & \textrm{Tr}_B\left\{ \hat{X} \right\} = \sigma_A,\\
    & \hat{X} \geq 0.
\end{aligned}
\end{equation}
In this context, $\hat{B}_0$ and $\hat{B}_1$ are operators related to the measurement of $v$ and $c$ and $\left(\sigma_A\right)_{k,l} := 1/4 \braket{\alpha_k}{\alpha_l}$ is given by the overlap between Alice's (coherent) signal states. The final constraint requires $\hat{X}$ to be positive semidefinite. The found optimum $Z^*$ can be used to calculate the Holevo quantity through standard techniques. 

The present optimization problem can be solved numerically with standard SDP solvers. However, the numerical implementation on a computer requires one to cutoff the Fock basis, which is used to formulate the optimization problem, at some number $N_c$. We discuss this cutoff within the course of the subsequent method, where a similar cutoff assumption had to be made. It is noteworthy that the method by~\cite{Ghorai2019} can be generalized to higher constellations. However, owing to the increased computational demands of the numerical optimization for a larger number of signal states, transitioning to more complex constellations is deemed infeasible with present-day technology.

Owing to the computational complexity of the numerical methods, the analysis of quadrature amplitude modulation (QAM) constellations with $64$ or more signal states currently appears to be beyond reach. Therefore, in what follows we explore analytical methods that cover this regime. The work ~\cite{Denys2021} adopted an approach akin to~\cite{Ghorai2019}, that uses a Gaussian optimality argument. However, it provided a complete analytic solution that is feasible for general modulation schemes. The main idea of the method is the following. While the first term in the Devetak-Winter formula is essentially known, the authors optimized the Holevo quantity over all quantum channels compatible with Bob's measurement results, which turns out to be the more challenging task. By applying Gaussian extremality, they upper bounded the optimization by a supremum of the Holevo quantity over all Gaussian states compatible with Bob's observations. Since Gaussian states are fully characterized by their covariance matrix, this simplifies the task significantly. The only unknown is the covariance $Z$. While this quantity can be calculated easily for Gaussian protocols, this is not the case for DM protocols. However, the authors formulated the objective optimization problem as a semidefinite program of the form
\begin{equation}
    \begin{aligned}
    \min & \Tr{\left(\hat{a}\hat{b}+\hat{a}^{\dagger}\hat{b}^{\dagger}\right) \hat{\rho}}\\
    \text{subject to }&\\
    & \Tr{\hat{C}_1 \hat{\rho}} = 2 c_1,\\
    & \Tr{\hat{C}_2 \hat{\rho}} = 2 c_2,\\
    & \Tr{\hat{N}_B \hat{\rho}} = n_B,\\
    & \textrm{Tr}_B\left\{ \hat{\rho} \right\} = \bar{\tau},\\
    & \hat{X} \geq 0,
\end{aligned}
\end{equation}
where $\hat{C}_1, \hat{C}_2$ and $\hat{N}_B$ are observables with expectations $c_1, c_2$, and $n_B$ and $\bar{\tau}$ is Alice's reduced density matrix given by the overlap between her signal states. Authors then provide analytical bounds for the value of $Z$, with the lower bound reading
\begin{equation}
    Z \geq 2 c_1 - 2 \sqrt{w\left(n_B - \frac{c_2}{\langle \hat{n} \rangle}\right)}.
\end{equation}
This lower bound is a function of the experimentally accessible quantities $c_1, c_2$ and $n_B$ (with the latter quantity being the received photon number on Bob's side) and $w$ and $\langle \hat{n} \rangle$ which quantify the modulation strength, thus are functions of protocol parameters. This finally allows the calculation of the Holevo quantity via symplectic eigenvalues, and therefore gives a lower bound on the secure key rate. For coherent-state protocols, the obtained bounds recover the rates known from Gaussian modulation and closely match the numerical bounds by~\cite{Ghorai2019} in most practical regimes. The authors stated that their proof was already essentially tight for modulations with $64$ QAM and higher.

The proof presented by~\cite{Kaur2021} follows a distinctive approach, in addition to incorporating a Gaussian extremality argument. The authors utilized a method to approximate Gaussian probability distributions using a finite-size Gauss-Hermite constellation, and they bounded the approximation error in the trace norm. Eve's information for this approximated constellation, which is bounded by the Holevo quantity, closely approaches the Holevo information for a Gaussian-modulated protocol. The error made can be bounded by a function of the approximation error in the trace norm, employing an entropic continuity bound. In contrast to~\cite{Denys2021}, the authors asserted that their method requires a constellation with about $5000$ states to achieve results close enough to Gaussian modulation, which suggests that the bounds provided by~\cite{Denys2021} are already tighter for lower constellation sizes and that bounds tend to be loose for smaller constellations. However, they essentially align with the results known from Gaussian modulation when dealing with sufficiently large constellations. While the Gaussian extremality argument is tight for Gaussian protocols, it can overestimate the adversary's knowledge about the key for DM protocols, particularly, if the modulation pattern is far from Gaussian. 
\subsubsection{Asymptotic security without Gaussian extremality}
\label{subsubsec:asympNG}
In the previous section, we saw that Gaussian-extremality-based arguments introduce looseness in the key rates of DM CV-QKD protocols, particularly for modulation patterns with a low number of signal states. However, technical challenges and implementation difficulties for Gaussian CV-QKD protocols that still persist motivated the development of techniques that directly evaluate the Devetak-Winter formula given in Eq. (\ref{eq:DWrate}) for the employed quantum states. This, however, is a notoriously difficult task that requires new techniques. 

\paragraph{Restriction to purely lossy channels.} 
Early security analyses not only focused on the asymptotic regime but also relied on the simplifying assumption of purely lossy (i.e., zero noise, $\nu=0$) channels. One of the pioneering works traces back to~\cite{Heid2006}. In this study a protocol with binary modulation in which Bob performs heterodyne measurements was analyzed. Under the simplifying assumption of a loss-only channel with transmittance $T$, the optimal attack is known to be the beam splitter attack, where Eve replaces the channel with a beam splitter and branches off $\ket{\pm \sqrt{1-T}\alpha}$, while Bob receives $\ket{\pm\sqrt{T}\alpha}$. Mathematically, Bob's heterodyne measurement is a projection onto coherent states, enabling him to confirm that he received a pure state. Eve's share has a tensor-product structure with Bob's pure state and is unitarily equivalent to $\ket{\epsilon_i}:=\ket{(-1)^{i} \sqrt{1-T}\alpha}$, where $i=0$ or $1$. In both direct and reverse reconciliation, this allows straightforward calculation of the Holevo quantity. Since the mutual information between Alice and Bob follows from purely classical considerations, we already hold all the ingredients to evaluate the Devetak-Winter formula to obtain secure key rates. In addition, with slight modifications this approach allows taking  postselection as well as trusted detector noise into account. Although the proof in this work was explicitly carried out for two states, the idea can be generalized to higher constellations, as demonstrated by~\cite{Sych2010} for two to eight states, for an arbitrary number of signal states as a side result by~\cite{Kanitschar2022} and, recently, even to the multiuser scenario~\cite{Kanitschar2024}.

\paragraph{Noisy channels and specific constellations.}
One of the pioneering approaches that considered excess noise was presented by~\cite{Zhao2009}. The authors analyzed a two-coherent-state CV-QKD protocol with homodyne detection in the asymptotic limit and proved security against collective attacks. They employed a method developed by~\cite{Rigas2006}, that estimates the largest eigenvalue and corresponding eigenvector, based on first- and second-moment measurements. Subsequently, they use Bob's homodyne measurement results to construct a statistical model of Eve's quantum state that needs to be compatible with Bob's measurements. In the worst-case scenario, where Eve gains the most information, this approach provides a lower bound on the secure key rate. The crucial steps rely on two properties of the binary Shannon entropy, namely the monotonicity and the concavity as a function of the absolute value of the overlap of the two states. While proving these properties for two states is feasible, generalizing them poses a considerable challenge. The main idea for~\cite{Bradler2018} was to establish both properties for the ternary Shannon entropy that is used to complete the security proof for a phase-shift-keying protocol with three signal states. This extended the proof idea to a three-state phase-shift-keying protocol.  However, generalizing monotonicity and concavity to the $N$-state case in a similar fashion is challenging, if not currently infeasible. As concluded by~\cite{Bradler2018}, it is likely that a more general argument is needed to establish similar properties for suitably generalized entropic quantities, thus extending this proof further to $N>3$. This suggests the necessity of a fundamentally different approach for a higher number of signal states. In addition, compared to the more recent proofs that we subsequently discuss, the key rates turn out to be pessimistic.

\paragraph{General discrete constellations.}
By the end of the 2010s, the incorporation of SDP for secure key rate calculations opened the door to a new generation of security proofs, with a representative not exploiting Gaussian extremality being~\cite{Lin2019}, which is part of the numerical framework by L{\"u}tkenhaus et al.~\cite{Coles2016, Winick2018}.  

The Devetak-Winter rate is reformulated as
\begin{align}
    R^{\infty} = \min_{\hat{\rho}_{AB}\in \mathcal{S}} D\left(\mathcal{G}\left(\hat{\rho}_{AB}\right) ||\mathcal{Z}\left( \mathcal{G}\left(\hat{\rho}_{AB}\right)\right) \right) - p_{\text{pass}} \delta_{EC},
\end{align}
where $D(\cdot||\cdot)$ denotes the quantum relative entropy, $\mathcal{G}$ is a completely positive trace nonincreasing map describing the classical postprocessing and $\mathcal{Z}$ is a pinching channel describing the key map, $\delta_{EC}$ denotes the leakage per signal and $p_{\text{pass}}$ is the sifting probability. The optimization is carried out over the set $\mathcal{S}$, which is defined by Bob's observations and the requirement that $\hat{\rho}_{AB}$ is the joint quantum state of Alice and Bob. Thus, the main idea of this proof method can be summarized as finding the worst-case density matrix that is physical and compatible with Bob's observations. This problem turns out to be a semidefinite program that then needs to be solved efficiently.

Assuming that Alice prepares one out of $N_{\textrm{St}}$ coherent states $\ket{\psi_i}$ with probability $p_i$ leads to the following optimization problem
\begin{equation} \label{eq:SDP}
\begin{aligned}
   \textrm{minimize } &D(\mathcal{G}(\hat{\rho}_{AB}) || \mathcal{G}(\mathcal{Z}(\hat{\rho}_{AB}))  )\\
   \textrm{subject to: } &\\
    &\Tr{\hat{\rho}_{AB} \left( \ketbra{k}_A \otimes \hat{x} \right) } = p_k \langle \hat{x} \rangle_k\\
    &\Tr{[ \hat{\rho}_{AB} \left( \ketbra{k}_A \otimes \hat{p} \right) } = p_k \langle \hat{p} \rangle_k\\
    &\Tr{ \hat{\rho}_{AB} \left( \ketbra{k}_A \otimes \hat{n} \right) } = p_k \langle \hat{n} \rangle_k\\
    &\Tr{ \hat{\rho}_{AB} \left( \ketbra{k}_A \otimes \hat{d} \right) } = p_k \langle \hat{d} \rangle_k\\
    &\textrm{Tr}_B\left\{ \hat{\rho}_{AB} \right\} = \sum_{i,j=0}^{N_{\textrm{St}}-1} \sqrt{p_i p_j} \langle \psi_j | \psi_i\rangle ~\ket{i}\!\!\bra{j}_A \\
    & \hat{\rho}_{AB} \geq 0,
\end{aligned}
\end{equation}
where $\hat{x}$ and $\hat{p}$ are the quadrature operators, $\hat{n}$ is the photon-number operator and $\hat{d}:=\hat{x}^2-\hat{p}^2$ is another second-moment operator. The objective function is convex and highly nonlinear. This problem is addressed by employing the numerical security proof framework by~\cite{Coles2016,Winick2018}, which involves a two-step process. In the first step, the nonlinear objective function is linearized and solved approximately. This can be accomplished, for instance, through an iterative first-order algorithm like the Frank-Wolfe algorithm~\cite{Frank1956}. As we cannot expect to find the exact minimum, the output of the first step serves only as an upper bound on the secure key rate. Therefore, in the second step, this approximate solution is transformed into a secure lower bound by combining another linearization and SDP duality theory. In addition, the application of a relaxation theorem~\cite{Winick2018} takes numerical imprecision into account. 
The authors of~\cite{Lin2019} demonstrated their approach for both a two-state protocol with homodyne detection and a four-state QPSK protocol with heterodyne detection. Other studies extended the method to higher constellations (8PSK, 12PSK, two-ring constellation, etc.)~\cite{Kanitschar2022, KanitscharT2021, Wang2023}, broadened the framework to include trusted detectors~\cite{Lin2020}, optimized key mapping for QPSK encoding to achieve higher noise tolerance~\cite{Liu2021}, and generalized the framework to the multiuser scenario~\cite{Kanitschar2024}. While the nonlinear objective function given by~\cite{Lin2019} necessitates an iterative approach and is numerically more intense and demanding than the method proposed by~\cite{Ghorai2019}, the key rates obtained by~\cite{Lin2019} are significantly higher, particularly for scenarios with medium to high loss. This highlights the advantages of DM CV-QKD security proofs without Gaussian extremality argument. An illustration can be found in Fig. 9 of ~\cite{Lin2019}. In addition, it is noteworthy that the method by~\cite{Lin2019} allows postselection, which can further increase key rates.

As in the case of the proof by~\cite{Ghorai2019}, the optimization has to be carried out numerically on a computer [for example, via the numerical packages based on those of~\cite{SDPT3a, SDPT3b, cvx1, cvx2}]. Consequently, the Fock basis used to describe the generally infinite-dimensional density matrix and operators must be truncated at a certain value $N_c$ to allow representation on a computer. While the computed rates numerically saturate when the value of the cutoff is increased, which suggests that the error introduced by the truncation becomes negligible, this potentially introduces a vulnerability in both security arguments. This concern was addressed by~\cite{Upadhyaya2021}, who eliminated this photon-number cutoff assumption. The authors develop a continuity-bound argument that allows the quantification of the error introduced when the evaluation of the objective function on the original infinite-dimensional state is substituted by a finite-dimensional cutoff version. Moreover, they described a method to calculate the weight of the original state outside of the cutoff space. This weight in turn leads to a correction term for the key rate that ensures security while eliminating the photon-number cutoff assumption. We note that the correction term was improved by~\cite{UpadhyayaT2021}.

\subsubsection{Finite-size security}

Thus far, our discussions have focused on security proofs in the asymptotic regime, i.e., operation under the simplifying assumption that Alice and Bob exchange an infinite number of signals. However, in practical scenarios, Alice and Bob are constrained by real-world limitations and must cease quantum signal exchanges after a certain duration to start generating keys. Consequently, it ultimately requires security proofs within the finite-size regime. In addition, the obtained key should be usable safely in any larger cryptographic protocol, which means that the key needs to be composable secure (see Sec. \ref{sec:Composable_Securty} for details). Compared to the asymptotic case, this poses several challenges for security analyses in the finite-size regime. First, the Devetak-Winter formula offers a secure key rate expression based on von Neumann entropy, an asymptotic measure. However, deriving an equivalent bound for the finite-size regime is more intricate. Second, while in the asymptotic scenario protocol observables are deterministically known, statistical fluctuations and the exclusion of testing rounds for key generation purposes must now be considered. In essence, while in the asymptotic regime the channel parameters (like loss and noise) were basically predetermined, they remain uncertain even after protocol execution in the finite-size regime. Third, we can no longer expect to prove our protocol secure with certainty but instead must accept a small error probability $\epsilon$. Furthermore, whereas in the asymptotic regime collective attacks were deemed to be effectively optimal~\cite{Christandl2009}, this is no longer necessarily the case in the finite-size regime. In summary, the need for more sophisticated security arguments is evident, and finite-size key rates are expected to be lower than those achievable in the asymptotic regime. However, the goal is to derive expressions that reconcile the asymptotic key rates in the limit of an infinite number of rounds, $N\rightarrow \infty$. We note that various researchers employ slightly different total $\epsilon$ parameters, that quantify the error probability of their protocols, when presenting secure key rates, thus making direct comparisons challenging. Nonetheless, given the prevalence of values around $\epsilon = 10^{-10}$ in the literature, qualitative comparisons between different studies remain feasible. Having addressed the challenges, we now present the status of finite-size security proofs for CV-QKD protocols.\\

\paragraph{Finite-size security of Gaussian CV-QKD protocols via Gaussian extremality.}
An extension of Gaussian CV-QKD security analysis to the finite-size regime was performed by~\cite{Leverrier2010} following the formalism developed by~\cite{Renner2006} and applied to DV protocols by~\cite{Scarani2008}, by taking into account the finite-size effects in the parameter estimation. This included a reduction of the key rate due to a use of the fraction $m$ of the overall dataset $N=m+n$ for parameter estimation, redefinition of the Holevo bound as $\chi_{\epsilon_PE}(E:y)$, which is the maximum channel capacity between the reference side of the postprocessing $y$ and the eavesdropper's system $E$ that is compatible with the measurement statistics, except with a given probability of failure $\epsilon_\text{PE}$, and the introduction of the correction term $\Delta(n)$ related to the security of the privacy amplification and dominated by the speed of convergence of the smooth min-entropy of an i.i.d. state toward the von Neumann entropy. The resulting modified Devetak-Winter bound then reads
\begin{equation}
    k=\frac{n}{N}\big[\beta I(x:y)-\chi_{\epsilon_\text{PE}}(E:y)-\Delta(n)\big],
\end{equation}
where the postprocessing (error-correction) efficiency $\beta$, which prevents the trusted parties that possess the data $x$ and $y$ from reaching the Shannon mutual information $I(x:y)$, as we discuss in Sec. \ref{Sec:Post_Processing}, is taken into account. Evaluation of the modified Holevo bound $\chi_{\epsilon_\text{PE}}$ is then based on the derivation of the equivalent two-mode entangled state, which is similar to the asymptotic case but expressed through the minimum value of transmittance and the maximum value of the channel noise, that is compatible with the sampled data except for having a predefined probability $\epsilon_\text{PE}/2$. The correction $\Delta(n)$ can be for the binary-encoded data well approximated by $7\sqrt{\log{(2/\bar{\epsilon})/n}}$, where $\bar{\epsilon}$ is a predefined smoothing parameter and is largely governed by the data size $n$. 

The described approach was the first step toward a finite-size security analysis of CV QKD and can be seen as the patch to the asymptotic security proofs, which is valid in the assumption that Gaussian attacks remain optimal in the finite-size regime. However, the approach demonstrated the importance of finite-size effects in the practical security of the protocols. This approach to finite-size security of Gaussian CV QKD was extended to arbitrary signal states and optimized by~\cite{Ruppert2014} and then applied to CV-MDI-QKD protocols by~\cite{Papanastasiou2017}.

\paragraph{Composable security of Gaussian CV-QKD protocols.}
Composable finite-size security against collective attacks was rigorously proven for the no-switching protocol (coherent states and heterodyne detection) by~\cite{Leverrier2015}, who took into account discretization of the continuous-variable parameters and estimation of the symmetrized covariance matrix of an equivalent entangled state. The leftover hashing lemma (LHL)~\cite{Tomamichel2012a} applied to the smooth min-entropy~\cite{Renner2008} is used to bound Eve's information, that is compatible with the estimated states and probabilities of failure of the protocol procedures. The key rate of the $\epsilon$-secure protocol with $\epsilon = 2\epsilon_{\textrm{sm}} + \bar{\epsilon} + \sqrt{\epsilon_{\textrm{PE}}+ \epsilon_{\textrm{cor}} + \epsilon_{\textrm{ent}}}$, composed of the smoothing parameters $\epsilon_{\textrm{sm}}$ and $\bar{\epsilon}$, parameter estimation failure probability $\epsilon_{\textrm{PE}}$, error-correction failure probability $\epsilon_{\textrm{cor}}$, and smooth-min-entropy bounding failure probability $\epsilon_{\textrm{ent}}$, is then upper bounded by
\begin{equation}
\begin{split}
    2n \left[ 2\hat{H}_\text{MLE}(U) - f(\Sigma_a^{\max}, \Sigma_b^{\max}, \Sigma_c^{\min}) \right]- \\
    -\mathrm{leak}_\text{EC} - \Delta_\text{AEP} - \Delta_\text{ent} - 2\log\frac{1}{2\bar{\epsilon}},
\end{split}
\end{equation}
where $\hat{H}_\text{MLE}(U)$ is the empirical entropy (the maximum likelihood estimator of the entropy) of the discretized data obtained by Bob; 
\begin{equation}
\begin{split}
 \Delta_\text{AEP} := \sqrt{2n} \Big[(d+1)^2 + 4(d+1) \log \frac{2}{\epsilon_\text{sm}^2} + \\
 + 2\log \frac{2}{\epsilon^2\epsilon_\text{sm}}\Big]
 - 4 \frac{\epsilon_\text{sm} d}{\epsilon} 
 \end{split}
 \nonumber
\end{equation} 
is the correction to the bound on the smooth min-entropy due to asymptotic equipartition property (AEP)~\cite{Tomamichel2009}; $\Delta_\text{ent} :=   \log{(1/\epsilon)}-  \sqrt{8n \log^2 (4 n) \log{(2/\epsilon)}}$; $f$ is the Holevo information between Eve and Bob's  data, computed from the Gaussian state with the covariance matrix parametrized by $\Sigma_a^{\max}, \Sigma_b^{\max}$, and $\Sigma_c^{\min}$ that are the worst-case values compatible with $\epsilon_\text{PE}$; and $\mathrm{leak}_\text{EC}$ is the size of the syndrome of Bob's data, sent to Alice during the error correction and defined by the preset error-correcting code. Note that the proof was constructed for the more practical RR scenario but can be directly applied to DR by swapping the roles of the trusted parties. Using the postselection technique~\cite{Christandl2009} in the additional symmetrization step of the protocol, the finite-size security against collective attacks for the no-switching protocol can be extended to general attacks~\cite{Leverrier2015}. 
The achievable key rate asymptotically reaches the bound obtained in the assumption of Gaussian attacks. However, the key rates reached using the postselection technique remained limited and the bound for the no-switching protocol in the case of the general attacks in the finite-size regime was improved by applying Gaussian de Finetti reduction~\cite{Leverrier2017} enabled by an energy test, which allows truncation of the infinite-dimensional Hilbert (Fock) space. Hence, it is sufficient to show finite-size security of the no-switching Gaussian CV-QKD protocol against collective Gaussian attacks, using the previously described technique, in order to attain security against general attacks.

Alternatively, composable finite-size security against general attacks was shown for the squeezed-state protocol with homodyne detection~\cite{Cerf2001}, using the entropic uncertainty relations~\cite{Furrer2012}, and extended to the RR scenario~\cite{Furrer2014}. The achievable key rates are, however, still sensitive to losses and do not comply with the asymptotic rates in the limit of large data ensemble size~\cite{Leverrier2015} suggesting that the obtained bounds are not tight.

More recently,~\cite{Pirandola2024} derived an alternative formula for the composable-secure secret key rate of Gaussian CV-QKD protocols that holds for implementations where parameter estimation is done before error correction. Let $N=n+m$ be the total states in a block, where $m$ are used for PE, while the remaining $n$ are for key generation. According to Eq.~(68) of~\cite{Pirandola2024}, one can achieve the key rate 
\begin{equation}
R \le \frac{p_{\text{EC}}[n \widehat{R}_{\infty}^{\text{PE}}  - n \delta_\text{ent}-\sqrt{n}\delta_\text{AEP} + \theta]}{N}, \label{key_rate_2024}
\end{equation}
where $p_{\text{EC}}$ is the probability of successfully correcting a block, $\widehat{R}_{\infty}^{\text{PE}}$ is the asymptotic key rate computed from the $m$-state parameter estimation (with error $\epsilon_\text{PE}$), $\delta_\text{ent}=\log(n)\sqrt{2 n^{-1} \ln(2/\epsilon_\text{ent})}$ accounts for the imperfect estimation of Bob's entropy (with error $\epsilon_\text{ent}$), $\delta_{\text{AEP}}\simeq4\log\left(2^{d/2}+2\right)  \sqrt{\log(2/\epsilon_{\text{s}}^{2})}$ accounts for the AEP (with digitalization $d$ and smoothing parameter $\epsilon_{\text{s}}$), while $\theta:=\log(2\epsilon^{2}_{\text{h}}\epsilon_{\text{cor}})$ depends on the hashing parameter $\epsilon_{\text{h}}$ and the correctness $\epsilon_{\text{cor}}$. This rate is valid against collective Gaussian attacks with secrecy $\epsilon_{\text{sec}}=\epsilon_{\text{s}}+\epsilon_{\text{h}}$ and total security $\epsilon =\epsilon_{\text{cor}}+\epsilon_{\text{sec}}+\epsilon_{\text{ent}}+2\epsilon_{\text{PE}}$ (note that this is the residual error affecting each block). Finally, using the previously described energy testing procedure, the key rate formula in Eq.~\eqref{key_rate_2024} can be extended to security against coherent attacks for the specific case of the no-switching protocol.

Composable finite-size security of Gaussian CV-MDI QKD against general attacks was initially discussed by~\cite{Lupo2018} who first considered the finite-size security against collective Gaussian attacks and then applied Gaussian de Finetti reduction. The security proof was later refined and improved by~\cite{Papanastasiou2023} who also provided the explicit methods.~\cite{Ghalaii2023} extended the security of Gaussian CV-MDI QKD to free-space channels that are generally affected by diffraction, pointing errors, and turbulence effects. More generally,~\cite{Ghalaii2022} extended the composable security to Gaussian quantum networks with untrusted relays. Finally, by generalizing the security proof techniques applied to the no-switching protocol, the composable finite-size security against general attacks was also shown by~\cite{Ghorai2019} for two-way Gaussian CV QKD with heterodyne detection.

\paragraph{Finite-size security of DM CV-QKD protocols via Gaussian extremality.}
Methods similar to asymptotic proofs for proving the security of DM CV-QKD protocols by relying on Gaussian extremality arguments emerged. ~\cite{Papanastasiou2021} undertook a thorough analysis of the composable finite-size security of a CV-QKD protocol using phase-encoded coherent states and heterodyne detection against Gaussian collective attacks. Their investigation assumes the presence of trusted thermal noise and does not consider postselection. By leveraging the LHL against quantum side information~\cite{Tomamichel2012a}, they derived a bound on the secure key rate.  Exploiting the tensor-product structure of the quantum state, which follows from the collective-attack assumption, and, using the AEP, they expressed the key rate formula in terms of the von Neumann entropy. This formulation allowed to establish a connection between finite-size key rates and asymptotic key rates that account for finite-size correction terms. Parameter estimation is conducted through maximum-likelihood estimators. The resulting key rate formula is then evaluated numerically, with the introduction of a cutoff to make the problem computationally feasible. However, this cutoff remains an assumption. The paper~\cite{Papanastasiou2021} focused on the case involving two and three signal states, as for this limited number of states, the computational demand already seems to be considerable specifically, achieving a higher number of significant digits necessary for low transmission (i.e., high distances). While the approach is general, it remains uncertain whether the method can be effectively extended to accommodate a larger number of signal states.

The study by~\cite{Lupo2022} analyzed the composable finite-size security of a DM CV-QKD protocol with coherent states against collective attacks. A key aspect of their approach involves consideration of the limitations inherent in real detectors, such as finite detection range and precision. The authors exploited this to establish a bound on the probability that the pulse Bob receives contains an extensive number of photons; this bound is used to define the numerical cutoff. Subsequently, Winter's gentle measurement lemma~\cite{Winter1999} allows the bounding of the trace norm between Alice and Bob's actual (infinite-dimensional) quantum state and a cutoff version thereof. Shirokov's continuity bound for the Holevo quantity~\cite{Shirokov2017} is utilized to quantify the effect of this cutoff on the Holevo information by introducing a penalty term. Finally, a Gaussian extremality argument and additional adjustments due to the application of the AEP~\cite{Tomamichel2009} lead to a semidefinite program for the secure key rate that can be solved using methods presented by~\cite{Ghorai2019}, all without relying on a cutoff. Following the idea proposed by~\cite{Leverrier2015}, the composable finite-size parameter estimation procedure takes place after error correction and is based on a Chernoff-bound argument. The authors reported nonzero key rates for block-sizes of $N=10^{10}$ and larger and for large $N$, their key rates converge to their own asymptotic key rates (including detector limitations) as well as to those reported by~\cite{Ghorai2019}. 

\paragraph{Finite-size security without Gaussian extremality.}
For reasons similar to those for the asymptotic regime, for DM CV-QKD protocols with low modulation patterns, methods that do not rely on Gaussian extremality arguments can yield significantly higher finite-size key rates. 

The work~\cite{Matsuura2021} followed a phase-error rate-based idea regarding composable security of a binary phase-modulated CV-QKD protocol with heterodyne detection in the finite-size regime. The central idea revolves around a new method used to estimate the fidelity between the received optical state and a coherent state and a conversion of the unbounded expectation values obtained by heterodyne measurements into bounded values. This allows for reducing the problem similar to earlier Shor-Preskill-style security proofs for DV-QKD protocols. Notably, heterodyne detection is utilized for testing purposes, while homodyne detection is employed for key generation. This requirement of both homodyne and heterodyne measurement was removed by~\cite{Yamano2024}. While this work marked the first general security proof for a DM CV-QKD protocol, their analysis does not include the option to trust parts of the noise or to perform postselection and the obtained key rates for realistic settings are limited. In addition, it is not clear whether the proof method can be generalized to higher modulations, posing a challenge for its application in more complex scenarios.

Thus, we shift our focus to finite-size security proofs for general modulation patterns. Both discussed methods  build upon the work~\cite{Lin2019}, which we discussed in Sec. \ref{subsubsec:asympNG}. Consequently, in principle, these methods are independent of the modulation pattern. However, owing to computational constraints, they are practically restricted to a low number of signal states. As a result, both studies showcase their secure key rates using the QPSK protocol.

The work~\cite{Kanitschar2023} established composable finite-size security for DM CV-QKD protocols using coherent states against collective attacks. The authors introduced a novel energy testing theorem, enabling the weight of Alice and Bob's state to be bounded outside a finite-dimensional Hilbert space and prove security within Renner's $\epsilon$-security framework~\cite{Renner2006}. They utilized an argument based on the finite-detection range of real detectors to derive bounds for Bob's observables and Hoeffding's inequality to define an acceptance set. For states passing both the energy test and the acceptance test (the composable finite-size version of parameter estimation), the LHL against infinite-dimensional side information~\cite{Tomamichel2012a} establishes a relationship between the achievable secure key length and the smooth min-entropy of the considered state. The application of the AEP allows the smooth min-entropy to be rewritten in terms of the von Neumann entropy. To rigorously account for the numerical cutoff, a generalized version of the dimension reduction argument of~\cite{Upadhyaya2021} that introduces an additional weight-dependent correction term in the key rate formula is applied. Subsequently, the secure key rate is numerically evaluated using an extended version of the framework proposed by~\cite{Coles2016, Winick2018, Lin2019}, in which optimization over the density matrices in the acceptance set is performed. The method accommodates postselection and considers nonideal and trusted detectors. They reported secure key rates for block sizes of $N=10^9$ and beyond under experimentally viable conditions. In addition, in the limit of $N\rightarrow \infty$, their secure key rates converge to the rates reported by~\cite{Lin2019, Upadhyaya2021}. Furthermore, they discussed and addressed realistic and practically relevant nonunique acceptance scenarios, coining the notion of expected key rates under honest channel behavior, and attributing the nonzero abort probability of real protocols.

Finally, we turn to applications of entropy accumulation arguments. A study by~\cite{Baeuml2023} established composable finite-size security for DM CV-QKD protocols using coherent states against general attacks, albeit they still relied on a photon-number cutoff. The authors applied the entropy accumulation theorem (EAT)~\cite{Dupuis2020, Dupuis2019}, which allows the lower bounding of the conditional smooth min-entropy of an unstructured (and thus general) state in terms of a so-called min-trade-off function. However, encountering challenges in satisfying a Markov condition necessary for the application of the EAT in the P\&M version of the analyzed DM CV-QKD protocol, the authors introduced an additional round of testing and invented a state-tomography procedure that allows correlations between Alice and Bob to be certified. Subsequently, the min-trade-off function is computed using the numerical method by~\cite{Coles2016, Winick2018, Lin2019}, which assume a numerical cutoff leading to numerical key rates. The authors reported composable-secure finite-size keys for block sizes of $N=10^{12}$ and larger under practically relevant conditions. The asymptotic behavior of this method closely resembles the secure key rates reported by~\cite{Lin2019}. A more recent version following this approach~\cite{PascualGarcia2025} employed the generalized EAT~\cite{Metger2022} as well as alternative advanced conic optimization routines~\cite{Lorente2024}. This improves the stability and precision of the numerical algorithm and, in certain scenarios, allows nonzero key rates for block sizes of $5\times 10^{8}$ (as opposed to $5\times 10^{12}$ in the earlier version). However, the work still assumes a cutoff.
This assumption was lifted by~\cite{Primaatmaja2024}, marking the first security proof against sequential coherent attacks without a photon-number cutoff assumption. The authors applied the generalized EAT to a DM CV-QKD protocol that estimates probability distributions rather than moments. They propose a novel reduction technique, taking care of numerical cutoffs rigorously, which lowers the secure key rate via comparably large finite-size correction terms and does not lead to convergence to the known asymptotic key rates. The authors noted that the main loss in key rate is due to working with probabilities rather than observables, which was necessitated by applying entropy accumulation, along with losses in key rate due to using low-order Gauss-Radau quadrature when applying the numerical method of~\cite{Araujo2023} to the evaluation of key rates.

Finally, note that arguments based on the generalized EAT usually rely on a sequential assumption~\cite{Metger2023}. This means that Eve can access only one of the quantum systems sent from Alice to Bob at the same time, which can easily be enforced by either requiring confirmation of arrival or a certain temporal schedule for sending the signals. Both, however, limit the number of signals sent per time interval severely, particularly for CV systems that usually operate at high repetition rates. While in principle this assumption can be lifted to the case where Eve has access to no more than $s$ subsequent signals at the same time, this leads to increased second-order corrections.

\subsubsection{Limits of CV QKD}
\label{subsec:limits}
Ideal models of CV QKD provide the highest achievable theoretical key rates, which are close to the secret key capacity of the bosonic lossy channel (the standard transmission model through an optical fiber). The exploration of the limits of CV QKD started with the definitions of direct and reverse secret key capacities of a quantum channel $K_{D}$ and $K_{R}$~\cite{Pirandola2009}. These capacities represent the maximum rates achievable via CV (and DV) QKD in direct and reverse reconciliation. They can be lower bounded by the coherent information $I_{\text{c}}$~\cite{Lloyd1996,Schumacher1996} and the reverse coherent information (RCI) $I_{\text{rc}}$~\cite{Patron2009}. In particular, for a bosonic lossy channel with transmittance $T$, we find that $K_{D} \ge I_{\text{c}}= \log{[T/(1-T)]}$ and that the RCI lower bound $K_{R} \ge I_{\text{rc}}= \log{[1/(1-T)]}$~\cite{Pirandola2009}. 

These rates can be achieved by specific CV-QKD protocols working in direct and reverse reconciliation, even if they are not implementable in practice since they require infinite two-mode squeezing and a quantum memory (so that Bob can always perform homodyne detection in the correct quadrature). That said, practical CV-QKD protocols based on coherent states and the heterodyne detection can reach half of $I_{\text{rc}}$ in the ideal asymptotic scenario~\cite{Usenko2016}.

An optimal CV-QKD protocol does not need to be restricted to one-way classical communication (as in direct and reverse reconciliation) but may exploit more general two-way classical communication and be adaptive, meaning that the classical information exchanged in a round may be used to improve the quantum communication in the next rounds. The optimization over two-way classically-assisted QKD protocols (both CV and DV ones) defines the secret-key capacity of a quantum channel $K \ge \max(K_{D}, K_{R})$. 

Following the initial studies on the one-way classically-assisted capacities $K_{D}$ and $K_{R}$~\cite{Pirandola2009}, upper bounds were later developed for two-way classically-assisted capacity $K$, with a preliminary approach based on squashed entanglement~\cite{Takeoka2017}. More recently,~\cite{Pirandola2017} adopted a completely different approach where the key rate of an arbitrary two-way classically-assisted QKD protocol is upper bounded by the channel's relative entropy of entanglement, suitably simplified via a technique of teleportation simulation. 

Using this approach,~\cite{Pirandola2017} showed that $K \le  \log{[1/(1-T)]}$, strictly bounding the rate of any QKD protocol implemented over a bosonic lossy channel with transmittance $T$, a result known as the Pirandola-Laurenza-Ottaviani-Banchi (PLOB) bound. Because the PLOB bound coincides with the RCI lower bound, it exactly establishes the secret-key capacity $K  = \log{[1/(1-T)]}$ of the bosonic lossy channel.

The PLOB bound represents the optimal QKD rate over a lossy channel without intermediate repeaters (repeaterless bound). Other versions exist for repeater-assisted communications and quantum networks~\cite{Pirandola2019}, as well as for free-space quantum communications~\cite{Pirandola2021b}.
 
While the limit of CV QKD (and all QKD) is clear on a lossy channel, the situation for a more general thermal-loss channel, whose structure is discussed in Sec.~\ref{sec:Gaussian_Channels} is different. This is a bosonic channel that is characterized by transmittance $T$ and thermal noise $\bar{n}_N$, the latter being the mean number of environmental photons such that $V_N=2\bar{n}_N+1$. This noise can equivalently be represented as excess noise via the formula $\nu = 2T^{-1}(1-T) \bar{n}_N$. In the presence of excess noise, we do not know the formula of the secret-key capacity $K=K(T,\bar{n}_N)$, but we can only write a sandwich $K_{\text{LB}} \le K \le K_{\text{UB}}$ using the bounds from~\cite{Pirandola2009,Pirandola2017}. The lower bound $K_{\text{LB}}$ has been slightly improved using optimized Gaussian-modulated CV-QKD protocols with engineered loss and thermal noise at the detector setups~\cite{Ottaviani2016} and through the engineering of an entanglement distribution protocol able to provide a positive rate where the reverse coherent information is zero~\cite{Mele2025}. The discovery of $K(T,\bar{n}_N)$ is a central question in CV QKD and quantum information theory.

It is evident from the expression of $K_{D}$ and $K_{R}$ that, while RR CV QKD can theoretically provide security for any pure channel attenuation (which is, however, limited by practical finite-size effects and device imperfections), DR protocols are fundamentally limited by $T=0.5$, and hence by 3~dB of channel loss. For the entanglement-in-the-middle protocol and the CV-MDI protocol with detection in the middle of the channel, the optimal key rates correspond to $K_{D}$ and $K_{R}$ for DR and RR respectively, up to the subtracted term $1/\ln{b}$, where $b$ is the information unit, and hence the base of the logarithms used to calculate the entropic quantities~\cite{Lasota2023}. The perfectly implemented entanglement-in-the-middle scheme is therefore limited by the channel transmittance of $T \approx 0.63$, and the symmetrical CV-MDI scheme is limited by $T \approx 0.73$. 

As mentioned, CV-QKD protocols are also limited by channel noise. Even in conditions of a perfectly transmitting channel ($T=1$), the maximum tolerable channel excess noise $\nu$ is limited by about 0.8 SNUs (in both DR and RR). Furthermore, it continuously decreases by decreasing the transmittance $T$~\cite{Usenko2016}. Two-way protocols make the dependence slower and improve the robustness to channel noise~\cite{Pirandola2008a}. For a generic P\&M CV-QKD protocol, the maximum tolerable channel noise $V_N$, as shown in Fig. \ref{fig:TL}(b), can be approximated through the Lambert W function as $1+2e^{1+W_{-1}(-T/e)}$~\cite{Lasota2017}.

\subsection{Practical implementation of CV QKD}

\label{sec:practimpl}

After several theoretical proposals of CV QKD~\cite{Ralph1999,Hillery2000,Cerf2001,Grosshans2002}, the first experimental proof-of-principle test was reported by~\cite{Grosshans2003} at 780~nm wavelength using Gaussian quadrature modulation (performed as an amplitude modulation with an electro-optical Mach-Zehnder interferometer and phase scanning via a piezoelectric transducer), homodyne detection and RR postprocessing. The tabletop experiment resulted in 1.7 megabits per second (Mbps) secret key rate (secure against individual attacks and considered in the asymptotic limit) in the loss-free channel and 75 Kbps upon 3.1~dB attenuation.

The first practical demonstration of coherent-state CV QKD was reported by~\cite{Lodewyck2007}, where 2 Kbps secret key rate, secure against collective attacks in the asymptotic regime, was obtained in a 25-km-long fiber channel (with the loss of 5.2~dB) with a system operating at the telecom wavelength of 1550~nm and while employing time multiplexing between copropagating signal and the LO. The modulation was performed and fine-tuned using automated electro-optical amplitude and phase modulators and an attenuator. The purification-based security analysis was adopted and incorporated detector imperfections. The work revealed the importance of error-correction efficiency, which limits the applicable signal modulation and the overall performance of the protocol. The efficiency $\beta$ obtained using low-density parity-check (LDPC) codes, discussed in Sec. \ref{Sec:Post_Processing_EC}, was approximately $0.9$.

A field test a coherent-state CV-QKD prototype at the telecom wavelength was reported by~\cite{Fossier2009} using time and polarization multiplexing for LO propagation and advanced error-correction codes with $90\%$ efficiency. The achieved secret key rate, which was secure against collective attacks in the asymptotic limit, was 8 Kbps (average after 3 days of operation), over an optical fiber with 3~dB loss (corresponding to 15~km distance). 

The secure distance of coherent-state CV QKD was extended to 80~km of optical fiber at the telecom wavelength by~\cite{Jouguet2013} using error-correcting codes with an efficiency of $95\,\%$. The achieved key rates ranged from 10 Kbps in a 25~km channel, a few Kbps over 53~km and around 200 bps in a 80~km channel, with security analyzed against collective attacks in the finite-size regime.

By controlling the noise of the setup (using high-sensitivity homodyne detectors tolerating copropagated LO attenuation, and high-precision phase compensation), the performance of coherent-state CV QKD was improved to an approximately 0.5 Kbps secret key rate (secure against collective attacks in the finite-size regime of $10^{10}$ block size enabled by a highly stable setup) over a 100-km-long fiber~\cite{Huang2016}, with an error-correction efficiency of $95.6\%$. Over the shorter distance of 25~km, the secret key rate of 1 Mbps was achieved using high-speed (1 GHz) homodyne detection and a 50 MHz clock rate~\cite{Huang2015}.

The record-breaking distance of over 200~km in ultralow-loss fiber, which reached a secret key rate of about 6 bps, was reported for coherent-state CV QKD by~\cite{Zhang2020} using a highly stable setup and error-correcting codes with $98\%$ efficiency. 

Composable security against collective attacks for the practical long-distance coherent-state CV QKD was demonstrated by~\cite{Jain2022} over 20~km fiber using phase compensation enforced with machine learning techniques, which enabled the secret key rate of 4.7 Mbps.

High-speed Gaussian-modulated coherent-state local local oscillator (LLO) CV QKD, as described in Sec. \ref{subsec:implsec}, was demonstrated by~\cite{Huang2015a}, and reached about 100 Kbps over a 25~km link with $97\%$ error-correction efficiency, while a long-distance implementation of the LLO protocol reported by~\cite{Hajomer2024} reached 25.4 Kbps over a 100~km telecom fiber, both implementations with the finite-size collective-attack security. 

Coherent-state DM CV QKD was implemented by~\cite{Tian2023}, who used 16-state constellations reaching 49.02 and 2.11 Mbps of asymptotic collective-attack secret key rates over 25 and 80~km optical fibers, respectively, corresponding to $67.4\%$ and $66.5\%$ of the Gaussian CV-QKD key rates in the same channels. A high-speed implementation operating at 10 GHz using 16, 32, and 64 coherent-state constellations was reported by~\cite{Hajomer2023} over 5 and 10~km reaching about 900 and 500 Mbps. The experiment by~\cite{Roumestan2024} reported secret key rates of 92 and 24 Mbps over 10 and 25~km, respectively, using a probabilistically shaped 64 QAM modulation based on the security proof of~\cite{Denys2021} and a worst-case finite-size security analysis. Composable security for collective attacks was achieved using a QPSK constellation over a 20~km fiber channel~\cite{Hajomer2025a} and a 3~dB free-space channel that was simulated in the lab~\cite{Jaksch2024}.

CV-MDI QKD was tested by~\cite{Pirandola2015} with 0.1 Kb of asymptotic secret key per relay use obtained at 4~dB loss. Another test of the MDI protocol, reported by~\cite{Tian2022}, reached 0.43 and 0.19 bits per symbol over 5 and 10~km fiber links, respectively. An experimental implementation operating at 20 MHz achieving a secure key rate of 2.6 Mbps over a 10~km fiber link operating in the finite-size regime against collective attacks was demonstrated by~\cite{Hajomer2025}. 

Entanglement-based protocols remain on the level of proof-of-principle tests, with a full-scale implementation yet to be performed. Entanglement-based CV QKD was first tested by~\cite{Su2009}, who reported 83 and 3 Kbps of key, secure against asymptotic collective attacks, upon $80\%$ and $40\%$ transmittance respectively. Modulation was reported to improve EB CV QKD over noisy channels and tested by~\cite{Madsen2012}. It reached $4 \cdot 10^{-3}$ bits per symbol in a highly noisy channel (where the coherent-state protocol would fail). An entanglement-based protocol with composable and one-sided device-independent security tested by~\cite{Gehring2015} reached 0.1 bits per symbol at 0.76~dB of channel attenuation (corresponding to 2.7~km telecom fiber).

Many more experimental tests and implementations were reported in addition to the aforementioned milestones. It is often difficult to compare different realizations because of the nonmatching security assumptions (types of attacks, finite-size versus asymptotic analysis, etc.). Practical CV QKD is still in the process of development toward higher key rates for the full-scale implementations in the most general security scenarios and over long-distance channels. In the following subsections we discuss the main state-of-the-art techniques and methods used in the practical realizations of CV QKD.

\subsubsection{Quadrature modulation}

The quantum states on which a CV-QKD protocol realization is based have certain physical properties. Among those are the wavelength or light frequency, the temporal shape, and the polarization. The physical properties define the mode in which the coherent or squeezed states reside and that has to be measured by the detector for optimal performance. Several different physical realizations have been investigated.

The basis of all implementations is light produced by a laser at a certain carrier angular frequency $\Omega$ and described by its electric field $E\exp(i\Omega t)$, where $E$ is the electric field amplitude and $t$ is the time. To implement the random quadrature displacement in P\&M coherent-state or squeezed-state protocols quadrature modulation is used. For this purpose the light is passed through an electro-optic modulator modulating the phase or amplitude of the light by changing the refractive index through an applied electric field.

An often-used implementation is based on a pulsed laser that, at a typical repetition rate in the MHz range, emits hundreds of nanoseconds-long pulses of light~\cite{Jouguet2013,Zhang2020} or laser pulses that are carved into a continuous-wave (cw) laser beam using a high-extinction intensity modulator that currently achieves higher repetition rates of up to 250 MHz~\cite{Laudenbach2019}. After optical attenuation to on average a couple of photons per pulse, the pulses are in a coherent state whose amplitude is randomly changed with an amplitude modulator and are moved to all four quadrants of the phase space using a phase modulator that has to be driven up to its half-wave voltage. 

\begin{figure}
\includegraphics{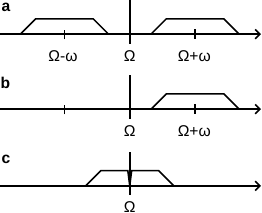}
\caption{Different possibilities for physical realizations of coherent-state modulation based on the displacement of vacuum states. (a) Double-sideband modulation. (b) Single-sideband modulation.
(c) Baseband modulation. The angular frequency of the laser is $\Omega$, while the center modulation sideband angular frequency is $\omega$.}
\label{fig:quadraturemodulation}
\end{figure}

In other implementations the vacuum state at a sideband frequency of the carrier is displaced instead of using the carrier itself~\cite{Pirandola2015,Kleis2017,Jain2022}. Those implementations use cw lasers, and the temporal shape of the coherent states is defined by the electrical waveform driving the modulators. Since a vacuum state and not an excited coherent state is displaced, smaller driving voltages can be used and the modulator is operated in its linear regime. Generally, three different sideband modulation techniques have been explored. They are sketched in Fig.~\ref{fig:quadraturemodulation} and outlined next.

Amplitude and phase modulation generates two frequency sidebands that are symmetric around the laser's carrier frequency. The rate at which the complex amplitude of the coherent states is changed is directly reflected in how much bandwidth is occupied in the Fourier domain with the roll-off given by the low-pass filter defining the temporal shape. The first modulation technique, shown in Fig.~\ref{fig:quadraturemodulation}(a), uses up-conversion of the modulation to an intermediate carrier frequency $\omega$, such that the electric field is $E(t) \propto \alpha(t)\exp(i(\Omega + \omega)t) + \alpha^*(t)\exp(i(\Omega-\omega)t)$, with $\alpha(t)$ describing the coherent-state amplitude. The receiver has to measure both sidebands, which is possible with homodyne and intradyne detection, as later discussed. To avoid having to measure both sidebands and to enable radio-frequency heterodyne detection, one of the sidebands can be suppressed, leading to single-sideband modulation as shown in Fig.~\ref{fig:quadraturemodulation}(b). This can be achieved via appropriate modulation using an in-phase and quadrature (IQ) modulator, which removes one of the sidebands through optical interference~\cite{Kleis2017,Jain2021}. The resulting electric field is given by $E(t) \approx \frac{\mu}{2}\alpha(t)\exp(i(\Omega + \omega)t) + \frac{\delta}{2}\alpha^*(t)\exp(i(\Omega-\omega)t)$ with $\mu$ the modulation index and $\delta < \mu$ the effective modulation index taking the sideband suppression ratio into account. While the suppressed sidebands usually have around 30~dB lower power than the desired sidebands, their existence leads to a potential side channel that has to be appropriately taken into account in the security analysis~\cite{Jain2021}. By instead using baseband modulation, which is shown in Fig.~\ref{fig:quadraturemodulation}(c), we can avoid this side channel~\cite{Hajomer2022}. Here the up-conversion of the modulation is removed, i.e.,\ $\omega=0$. However, to avoid the strong carrier and the laser noise close to the carrier, a high-pass filter has to be introduced. This high-pass filter has to be optimized to reduce correlations between quantum states to the largest possible extent~\cite{Hajomer2022}. An advantage of this modulation scheme is that it is the most transmitter bandwidth efficient, which is important for CV QKD at high speed~\cite{Hajomer2023}.

Besides the coherent states for the CV-QKD protocol, auxiliary signals have been generated that are time or frequency multiplexed. Those signals usually aid digital signal processing (DSP) procedures in the receiver, for instance, for phase recovery and clock synchronization~\cite{Chin2022}. Typical signals are pilot tones and signals of higher power than the quantum states using, for instance, QPSK modulation formats.

\subsubsection{Coherent detection}

\begin{figure}
    \centering
    \includegraphics{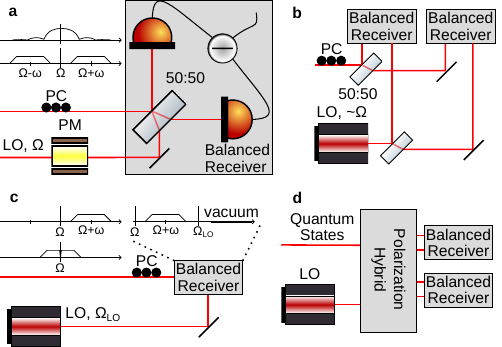}
    \caption{CV-QKD receiver schemes. (a) Homodyne receiver. (b) Heterodyne (phase-diverse) receiver. (c) rf heterodyne receiver. (d) Polarization-diverse rf heterodyne receiver. LO, local oscillator; PM, phase modulator; and PC, polarization controller.}
    \label{fig:receiver}
\end{figure}

At the receiver the quantum states are typically measured with quadrature detectors that can be divided into two types: receivers measuring one quadrature at a time, i.e., homodyne detection, and receivers measuring the amplitude and phase quadratures simultaneously, which are often called heterodyne-detection receivers in the quantum optics community.

In a homodyne receiver, the LO beam interferes with the quantum states at a balanced beam splitter whose outputs are detected by \textit{p-i-n} photodiodes, as shown in Fig.~\ref{fig:receiver}(a). Having the same frequency as the carrier of the coherent states, the phase of the LO determines the measured quadrature. To follow the protocol, the measured quadrature has to be chosen randomly for each received quantum state that can be implemented most practically when pulsed lasers or pulse-carved laser beams are used. Nevertheless, the switching speed is limited, and therefore it has been implemented only with state repetition rates up to 5 MHz~\cite{Zhang2020}. To achieve a stable phase relation, the LO is usually derived from the transmitter and sent to the receiver through a fiber. Recently, remote homodyne detection of a squeezed state was achieved with a separate laser for the LO using an optical phase-locked loop~\cite{Suleiman2022}.

Heterodyne detection avoids active basis choice. A common implementation is to split the quantum states at a balanced beam splitter and to detect its outputs with two homodyne detectors measuring orthogonal quadratures. This is shown in Fig.~\ref{fig:receiver}(b). The orthogonality of the measured quadratures can automatically be ensured using $90^\circ$ optical hybrids that utilize polarization optics to turn the phase of the LO in one arm by $90^\circ$ with respect to the other. The introduction of optical hybrids into CV-QKD systems has enabled the development of LLO schemes~\cite{Qi2015,Huang2015a,Soh2015} that aim to remove the attacks on the LO, as described in Sec. \ref{subsec:implsec}. For them the LO is derived from a laser located at the receiver that is independent of the laser used at the transmitter. The frequency of the LO laser can be tuned close to that of the transmitter laser, which yields a phase drift proportional to the frequency difference that is corrupted by phase noise of the lasers. The phase drift and noise have to be measured and can be compensated for in postprocessing by applying simple rotations in the phase space spanned by the measured orthogonal quadratures. The detection scheme, which is called intradyne detection, works well for frequency differences of lasers lower than half the bandwidth of the quantum states. The scheme can be implemented for modulation schemes based both on displacing coherent states and on baseband modulation, i.e., the displacement of vacuum states.

A second implementation approach to heterodyne detection known as radio-frequency (rf) heterodyne detection is depicted in Fig.~\ref{fig:receiver}(c). Here the LO is detuned from the frequency band occupied by the quantum states. Since a detuned LO also interferes with the image band to the frequency band occupied by the quantum states, a simple balanced receiver setup with one detector is sufficient to measure the two quadratures simultaneously. As a single balanced detector is enough, advantages in comparison to an optical hybrid are that the splitting ratio of the quantum states is exactly 50:50 and that the quantum efficiency is the same for both quadratures. rf heterodyne detection is best used with single-sideband or baseband modulation schemes.

Another dimension is polarization. For balanced reception to work efficiently, the polarization of the LO and the quantum states have to match. Most often this is achieved via polarization controllers optimizing the polarization of the quantum states after transmission through the channel, as shown in Fig.~\ref{fig:receiver}. Recently, polarization hybrids splitting the polarization into two components as depicted in Fig.~\ref{fig:receiver}(d)~\cite{Pareira2023,Chin2024}, have also been introduced. The components are then measured independently and combined using DSP, as discussed further.

\subsubsection{Digital signal processing}
With the introduction of LLO schemes and the use of some form of heterodyne detection, DSP has become increasingly important to experimental implementations, as it allows one to compensate for impairments digitally instead of in the analog domain. Some of the impairments DSP deals with are phase noise, polarization mismatch, and clock desynchronization. DSP is also sometimes used to achieve mode matching between the received and the transmitted quantum states.

One of the most challenging tasks is carrier frequency and phase recovery, which compensates for phase noise originating from the free-running transmitter and receiver lasers and -- to a lesser extent -- from the channel. Phase noise from the lasers is generally the largest source of excess noise in experimental implementations~\cite{Hajomer2024}, particularly when compared to implementations using transmitted LO~\cite{Zhang2020}. To achieve phase matching, the phase of the quantum signal with respect to the LO has to be measured, estimated, and compensated for. For this purpose, usually either frequency-multiplexed (and sometimes polarization-multiplexed) pilot tones or time-multiplexed reference symbols of higher power are used. In the case of pilot tones, the measured signal can be described by
\begin{equation*}
A\sin\left(i\Omega_\text{LO-pilot-beat}(t) t + \phi(t)\right) + X_\text{meas}(t) ,
\end{equation*}
where $\Omega_\text{LO-pilot-beat}(t)$ is the time varying beat frequency of the pilot tone with the LO, $\phi(t)$ is the phase noise, $t$ is the time, and $(A/\sqrt{2})^2 / \text{var} (X_\text{meas})$ is the SNR given by the pilot tone power $A^2/2$ and the power of the quadrature noise, $\text{var} (X_\text{meas})$, induced by the measurement (often shot noise and electronic noise). After the measurement, the phase $\phi(t)$ has to be estimated from the measurements. This can be achieved by taking either the arctangent of the two measured quadrature components (phase-diverse reception) or the measured signal and its Hilbert transform (rf heterodyne reception). However, while computationally inexpensive, this estimation method is not statistically optimal. Statistically optimal  methods are based on Bayesian inference~\cite{Saerkkae2013}. Kalman filters produce estimates of unknown time varying variables in the presence of noise. Their extended and unscented versions can deal with the nonlinearity of the measurement equation with respect to phase and have shown exceptional performance down to much lower SNRs than the arctangent-method~\cite{Chin2021,Kleis2019,Su2019}. The required prior model in Bayesian inference is usually taken from a Wiener model of phase noise whose linewidth parameters are determined from the data. Alternative approaches have used neural networks for phase estimation~\cite{Long2023,Xing2022}.

Since eventually the SNR of the pilot tone or reference symbols is the figure of merit for achieving high-quality phase estimation, some experiments have used polarization multiplexing to accommodate higher power pilot tones and low-power quantum states~\cite{Laudenbach2019}. A recent study showed that 10 kHz linewidth lasers can be used with SNRs above 28~dB; however, higher linewidth lasers require higher SNR, which are experimentally challenging to achieve without impacting the linearity of the receivers.

In transmitted LO implementations, clock synchronization can be achieved simply by measuring a fraction of the LO pulse~\cite{Jouguet2014}. Modern DSP-based systems often use oversampling by an asynchronous analog-to-digital converter and a timing error detector based on one or more auxiliary signals of known frequency difference to determine the correct sampling points~\cite{Kleis2017,Chin2022,Wang2022}.

DSP has also been used to compensate for imperfect modulator~\cite{Hajomer2023} or receiver characteristics~\cite{Jain2022}. Machine learning techniques such as neural-network-based autoencoders have also been used to compensate for nonlinear distortions~\cite{Zhang2022,Long2023}. It has also been the focus of an open-source modular software platform benchmarked on multiple CV-QKD configurations and developed to stimulate further explorations~\cite{Pietri2024}. Indeed, although the exploration of DSP for CV QKD is still at its early stages, it represents a promising path toward simpler photonic circuits with lower-quality components, which are crucial for the envisioned cost reduction of QKD systems.

\subsubsection{Squeezed signal states}
\label{subsec:sqz_sig}

A few proof-of-concept implementations of QKD based on squeezed states have been demonstrated to date. These schemes, which are detailed in Sec. \ref{sec:protocols}, introduce promising advancements for CV QKD, including a modulation-free approach~\cite{Su2009}, extended operational range~\cite{Madsen2012}, the enabling of composable security~\cite{Gehring2015}, one-sided device independence~\cite{Walk2016} and the elimination of information leakage to eavesdroppers~\cite{Jacobsen2018}, which reduces postprocessing requirements. We now delve into the experimental tools used in these implementations and highlight their key findings.

\textit{Parametric down-conversion} (PDC) has been the dominant method for generating squeezed light~\cite{Andersen2016} since its first demonstration by~\cite{Wu1986}, where noise suppression of 3.5~dB beyond the shot-noise limit was achieved. Since that time, squeezed light with noise suppression of up to 15~dB has been realized using the PDC process~\cite{Vahlbruch2016}. In PDC a pump wave with frequency \( \omega_{\mathrm{p}} \) interacts with a 
 \(\chi^2\)-nonlinear medium to produce two quantum-correlated (entangled) waves known as the signal and the idler~\cite{Boyd2020}. When these waves are degenerate across all degrees of freedom -- spatial, polarization, frequency, and temporal modes -- their entanglement translates into quadrature squeezing of the resulting single mode. Efficient squeezing generation relies on a pair of key conditions: high power density in the crystal, which can be achieved using pulsed lasers, by placing the crystal in a cavity, or by employing an optical waveguide, and precise phase matching of the parametric process, which can be achieved through methods such as temperature tuning of the crystal, adjusting the pump wavelength, or periodically poling the crystal. Currently, the most effective strategy for generating quadrature squeezed light is to use a cw pump beam and integrate a periodically poled crystal [for example, potassium titanyl phosphate (KTP), or LiNbO$_3$] inside a low-loss cavity that is resonant at the signal-idler wavelength (and potentially also at the pump wavelength). This method has been employed in all existing demonstrations of squeezed-light-based CV QKD.

While many other methods exist for generating squeezed light~\cite{Andersen2016}, it is crucial to select a process that produces squeezed light suitable for QKD. For fiber-based QKD, the wavelength of the squeezed light should fall within the telecom band to minimize fiber transmission losses. For free-space QKD, suitable wavelengths are 800-850,, 1064 and 1550~nm, where atmospheric attenuation is relatively low depending on the weather and where laser sources are commonly available. For example, a KTP-crystal-based PDC process can generate squeezed light at all of those wavelengths, whereas atomic-based squeezed-light sources are often limited to specific atomic transition wavelengths, such as cesium at 852~nm and rubidium at 795~nm. Another advantage of the PDC process is its bandwidth, ideally constrained only by the frequency band where phase matching occurs, which can extend into the terahertz range. However, as noted, the bandwidth of the PDC process is often limited by the cavity used to enhance it, narrowing it to the cavity's linewidth -- which is often still considerable (several GHz). Another important factor is the temporal mode of the squeezed state. To improve the efficiency of homodyne detection and to mitigate issues with dispersion, it is beneficial to produce the squeezed state in a cw or quasi-cw mode. cw operation also provides considerable flexibility in digitally defining the temporal mode via DSP. This preference for cw operation makes the cavity-enhanced PDC process the most obvious choice, although recent reports on cw squeezed light produced in a waveguide are also encouraging~\cite{Kashiwazaki2023}.

In all experiments on CV-EB QKD conducted to date, squeezed light was generated using cavity-enhanced PDC with periodically poled KTP crystals housed in high-quality linear-optical cavities. In a few experiments~\cite{Su2009,Wang2018}, the CV entanglement -- also known as a TMSV state -- was directly produced by operating the PDC process in a nondegenerate mode, whereas in the other experiments~\cite{Madsen2012,Gehring2015,Walk2016}, entanglement was achieved by combining two single-mode squeezed states at a balanced beam splitter. In each setup one mode of the two-mode squeezed state was measured with high-efficiency homodyne detection at the transmitting station, while the other mode was transmitted through free space to the receiving station, where homodyne or heterodyne measurements were performed. The transmitter's measurements influenced the state observed at Bob's station, resulting in a Gaussian distribution of amplitude or phase quadrature squeezed states, effectively simulating a single-mode P\&M squeezed-state protocol.
The noise suppression achieved below the shot-noise level was measured at 3.0~dB at 1080~nm~\cite{Su2009}, 3.5~dB at 1064~nm~\cite{Madsen2012}, and 10.5~dB at 1550~nm~\cite{Gehring2015}. Although the theoretical clock rate in these experiments could approach the bandwidths of the cavity-enhanced parametric amplifiers (20, 21, and 63 MHz, respectively), the effective operational bandwidth was restricted to a few kHz via electronic filtering in postprocessing. 

QKD schemes based on two-mode squeezed states do not require electro-optical modulators to define the alphabet. As described above, the result of a quadrature measurement on one mode of a two-mode squeezed state directly produces a Gaussian alphabet of squeezed states, thereby eliminating the need for physical electro-optical modulators. In this approach the modulation arises naturally from the intrinsic quantum randomness of measuring the two-mode squeezed state, meaning that the randomness used for key generation is inherent to the measurement process itself rather than relying on an external quantum random number generator. The preparation basis choice is then performed by measuring on one of the modes of an EPR state~\cite{Silberhorn2002}. A proof of concept for such a modulation-free scheme was first demonstrated by Su et al.~\cite{Su2009}, where an early-stage security analysis showed secret key rates of 84 and 3 kbps for transmission efficiencies of 80\% and 40\%, respectively, under the assumption of Gaussian collective attacks.

In systems using squeezed states, the alphabet size is typically constrained by the degree of antisqueezing, limiting the potential to maximize the key rate based on modulation depth. However,~\cite{Madsen2012} demonstrated that expanding the Gaussian alphabet through conventional amplitude and phase modulation at Alice's station enhances the system's resilience to excess noise, thus allowing for secure communications over greater distances. This protocol can equivalently be realized using the EB scheme, which leaves Alice with two sets of data: the homodyne measurement results $X_{HD}$, and the Gaussian modulation values $X_M$. To optimize the protocol's resilience to excess noise, these two sets of data were combined with an optimal weighting $X_M+gX_{HD}$, where the factor $g$ depends on the strength and purity of the squeezed states. For large, pure squeezing, the two datasets should be equally weighted ($g=1$), whereas for lower squeezing a greater weight should be assigned to the modulation data, with $g\rightarrow0$ as the squeezing vanishes and the protocol changes to the coherent-state one. Using such a weighting strategy, the squeezed-state scheme outperforms the coherent-state scheme for all levels of squeezing.~\cite{Madsen2012} produced two-mode squeezed states with 3.5~dB of noise suppression and 8.2~dB of antisqueezing, with coherent modulation levels varying from 0 to 15~dB. Specifically, with a modulation of 23.4~SNUs, channel transmission of 95\%, and a channel excess noise of 0.45~SNU, a key rate of 0.004 bit per state was achieved in the asymptotic limit under collective attacks (and perfect postprocessing $\beta=1$). Notably, neither a coherent-state protocol nor a nonmodulated squeezed-state protocol could generate a key under these conditions. In addition, they identified the tolerable levels of excess noise for secure key generation in a 10\% transmissive channel across various modulation strengths, and found that modulated squeezed-light realizations consistently outperformed the coherent-state protocol under all tested conditions. Although these experimental tests assumed ideal postprocessing efficiency,~\cite{Madsen2012} also demonstrated that the squeezed-state protocol is inherently more robust than the coherent-state protocol in the presence of postprocessing inefficiencies.    

In these initial demonstrations of EB CV QKD, the security proofs typically addressed only collective attacks and did not guarantee composability against coherent attacks. However, 
\cite{Gehring2015} reported the first implementation of a CV-QKD protocol that achieves composable security against coherent attacks while being 1sDI. Using highly squeezed EPR-entangled states, they achieved a composable-secure key generation rate of 0.1 bit per sample for a channel loss of 0.76~dB (equivalent to 2.7~km of fiber) and a reconciliation efficiency of 94.3\%. The experiment demonstrated robustness against intensity fluctuation attacks on the LO and Trojan-horse attacks, with the EPR source located at Alice's station ensuring security even when Bob's devices were untrusted. Building on this work,~\cite{Walk2016} extended the scope of CV-1sDI-QKD by implementing a range of Gaussian protocols, including both P\&M and EB schemes using TMSV states. In their experiment with TMSV states (operated at 1064~nm), they demonstrated secure key generation over distances of up to 7.5~km of effective optical fiber transmission using reverse reconciliation, highlighting the practicality of 1sDI QKD with current technology.

While most squeezing-based QKD implementations leverage two-mode squeezed states, as previously discussed, there have also been successful demonstrations utilizing single-mode squeezed states. These approaches reduce experimental complexity by simplifying the optical setup while maintaining compatibility with QKD protocols. One such implementation by~\cite{Eberle2013} employed a single-mode squeezed state of 11.1~dB quantum noise suppression at 1550~nm to generate asymmetric entanglement by having the squeezed state interfere with the vacuum on a balanced beam splitter. This setup established quantum correlations between one pair of quadratures (for example, the amplitude quadrature for an amplitude-squeezed state), while the conjugate quadratures (for example, the phase) remained uncorrelated. Despite this asymmetry the source was proven to be suitable for QKD with composable security, supporting secure key distribution over distances of up to 10~km. 

Another notable experiment~\cite{Jacobsen2018} used single-mode squeezed states to encode information into displacements along a single quadrature, thereby creating a unidimensional Gaussian alphabet~\cite{Usenko2015}. For instance, an amplitude-squeezed state was solely displaced in the amplitude quadrature direction using an amplitude modulator. By aligning the displacement with the squeezed quadrature, this approach achieved a significant security advantage. Specifically, it was demonstrated that for a purely lossy transmission channel, an eavesdropper's information can be completely eliminated when the variance of the squeezed alphabet matches the variance of the vacuum state. In the experiment a 3dB squeezed state at 1064nm was displaced with an amplitude modulator with varying modulation depths. It was observed that for a specific modulation depth where the alphabet size coincided with the width of the vacuum noise, Eve's Holevo information dropped to exactly zero. This effect was tested across four different transmission coefficients of a simulated lossy channel (using a variable beam splitter). In all cases complete elimination of the Holevo information was achieved. This suppression of eavesdropper information arises because the amplitude quadrature correlations between Eve and Bob vanish entirely when the variance of the alphabet equals the variance of the vacuum state (corresponding to the noise floor assumed in Eve's beam splitter attack). By eliminating Eve's ability to gain information, this protocol not only ensures security but also significantly reduces the computational complexity of postprocessing, making it an attractive option for future CV-QKD systems, particularly in scenarios where low computational overhead is critical.

All the aforementioned proof-of-concept experiments with squeezed light employed a LO derived from the same laser used to generate the squeezed light. This approach simplifies the experimental setup due to the inherent frequency matching and phase stability between the squeezed light and the LO. However, in practical QKD scenarios, transmitting the LO alongside the quantum signal poses significant challenges. Transmitting high-power LOs over long distances not only creates issues with nonlinear induced mode distortion and light scattering but also opens the door to eavesdropping strategies that exploit the transmitted LO to compromise the coherent detection at the trusted receiver. To address these vulnerabilities, it is preferable to generate the LO locally at the receiving nodes. Generating the LO locally introduces its own challenges, particularly the need for real-time control of the phase difference between the signal and the LO. Without frequency locking between the two independent lasers, maintaining stable phase coherence can be difficult. ~\cite{Suleiman2022} reported on the generation, transmission, and homodyne detection of single-mode squeezed states of light at 1550~nm using a locally generated LO with real-time phase control. By implementing a real-time feedback system, they achieved precise control over the frequency and phase of the LO. This enabled homodyne detection of a squeezed state with a phase uncertainty of only 2.51$^o$, even after the squeezed state had propagated through up to 40~km of telecom fiber. This demonstration represents a crucial step toward practical implementations of squeezed-light-based CV QKD. By overcoming the challenges of locally generated LOs and maintaining stable phase control, this work lays the foundation for future QKD schemes that combine the security advantages of squeezed light with realistic deployment in telecom networks.

\subsubsection{Measurement-device-independent protocols}

CV-MDI QKD eliminates all side-channel attacks on detectors by performing a CV Bell measurement at an untrusted relay. While this feature provides strong security, the protocol's practical implementation is more complex than one-way CV QKD, leading to relatively few experimental demonstrations, despite significant theoretical progress. We now review recent advancements in CV-MDI-QKD implementation.

The first proof-of-concept experiment, conducted in 2015~\cite{Pirandola2015}, used a free-space optical setup with a shared 1064~nm cw laser modulated to prepare coherent states. Phase locking between Alice's and Bob's signals was achieved using piezoelectric transducers, while the CV Bell measurement was performed with a balanced beam splitter and homodyne detectors. Although this experiment demonstrated the feasibility of CV-MDI QKD, it lacked critical practical features, such as telecom-compatible wavelengths, independent lasers, and real fiber channels, limiting its applicability for in-field deployment.

To address these limitations, a more practical CV-MDI-QKD implementation was reported in 2022~\cite{Tian2022}. This system employed a 10~km standard single-mode fiber channel and independent lasers operating at the telecom wavelength of 1550~nm. Frequency locking was achieved via phase-locked loops, and phase drift was compensated for using a combination of pilot and phase-reference pulses. While the system achieved a low excess noise of 0.0045 SNU and a secret key rate of 0.19 bits per relay use, it remained complex due to the need for heterodyne receivers for phase locking, additional modulators for pulse carving, and polarization and time multiplexing.

A significant step toward simplifying CV-MDI QKD was achieved with a system design based on DSP~\cite{hajomer2022high}. This design eliminated the need for phase locking by sharing a single laser source and replaced traditional pulse carving with digital pulse shaping using root-raised cosine filters. These innovations enabled a tenfold increase in the symbol rate, reaching 5 MBd. Furthermore, the CV Bell measurement was implemented with a simplified relay structure based on a polarization-based $90^\circ$ hybrid. These advancements resulted in a key rate of 0.12 bit per relay use over a 2~dB channel loss. However, further work is required to integrate clock synchronization and adapt the system for deployment over fiber channels.

Recent advancements have focused on practical integration and achieving higher key rates. A notable system design~\cite{Hajomer2025} addressed challenges such as clock synchronization and coexistence with classical communication channels. By colocating the relay at Alice's station and using a digital signal processing module for quantum state preparation, the system achieved CV Bell measurement at 20 MBd. Synchronization was achieved with a 1310~nm optical clock copropagating with the quantum signal, complemented by real-time phase locking. This setup demonstrated a secure key rate of 2.6 Mbps against collective attacks in the finite-size regime over a 10~km fiber link. Despite these advancements, further improvements are needed, particularly in the postprocessing phase.

\subsubsection{Fiber and free-space channels}
Most of the tests and realizations of CV QKD were performed over fiber channels, which are the main type of links for optical communications due to high and stable transmittance at the telecom wavelengths (typically 1550~nm); low noise, especially in single-mode fibers if free from other optical signals; and availability due to the existing telecom infrastructure. A fiber channel is characterized by transmittance $T$ and is modeled by a beam splitter, that couples the quantum signal to a vacuum mode.   The telecom fibers, usually made of silica, nominally have a transmittance of -0.2~dB/km (which can also be referred to as 0.2~dB of channel loss), but deployed fiber commonly has higher loss due to splices and connectors. Ultralow-loss telecom fibers with transmittance of -0.15 to -0.16~dB/km are also available and were used in laboratory experiments to demonstrate long-distance CV QKD~\cite{Zhang2022,Hajomer2024}. Fiber attenuation is caused mainly by scattering and absorption and dominantly contributes to the signal loss. In any practical implementation, the channel transmittance has to be precisely estimated and controlled and should be clearly distinguished from the losses in the trusted devices. 

Typical noise sources introduced by fiber channels are phase and polarization fluctuations that have to be suppressed by the QKD system~\cite{Laudenbach2018,Liu2020}. The excess noise created by phase fluctuations not only depends on the amount of residual phase noise after phase compensation, but also on the modulation variance~\cite{Marie2017}. Therefore the QKD system not only has to excel in phase compensation, but also has to optimize the modulation variance to achieve best performance~\cite{Hajomer2024}. The amount of polarization fluctuations stemming from the fiber channel depends on whether a fiber is on a spool in a laboratory, is deployed underground, or is aerial~\cite{Jain2024}. Noise also arises when the quantum states coexist in a fiber with classical telecom signals through wavelength division multiplexing~\cite{Eriksson2019}, especially through Raman scattering. However, amplified spontaneous emission from the lasers used for high-rate data transmission also creates noise, which can be suppressed with optical filters before the light is multiplexed with the quantum states. The precise estimation of channel noise is crucial for system performance and imperfect estimation in the finite-size regime implies pessimistic assumptions on the noise levels. As with the channel transmittance, the noise has not only to be carefully characterized but also distinguished from the noise in the trusted devices.

The situation is different in free-space atmospheric channels, where turbulence effects and background light drastically affect optical quantum communication. CV QKD, however, can be particularly robust against background light, because the narrow-band LO acts as a filter such that the homodyne detector does not register light, that does not match the LO wavelength. This makes CV QKD particularly promising for daylight operations, where DV protocols would need additional filtering, which introduces extra loss to the overall optical budget of the link. Free-space implementations define the suitable wavelengths of 800-850, 1064, and 1550~nm, on which the optical signals are better transmitted through the atmosphere. Atmospheric turbulence, however, complicates things for CV QKD, leading to effective transmittance fluctuations, also referred to as channel fading. Fading is caused mostly by the beam deformation and wander effect, which dominate the channels with weak turbulence, where transmittance fluctuations are well described by the log-negative Weibull distribution, and additional broadening and deformation of the beam in the strong turbulence with smooth transmittance probability distributions~\cite{Vasylyev2016}. Fading channels are non-Gaussian~\cite{Dong2010} and result in the fading-related channel excess noise $\nu_f=\text{var}(\sqrt{T})(V_S+V_M-1)$, which is governed by the variance of the coupling to vacuum (square root of transmittance) $\text{var}(\sqrt{T})$ and signal and modulation variances in the given quadrature $V_S$ and $V_M$, respectively~\cite{Usenko2012}. Such noise has to be assumed to be untrusted, as the transmittance fluctuations can be controlled by Eve, limiting the applicability of CV QKD over fading channels, and requiring additional optimization of the signal and modulation. It was shown that signal squeezing can be helpful in fading channels, but has to be optimized owing to the fading-related noise in the antisqueezed quadrature, where the variance of the signal is significantly above the shot noise~\cite{Derkach2020}. 

The negative impact of transmittance fluctuations in CV QKD can be partially compensated for using adaptive optics, for example, by beam expansion~\cite{Usenko2018a} or wave-front correction~\cite{Chai2020,MarulandaAcosta2024}. Alternatively, the transmittance distribution can optimally be divided into clusters where the variance of the fluctuations is lower, and QKD security can be analyzed for those clusters separately, which increases the overall secret key rate~\cite{Ruppert2019}. This technique is promising for realizations of CV QKD over satellite channels using coherent~\cite{Dequal2021} and squeezed states~\cite{Derkach2020a}. A comprehensive analysis of free-space CV QKD in the composable finite-size security framework was performed by~\cite{Pirandola2021b}, who established the bounds on free-space CV QKD and showed that the coherent-state protocol can perform close to those bounds. Composable finite-size security of the no-switching protocol in free-space channels was studied by~\cite{Hosseinidehaj2021} who confirmed the advantage of channel clusterization. 

CV QKD was tested in free-space atmospheric channels using single-quadrature encoding with a 0.152~Kbps secret key rate that was secure against collective attacks in the asymptotic limit. The results were reported over a 460 m channel~\cite{Shen2019}.

\subsection{Implementation security}

\label{subsec:implsec}

Practical implementation of QKD is never ideal because of device imperfections, which have to be taken into account in the security analysis. Usually,  practical imperfections are distinguished between normal device operation in accordance with their specification, such as finite detection efficiency and electronic noise of the detectors, which can be observed by the trusted parties and directly accounted for, and undocumented features or their combinations, which can lead to security loopholes, when the trusted parties are not even aware of the possible information leakage. The well-known example of this is detector blinding in DV QKD~\cite{Lydersen2010}. Therefore, the practical security of QKD relies on the best existing knowledge of the setup and its components and requires a thorough investigation of the component specification as well as their study, which aims to find possible quantum exploits. We start the review of the practical security of CV QKD with a description of the major known device imperfections that directly affect the performance of the protocols. We base our discussion on the Gaussian protocols in the collective-attack scenario while assuming asymptotic data sizes, which can be directly extended to a finite-size Gaussian collective-attack analysis for all Gaussian protocols following the techniques from~\cite{Leverrier2010} and to composable security against Gaussian collective and general attacks using the proof~\cite{Leverrier2015} and the Gaussian de Finetti reduction~\cite{Leverrier2017} for the no-switching protocol. This most general CV-QKD security in the case of the major practical imperfections was recently analyzed by~\cite{Pirandola2021a}. The comprehensive analysis of practical imperfections in the DM CV-QKD security has yet to be performed; therefore, we limit our review to Gaussian protocols. 

Before we go into details of the practical security of CV QKD, it is important to introduce the \textit{trust assumption}. The trusted parties decide which parts of the setup they assume to be out of control of an eavesdropper, refer to these parts as trusted and adapt their security analysis accordingly. The typical trust assumption is that the devices located in Alice's and Bob's laboratories are properly characterized and isolated so that no other party can control the noise that is introduced by those devices or receive the information leakage from the trusted devices. Those trusted parts of the protocol are shown in dashed boxes in Fig. \ref{fig:P_and_M} and typically include the source of the signal states, the modulator, and the detector. The trust assumption on the detectors can be waived in the MDI protocol shown in Fig. \ref{fig:MDI}. Assuming that a device is untrusted typically means that the related noise is attributed to the channel, which strongly limits the performance of the protocols. Therefore, in this review, we keep to the standard assumption of trusted sender and receiver stations, unless otherwise specified. 

\subsubsection{Source and detection imperfections}
The generated and modulated states can contain excess noise, originating either from the signal source (the laser or source of squeezed states in the case of coherent- or squeezed-state protocols, respectively) or from the imperfect modulation. If the trusted parties can characterize this noise separately from the channel noise and localize it to the preparation part of their setup (by monitoring the signal in front of the channel using a tap-off or optical switch for back-to-back measurements in the sending station), they can consider this noise to be trusted. Such excess noise, referred to as \textit{preparation noise}, was studied by~\cite{Filip2008} for purely lossy (noiseless) channels and extended to noisy channels by~\cite{Usenko2010a}. In the context of purification-based theoretical security analysis, the trusted preparation noise is then purified using an entangled source in modes $P_1$ and $P_2$ that is similar to the untrusted noise purification by the entangling cloner $E_1$ and $E_2$, which is shown in Fig. \ref{fig:TL}(c). One of the entangled modes is coupled to the signal mode prior to the channel on a beam splitter with transmittance $T_P$, and the variance of the entangled state is set to $V_P/(1-T_P)$, where $V_P$ is the variance of the preparation excess noise. For $T_P \to 1$ the noise addition is then performed approximately losslessly (without attenuating the signal). The modes $P_1$ and $P_2$ are not available for the measurement either to the trusted parties, or to the eavesdropper and are used to keep the trusted preparation noise out of Eve's purification. The Holevo bound is then obtained as $S(A,B,P_1,P_2)-S(A,P_1,P_2|B)$ from the entropy of the four-mode state in modes $A,B,P_1,$ and $P_2$ and the entropy of the conditional state after measurement on Bob's side in the RR scenario (or from the entropy $S(B,P_1,P_2|A)$ of the conditional state after measurement on Alice's side in the case of DR). While the trusted preparation noise does not directly contribute to Eve's information (as it is kept out of Eve's purification of the channel noise), it reduces the mutual information between the trusted parties, and therefore the resulting key rate as well. It was shown that the bound on the trusted preparation noise in RR in the purely lossy channels with transmittance $T$ in the asymptotic limit and in the limit of infinite modulation $V \to \infty$ reads $1/(1-T)$ for the coherent-state protocol~\cite{Filip2008} and generalizes to $(2-T)/(1-T)-V_S$ for an arbitrary signal state variance $V_S$~\cite{Usenko2016}. Controllable attenuation was suggested to compensate for the trusted preparation noise in RR~\cite{Filip2008}. Conversely, the trusted preparation noise is harmless in the DR scenario~\cite{Weedbrook2010}, and it even improves the security of the protocol in the noisy channels~\cite{Weedbrook2012}, which potentially enables CV QKD in the wavelengths beyond the optical domain up to the microwave regime.

On the receiver side, the homodyne detectors are inclined to losses (owing to the limited detection efficiency) and electronic noise, which, in the purification-based theoretical security analysis, is modeled similarly to the preparation noise by introducing a two-mode entangled state with variance $V_D/(1-T_D)$, where $V_D$ is the electronic noise and $T_D$ is the detection efficiency~\cite{Lodewyck2007}. As with the preparation noise, this implies a Holevo bound evaluation from a four-mode state (or from a six-mode state if both preparation and detection noise are to be taken into account). The trusted \textit{detection noise} then only limits the mutual information between Alice and Bob, but, unlike the preparation noise, it limits the applicability of the protocols in DR. Robustness of CV QKD with DR and noisy detection is bounded by $(2T-1)/(1-T)$ SNUs of detection noise regardless of the signal variance~\cite{Usenko2016} already in the purely lossy channels in the asymptotic limit. However, in the case of RR, the detection noise can be even helpful for the protocols in noisy channels~\cite{Garcia2009}. This detection-noise RR counterpart of the positive effect of the preparation noise in DR protocols is known as \textit{fighting noise with noise}, when the trusted noise on the reference side of the error-correction protocols (Alice in DR and Bob in RR) helps the trusted parties to decouple Eve, who performs active attacks in the noisy channels, from the key data~\cite{Usenko2016}. The trusted noise addition can be optimized to maximize the improvement to the security of CV QKD over noisy channels~\cite{Madsen2012}.

Besides the excess noise and signal attenuation, the imperfections of Alice's and Bob's devices may lead to other unwanted effects, such as imperfect shot-noise calibrations or phase noise, with results in random phase rotations between the sent and received data. Those imperfections were studied by~\cite{Jouguet2012} for coherent-state CV QKD taking into account finite-size effects for Gaussian collective attacks. It was shown that the phase noise degrades the correlations between the trusted parties, which also limits the performance of the protocols. By properly characterizing the phase noise, the trusted parties can, however, put tighter bounds on Eve's information~\cite{Jouguet2012}.

Systematic analysis of trusted noise in realistic coherent-state CV QKD was performed by~\cite{Laudenbach2018} who developed models for various sources of noise, including source and LO intensity fluctuations, imperfect modulation, phase noise, common-mode rejection ratio in a homodyne detector, detection noise, and quantization noise of the analog-to-digital converter. Those models allow one to efficiently predict the performance of practical CV-QKD systems. However, the security of a particular implementation can be based only on the actually measured data and subsequent characterization of trusted and untrusted noise to be incorporated to the security analysis. Analysis of practical imperfections in the composable finite-size security framework was performed for the no-switching protocol by~\cite{Pirandola2021a} who considered different trust levels of the setup components.

\subsubsection{Security side channels and loopholes}
Besides the losses and noise in the sending and receiving stations, which can be assumed to be either trusted or untrusted, the security of CV QKD can be affected by \textit{side channels}, providing an eavesdropper with partial knowledge on the distributed key data or on the noise that is added to the key data. Side-channel leakage from the sending station and side-channel noise addition in the receiving station were theoretically analyzed by~\cite{Derkach2016} and were shown to be potentially harmful for RR and DR protocols, respectively. Besides isolation of the trusted stations to prevent signal leakage and noise addition, the trusted parties may implement controllable noise insertion to the leaking mode or noise monitoring and subtraction of the side-channel noise. Leakage of the generated signal prior to its modulation can be harmful to the squeezed-state protocols, while coherent states are immune to such leakage, providing no additional correlations with the leaking mode~\cite{Derkach2017}. However, leakage of the modulated signal can directly provide an eavesdropper with the modulation data and can remain unknown to the trusted parties. This was shown for the CV-QKD system using an optical in-phase and quadrature modulator with imperfect suppression of the sidebands, which are modulated but not measured by the receiver~\cite{Jain2021}. This emphasizes the importance of good single sideband suppression in the sending station and careful characterization of the transmitted modes. The negative role of imperfect state preparation in parameter estimation, potentially leading to overestimation of the secret key rate, was studied by~\cite{Liu2017} who also suggested the methods for proper calibration of imperfect CV-QKD setups.

In addition to obtaining information on the key from side-channel leakage, Eve can send a light pulse to the sending station to learn the modulator settings in the so-called \textit{Trojan horse attack}, which is well known for DV-QKD protocols~\cite{Gisin2006}. Such attacks can threaten the security of CV QKD~\cite{Pereira2018} by providing Eve with side information, which is equivalent to worsening the channel parameters. Such attacks can be ruled out by an additional sensing system in the sending station for monitoring the incoming and outgoing light~\cite{Pereira2018}. 

Besides attempting to learn the modulator settings using light probes, Eve may temper the state preparation in the quantum hacking attacks on the sending station. In particular, Eve can perform an \textit{attack on the optical attenuator} used for state preparation in the sending station~\cite{Zheng2019}. By sending bright light, Eve can damage the variable optical attenuator and reduce its attenuation level. That way, Eve can also make trusted parties to overestimate the channel transmittance (and hence underestimate Eve's attack), which can result in a security breach. The attack on the optical attenuator can be ruled out by optical isolation or, more efficiently, by continuous monitoring of signal modulation prior to sending the states to the channel~\cite{Zheng2019}. Eve can also perform a \textit{laser seeding attack} by injecting controllable cw laser light into a semiconductor laser source in order to manipulate the intensity of the laser used on the trusted sender side
for signal state generation. That way, Eve can scale up the displacements of the signals sent to the channel and force the trusted parties to underestimate the channel attenuation and overestimate the bounds on the key rate, hence leading to a security breach~\cite{Zheng2019a}. The attack was shown to be particularly harmful in CV-MDI QKD but can also break the security of conventional CV-QKD schemes. An optical isolator can be used to block the injected laser seeding light, but it can also be laser damaged by Eve. Alternatively, a real-time monitoring of LO intensity can be used to reveal and compensate for a laser seeding attack in CV-QKD systems with LO generation on the sending side. 

On the receiver side, additional detection noise may originate from imperfect mode matching in the case of multimode signals~\cite{Usenko2014} and should be assumed to be untrusted since they are potentially under Eve's control owing to propagation through the channel. This noise can be reduced or removed by mode filtering or increasing the brightness of the LO~\cite{Kovalenko2019}. Alternatively, the impact of the electronic noise of the detectors on the security of CV QKD can be reduced by using multimode signals~\cite{Kumar2019}, which improve the signal-to-noise ratio and the resulting secret key rate. Insufficient bandwidth in a balanced homodyne detector can lead to incorrect parameter estimation in a CV-QKD system and can affect its performance, which requires optimization of the signal repetition rate and modulation pattern~\cite{Liu2022a}.

The homodyne detection based on the LO beam transmitted through the untrusted channel can be a target of numerous hacking attacks when Eve can obtain information on the key while hiding her attack by forcing the trusted parties to incorrectly estimate the channel parameters. In particular, Eve can carry out an \textit{LO intensity attack}~\cite{Haeseler2008}, an intercept-resend attack with an optimized detection and the same amplitudes for the resent signal and the LO, leading to lower variances obtained from the measurements on the receiving side and hence lower estimated channel noise. Another possibility for Eve is to introduce \textit{fluctuations to the LO intensity}~\cite{Ma2013a}, which allows Eve to compromise channel noise estimation and force the trusted parties to underestimate the collective attack in the quantum channel, particularly in the RR scenario. The intensity attack on the LO using introduced fluctuations was also recently reported for the DM CV-QKD protocols~\cite{Fan2023}. The intensity attacks on the LO can be prevented by LO monitoring, which, however, can be challenging for detecting the simulated fluctuations~\cite{Ma2013a}. The trusted parties can also vary the LO intensity in the sending station to prevent the LO intensity attacks and even improve the robustness of the protocols to the channel noise~\cite{Ma2014}. In addition to manipulating the LO intensity, Eve can perform the \textit{wavelength attack} on the LO by intercepting both the signal and the LO and then reemitting the signal and the LO at different wavelengths so that the wavelength-dependent beam splitter in Bob's heterodyne detector becomes properly unbalanced~\cite{Ma2013,Huang2013}. This allows Eve to impose required outcomes on Bob's detector and hide her attack under the shot-noise contribution of the heterodyne detector. The wavelength attack can also be implemented against protocols with homodyne detection~\cite{Huang2014} by affecting the shot-noise measurements, leading to wrong estimation of the channel noise. Such attacks can be compensated for by randomly applying wavelength filters or controllable attenuation in the receiver and advanced real-time shot-noise monitoring~\cite{Huang2014}. 

Estimation of shot-noise can be affected by Eve using a \textit{calibration attack} on the LO~\cite{Jouguet2013a}. This is done by attenuating the beginning of an LO pulse, thereby introducing a delay to the trigger signal for the homodyne detector, which decreases the detection response slope. The trusted parties then overestimate the value of the shot noise and subsequently underestimate the channel noise, which hides Eve's attack in the quantum channel and leads to a security breach. This type of quantum hacking attack can be repelled by real-time shot-noise measurements, such as random strong attenuation applied to the incoming light in the receiver or introduction of a beam splitter and a subsequent calibration of the relative sensitivity of the two homodyne detectors~\cite{Jouguet2013a}. Calibration in the CV-QKD measurement can also be exploited by Eve using a \textit{polarization attack}~\cite{Zhao2018} on systems using polarization multiplexing of the signal and the propagated LO. In such schemes the polarization drift of the incoming light is compensated for using polarization monitoring of the LO and real-time feedback control. However, only part of the pulses is measured owing to the limited detection rate. It can even change the polarization of the remaining LO pulses, leading to the leakage of the LO into the signal after demultiplexing (without interference with the signal due to time delay) and to reduction of the LO intensity. The measured shot noise is then reduced, which again leads to underestimation of the channel noise. Such an attack can be removed via polarization monitoring on all the pulses~\cite{Zhao2018} or via real-time shot-noise monitoring~\cite{Kunz2015}. 

Attacks on the transmitted LO and the subsequent parameter estimation can be removed by the local generation of the LO in the LLO schemes. In this case Alice generates reference pulses (pilot tones) that are used by Bob to phase lock his local strong coherent LO with Alice's prepared signal~\cite{Soh2015,Qi2015,Marie2017}. The residual phase drift between the measurement and preparation bases should then be removed by the data postprocessing as discussed in Sec. \ref{Sec:Post_Processing}. Recently, the time-variant parameter estimation and compensation was proposed by~\cite{Xu2024} to improve robustness of LLO-based schemes. 

However, the LLO systems can be vulnerable to an attack on the pilot tones referred to as the \textit{reference pulse attack}~\cite{Ren2019}. Within the trust assumption on the phase noise due to the basis mismatch, Eve can identify the reference pulses and propagate those through an ideal channel, thereby reducing the phase noise. Then, within the total estimated excess noise at the receiver, Eve can increase the fraction related to the channel noise by performing a stronger eavesdropping attack in the quantum channel, which remains hidden and can lead to a security breach. To prevent these attacks, the trusted parties can assume perfect transmission of the reference pulses while keeping the trust assumption on the phase noise. A more efficient countermeasure for the trusted parties is to continuously monitor the intensity of the pilot tones and calibrate the phase noise accordingly~\cite{Ren2019}. LLO schemes can also be vulnerable to the polarization attack as in the conventional protocols with the shared LO~\cite{Shao2022}. To avoid attacks on a shared phase reference, be it LO or pilot tones, ~\cite{Usenko2025} recently proposed the shared-reference-frame-independent CV-QKD protocol based on UD encoding, heterodyne detection, and subsequent basis alignment on the receiver side by a suitable phase-space rotation of the measured data.

As an alternative to the attacks on the phase reference, Eve can perform the \textit{saturation attack} on CV QKD~\cite{Qin2016} by strongly displacing the incoming signals in order to reach the nonlinear response regime of the homodyne detector. The measured quadrature variance then becomes reduced, which leads to an underestimation of the channel excess noise and to a security breach of the protocol. To prevent this type of attack, the trusted parties should ensure the linear regime of the detector, for example, by postselection tests or by introducing a variable attenuator before the detection (the attenuator has to be wavelength insensitive to avoid opening another loophole). Furthermore, the homodyne detection in CV QKD can be compromised by the \textit{blinding attack}~\cite{Qin2018} similarly to the well-known hacking attack on the single-photon detectors in DV QKD~\cite{Lydersen2010}. By sending bright light to the homodyne receiver with an imbalance between two photodetectors, Eve can produce the signal current displacement, resulting in saturation of the electronic amplifier in the receiver. As with the saturation attack, the detection becomes nonlinear, leading to a security breach. Unlike the saturation attack, however, Eve does not need to phase lock the injected light with the signal in order to carry out the blinding attack, which makes this type of quantum hacking more feasible than the saturation attack. This type of attack can also be ruled out by verifying the linearity of the detection~\cite{Qin2018}. Both attacks were experimentally studied in the context of the attack rating, with the aim of quantifying the risks for practical CV-QKD systems~\cite{Kumar2021}.

The MDI protocols, which waive the trust assumption on the measurement in CV QKD, are immune to any of the previously described types of detector attacks but can be vulnerable to attacks on the sources, such as the Trojan horse attack.

\subsubsection{Security certification}

As CV QKD is approaching technological maturity--yet still the subject of active research and development--its practical use is closely related to certification of its security. In this process the QKD products undergo rigorous tests and assessments to confirm that they meet defined security standards and requirements. Despite commercial QKD products having been available for more than 20 years, no independent security evaluation or certification has yet been performed owing to the absence of required standards and of a complete ecosystem. 

Security certification is characterized by evaluation assurance level (EAL), which ranges from 1 (functionally tested) to 7 (formally designed and verified). Achieving a high-assurance security certification with EAL above 4 for QKD systems will represent a significant milestone, as it will establish a high level of trust in QKD products and unlock access to strategic and highly regulated sectors that require certification for secure communication products, such as government and defense, finance and banking, health care, telecommunication, and critical infrastructures.
Increasing amounts of effort are have been invested toward this goal, with significant progress in recent years.

The methodology for achieving high-assurance security certification in QKD systems in the Common Criteria framework \footnote{see https://www.commoncriteriaportal.org/cc/index.cfm} involves a collaborative process among various entities that is based on several steps and documents:
\begin{itemize}
\item \textbf{Protection profile} (PP). A document that defines the target of evaluation and the security requirements that will need to be tested and validated. It is typically developed by experts and may be endorsed by national or international bodies to ensure consistency and relevance across sectors.
The PP outlines criteria for evaluating the security properties of a system (here QKD) covering both the functional requirements and threat mitigation strategies. 

\item \textbf{Security target.} The manufacturer of the QKD system is responsible for designing and producing the product in alignment with the PP requirements. To illustrate this compliance, the manufacturer provides  documentation called the security target on the product's architecture, security features, and implementation details. It is used by the evaluation lab during testing.

\item \textbf{Evaluation report.} An accredited third-party evaluation lab, helped by the security target document, performs a thorough assessment of the QKD product against the PP. The lab conducts tests to verify that the product meets the specified security standards and that it is resistant to the types of attacks outlined in the PP. Following the evaluation the lab produces an evaluation report that details the findings, including any security weaknesses or compliance issues. 

\item \textbf{Certification by national authority.} The evaluation report is submitted to a recognized certification body, typically a national authority or a standards organization. This body reviews the report and ensures that the QKD product meets all required security criteria for the targeted EAL. Upon satisfactory review the certification body issues an official certificate, endorsing that the QKD system has met the PP requirements for the targeted EAL.
\end{itemize}

Note that certification of a QKD system is a risk-based process that differs in nature from the derivation of a security proof. Yet the two approaches are compatible and complementary. Certification provides the assurance of a lower bound on the practical security of a QKD system. This requires the complexity of the possible attacks to be assessed, using, for instance, attack ratings~\cite{Kumar2021}. An important effort is also needed to standardize some key aspects (tackled in different sections of this review), such as  security proofs (\ref{sec:theosec}), practical implementations including parameter estimation procedures (\ref{sec:practimpl}), vulnerability analysis (\ref{subsec:implsec}), and the cryptographic use cases, i.e., how QKD is used and combined with other cryptographic primitives to achieve a particular security service.

From the viewpoint of security evaluation and certification, CV-QKD systems have several specificities: 
\begin{itemize}
    \item Calibration. The parameter estimation phase requires, for most CV-QKD protocols, some quantities that can be considered trusted (see \ref{sec:practimpl}) to be calibrated. Overestimating trusted noise can lead to attacks~\cite{Huang2014} while its underestimating leads to reduced performance. The precision with which these calibrations can be achieved depends on the number of samples used for the calibration procedure, and also on the sampling dynamic and the stability of the processes. This is notably the case for electronic noise~\cite{brunner2020precise} and shot noise~\cite{van2023receiver}, which are crucial procedures for which one can foresee the need for a standardized approach.
\item Digital signal processing. Modern CV-QKD systems can leverage high-bandwidth acquisition electronics to digitally track and correct noise occurring at the optical channel level (such as frequency and phase noise, dispersion, and polarization drift). DSP has allowed high key rates~\cite{roumestan2022} as well as high spectral efficiency~\cite{Eriksson2020} to be reached, but also requires one to revisit calibration procedures and how the logical channel is defined~\cite{chen2023continuous} as a function of analog and digital filters.
\item Integration and system modularity. A specificity of the CV-QKD connection is that it can essentially be operated over a single mode, with hardware subsystems (laser, modulators, coherent receivers, etc.) that are extremely close to those used by classical optical coherent communication systems. The single-mode feature reduces vulnerability to optical side channels and moreover facilitates key aspects of network integration such as joint operation~\cite{aymeric2022symbiotic} and coexistence with classical channels~\cite{Kumar2015} and point-to-multipoint operation~\cite{hajomer2024continuous}.
\end{itemize}

\subsection{Postprocessing}\label{Sec:Post_Processing}

In general, for a typical P\&M QKD protocol, after Alice has prepared a sequence of quantum states and Bob has measured these states, they jointly convert their classical information on those quantum states (for example, preparation bases and bit values, measurement basis, and results) into a final secure shared key. This conversion process is commonly called postprocessing (see~\cite{Luo2024} for a recent comprehensive review of QKD postprocessing) and consists of several separate steps, namely, sifting, parameter estimation, information reconciliation, and  privacy amplification, that must be implemented according to the security proof. To accomplish these tasks, Alice and Bob exchange messages over an authenticated channel on which Eve can learn the content of the messages but cannot modify them.

\subsubsection{Postprocessing performance}

We start with discussing the important figures of merit, characterizing the performance of postprocessing implementations. General performance characteristics for all postprocessing steps are (i) achievable throughput of the secure key generated and (ii) computing resources (computing power, memory, etc.) needed to generate one secure key bit.
 
We observe that in CV-QKD protocols all quantum states that Alice prepares and transmits are detected and measured by Bob. This is in contrast to DV QKD, where the detection (click) rate of the single-photon detector(s) is much smaller than the number of states prepared.
Therefore, if postprocessing should occur in real time (such that no quantum states remain unused) after the sifting step, a CV-QKD protocol still has to process information at a rate that is of the same order of magnitude as the state generation rate $r_\text{gen}$. In comparison, a DV-QKD protocol processes state information after sifting only at a rate that is orders of magnitudes lower, i.e., at $r_\text{gen}\times \mu \times T\times \eta$, where $\mu$ (typically, $\mu<1$) is the mean photon number of the states prepared, $T\ll 1$ is the channel transmission, and $\eta<1$ is the detection efficiency.

For the information reconciliation step, the \textit{error-correction efficiency} $\beta$ is an important parameter that can be decisive for secure key generation by a QKD protocol. For reconciliation of a block of $n$ bits, that were transmitted over a noisy quantum channel, by exchanging $n_{leak}$ bits on the classical error-free channel, $\beta$ is given by 
\begin{equation}
    \beta=\frac{n-n_{leak}}{nC},
\end{equation}
where $C$ is the channel capacity, which refers to the maximum rate (per channel use) at which information can be reliably (with a negligible error rate) transmitted over a noisy channel. If reconciliation is performed by transmitting syndromes of linear block codes with a block size $n$, a syndrome length $s$, and a corresponding code rate $R=1-s/n$, then
\begin{equation}
    \beta=\frac{R}{C},
\end{equation}
since the number of leaked bits $n_{leak}$ is given by the length of the syndrome $s$.

The \textit{frame error rate} (FER, also sometimes referred to as the word error rate), which is the ratio of the number of data frames that cannot be successfully reconciled with the total number of frames, is another important parameter that influences the secure key rate. Recently, the FER has been taken into account in a security proof in the finite-size regime against collective attacks~\cite{Hajomer2025a}, scaling the typically dominating component of the key rate by the ratio of
successfully reconciled frames -- hence, by the value of FER
subtracted from 1\footnote{This follows from combining Eqs.(2) and (3) by~\cite{Hajomer2025a}.}. Note that the channel capacity is defined for a negligible FER. If and only if reconciliation must have a negligible FER, $R<C$; i.e., $\beta<1$ must hold.

Owing to imperfections in codes and decoders and limited computational decoding resources, the efficiency $\beta$ is typically lower than $1$ even if a non-negligible FER is accepted. Values of $\beta$ up to $98\%$ were reported in CV-QKD experiments (see Table \ref{tab:tests}). However, accepting a FER of $99.99\%$~\cite{Johnson2017} demonstrated that an LDPC code can reach an efficiency of $\beta=1.09$.

Often, information reconciliation uses an iterative decoding algorithm, and the number of iterations then used is another important parameter because the time needed to decode is directly proportional to the number of iterations performed. In general, using a higher number of iterations achieves a higher value of $\beta$ for the same value of FER. 

For the privacy amplification step, we observe that finite-size proofs typically demand a data block size of $10^{10}$ or even much larger, which can result in the need for large memories.  We now further discuss the main steps and respective methods for postprocessing in CV QKD.

\subsubsection{Sifting}
As mentioned, two detection schemes are typically used in CV QKD: homodyne detection, where Bob measures a single quadrature (either $\hat{x}$ or $\hat{p}$), and heterodyne detection, where Bob measures the two quadratures simultaneously. In the homodyne-based CV-QKD protocols, Bob therefore has to randomly switch between noncommuting quadratures for each received quantum state. This is typically accomplished by adjusting the LO phase between 0 and $\pi/2$, often with the aid of a quantum random number generator. Random switching is essential for security, as it means that, in a direct attack, Eve would match Bob's basis only half the time. Even if she could retain quantum states in memory and wait until Bob reveals his basis, she would still not be able to influence the measurement outcome.
This switching process, however, requires a ``sifting'' step where Bob communicates with Alice his choice of quadrature for each received quantum state over an authenticated public channel, allowing Alice to retain only the data  corresponding to Bob's selected quadrature, ensuring that they are in agreement. 

\subsubsection{Postselection}
\label{Subsec:postselection}

As discussed in Sec.\ref{Sec:Advances_Post_Selection} in more detail, discarding information of selected states from further processing conditioned on measurement outcome can be utilized in some CV-QKD protocols to increase the secure key rate. Another advantage of postselection relevant in the context of postprocessing is that the reduced number of states leads to a reduction in the computing resources needed. During the postselection Bob has to decide which states will be discarded.

\subsubsection{Parameter estimation}
After the quantum phase, where Alice and Bob prepare and measure quantum states, they move on to parameter estimation. This step enables them to estimate system parameters essential for calculating the secret key rate according to a specific security proof. Traditionally, parameter estimation involves Alice and Bob disclosing a portion of their modulated and measured quadrature values. However, this approach limits the length of the generated secret key. To address this, most recent CV-QKD implementations have reordered the postprocessing steps such that information reconciliation occurs before parameter estimation~\cite{Leverrier2015,Hajomer2024,Jain2022}. This adjustment allows all measurements to contribute to both parameter estimation and secret key generation.  

The parameter estimation process in DM CV QKD differs from that of traditional Gaussian-modulated protocols. To clarify, we examine this step using the traditional Gaussian-modulated CV-QKD protocol as an example.  In such a protocol, calculating the secret key rate requires one to construct the covariance matrix of the P\&M scheme, which can then be translated into an equivalent EB scheme for security analysis. To build this covariance matrix, Alice and Bob must estimate the channel parameters, specifically, the channel transmittance $T$ and the excess noise $\nu$, which they can achieve through the following measurements: 

\begin{enumerate}
    \item Electronic noise. Measured by turning off the LO and blocking the signal input port.
\item Vacuum noise (shot noise). Measured by blocking the signal input port while keeping the LO port open.
\item Modulated quantum states. Measured by keeping both the signal and LO ports open. 
\item Back-to-back measurement. Conducted to monitor and calibrate the modulation variance at the channel input. This is typically done by connecting the transmitter and the local receiver on the sender side through a short channel (an optical switch between the local receiver and the quantum channel) or, alternatively, by tapping a small portion of the signal at the transmitter side and measuring it locally. In this setup it is assumed that the channel transmittance is equal to 1 and that the excess noise is 0 (hence, the channel is perfect). Given the quantum efficiency of the receiver $\eta$, the modulation variance $V_m$ can be estimated as 
\begin{equation}
    V_m = \frac{V_{B2B}-V_\text{elec}-1}{\eta \mu},
\end{equation}
where $V_{B2B}$ is the variance of the measured quadrature in shot-noise units, $V_\text{elec}$ is the electronic noise variance, and $\mu$ is the parameter equal to 1 for homodyne detection  or  2 for heterodyne detection.    
\end{enumerate}

After completing these measurements, Bob normalizes his measurement using the variance of the vacuum noise. Alice and Bob can then use their shared quadrature values to estimate the channel parameters as  
\begin{equation}
    T = \frac{z^2}{V_m \eta}\ , 
    \nu_\text{out} =TV_m- \frac{V_B-V_\text{elec}-1}{\eta \mu}\ ,
\end{equation}
where $z^2$ is the covariance (correlation) between Alice's and Bob's quadrature values.  

This estimation is based on the assumption that Alice and Bob exchange an infinite number of quantum states, allowing the calculation of the secret key rate in the asymptotic regime. However, in practical systems, the number of exchanged quantum states is finite, leading to fluctuations in parameter estimates. This finite-size effect is mitigated by applying a confidence interval based on the number of exchanged states to ensure a worst-case estimation~\cite{Leverrier2010} and is further discussed in Sec. \ref{sec:theosec}.   
 
\subsubsection{Information reconciliation (error correction)}\label{Sec:Post_Processing_EC}

Information reconciliation (also called error correction) is always needed to achieve a \textit{correct} protocol, i.e., one that either produces identical keys for Alice and Bob or aborts with a probability close to 1. This process involves mapping Alice's and Bob's quadrature values into binary data and then applying an error-correction algorithm, which requires information, such as syndromes, to be exchanged over an authenticated public channel. 

In Gaussian-modulated CV QKD, there are two main reconciliation approaches, depending on how the quadrature values are mapped to binary strings: \textit{sliced reconciliation}~\cite{van2004reconciliation} and \textit{multidimensional} (MD) \textit{reconciliation}~\cite{leverrier2008multidimensional}. 
 
Sliced reconciliation uses a multilevel encoding and multistage decoding process that applies different error-correction codes, such as LDPC codes or polar codes, as the SNR varies across levels of coding. This approach allows encoding of more than 1 bit per quantum state and is typically applied at SNR values above 0~dB, making it effective for moderate distances (of up to around 30~km). This distance limitation occurs because slicing is effective only when the received SNR at Bob's end is sufficiently high. With independently chosen code rates for each $m$ slice [$R_i(1\le i \le m) $], the error-correction efficiency $\beta$ can be computed as~\cite{jouguet2014high} 
\begin{equation}
\beta=\frac{H[Q(B)]-m+\Sigma^m_{i=1} R_i}{I(A;Q(B))},
\end{equation}
where $H$ is the binary entropy function and $Q$ is the quantizing function.

MD reconciliation is an alternative method designed for low-SNR regimes (below 0~dB). The core idea in MD reconciliation is to rotate Gaussian variables such that they lie on a unit sphere in a 
$d$-dimensional space, using their norm to avoid issues with small absolute values. This approach reduces the high bit error rate observed in slicing protocols for low-SNR regimes. MD reconciliation has shown superior performance over slicing at longer transmission distances, supporting ranges up to 200~km~\cite{Zhang2020}. Like slicing, MD reconciliation can be paired with various error-correcting codes; however, it particularly benefits from multiedge-type LDPC codes, which are optimized for MD use at a low received SNR~\cite{mani2021multiedge}. In this regime state-of-the-art decoding can achieve throughputs of 1.44 and 0.78 Gbps for code rates of 0.2 and 0.1, respectively, enabling real-time secret key generation at 71.89 Mbps at 25~km and 9.97 Mbps at 50~km~\cite{Luo2024}. 

The efficiency of MD reconciliation, defined as $ \beta (SNR) = R/C_\text{AWGN}(SNR)$, where $R$ is the code rate and $C_\text{AWGN}$ is the capacity of the additive white Gaussian noise (AWGN) channel, depends on the intrinsic efficiency of the error-correcting codes, the received SNR, and efficiency of mapping. This becomes evident when the overall efficiency is expanded in terms of the code efficiency, $\beta_{\text{code}}$, and the channel efficiency, $\beta_{\text{channel}}$, which depend on the dimension of the mapping. The overall efficiency can be expressed as~\cite{Mario2018}:
\begin{equation}
   \beta =  \beta_\text{code}\times\beta_\text{channel}=\beta_{code}\times\frac{C_{d}(SNR)}{C_\text{AWGN}(SNR)}, 
\end{equation}
where $C_{d}(SNR)$ is the ergodic capacity of a multidimensional scheme with dimension $d =$ 2,4, or 8. Consequently, rate-adaptive error correction, which uses techniques such as puncturing and shortening to adjust the rate of error-correcting codes, has been introduced to improve efficiency in practical implementations where SNR fluctuates over the measurements. For a comprehensive comparison of the performance of different reconciliation methods and error-correcting codes, see~\cite{Luo2024}.

After information reconciliation Alice and Bob perform error confirmation to ensure that the corrected key bit strings match on both sides. This is done by Alice and Bob randomly selecting a hash function from a family of universal hash functions. They then use this function to calculate the hash values of their corrected key bits and compare those values to determine whether the error correction was successful. If the hash values are identical, the keys are further processed in the next step. If the hash values do not match, the corresponding keys are deemed to be incorrect. Those keys are either discarded entirely or corrected by one party sending its key in plaintext to the other. The choice of method depends on the security proof. For a recent review on information reconciliation in CV QKD, see~\cite{Yang2023}. 

\subsubsection{Privacy amplification}\label{Sec:Post_Processing_PA}
Privacy amplification is a crucial step in QKD that removes information leaked about key bits that Eve might have gained during earlier stages of the QKD protocol. It is typically carried out by first choosing a hash function randomly from a so-called \emph{family of universal hash functions}. The value that chooses which hash function from the family (set) of hash functions is used is called the seed and is assumed to be sampled from a uniform distribution. Then this hash function is used to map the corrected bit strings of length $N$ at both Alice and Bob to a shorter secure key of length $L$~\cite{bennett1995generalized}. 
 
Among various families of universal hash functions, multiplication with a randomly chosen \textit{Toeplitz matrix} is the most commonly used method for privacy amplification in QKD. In a Toeplitz matrix all elements along each diagonal (from the upper left to the lower right) are identical. This property allows the matrix to be efficiently constructed using only the elements of its first row and first column. Privacy amplification using a Toeplitz matrix $\mathbb{T}$ can be directly implemented through matrix multiplication, $k_{\textrm{sec}} = \mathbb{T}k_{\textrm{corr}}$, where $k_{\textrm{sec}}$ is the final secure key with length $L$ and $k_{\textrm{corr}}$ is the key after error correction with length $N$. 
 
However, in the finite-size CV-QKD regime, $N$ is typically  large -- on the order of at least $10^{10}$. This renders direct matrix multiplication infeasible, as the direct implementation of Toeplitz matrix multiplication has a time complexity of $O(N^2)$, which would significantly slow down the privacy amplification process, especially in high-speed QKD systems.

To address this, the fast Fourier transform (FFT)-based Toeplitz method has been used in several works~\cite{tang2019high}. This approach reduces the time complexity to $O(N\log{N})$ by converting the Toeplitz matrix into a circulant matrix, enabling more efficient computation. Using this method, privacy amplification has been implemented in various hardware platforms, including graphics processing units, field-programmable gate arrays, and coprocessors~\cite{weerasinghe2024practical}. These implementations have achieved execution speeds ranging from Mbps to Gbps, making them suitable for real-time QKD systems.

The state-of-the art method for accelerating the implementation of the privacy amplification for CV QKD is the number-theoretical transform~\cite{takahashi2016high}, which, like the FFT, reduces the time complexity from $O(N^2)$ to $O(N \log{N})$. Unlike the FFT, the number-theoretical transform operates entirely in finite fields using modular arithmetic, eliminating the need for floating-point and complex variables. This reduces the memory needed and, at the same time, avoids any potential round-off errors of floating-point operations. This approach throughput similar to the FFT approach, reaching rates ranging from Mbps to Gbps~\cite{takahashi2016high,yan2022efficient}.

Alternatives to Toeplitz hashing that are based on multiplication with either (i) a modified Toeplitz matrix or (ii) circulant matrix and that are slightly faster and need a shorter seed were discussed by~\cite{Hayashi2016}. These alternatives also have a computational complexity of $O(N \log N)$. 

\subsubsection{Message authentication}\label{Sec:Post_Processing_Auth}
As mentioned, QKD protocols consist of two main phases, namely the quantum phase, where quantum states are transmitted and detected, and the postprocessing phase, which includes classical communication and data processing. To achieve information-theoretic security, the postprocessing phase requires a public authenticated channel to prevent man-in-the-middle attacks. 

Authentication in QKD typically relies on either preshared symmetric seed keys or postquantum cryptographic (PQC) keys for encrypting (signing) and decrypting (verifying) the hash values of classical messages~\cite{fregona2024authentication}. While symmetric key-based authentication provides information-theoretically secure solutions, it faces significant scalability challenges in large networks. The number of symmetric key pairs grows quadratically with the number of users, leading to increased overhead in storage, synchronization, and management. Trusted relays can alleviate this by forming star-type networks~\cite{frohlich2013quantum} where each user shares keys only with the relay. However, this architecture limits direct communication between users.

In contrast, PQC-based methods utilize a public key infrastructure~\cite{wang2021experimental} where users receive digital certificates signed by a trusted authority. This approach enables an efficient and scalable authentication for large networks. Notably, using PQC in authentication, rather than for confidentiality or key distribution, ensures that, as long as the algorithm remains secure during the authentication process, the QKD-generated keys are also secure, even if the PQC algorithm is later compromised.

Authentication protocols in QKD include approaches such as encode-decode methods~\cite{gilbert1974codes}, key recycling~\cite{wegman1981new}, and ping-pong delayed authentication~\cite{kiktenko2020lightweight}. Encode-decode methods provide strong security by using each key only once, but their high key consumption makes them impractical. Key recycling protocols address this limitation by reusing keys cyclically, such as with encrypting tags with one-time pads, thus significantly reducing key usage. Ping-pong authentication further optimizes this process by consolidating multiple rounds of postprocessing into a single authentication step. This method employs bidirectional authentication, where both parties alternately verify each other. 

For authentication universal hash functions such as $\epsilon$-AU$_2$ (almost universal)~\cite{abidin2012new}, $\epsilon$-AXU$_2$ (almost XOR universal)~\cite{rogaway1995bucket}, and $\epsilon$-A$\Delta$U (almost $\Delta$ universal)~\cite{stinson1996connections} are integral, with advancements focusing on minimizing key consumption while maintaining security. 

In practical implementations of QKD, the three following key aspects of authentication must be addressed to ensure overall security:
\begin{enumerate}
    \item \textit{Security of preshared authentication keys}. Preshared keys may be distributed out of band, for example, by using couriers. PQC methods have alternatively been proposed to secure those keys~\cite{wang2021experimental}.
    \item \textit{Key reuse}. Reusing keys for multiple authentications could potentially compromise security. Encrypting all tags with one-time pads can mitigate this risk.
    \item \textit{Security of QKD-generated keys}. After the initial round of authentication using preshared keys, QKD-generated keys are used for subsequent authentications. 
\end{enumerate}

For an in-depth analysis of authentication security under various nonideal conditions, see Table~9 of~\cite{Luo2024}.

\subsection{Advances in CV QKD}
CV QKD remains an actively developing field of research and many methods for improving the efficiency and robustness of  protocols have been suggested during the past decades. Development is ongoing toward quantitative improvement of the figures of merit, such as key rate, secure distance (or, equivalently, tolerable attenuation), tolerable channel or untrusted preparation and detection noise (which is particularly relevant for protocol implementations beyond the optical domain), as well as cost reduction by means of the simplification and integration of devices. In addition, effort is being directed toward qualitative improvements, such as the reduction of security assumptions, the generalization of security models, and the closing of practical loopholes. 

The quantitative improvements of the protocol performance can first be achieved by optimizing the parameters of the protocol. For example, it is already  known from the early implementations of CV QKD that modulation upon the imperfect error correction has to be optimized~\cite{Lodewyck2007}. Controllable trusted noise, which can be added to compensate for the noise in the quantum channel~\cite{Garcia2009} should also be optimized to provided maximum robustness to noise~\cite{Madsen2012}. Optimized parameter estimation can also improve the performance~\cite{Ruppert2014}, especially when the channels are fluctuating~\cite{Ruppert2019}. 

In addition to parameter optimization, CV QKD can benefit from signal state engineering, with the most typical example being the use of squeezed signal states instead of coherent signals, which allows the performance and robustness of the protocols to be improved~\cite{Navascues2005,Garcia2009,Madsen2012}. Another example involves fluctuating free-space channels~\cite{Derkach2020}, where squeezing has to be limited (optimized). Beyond parameter and signal state optimization, the applicability and security of CV QKD can be improved by various methods, which we discuss below.

\subsubsection{Quantum amplifiers and repeaters}

Amplification is well known and widely used in classical communications, where signal amplifiers (particularly laser amplifiers in the optical domain) increase the magnitude of the signal in order to compensate for the transmission losses. However, quantum features are typically lost in such a process, and quantum communication should be performed on the dedicated fibers free from standard classical networking equipment. Quantum states can also be amplified, which leads to enhancement of the signal but comes at the cost of increased noise or probabilistic outcomes. 

Quantum amplifiers in the phase-space description are usually distinguished between \textit{phase-insensitive amplifiers} (PIAs), which upscale both quadratures at the cost of excess noise in both, and \textit{phase-sensitive amplifiers} (PSAs), which amplify one of the quadratures while deamplifying another one, with a typical example of such process being quadrature squeezing. The PIA is therefore a nondegenerate optical parametric amplifier (OPA) that transforms the signal ($S$) and idler ($I$) modes with quadratures $\hat{x}_S,\hat{p}_S$ and $\hat{x}_I,\hat{p}_I$ as
\begin{equation}
\nonumber
\begin{split}
    \hat{x}_{S/I} \to \sqrt{g}\hat{x}_{S/I}+\sqrt{g-1}\hat{x}_{I/S},\\
    \hat{p}_{S/I} \to \sqrt{g}\hat{p}_{S/I}-\sqrt{g-1}\hat{p}_{I/S},
    \end{split}
\end{equation}
where $g>1$ is the amplification gain. The idler mode is ideally in a vacuum state or, more realistically, in a thermal state, that results in the excess noise added to the signal quadrature variances. The PSA, alternatively, ideally results in the signal transformation of the form $\hat{x}_S\to\sqrt{g}\hat{x}_S$ and $\hat{p}_S\to\sqrt{1/g}\hat{p}_S$ such that the amplified quadrature $\hat{x}$ becomes more noisy and the deamplified quadrature $\hat{p}$ becomes squeezed. It was shown that an optimal PSA can compensate for the practical imperfections (losses and noise) of a homodyne detector, while an optimal PIA compensates for the imperfections of a heterodyne deceiver~\cite{Fossier2009a}, which is valid even for practical noisy amplifiers, provided that their noise is limited. The positive role of PSAs in CV QKD was recently demonstrated experimentally by~\cite{Liao2025}, where it improved the photodetector clearance, resulting in the homodyne-detection efficiency of $96\%$.

The possibility of using PIAs and PSAs in coherent-state CV QKD over multispan links was recently studied by~\cite{Notarnicola2024}. It was shown that the key rate can be increased using multiple PSAs placed between the link segments if all the segments are considered untrusted, while both PIAs and PSAs can be helpful for only one untrusted link segment. 

To avoid noise addition by the PIA, postselection can be used, resulting in the probabilistic \textit{noiseless linear amplifier} (NLA)~\cite{Xiang2010}, which ideally is free from the idler-mode noise and can distill continuous-variable entanglement, that is a resource for CV QKD. By considering NLAs after the quantum channel and building an effective channel model, it was shown that NLAs can extend the secure distance of Gaussian CV QKD or, equivalently, improve the robustness to noise at a given channel attenuation~\cite{Blandino2012}. The operation should be optimized due to a trade-off between the probability of success and the amplification gain. 

The main building block of an NLA is typically the so-called \textit{quantum scissors} scheme~\cite{Pegg1998}, which is based on the input beam (coming from a multiport splitting of the signal entering the NLA) balanced coupling to a single-photon state, mixed with vacuum on a variable coupler. The output of the variable coupler, conditioned on the click on the proper output of the balanced coupler, then results in the scissors output. The outputs are recombined with the other outputs from other parallel scissors schemes to produce the probabilistically noiselessly amplified signal~\cite{Dias2020}. The quantum scissors scheme in the receiving station of Gaussian CV QKD was studied by~\cite{Ghalaii2020} who incorporated the essentially non-Gaussian single-photon coupling and filtering operations into security analysis. The obtained bounds on the secret key rate demonstrate the improvement of the secret distances of the protocol in certain regimes of nonzero channel excess noise, suggesting perspectives of using quantum scissors as a building block for CV quantum repeater schemes~\cite{Dias2017} that are useful for CV QKD. The result was extended to DM CV QKD by~\cite{Ghalaii2020a}. The quantum scissors scheme can be generalized to $N$-photon Fock states~\cite{Winnel2020}, which was shown to be more efficient for NLA than multiple applications of single-photon scissors and can be further improved by optimizing the general interferometric coupling~\cite{Fiurasek2022}. 

Another CV quantum repeater scheme, based on the non-Gaussian entanglement distillation (enabled by the single-photon states and detections)~\cite{Lund2009,Fiurasek2010} with subsequent Gaussification~\cite{Campbell2012}, postponed to the end of the repeater protocol by means of non-Gaussian entanglement swapping, was proposed and analyzed for CV QKD~\cite{Furrer2018}, showing the possibility of improving the rates and distances of the protocols. It was also shown that losses can be noiselessly suppressed by a combination of noiseless attenuation (heralded by single-photon detection on the residual port of an attenuating beam splitter) prior to transmission and noiseless amplification of the transmitted state~\cite{Micuda2012}. As an alternative to the quantum scissors scheme, which has to be multiply implemented on the split signal, NLAs can be efficiently realized by using multiple single-photon additions and subtractions~\cite{Zavatta2011} or two-photon addition and subtraction~\cite{Neset2024}, using weak measurements based on strongly nonlinear optical interactions~\cite{Menzies2009}, or combining Gaussian thermal noise and photon counting~\cite{Usuga2010}, which allows concentration of the phase information of the coherent states.

The practical implementation of NLAs and other CV quantum repeater schemes, which are potentially useful for CV QKD, is, however, challenging as it requires single-photon (or $N$-photon) state generation and efficient photodetection or high optical nonlinearities. Nevertheless, the NLA scheme can be implemented in the form of postselection, as we discuss next.

\subsubsection{Postselection}\label{Sec:Advances_Post_Selection}

Classical postselection in the sense of conditioning on a particular measurement outcome was first suggested for DR CV-QKD protocols to beat the 3~dB channel loss limit~\cite{Silberhorn2002a} in the absence of channel noise. Alternatively, a quantum postselection that is equivalent to a physical operation on the state can be applied. NLAs, that were virtually implemented in the form of Gaussian postselection were shown to be compatible with the Gaussian security proofs and helpful for CV QKD by~\cite{Fiurasek2012}, especially if they were combined with the virtual noiseless attenuation on the sender side (even without the need of preselection on this stage). It was shown that such virtual NLAs (also referred to as measurement based) can improve secure distance or tolerable channel noise of the protocols~\cite{Walk2013} and restore the key rate in the EB realization of CV QKD~\cite{Chrzanowski2014}. The quantum filtering, which is equivalent to NLA operation, should also be optimized owing to a trade-off between the postselection region and reduction of the raw key data. The filter implementation requires a cutoff on the data filter that should be large enough to result in Gaussian statistics. In the finite-size regime, this cutoff has to be optimized to provide sufficient improvement of the effective transmittance and, simultaneously, to preserve Gaussian statistics and not worsen the parameter estimation~\cite{Hosseinidehaj2020}. The equivalence between physical and virtual realizations of NLA was studied in detail by~\cite{Zhao2017} who showed the lower success rate of the measurement-based virtual realizations. Practical realizations of NLAs using quantum scissors and single-photon catalysis [photon addition or subtraction operation, which can reportedly act as NLA~\cite{Zhang2018a}] were theoretically studied by~\cite{Notarnicola2023} who took limited detection efficiency into account, and compared to the ideal NLA analyzed by~\cite{Blandino2012}, resulting in the extended secure distance for both practical realizations. Classical postselection was analyzed and shown to be helpful for DM CV QKD with phase-shift keying~\cite{Kanitschar2022}, where optimal postselection strategies can improve the performance of protocols with fewer signal states and reduce the performance gap between trusted and untrusted detector models. 

\subsubsection{Photonic integration}
The development of quantum integrated photonics is a major challenge and key enabler across all quantum technologies as it paves the way to systems with reduced cost and complexity and increased scalability, stability, and reproducibility~\cite{Moody2022}. In particular, for quantum communication and QKD, the use of photonic integrated circuits (PICs) will facilitate their deployment in practical infrastructures, notably in heavily constrained environments like data centers or satellite payloads. Significant progress has been achieved in recent years on all major PIC platforms, including silicon, III-V compound semiconductors, and lithium niobate~\cite{Aldama2022}. Many of these developments have focused on DV technology, which, however, faces the challenge of integrating single-photon detectors~\cite{Liu2022}. Coherent detectors used in CV QKD are more amenable to photonic integration, leading to several demonstrations in this direction, although the field is in general relatively young and more progress is expected in the following years. Most demonstrations leverage the silicon (Si) platform, which offers compatibility with mature manufacturing processes and low-loss components, but indium phosphide (InP) opens the prospect of monolithic integration, as it allows for laser integration in addition to high-speed modulators and efficient photodiodes.

At the component level, early experiments focused on Si-integrated homodyne detectors. by~\cite{Raffaelli2018}, an electronic-to-shot-noise clearance of 11~dB at 150~MHz was achieved, while the device was also used to produce random numbers at 1.2 Gbps. The experiment by~\cite{Bruynsteen2021} combined a Si optical front end and a custom integrated transimpedance amplifier designed with a 100~nm GaAs pseudomorphic high-electron-mobility transistor technology, showing enhanced stability, bandwidth, and noise performance for the developed photonic-electronic balanced homodyne detector. The clearance was 28~dB, and the shot-noise-limited bandwidth exceeded 20 GHz. ~\cite{Jia2023} used the silicon-on-insulator  platform for a time-domain balanced homodyne detector, that featured a common-mode rejection ratio of 86.9~dB, a clearance of 19.42~dB, and a quantum efficiency of 38\%. This performance enabled CV QKD with a secret key generation rate of 0.01~bits/pulse at 50~km. Still on the detection side,~\cite{Milovancev2021} showed first the use of commercially available die-level components for balanced homodyne detection, with a clearance exceeding 20~dB and a common mode rejection ratio of 50~dB at 1~GHz. Then~\cite{Milovancev2024} presented a receiver combining Si and BiCMOS and featuring much lower noise than wire-bonded counterparts at an equivalent bandwidth. This performance was compatible with a key rate of up to 30~Mbps at 10~km when considering CV QKD with Gaussian modulation and an untrusted receiver assumption.

In addition to the aforementioned developments at the component level,~\cite{Bian2024} demonstrated improved system stability reducing the standard deviation of the secret key rate by an order of magnitude using a biased Mach-Zehnder interferometer integrated on Si. The experiment showed a stable rate of 1.97~Mbps at 60~km with fluctuations on the order of 1\%, assuming a finite-size regime. Furthermore,~\cite{Li2023} used an InP reflective semiconductor optical amplifier coupled to a low-loss silicon nitride cavity to develop on-chip low-linewidth lasers (with linewidth ranging from 1.6 to 3.2~kHz), hence addressing the laser integration issue inherent in the Si platform. These lasers were used as signal and LO sources in a CV-QKD experiment reaching a secret key generation rate of 0.75~Mbps at 50~km.

Operation of PIC-based CV-QKD subsystems, receivers, or transmitters has been demonstrated only recently. A Si-integrated photonic-electronic receiver by~\cite{Hajomer2023} implementing phase-diverse heterodyne detection (with two balanced detectors) showed CV-QKD operation at a record-high 10~GBd rate with advanced DSP techniques. A Si-integrated receiver with germanium photodiodes implementing rf heterodyne detection (with one balanced detector) was used by~\cite{Pietri2024a} with the optimal coupling efficiency reaching 26\%, a clearance of 10~dB at 150~MHz bandwidth, and an achieved secret key generation rate of 2.4~Mbps at 10~km and 220~kbps at 23~km, also with advanced DSP. The experiment of~\cite{Bian2024a} used Si photonics for a receiver featuring a clearance of 7.42~dB at a bandwidth of 1.5~GHz, which is compatible with CV QKD in a local LO configuration with a secret key generation rate of 1.38 or 0.24 Mbps in the asymptotic or finite-size regime, respectively, at 28.6~km with a trusted receiver. Progress has been achieved on the transmitter side as well, with a pulsed experiment with Gaussian modulation and pulse shaping using an InP-integrated transmitter that includes an electro-optic modulator, an IQ modulator equipped with thermo-optic and current-injection phase shifters, and a variable optical attenuator featuring a 1~GHz bandwidth~\cite{Aldama2023}. The experiment achieved a rate of 2.3~Mbps in a back-to-back configuration or 0.4~Mbps at 11~km.

Full on-chip CV-QKD system operation remains challenging. The first PIC used for CV QKD~\cite{Ziebell2015} was in fact a transceiver including modulators based on carrier depletion, injection and thermal effects, and germanium photodiodes in a homodyne detector. However, although this work demonstrated the possibility of integrating all key components, there was no performance assessment as a full system. More recently,~\cite{Zhang2019} showed a Si-integrated system with both transmitter and receiver chips, including all key components. The system operated in a pulsed, transmitted LO configuration, with a homodyne detector and Gaussian modulation. The laser was not integrated and the bandwidth was limited to 10~MHz owing to to the detector transimpedance amplifier. The detector featured a clearance of 5~dB and a quantum efficiency of~49.8\%, and the achieved secret key generation rate was 0.14~kbps over a simulated 100~km fiber.

With the increasing importance of photonic integration in several technological fields, significant progress is expected in the coming years, with a direct positive impact on quantum communication. Further advances will undoubtedly include advanced and efficient packaging techniques, optimized coupling methods, and the cointegration of photonic and electronic chips. In particular, for CV schemes, in addition to low-linewidth laser integration, a crucial challenge remains regarding how to address the inherent trade-offs involved in high-sensitivity, low-noise and high-bandwidth detection in the presence of multiple photonic components on the same chip. This calls for precise control and optimization procedures in the entire chip and surrounding circuit that need to be validated in deployed quantum communication systems, hence confirming the expected advantages of the use of integrated photonics.

\subsubsection{Channel multiplexing and coexistence}
In the first tests of CV QKD, mode multiplexing was already used to simultaneously transmit quantum signals and LO~\cite{Lodewyck2007,Fossier2009}. However, signal multiplexing can also be used to improve the performance of CV-QKD protocols. In particular, three-dimensional multiplexing (in wavelength, polarization, and orbital angular momentum) was shown to increase the rates of DM CV QKD over free-space channels~\cite{Qu2017}. Multiplexed quantum teleportation was shown to be a promising tool for enhancing quantum communication rates~\cite{Christ2012}. The possibility of upscaling the secret key rate of Gaussian CV QKD by means of mode multiplexing in the multimode fibers was demonstrated by~\cite{Sarmiento2022}. It was also shown theoretically and verified in a proof-of-principle test on a frequency-multiplexed source of entanglement by~\cite{Kovalenko2021}, revealing the negative role of intermode crosstalk, as well as suggesting crosstalk removal by means of optimized data processing (equivalent to a passive linear-optical network used to comply with Gaussian security proofs). Multiplexed CV QKD over 194 channels with a total key secret rate of 172.6 Mbps over 25~km was reported by~\cite{Eriksson2020}, revealing the power of mode multiplexing for increasing the capacity of the coherent-state protocol.

Quantum signals can also be multiplexed with classical communication channels, which is referred to as coexistence between QKD and classical communication and is essential for practical implementation of CV QKD over existing fiber-optic networks. The possibility of performing CV QKD over dense-wavelength-division-multiplexing networks was experimentally studied by~\cite{Kumar2015} in a 25-km-long fiber link. It was shown that Gaussian coherent-state CV QKD operating at 1500~nm can coexist with noise from up to 11.5~dBm classical channels at the same wavelength in the forward direction (9.7~dBm in the case of backpropagation). In the 75~km channel, the tolerable classical channel power values were 3 and 9~dBm, respectively. An experimental test of coherent-state CV QKD together with 100 wavelength-division-multiplexed (WDM) channels with a total data rate of 18.3 Tbps in a 24-hour operation was reported by~\cite{Eriksson2019a}. Crosstalk from the classical channels was studied in its impact on coherent-state CV QKD by~\cite{Eriksson2019}, showing that in-band crosstalk from 30 WDM channels can stop the protocol, but when CV QKD is placed in a wavelength not used by the classical channels, the protocol can be realized even in the presence of amplified spontaneous emission noise. Impact of four-wave-mixing noise from dense-WDM systems on EPR-based CV QKD was studied by~\cite{Du2020}. A demonstration of CV QKD in the existing telecommunication infrastructure was recently reported by~\cite{Jain2024} who combined CV-QKD channels with 100 G optical transceivers operating at 1300~nm.

Further directions of ongoing development in CV QKD include advancement of the detection schemes, particularly in a hybrid Gaussian CV-QKD protocol with photon-number-resolving detectors, that can improve the secret key rate in the purely lossy channels and under the assumption that Eve employs homodyne detection~\cite{Cattaneo2018}. Recent results of the optimized state-discriminating receivers report extension of the security distance of CV QKD in wiretap (noiseless) channels~\cite{Notarnicola2023a}. Protocol simplification by so-called \textit{passive preparation} using splitting of a thermal state can reduce the cost and potentially remove some of the side channels in the preparing station, as suggested theoretically~\cite{Qi2018} and recently studied experimentally~\cite{Qi2020}. 

\section{CV quantum communication beyond QKD}
\label{sec:cvbeyond}
While QKD remains the most mature application of CV quantum communication, there are a few more protocols that can  potentially be realized using CV systems, as we discuss in this section.

\subsection{Quantum secure direct communication}
CV QSDC was proposed by~\cite{Pirandola2008b} and consists of two major phases, message mode and control mode, which relate to the message transmission and verification of its security, respectively. The phase space is first discretized via a square lattice such that any pair of bits is specified by the parity of the phase-space coordinates of the lattice cell. In the \textit{message} mode, for each pair of message bits $(u,u')$, the sender Alice computes the message amplitude $\alpha_{uu'}$, which points to the center of the lattice cell. Alice then adds a mask amplitude $\alpha_M$ to each of the message amplitudes such that the generated coherent signal states with amplitudes $\bar{\alpha}=\alpha_{uu'}+\alpha_M$ are continuously Gaussian distributed across the phase space. Each signal state is measured by the remote receiving party Bob using a heterodyne detector, resulting in the outcome $\beta \backsimeq \bar{\alpha}$. The CV-QSDC setup is, therefore, conceptually similar to the P\&M CV-QKD scheme shown in Fig. \ref{fig:P_and_M}. For each pair of bits, Alice classically communicates the mask $\alpha_M$ to Bob, who can then unmask the signal by computing $\beta-\alpha_M \backsimeq \bar{\alpha}-\alpha_M=\alpha_{uu'}$ and retrieve the message bits $(u,u')$ through lattice decoding. Security analysis of the scheme relies on the assumption of individual attacks, in which it is optimal for an eavesdropper Eve to apply a universal Gaussian quantum cloner, which would unavoidably introduce noise to the signal. Therefore, in the \textit{control} mode of the protocol, Alice prepares the Gaussian-distributed signals $\bar{\alpha}$ only and, after the transmission and detection, communicates their amplitudes to Bob, who extracts a test variable as $\beta-\bar{\alpha}$, from which the total channel noise can be inferred.

By setting a noise threshold value, the trusted parties can verify whether they should continue or abort the protocol. In principle, the threshold can even be set to zero noise, and the security of the scheme was analyzed in this strict case~\cite{Pirandola2008b}. Trusted parties can optimize the lattice size and the fraction of the control mode rounds of the protocol, which should be randomly switched from the message mode, while Eve can optimize the added noise to obtain the maximum information on the message bits. Using the optimized settings, it was shown that Alice can transmit 630 bits using $2.2 \times 10^4$ systems, while Eve can steal 80 bits. The amount of leaking information can be decreased by increasing the fraction of the control mode runs (making the protocol less efficient) or by employing classical error-correction codes, which make the decoding more sensitive to the presence of channel noise. In the aforementioned model case, the optimal code allows the stolen bits to be decreased to 10. 

A CV-QSDC protocol based on Gaussian mapping of the classical data encoded in the quadratures of TMSV states shared between Alice and Bob was proposed and analyzed by~\cite{Cao2021} who built upon previous theoretical works~\cite{Chai2019,Srikara2020}. In this scheme Alice prepares a TMSV state, randomly measures the quadratures of one of the modes, communicates the time slots and measurement outcomes to Bob, and sends both modes to Bob. Bob performs the same measurements on the other mode and verifies the nonseparability of the obtained TMSV. The trusted parties check the error rate between their datasets to be below a pre-agreed-upon threshold and proceed to the Gaussian mapping stage once the condition is fulfilled. The secret message is divided into blocks and mapped onto a sequence of Gaussian random values $G_1$. The values are used to modulate one of the quadrature observables of Alice's mode, another Gaussian sequence $G_2$ is used to modulate the complementary quadrature and the modulated mode is sent to Bob. Alice also classically communicates the sequence $G_2$ to Bob, who couples the modulated mode with the residual mode of TMSV and measures either of the quadratures on the two output modes. After that, the trusted parties perform the parameter estimation based on Bob using the measurement outcomes of the quadrature to which the sequence $G_2$ was encoded with the $G_2$ received from Alice. Alice and Bob perform the error correction on the data sequence $G_1$ (as modulated by Alice), and the version of it obtained by Bob after the channel. Finally, the trusted parties perform privacy amplification to eliminate the information introduced by the Gaussian mapping (which has to be optimized to account for the nonuniformity of the messages), as well as the information leaked from the quantum channel. The security of the scheme is analyzed against collective attacks following the techniques for Gaussian CV QKD, strictly assuming that the second mode of TMSV, which is used by Bob in the detection stage, is not accessible to an eavesdropper (while being transmitted through the same quantum channel). 

The CV-QSDC protocol using single-mode squeezed states was proposed and experimentally tested by~\cite{Paparelle2025} who assumed beam-splitting eavesdropping attacks. In the two-way quantum communication scheme, squeezing was considered on one side or on both sides of the channel (referred to as the asymmetric and symmetric protocols, respectively). Using an optical switch, Alice randomly chooses between the control and message states, received by Bob, measuring the former with a homodyne detector and storing the latter in a quantum memory or a delay line. By comparing the measurement outcomes to the amplitudes, classically communicated by Bob for the designated control states, and verifying a sufficiently low amount of errors, Alice then encodes the message by controllably attenuating part of the message states [as opposed to applying more technically demanding displacements, initially suggested by~\cite{Srikara2020}] and sends them back to Bob along with nonattenuated decoy signal states that are used to further bound the eavesdropping. It was shown that squeezing is advantageous for such a CV-QSDC protocol even in the asymmetric configuration~\cite{Paparelle2025}.

\subsection{Quantum dense coding}
\begin{figure}
    \centering
    \includegraphics[width=0.9\linewidth]{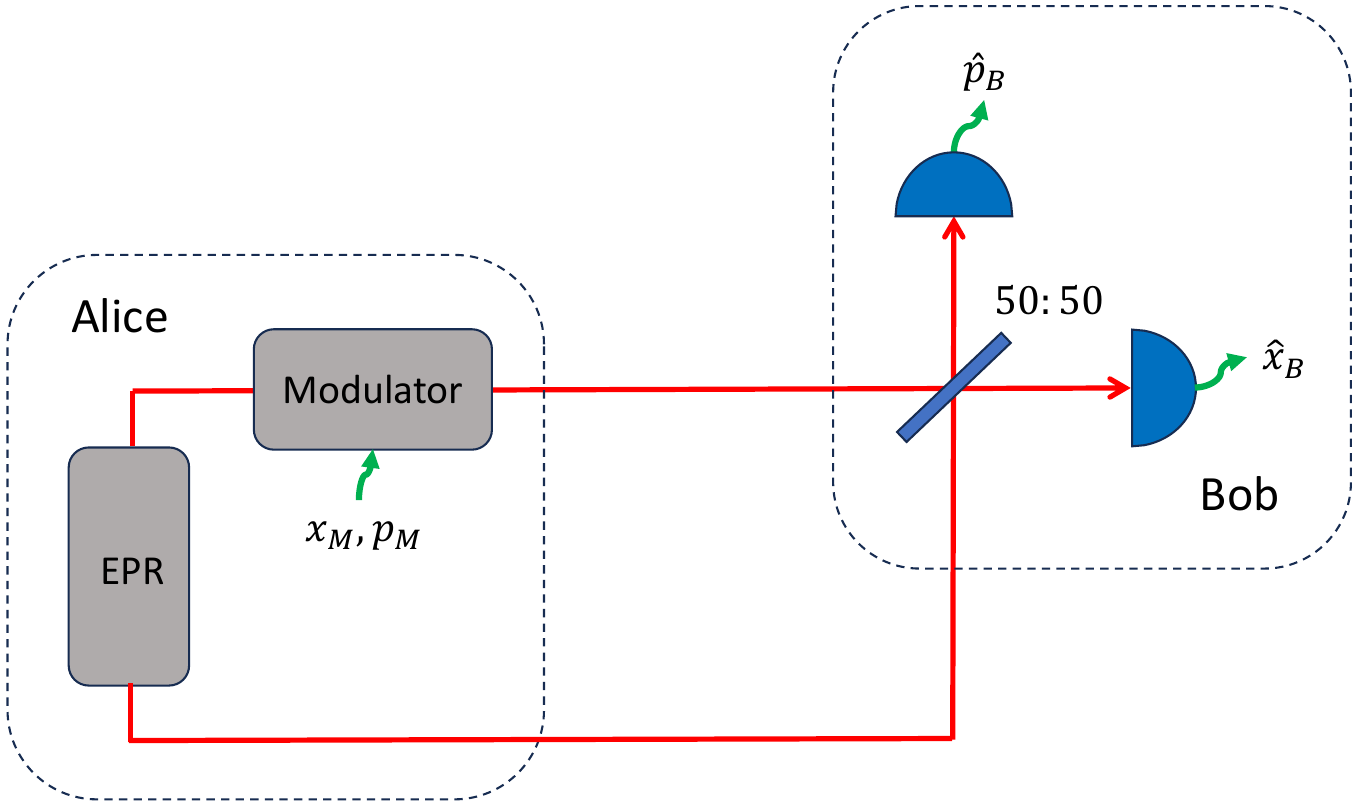}
    \caption{Generic CV-QDC scheme. Alice generates a TMSV  (EPR) state, applies the displacements $x_M$ and $p_M$ to one of the modes, and sends both modes to Bob. He then combines them on a balanced beam splitter and measures $\hat{x}_B$ and $\hat{p}_B$ quadratures on the output to obtain the encoded message (encoding and decoding parts are not depicted and are explained in the text).}
    \label{fig:dense_coding}
\end{figure}

CV QDC was first proposed by~\cite{Ban1999} and~\cite{Braunstein2000a} on the basis of highly entangled TMSV states, as shown in Fig. \ref{fig:dense_coding}. One of the modes is modulated by the sender Alice, who performs encoding of a message by applying quadrature displacement with an amplitude $\alpha_\text{in}=x_M+ip_M$. The displaced mode and the second mode of the TMSV are then sent to the receiver Bob, who combines the modes on a balanced beam splitter and measures the quadratures $\hat{x}_B$ and $\hat{p}_B$ on the outputs. The resulting outcomes are normalized and combined to obtain the output amplitude $\alpha_\text{out}=x_B+ip_B$, which constitutes the message received by Bob. In the limit of infinitely strong TMSV squeezing (or, equivalently, mean photon number $\bar{n}\to\infty$) $\alpha_\text{out}=\alpha_\text{in}$, meaning that the message is perfectly recovered. However, even upon limited squeezing (and finite $\bar{n}$), CV entanglement of a TMSV state enables dense coding by achieving larger channel capacity than coherent- or squeezed-state communication with the same mean photon number. Indeed, the dense coding capacity of the CV-QDC scheme in the case of Gaussian-distributed signals $\alpha_\text{in}$ reads $C_\text{QDC}=\log{(1+\bar{n}+\bar{n}^2)}$, which for the strong squeezing is approximated through the squeezing parameter as $4r$. Alternatively, P\&M communication using coherent states can achieve $C_\text{coh}=\log{(1+\bar{n})}$, which is always less than $C_\text{QDC}$. Using squeezed states, a capacity of $C_\text{sq}=\log{(1+2\bar{n})}$ can be achieved, which is less than $C_\text{QDC}$ for any $\bar{n}>1$~\cite{Braunstein2000a}.
The CV-QDC protocol, operating at high parametric gains of the TMSV sources, allows for highly efficient deterministic transmission, unlike the DV QDC protocols, which employ weak sources and perform probabilistically~\cite{Braunstein2000a}. The analysis of CV QDC was extended to lossy channels by~\cite{Ban2000} who showed that CV QDC remains advantageous at channel transmittance levels above approximately $0.63$. 

An experimental test of CV QDC was reported by~\cite{Li2002}, with simultaneous sub-shot-noise measurements of signals encoded in the amplitude and phase quadratures that were enabled by the CV entanglement produced using a nondegenerate OPA.

\subsection{Quantum digital signatures}
\begin{figure}
    \centering
    \includegraphics[width=0.9\linewidth]{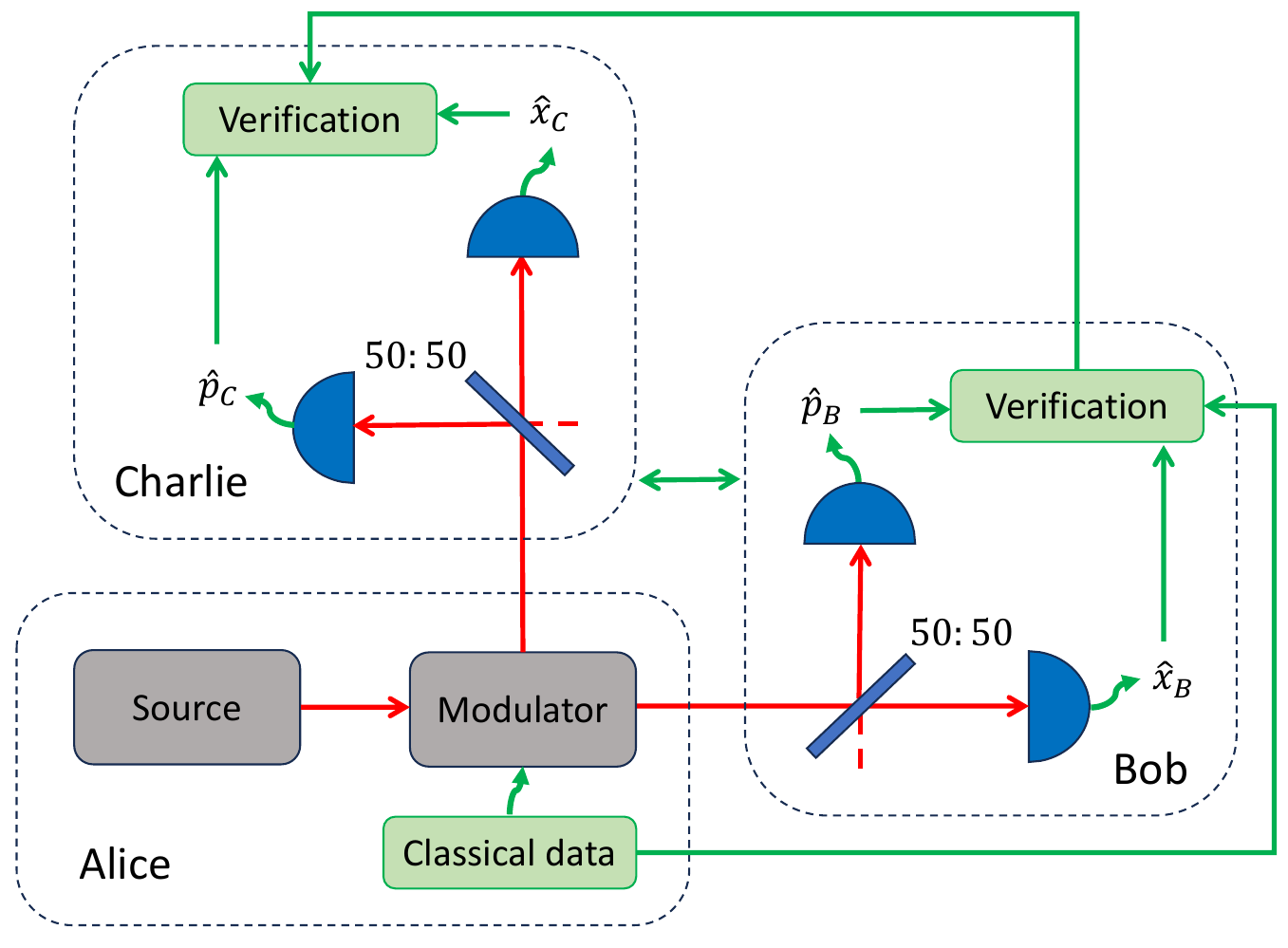}
    \caption{CV-QDS scheme. Alice uses a laser source and a quadrature modulator to prepare the public key sequences of quantum states, and sends them to the remote parties Charlie and Bob. They perform heterodyne measurements of quadratures $\hat{x}_B,\hat{p}_B$ and $\hat{x}_C,\hat{p}_C$ respectively, feeding the outcomes to the verification procedures as well as exchanging the measurement settings and outcomes. Quantum links are given in red, while classical links are given in green.}
    \label{fig:QDS}
\end{figure}

A QDS scheme using CV heterodyne measurements suggested by~\cite{Croal2016} consists of two stages, namely, the distribution stage and the messaging stage. In the \textit{distribution} stage, the sender Alice selects sequences of nonorthogonal quantum states for encoding a message bit that forms the public keys. The classical information on the state sequences forms the private keys. Alice sends the public keys to two remote parties Bob and Charlie over the respective quantum channels, as shown in Fig. \ref{fig:QDS}. Bob and Charlie perform heterodyne measurements on the incoming quantum signals, hence obtaining the partial information on the public keys (eliminating the less probable quantum states according to the measurement results), which concludes the distribution stage. In the \textit{messaging} stage, Alice sends a message to Bob or Charlie along with the corresponding private key. The message recipient determines whether the private key sufficiently matches the previously obtained public key, which excludes the message forgery (when one of the remote parties fakes a message as originating from Alice). The protocol also enables message forwarding when the new recipient checks for mismatches with his measurement records and involves symmetrization to exclude repudiation by Alice (when Bob and Charlie disagree on the authenticity of a message), which relies on QKD established between the recipients. The mismatch threshold for accepting a forwarded message is then less strict compared to the direct messages received from Alice. 
Assuming collective attacks by dishonest parties and an otherwise authenticated quantum channel, the probability of a successful repudiation or forgery, as well as the probability of a protocol failure, can be bounded from the protocol parameters. The probability of repudiation is bounded as 
\begin{equation}
p_\text{rep} \leq 2\exp\Big[-(S_C-S_B)^2\frac{L}{4}\Big],
\label{eq:qsc_prep}
\end{equation}
where $S_C$ and $S_B$ are Charlie's and Bob's acceptance thresholds, respectively. The probability of forgery is bounded as 
\begin{equation}
p_\text{forg} \leq \exp[-(C_\text{min}-S_C)^2L],
\end{equation}
where $C_\text{min}$ is the minimum probability that an honest party will detect an error in an individual signature element coming from the forger. Finally, the probability of the protocol failure is bounded as
\begin{equation}
p_\text{fail}\leq 2\exp\Big(-\frac{g^2}{16}L\Big).
\end{equation}
where $g=C_\text{min}-p_\text{err}$, with probability of error $p_\text{err}$, the probability that an honest recipient, properly following the protocol, will eliminate the state actually sent by Alice. The figure of merit for the CV-QDS protocol is the signature length $2L$, required to sign a 1 bit message with a fixed probability of failure. 

The previously described protocol was tested using sequences of four coherent states, $|\alpha\rangle,|i\alpha\rangle,|-\alpha\rangle$, and $|-i\alpha\rangle$, to encode single-bit messages, assuming authenticated quantum channels between Alice, Bob, and Charlie and achieving much larger signature lengths per quantum state compared to the DV QDS protocols~\cite{Croal2016}. 

The CV-QDS scheme was extended to insecure quantum channels by~\cite{Thornton2019} who assumed collective attacks modeled by an entangling cloner. The probability of failure in a purely lossy channel with transmittance $T$ then reads $p_\text{err}=(1/2)\text{erfc}[\alpha\sqrt{T/2}]$. It was shown that the quantum encoding alphabet can be optimized in terms of the number of states and the displacement $\alpha$. This allowed  the practical performance of the protocol to be assessed, predicting the possibility of signing a single-bit message on the timescale of 1 ms at a distance of 20~km of standard telecom fiber, when one assumes a repetition rate of 100 MHz~\cite{Thornton2019}.

While the previously described protocols aim at iteratively signing single-bit messages, CV QDS for multibit messages was developed by~\cite{Zhao2021}, also providing stronger security against the dishonest Bob. Furthermore, the security of multibit CV QDS was extended to coherent attacks~\cite{Zhang2024} via the employment of one-time universal hashing and advanced DM CV-QKD security proofs~\cite{Matsuura2021}.

\subsection{Quantum authentication}
The protocol for CV QA of the physical keys with the aim of entity authentication was proposed by~\cite{Nikolopoulos2017} and based on probing the keys using coherent states of light and their subsequent homodyne detection. The physical keys are represented by an optical multiple-scattering random medium and are supposed to be physically unclonable as their full characterization involves multiple degrees of freedom. The typical optical entity authentication is then done in two stages, namely, enrollment and verification. In the \textit{enrollment} stage, the authority that distributes the keys is characterizing the keys by interacting them with various probes (challenges) and recording responses to the challenges along with the parameters of respective probes. In the \textit{verification} stage, the verifier checks the authenticity of the key by challenging the key with randomly chosen probes and comparing the obtained responses to the recorded ones. It was shown that by using quantum probes, the security of the scheme is considerably improved~\cite{Nielsen2012}, while CV realization offers an advantageous homodyne-detection technique~\cite{Nikolopoulos2017}. 

The key in the CV quantum entity authentication scheme is represented by a disordered multiple-scattering medium, and the setup is largely based on the existing wave-front-shaping techniques for controlling the scattered light~\cite{Mosk2012}. The same setup is used to test the enrollment and the verification so that the authority generates a coherent state and uses its smaller fraction as a probe that is shaped, directed to the key and, after scattering, collected into a single-mode fiber. The stronger fraction of the initial coherent state is then used as an LO for the quadrature measurement of the collected scattered light on a homodyne detector. Provided that the verification setup is secure (hence, an attacker has no access to it), the protocol is reported to be collision resistant, meaning that it can distinguish between two randomly chosen keys and is sensitive to cloning of the keys (which is assumed to be imperfect), hence being able to distinguish between a cloned key and the original~\cite{Nikolopoulos2017}. 

The CV protocol for quantum identity authentication was proposed by~\cite{Huang2011} using Gaussian modulation of squeezed states and homodyne detection in the same setup as Gaussian P\&M CV QKD shown in Fig. \ref{fig:P_and_M}. Furthermore, the proposed scheme can be combined with QKD to provide authentication and prevent man-in-the-middle attacks~\cite{Zeng2000}. The scheme starts with a preshared key (possibly a residual key from the previous QKD session), and Alice, acting as a certifying authority, modulates the squeezed signal states, thus encoding the key bits, and sends them to Bob. Bob measures the corresponding quadratures and publicly announces certain outcomes as defined by Alice, who evaluates the fidelity between her and Bob's data. Based on the fidelity, Alice verifies Bob's identity as well as the presence of an eavesdropper. 

\subsection{Quantum oblivious transfer}
CV one-out-of-two randomized QOT protocol was suggested in the noisy storage model, i.e., assuming limited memory capacity of an attacker~\cite{Furrer2018a}. In this protocol Alice performs the bitwise XOR operations $x_0 \oplus s_0$ and $x_1 \oplus s_1$ between the input strings $x_0$ and $x_1$ and independent and uniformly distributed outputs $s_0$ and $s_1$. She then sends the results to Bob. Bob chooses a bit from $\{0,1\}$, specifying the bit string to be learned,  obtains the bit string $\tilde{s}$, and can learn the string $x_t$ by adding $\tilde{s}$ to $x_t \oplus s_t$. 

The protocol can be realized using Gaussian-modulated squeezed states of light measured with homodyne detectors in the setup shown in Fig. \ref{fig:P_and_M} (with $T_\text{hetB}\in\{0,1\}$) but was studied and tested in the EB version as shown in Fig. \ref{fig:EPR} (with, additionally, $T_\text{hetA}\in\{0,1\}$). Alice distributes $n$ EPR states, each measured by Alice and Bob using homodyne detectors in one of two randomly chosen orthogonal quadratures. Both parties discretize the measurement outcomes (Bob performs scaling first to compensate for the channel losses), obtaining the $n$-bit strings $Z$ and $Y$, respectively. Then the parties wait for some predefined time to impose a coherence requirement on Bob's memory in case he is malicious and proceed to the classical postprocessing, sifting the data (Bob stores the results obtained in compatible and incompatible bases independently) and splitting their strings to obtain correlated and anticorrelated substrings. The parties then perform one-way error correction and Bob, after applying the hash function as described by Alice, obtains the bit string $\tilde{s}$. 

The correctness of the protocol, imposing a uniform distribution of $s_0$, $s_1$, and the fact that Bob learns the desired string $s_t=\tilde{s}$, is analyzed in the composable security framework. The security of the scheme is shown for honest Bob against malicious Alice without any assumptions about the power of the latter. In the opposite scenario of dishonest Bob, the composable security of the protocol requires additional assumptions on the power of malicious Bob to store quantum states, which is limited in the time delay between the state distribution and the data postprocessing. Bob's quantum memory is then modeled as the lossy channel. The security is obtained using the specially derived entropic uncertainty relations for the discretized quadrature observables which limit Bob's knowledge on Alice's outcomes and assuming general as well as Gaussian storage operations. 

The protocol was tested using TMSV states produced by a balanced coupling of two orthogonally squeezed states generated via parametric down-conversion at 1550~nm and low-noise, highly efficient homodyne detectors operating at 100 kHz. The signal and the LO were copropagated through the variable attenuator simulating the quantum channel. Using optimized data processing and reconciliation, it was shown that in a highly transmitting quantum channel, the rate of the order of 0.1 bit per symbol can be achieved, while for stronger losses the results depend largely on the assumptions about the malicious Bob's storage~\cite{Furrer2018a}.

\subsection{Quantum teleportation}
\label{sec:cv-qt}
\begin{figure}
    \centering
    \includegraphics[width=0.9\linewidth]{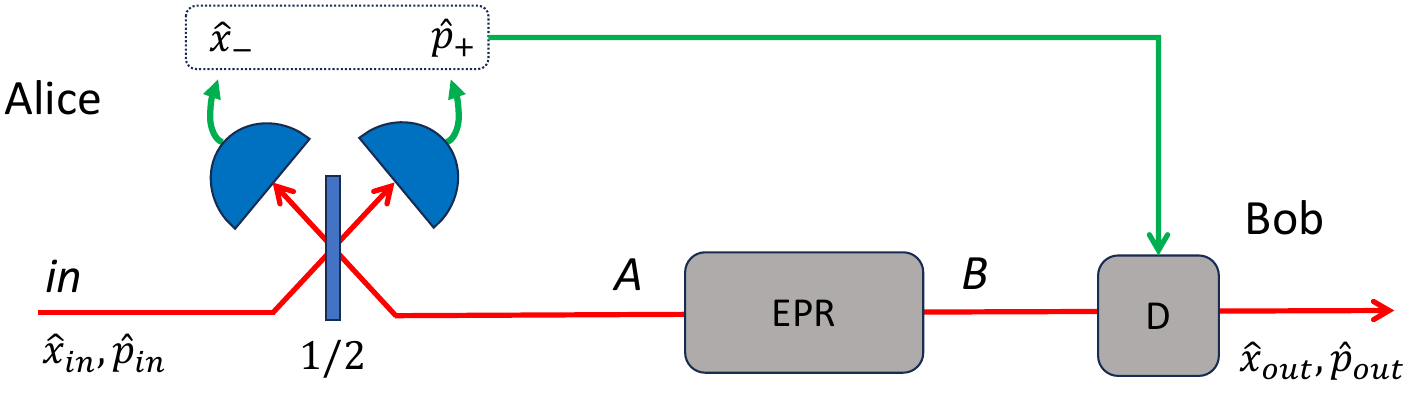}
    \caption{Generic CV-QT scheme. Alice and Bob share an entangled state in modes $A$ and $B$ (ideally, this is a CV EPR state with perfect quadrature correlations; in practice, this is a TMSV state). Alice couples the input state in mode $in$ on a balanced beam splitter with transmittance $1/2$ to the mode $A$ of the entangled state. The outputs $\hat{x}_-$ and $\hat{p}_+$ are measured and results are classically communicated to Bob, who accordingly performs displacement $D$ on mode $B$ of the entangled source to obtain the output state.}
    \label{fig:QT}
\end{figure}

CV QT was proposed by~\cite{Vaidman1994} who used perfect EPR-type entanglement, and was extended to finite correlations by~\cite{Braunstein1998}, in order to avoid qubit Bell measurements, which are very challenging to implement~\cite{Luetkenhaus1999}. In the CV-QT scheme~\cite{Braunstein1998} shown in Fig. \ref{fig:QT}, Alice and Bob share an entangled state in modes $A$ and $B$, while the incoming state in mode ``in'', which is characterized by the quadrature vector $\hat{\mathbf{r}}_\text{in}=\{\hat{x}_\text{in},\hat{p}_\text{in}\}^T$, is to be teleported. Alice couples modes $in$ and $A$ on a balanced beam splitter and measures quadratures $\hat{x}_-$ and $\hat{p}_+$, performing a CV Bell measurement. The measurement outcomes are classically communicated to Bob, who performs displacement $D$ on the mode $B$ of the shared entangled state and obtains the output state with the quadrature vector $\hat{\mathbf{r}}_\text{out}=\{\hat{x}_\text{out},\hat{p}_\text{out}\}^T$. 
In an infinitely strong CV EPR entangled state, the mode quadratures are fully correlated or anticorrelated: $\hat{x}_A=\hat{x}_B,\hat{p}_A=-\hat{p}_B$. Then, in a perfect implementation of the scheme, after values $x_-$ and $p_+$ of the measured quadratures $\hat{x}_-=(\hat{x}_A-\hat{x}_\text{in})/\sqrt{2}$ and $\hat{p}_+=(\hat{p}_A+\hat{p}_\text{in})/\sqrt{2}$ are sent to Bob, Bob's displacement is simply $\{-x_-\sqrt{2},p_+\sqrt{2}\}$ to obtain the output state $\hat{\mathbf{r}}_\text{out}\equiv\hat{\mathbf{r}}_\text{in}$, and hence perform perfect teleportation with a fidelity $F=1$. Such a unit fidelity cannot be exactly reached in practice, since it would require infinite entanglement or, equivalently, infinite entropy at the local stations, therefore violating holographic bounds~\cite{Pirandola2024b}. Thus, finite entanglement resources and setup imperfections limit the quality of CV QT. Nevertheless, the feasibility of the CV Bell measurement allows CV QT to be performed unconditionally.

In practice, the entangled resource state shared by the parties is a TMSV state. Bob's displacements should be optimized by applying a gain factor to the measurement outcomes communicated by Alice. This helps compensate for the imperfect resource state, increasing the teleportation fidelity in practical scenarios. Gain tuning was studied and optimized for coherent, vacuum, and single-photon inputs~\cite{Ide2002}, with further studies for DV input states under realistic experimental conditions~\cite{Takeda2013}. The limited fidelity of CV QT was shown to be equivalent to attenuation of the input signal~\cite{Hofmann2000} and is related to information about the signal, obtained from Bell measurement, which should ideally be absent~\cite{Hofmann2001}.

Generally, in the case of a mixed input state, the fidelity can be obtained as $F=\overline{\mathrm{Tr}(\sqrt{\hat{\rho}_\text{in}}\hat{\rho}_\text{out}\sqrt{\hat{\rho}_\text{in}})}$~\cite{Pirandola2006}, averaged over outcomes of the Bell measurement. The expression simplifies to $F=\overline{\langle\psi_\text{in}|\hat{\rho}_\text{out}|\psi_\text{in}\rangle}$ for the pure input state $\hat{\rho}_\text{in}=|\psi_\text{in}\rangle\langle\psi_\text{in}|$. If the teleportation channel is a Gaussian state with a covariance matrix $\gamma_{AB}$ having the form of Eq. (\ref{eq:CovarianceMatrix}) and the input state is a pure single-mode Gaussian state with covariance matrix $\gamma_\text{in}$, the fidelity of teleportation with a gain of $\sqrt{2}$ can be obtained as an overlap integral between the input and output Wigner functions~\cite{Fiurasek2001},
\begin{equation}
F=2\pi\int_{-\infty}^{\infty}{W_\text{in}(x,p)W_\text{out}(x,p)dxdp},
\end{equation}
which can be expressed as $F=2/\sqrt{det\Gamma}$, where the matrix $\Gamma$ reads
\begin{equation}
\Gamma=2\gamma_\text{in}+\gamma_B+\sigma_z\gamma_A\sigma_z-\sigma_z\sigma_{AB}-\sigma_{AB}^T\sigma_z.
\end{equation}

For a coherent-state input and a TMSV resource state with squeezing $r$ and unity gain, the fidelity is given by $F_\text{coh}=1/[1+\exp(-2r)]$~\cite{Braunstein2001a}. While in the case of infinitely strong entanglement ($r \to \infty$) the fidelity approaches 1, in the absence of entanglement ($r=0$), the fidelity is $F_\text{coh}^{r=0}=1/2$, which gives the classical bound on the coherent-state fidelity. Alternatively, quantum cloning (also called quantum duplication) can achieve a fidelity of $F=2/3$ that is referred to as the no-cloning limit~\cite{Grosshans2001}. Therefore, to not only show the advantage above the classical strategies but also make sure that the teleported state is better than any cloned copy, QT should overcome the no-cloning limit. Beating the no-cloning limit also indicates that a quantum teleporter is able to preserve the nonclassicality (Wigner negativity) of the teleported states~\cite{Ban2004}.

When the teleported state is a two-mode entangled state itself and the communicating parties do not initially share an entangled resource, but, instead, each keeps a mode of its entangled mode pairs while sending the two residual modes to a third party who performs the Bell quadrature measurement and broadcasts the outputs, the QT scheme changes to CV entanglement swapping~\cite{Tan1999,Loock1999}. The state shared between Alice and Bob and conditioned on the broadcast Bell measurement outcomes is now entangled, even though the entangled modes never physically interact. Importantly for quantum communication, entanglement swapping can be a building block for a CV quantum repeater~\cite{Dias2017}. Note that CV entanglement swapping represents the EB version of the CV-MDI protocol, which was described in Sec. \ref{sec:protocols}. It is therefore an ideal setting to study the security of CV-MDI QKD, which is similar to the role that DV entanglement swapping plays in the security of DV-MDI QKD~\cite{Braunstein2012}.

While QT supposes the teleportation of a quantum state, its experimental test must also include verification, performed by a verifier (often referred to as a third party Victor) who prepares and characterizes the input state and measures the output state in order to assess the resulting fidelity. Experimental realization of unconditional CV QT was first reported by~\cite{Furusawa1998} who used two-mode entanglement produced by parametric down-conversion in an optical parametric oscillator (OPO). The experimental fidelity of coherent-state teleportation of $0.58 \pm 0.02$ was obtained, thus confirming the advantage of quantum teleportation over the classical benchmark. Fidelity of $0.61 \pm0.02$ (or $0.62$ after correcting for the efficiency of the verifying detector) was reported by~\cite{Zhang2003}. Teleportation of modulated coherent states was demonstrated by~\cite{Bowen2003} in an experiment with an entangled channel state, produced by coupling of two -4.8~dB squeezed states from two OPAs, showing the optimized fidelity of $0.64 \pm 0.02$. 

The benchmark of no-cloning limit was surpassed by~\cite{Takei2005} who reported the fidelity of $0.7 \pm 0.02$ and also demonstrating CV entanglement swapping, which was verified through confirmation of the sub-shot-noise fluctuations of the quadrature differences of the two residual modes. The experimental scheme was based on two sources of entanglement (by coupling pairs of orthogonally squeezed states produced by four OPOs), measured with four homodyne detectors. Teleportation of a squeezed state with confirmed sub-shot-noise fluctuations on the output was reported by~\cite{Takei2005a}  with fidelity $0.85 \pm 0.05$, which is above the classical limit for such states that has been experimentally verified as $0.73 \pm 0.04$. A coherent-state teleportation fidelity of $0.83 \pm 0.01$ was reported by~\cite{Yukawa2008} who used high degrees (-6~dB) of OPO squeezing to produce strong two-mode entanglement, which also suggests sequential teleportation toward advanced quantum information processing.

Teleportation of a nonclassical Schr\"{o}dinger's cat state of superposed coherent states $|\alpha\rangle \pm |-\alpha\rangle$ was reported by~\cite{Lee2011} and confirmed by negativity of the Wigner function of the input and output states that is characterized by state tomography. Similarly, noiseless teleportation of a single-photon state was confirmed by the Wigner function negativity, as reported by~\cite{Fuwa2014} who used loss suppression based on conditioning on the Bell quadrature measurements, resulting in a $53\%$ success rate. Long-distance teleportation of a coherent state over 6~km of optical fiber was demonstrated by~\cite{Huo2018} and reached a fidelity of $0.62 \pm 0.03$ by means of OPO squeezing of approximately -5.3 dB. The milestone realizations of CV QT are summarized in Table \ref{tab:cvqt}.

\begin{table}[!]
    \centering
    \begin{tabular}{|p{3.5cm}|p{2cm}|p{1cm}|p{1cm}|}
    \hline
     \leavevmode\centering\textbf{Reference} & \leavevmode\centering\textbf{Input} & 
     \leavevmode\centering\textbf{F} &
    \leavevmode\centering\arraybackslash\textbf{Prob.}    \\
    \hhline{|=|=|=|=|}
     ~\cite{Furusawa1998} & \centering coherent & \centering 0.58 & \centering\arraybackslash 1 \\
      \hline
   ~\cite{Zhang2003} & \centering coherent & \centering 0.62 & \centering\arraybackslash 1 \\
    \hline
     ~\cite{Bowen2003} & \centering coherent & \centering 0.64 & \centering\arraybackslash 1 \\
        \hline
   ~\cite{Takei2005a} & \centering squeezed & \centering 0.85 & \centering\arraybackslash 1 \\
    \hline
     ~\cite{Takei2005} & \centering coherent & \centering 0.7 & \centering\arraybackslash 1  \\
      \hline
       ~\cite{Yukawa2008} & \centering coherent & \centering 0.83 & \centering\arraybackslash 1 \\
    \hline 
   ~\cite{Lee2011} & \centering cat & \centering - & \centering\arraybackslash 1 \\
    \hline
   ~\cite{Fuwa2014} & \centering single photon & \centering - & \centering\arraybackslash 0.53 \\
    \hline
   ~\cite{Huo2018} & \centering coherent & \centering 0.62 & \centering\arraybackslash 1 \\
    \hline
   ~\cite{Zhao2023} & \centering coherent & \centering 0.92 & \centering\arraybackslash $10^{-5}$ \\  
     \hline
    \end{tabular}
        \caption{Milestone bipartite CV-QT demonstrations (in chronological order), input states (coherent, squeezed, single-photon, or Schr\"{o}dinger's cat states), achieved fidelities $F$, and success probabilities (unity meaning unconditional, deterministic teleportation).}
    \label{tab:cvqt}
\end{table}

Several proposals were made for improving the fidelity of CV QT with finite resources, such as dividing the input state to multiple CV qubit teleporters and recombination of the outputs as suggested by~\cite{Andersen2013}. Equivalence between CV QT with finite resources and single-mode phase-insensitive Gaussian channels was analyzed by~\cite{LiuzzoScorpo2017} and used to theoretically optimize the teleportation of Gaussian-modulated coherent states. It was shown that optimal linear-optical operations on the signal and channel modes improve the fidelity and allow it to overcome the classical threshold at any degree of squeezing~\cite{Fiurasek2001}. However, CV QT can be used for efficient noiseless linear amplification of coherent states without the need for a single-photon addition~\cite{Fiurasek2022a}. Recently, a heralded quantum teleporter with $-6.5$~dB of squeezing and measurement-based NLA was suggested by~\cite{Zhao2023}, demonstrating the fidelity of $92\%$ at the cost of success rate of $10^{-5}$. Limitations and possibilities of the NLA-based scheme were analyzed by~\cite{Fiurasek2024}. Beyond the optical domain, CV QT and entanglement swapping were extended to mechanical modes such as the vibrational modes of a mirror~\cite{Pirandola2003,Pirandola2006a}. 

As an alternative to employing measurement and feedforward using, respectively, homodyne detectors and electro-optical modulators, CV QT can be performed in an all-optical setup using a PIA fed by an EPR state or properly pumped PSAs in an interferometric scheme~\cite{Ralph1999a}, enabling multiplexed realizations of CV QT~\cite{Liu2020a, Liu2024}  and entanglement swapping~\cite{Liu2022b}.

CV QT was extended in yet other ways. An important multipartite extension was the theoretical development of CV quantum teleportation networks, as proposed by~\cite{Loock2000} and as experimentally tested by~\cite{Yonezawa2004,Yan2024,Wu2024}. Another interesting variant is port-based teleportation, which was originally developed for DV~\cite{Ishizaka2008,Ishizaka2009} and was later extended to CV~\cite{Pereira2023}. In this variant the parties exploit a multicopy entanglement source and a multisystem measurement at Alice's side. The scheme is built in a way that Bob does not need to apply any unitary correction but instead needs to choose a system (or port) for the output state.   

As described by~\cite{Pirandola2015b}, one can consider four conditions to be met to have an ideal teleportation system: high efficiency (related to the practical success probability of Alice's detection), high average fidelity of the teleported states, distance of teleportation, and, finally, the availability of a quantum memory (to store the teleported states). In this context CV QT with optical modes represents one of the best choices in terms of efficiency ($100\%$) and fidelity ($>80\%$). However, challenges remain in terms of extending distance and providing an efficient interface into a quantum memory.

\subsection{GKP encoding}
CV quantum states and coherent detection can also be used for encoding and manipulating DV states, hence allowing CV-based operations with DV quantum information. The Gottesman-Kitaev-Preskill (GKP) encoding, which encodes a qudit (state of a finite $d$-dimensional quantum system; a qubit in the case of $d=2$) into a state of an infinite-dimensional quantum oscillator described by the field quadratures, was suggested by~\cite{Gottesman2001a}. It was shown that such encoding, which is possible with the CV states with finite squeezing, enables efficient error correction based on conversion of a photon loss by means of a PIA to a correctable Gaussian noise~\cite{Albert2018}. This makes the GKP codes particularly useful for fault-tolerant quantum computation.

In the context of quantum communication, the GKP codes were suggested as the basis for efficient quantum repeaters for qubit~\cite{Fukui2021, Rozpedek2021} and qudit states~\cite{Schmidt2024} using either one- or two-way quantum communication and possibly combining CV and DV error-correction codes. Particularly, it was shown that PSA is sufficient for converting losses to Gaussian noise in a two-way protocol and can be applied in the postprocessing~\cite{Fukui2021} Furthermore, it was shown that amplification is not needed for correcting channel losses with GKP codes and can be harmful in the relevant parameter regimes~\cite{Hastrup2023}. Recently, GKP encoding was realized on the propagating light at the telecom wavelength~\cite{Konno2024}, which opens the pathway to implementation of GKP-based CV quantum repeaters for DV quantum communication.

\section{Summary and outlook}
This review covers the fascinating field of continuous-variable quantum communication, which develops and studies methods for information transfer using quantum systems defined on the infinite-dimensional phase space. 

    \begin{table*}[!]
    \centering
    \begin{tabular}{|p{4cm}|p{2cm}|p{2cm}|p{2.2cm}|p{2.2cm}|p{5cm}|}
    \hline
     \leavevmode\newline\centering\textbf{Protocol} & \leavevmode\newline\centering\textbf{Signal} & \leavevmode\newline\centering\textbf{Modulation} & \centering\vspace*{\fill}
\textbf{Bob's}\newline\textbf{measurement}\vspace*{\fill} & \centering\textbf{Equivalent}\newline\textbf{Alice's}\newline\textbf{preparation} & \leavevmode\newline\centering\arraybackslash\textbf{Best known security proof(s)} \\
    \hhline{|=|=|=|=|=|=|}
       
       Squeezed-state \newline~\cite{Cerf2001}  & squeezed & Gaussian & homodyne & homodyne & composable \newline~\cite{Furrer2012,Furrer2014} \\ 
       \hline
       
       GG02 \newline~\cite{Grosshans2002}  & coherent & Gaussian & homodyne & heterodyne & collective composable \newline~\cite{Pirandola2024} \\
       \hline

              No-switching \newline~\cite{Weedbrook2004} & coherent & Gaussian & heterodyne & heterodyne & composable \newline~\cite{Leverrier2015} \\
       \hline

       Thermal \newline~\cite{Filip2008} & thermal & Gaussian & homodyne & heterodyne + noise & 
	asymptotic collective \newline~\cite{Filip2008} \newline finite-size collective linear \newline~\cite{Leverrier2010} \\
       \hline

        Two-way \newline~\cite{Pirandola2008a} & coherent/\newline squeezed & Gaussian &  homodyne/\newline heterodyne & homodyne/\newline heterodyne & 
	asymptotic collective \newline~\cite{Pirandola2008a} \newline composable \newline~\cite{Ghorai2019} \\
       \hline

      ~\cite{Garcia2009} & squeezed & Gaussian & heterodyne & homodyne & 
	asymptotic collective \newline~\cite{Garcia2009} \newline finite-size collective linear \newline~\cite{Leverrier2010} \\
       \hline

       DM \newline~\cite{Ralph1999,Leverrier2011} & coherent & discrete & homodyne/\newline heterodyne & \~heterodyne & 
	collective composable \newline~\cite{Kanitschar2023} \newline composable for QPSK+heterodyne \newline~\cite{Baeuml2023}  \\

       \hline

      ~\cite{Weedbrook2012} & thermal & Gaussian & heterodyne & heterodyne + noise & 
	asymptotic collective \newline~\cite{Weedbrook2012} \newline finite-size collective linear \newline~\cite{Leverrier2010} \\
       \hline

     ~\cite{Madsen2012} & squeezed & Gaussian & homodyne & homodyne + modulation & 
	asymptotic collective \newline~\cite{Madsen2012} \newline finite-size collective linear \newline~\cite{Leverrier2010} \\
       \hline

      ~\cite{Fiurasek2012} & coherent/\newline squeezed & Gaussian & homodyne/\newline heterodyne \newline + Gaussian \newline postselection & homodyne/\newline heterodyne & 
	asymptotic collective \newline~\cite{Fiurasek2012} \newline collective for coherent-state \newline \& heterodyne \newline~\cite{Hosseinidehaj2020} \\

       \hline

        Entanglement-in-the-middle \newline~\cite{Weedbrook2013} & entangled & - & homodyne/\newline heterodyne & - & finite-size \newline~\cite{Guo2019} \\
       \hline

       Unidimensional \newline~\cite{Usenko2015} & coherent & Gaussian 1D & homodyne & heterodyne + squeezing & 
	asymptotic collective \newline~\cite{Usenko2015} \newline finite-size collective linear \newline~\cite{Leverrier2010} \\
       \hline

      ~\cite{Usenko2018} & squeezed & Gaussian 1D & homodyne & homodyne + squeezing & 
	asymptotic collective \newline~\cite{Usenko2018} \newline finite-size collective linear \newline~\cite{Leverrier2010} \\
       \hline

    Thermal DM \newline~\cite{Papanastasiou2021} & thermal & discrete & heterodyne & \~heterodyne +noise & 
	collective composable \newline~\cite{Papanastasiou2021} \\

       \hline

    \end{tabular}
    \caption{Major trusted-device (non-MDI) CV-QKD protocols and their modifications (in chronological order), used signal states, modulation profiles, receiver (Bob's) measurements, equivalent sender (Alice's) state-preparing measurements in the EB representation plus additional operations on Bob's mode if needed, and the best known security proofs (asymptotic or finite-size unless specified, collective or general attacks unless specified, assumption of channel linearity, proof composability)}
    \label{tab:protocols}
\end{table*}

Using information encoding into the continuous quadrature observables and its efficient retrieval by means of coherent detection, continuous-variable quantum communication has developed into a widespread quantum technology that complements discrete-variable methods with high efficiency at the cost of higher sensitivity to effects of the environment. Furthermore, the elegant covariance-matrix formalism, which is directly applicable to the large class of Gaussian states typically produced in the quantum-optical laboratories, allows for efficient characterization of continuous-variable states and operations as well as the entropic properties that are relevant for communication tasks. These features make continuous-variable quantum communication a great tool in numerous applications, starting with the well-known secure quantum key distribution. The review is largely focused on this area of quantum technology, presenting the major protocols, their security proofs, and methods of their practical realization and discussing security aspects of the implementations, which differ from the idealized theoretical models. 

\begin{figure*}
    \centering
    \includegraphics[width=0.9\linewidth]{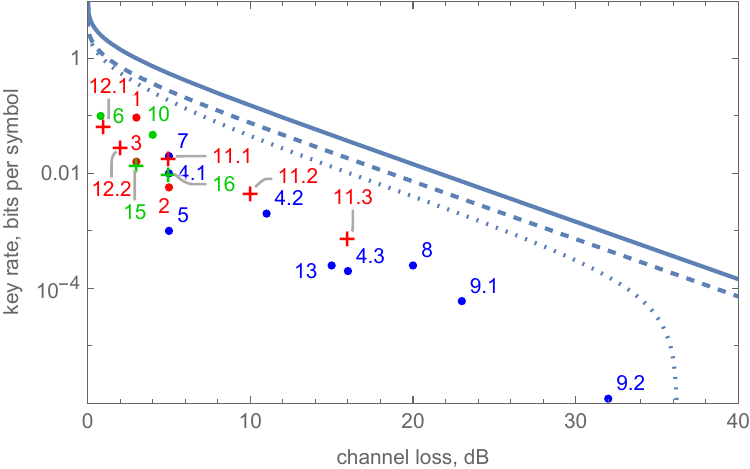}
    \caption{Secret key rates in bits per symbol (bits per channel use) vs respective channel loss in decibels achieved in the milestone CV-QKD demonstrations as summarized in Table \ref{tab:tests} and numbered accordingly. Red points, asymptotic security; blue points, finite-size security; green points, composable finite-size security. Circles, Gaussian modulation; crosses with call-out labels, discrete modulation. Solid line, PLOB bound~\cite{Pirandola2017}; dashed line, key rate for GG02 protocol~\cite{Grosshans2002} with perfect postprocessing ($\beta=1$) and strong modulation ($V_M=10^3$ SNUs) in a purely lossy (noiseless) channel; and dotted line, GG02 protocol with practical error-correction efficiency $\beta=95\%$, optimized modulation (of the order of few SNUs), and excess noise on the channel output $\nu_\text{out}=10^{-5}$ SNUs.}
    \label{fig:KRplot}
\end{figure*}

We summarize the quantum key distribution protocols based on continuous variables in Table~\ref{tab:protocols}, which covers state generation, modulation, and measurement techniques; equivalent EB representations for security analysis; and the best-known security proofs. Similarly, we summarize the milestone realizations of the protocols in Table~\ref{tab:tests}, presenting the protocol types, security proofs, achieved key rates, system clock rates, and error-correction efficiencies. Table~\ref{tab:tests} illustrates advances in practical CV QKD since its first experimental test in 2003, in terms of both achievable key rates and distances. The achieved key rates, normalized to the clock rates, are  plotted in Fig.~\ref{fig:KRplot} along with the fundamental PLOB bound (the solid line), providing the secret key capacity of the lossy channel and the asymptotic key rates for GG02 in a perfect noiseless realization (the dashed lines) and with finite postprocessing efficiency, optimized modulation, and the presence of small channel noise (the dotted line). It is evident from Fig.~\ref{fig:KRplot} that progress in the experimental techniques (particularly high clock rates providing large data ensembles and high system stability and efficient data processing resulting in lower excess noise levels) allows the security assumptions (from asymptotic to composable finite-size security) to be strengthened without the overall performance being degraded. However, there is still an evident gap between the practical performance of the protocols and the theoretical bounds, especially at strong channel loss levels, which requires experimental efforts toward even higher stability of the systems and efficient noise compensation methods, as well as theoretical developments, particularly for the tight practical security of discrete-modulation protocols. It is now clear that the development of practical continuous-variable quantum key distribution protocols will largely continue in the direction of discrete-modulation schemes, allowing for much higher repetition rates, while Gaussian-modulated protocols will remain important benchmarks as well as test beds for advanced tools. Yet, the choice of the best practical continuous-variable quantum key distribution protocol remains an open question, especially as its performance strongly depends on the practical security proofs and postprocessing techniques that are being developed~\cite{Mario2018,Mountogiannakis2022}.

    \begin{table*}[!]
    \centering
    \begin{tabular}{|p{4cm}|p{2cm}|p{2cm}|p{2.2cm}|p{2.2cm}|p{2.2cm}|p{1.5cm}|}
    \hline
     \leavevmode\centering\textbf{Reference} & \leavevmode\centering\textbf{Protocol} & 
     \leavevmode\centering\textbf{Security} 
    & \leavevmode\centering\textbf{Clock}
     & \leavevmode\centering\textbf{Keyrate} & \leavevmode\centering\textbf{Channel} 
     & \leavevmode\centering\arraybackslash$\beta$ \\
    \hhline{|=|=|=|=|=|=|=|}

       1.~\cite{Grosshans2003}  & GG02 & individual, asymptotic & 0.8 MHz & 75 Kbps & - / 3~dB & - \\ 
       \hline

       2.~\cite{Lodewyck2007}  & GG02 & collective, asymptotic & 0.35 MHz & 2 Kbps & 25~km / 5~dB & 90\% \\ 
       \hline

       3.~\cite{Fossier2009}  & GG02 & collective, asymptotic & 0.5 MHz & 8 Kbps & 15~km / 3~dB  & 90\% \\ 
       \hline

       4.~\cite{Jouguet2013}  & GG02 & collective, finite-size & 1 MHz & 
		10 Kbps \newline 2 Kbps \newline 0.2 Kbps & 
		25~km / 5~dB \newline 53~km / 11~dB \newline 80~km / 16~dB
	& 95\% \\ 
       \hline

       5.~\cite{Huang2015a}  & GG02+LLO & collective, finite-size & 100 MHz & 
		100 Kbps & 25~km / 5~dB 
	& 97\% \\ 
       \hline

       6.~\cite{Gehring2015} & squeezed & coherent, composable, finite-size & 100 kHz & 10 Kbps & - / 0.8~dB & 94.3\% \\
       \hline

       7.~\cite{Huang2015}  & GG02 & collective, finite-size & 50 MHz & 
		1 Mbps & 25~km / 5~dB & 93\% \\ 
       \hline

       8.~\cite{Huang2016}  & GG02 & collective, finite-size & 2 MHz & 
		0.5 Kbps & 100~km / 20~dB & 95.6\% \\ 
       \hline

       9.~\cite{Zhang2020}  & GG02 & collective, finite-size & 5 MHz & 
		0.3 Kbps \newline 6 bps & 140~km / 23~dB \newline 200~km / 32~dB 
	& 96\% \newline 98\% \\ 
       \hline

        10.~\cite{Jain2022}  & GG02+LLO & collective, composable, finite-size & 100 MHz & 
		4.7 Mbps & 20~km / 4~dB 
	& 94.3\% \\ 
       \hline

       11.~\cite{Tian2023}  & DM+LLO & collective, asymptotic & 2.5 GHz & 
		49 Mbps \newline 12 Mbps \newline 2 Mbps & 25~km / 5~dB \newline 50~km / 10~dB \newline 80~km / 16~dB
	& 95\% \\ 
       \hline

       12.~\cite{Hajomer2023}  & DM+LLO & collective, asymptotic & 10 GHz & 
		0.7 Gbps \newline 0.3 Gbps & 5~km / 1~dB \newline 10~km / 2~dB 
	& 95\% \\ 
       \hline

       13.~\cite{Hajomer2024}  & GG02+LLO & collective, finite-size & 100 MHz & 
		25 Kbps & 100~km / 15~dB 
	& 92.5\% \\ 
       \hline

       14.~\cite{Roumestan2024}  & DM & collective, finite-size & 1 GHz & 
		92 Mbps \newline 24 Mbps & 10~km / 2~dB \newline 25~km / 5~dB 	& - \\
       \hline

       15.~\cite{Jaksch2024} & DM & collective, composable, finite-size & 25 MHz & 375 Kbps & - / 3~dB & 89\%\\
       \hline

       16.~\cite{Hajomer2025a} & DM+LLO & collective, composable, finite-size & 125 MHz & 1.3 Mbps & 20~km / 5~dB & 87.8\% \\
        \hline
 
       \hline

    \end{tabular}
    \caption{Milestone trusted-device (non-MDI) CV-QKD tests and implementations (in chronological order), used protocols as referred to in Table \ref{tab:protocols} (LLO indicated where used), used security proofs, system clock rates, achieved key rates, channel distances in kilometers / loss levels in~dB, and error-correction efficiency $\beta$.}
    \label{tab:tests}
\end{table*}

Despite the attention given to quantum key distribution, continuous-variable quantum communication goes well beyond this field and includes such specific quantum communication tasks as quantum direct communication, quantum superdense coding, quantum digital signatures, quantum authentication, and quantum oblivious transfer. These tasks, although less mature than quantum key distribution, are only part of the plethora of novel communication methods offered by continuous-variable quantum technology. Many more tasks will likely be developed with the advances of quantum-optical technologies, while the existing tasks will proceed to full-scale implementations, in-field tests, and possible further industrialization, as in the case of quantum key distribution. Continuous-variable quantum teleportation, which has demonstrated impressive theoretical and experimental progress in the past 25 years, will benefit from long-distance implementation and hybrid integration with other systems capable of quantum storage, for example, atomic ensembles~\cite{Sherson2006}.

In our review we have limited the discussion to bipartite quantum communication (including third parties only in the protocol design, as in the case of quantum digital signatures). However, continuous-variable quantum communication is naturally developing toward quantum networking, for example, with the recent results in multiuser quantum key distribution~\cite{hajomer2024continuous} and the generation of multipartite entanglement~\cite{Asavanant2024} for continuous-variable teleportation networks. In parallel with this development, continuous-variable quantum communication will benefit from methods aimed at improving its efficiency, such as quantum repeaters and multiplexing, as well as from technological advances such as photonic integration and hybrid and satellite-based realizations aimed at reducing the cost and making the technology more applicable and broadly available. Continuous-variable quantum communication remains an active field of research, despite its immense progress in the recent decades, and we hope that this review will encourage and stimulate this development.

\section*{Acknowledgments}
We acknowledge our insightful discussions with Radim Filip, Anthony Leverrier, Norbert L\"utkenhaus, and Christoph Marquardt and thank Jarom{\'\i}r Fiur{\'a}{\v{s}}ek, Cameron Foreman, Andrew Lance, Gui-Lu Long, Andrey Rakhubovsky, Lev Vaidman, and Qi Wu for their feedback on the manuscript. 
This project has received funding from the European Union's Horizon Europe research and innovation programme under the project Quantum Secure Networks Partnership (QSNP, grant agreement No 101114043). V.C.U. acknowledges project No. 21-44815L of the Czech Science Foundation and project No. CZ.02.01.01/00/22\_008/0004649 (QUEENTEC) of the Czech Ministry of Education, Youth and Sports (MEYS). A.A.E.H., T.G., and U.L.A. acknowledge projects CryptQ No. 0175-00018A and CyberQ No. 3200-00035A of Innovation Fund Denmark, projects bigQ and DNRF142 of DNRF, and project No. 0171-00055B of Danmarks Frie Forskningsfond. V.C.U., A.A.E.H., T.G., U.L.A., C.P., and F.K. acknowledge QuantERA project CVSTAR, which has received funding from the European Union's Horizon 2020 Research and Innovation Programme under Grant Agreement No. 731473, the Czech MEYS under project No. 8C22002; Innovation Fund Denmark under Grant Agreement No. 101017733, and the Austrian Research Promotion Agency under project No. FO999891361. E.D. and R.A. acknowledge support from the PEPR integrated project QCommTestbed (grant No. ANR-22-PETQ-0011), which is part of Plan France 2030. A.A. acknowledges support from the Government of Spain (Severo Ochoa grant No. CEX2019-000910-S, NextGenerationEU grant No. PRTR-C17.I1 and FUNQIP), Fundaci{\'o} Cellex, Fundaci{\'o} Mir-Puig, Generalitat de Catalunya (CERCA program),  the ERC AdG CERQUTE, and the AXA Chair in Quantum Information Science. S.P. acknowledges support from EPSRC and UKRI via grants No. EP/M013472/1 and No. EP/T001011/1 and from the Integrated Quantum Networks Research Hub (IQN, grant No. EP/Z533208/1).

\bibliographystyle{apsrmp}
\bibliography{CV_QC-review}

@Article{Acin2007,
  Title                    = {Device-independent security of quantum cryptography against collective attacks},
  Author                   = {Ac{\'\i}n, Antonio and Brunner, Nicolas and Gisin, Nicolas and Massar, Serge and Pironio, Stefano and Scarani, Valerio},
  Journal                  = {Physical Review Letters},
  Year                     = {2007},
  Number                   = {23},
  Pages                    = {230501},
  Volume                   = {98},
  Publisher                = {APS}
}

@InProceedings{Bennett1984,
  Title                    = {Quantum cryptography: public key distribution and coin tossing},
  Author                   = {Bennett, Charles H and Brassard, Gilles},
  Booktitle                = {Proceedings of International Conference on Computers, Systems and Signal Processing},
  Year                     = {1984},

  Address                  = {Bangalore, India},
  Organization             = {IEEE},
  Pages                    = {9},
  Volume                   = {175}
}

@Article{Blandino2012,
  Title                    = {Improving the maximum transmission distance of continuous-variable quantum key distribution using a noiseless amplifier},
  Author                   = {Blandino, R{\'e}mi and Leverrier, Anthony and Barbieri, Marco and Etesse, Jean and Grangier, Philippe and Tualle-Brouri, Rosa},
  Journal                  = {Physical Review A},
  Year                     = {2012},
  Number                   = {1},
  Pages                    = {012327},
  Volume                   = {86},
  Publisher                = {APS}
}

@Article{Braunstein1998,
  Title                    = {Teleportation of continuous quantum variables},
  Author                   = {Braunstein, Samuel L and Kimble, H Jeff},
  Journal                  = {Physical Review Letters},
  Year                     = {1998},
  Number                   = {4},
  Pages                    = {869},
  Volume                   = {80},
  Publisher                = {APS}
}

@Article{Braunstein2012,
  Title                    = {Side-channel-free quantum key distribution},
  Author                   = {Braunstein, Samuel L and Pirandola, Stefano},
  Journal                  = {Physical Review Letters},
  Year                     = {2012},
  Number                   = {13},
  Pages                    = {130502},
  Volume                   = {108},
  Publisher                = {APS}
}

@Article{Braunstein2005,
  Title                    = {Quantum information with continuous variables},
  Author                   = {Braunstein, Samuel L and Van Loock, Peter},
  Journal                  = {Reviews of Modern Physics},
  Year                     = {2005},
  Number                   = {2},
  Pages                    = {513},
  Volume                   = {77},
  Publisher                = {APS}
}

@Article{Cerf2001,
  Title                    = {Quantum distribution of Gaussian keys using squeezed states},
  Author                   = {Cerf, Nicolas J and Levy, Marc and Van Assche, Gilles},
  Journal                  = {Physical Review A},
  Year                     = {2001},
  Number                   = {5},
  Pages                    = {052311},
  Volume                   = {63},
  Publisher                = {APS}
}

@Article{Chin2021,
author={Chin, Hou-Man
and Jain, Nitin
and Zibar, Darko
and Andersen, Ulrik L.
and Gehring, Tobias},
title={Machine learning aided carrier recovery in continuous-variable quantum key distribution},
journal={npj Quantum Information},
year={2021},
month={Feb},
day={04},
volume={7},
number={1},
pages={20},
doi={10.1038/s41534-021-00361-x}
}

@article{Chin2022,
doi = {10.1088/2058-9565/ac7ba2},
year = {2022},
month = {jul},
volume = {7},
number = {4},
pages = {045006},
author = {Hou-Man Chin and Nitin Jain and Ulrik L Andersen and Darko Zibar and Tobias Gehring},
title = {Digital synchronization for continuous-variable quantum key distribution},
journal = {Quantum Science and Technology}
}

@Article{Christandl2009,
  Title                    = {Postselection technique for quantum channels with applications to quantum cryptography},
  Author                   = {Christandl, Matthias and K{\"o}nig, Robert and Renner, Renato},
  Journal                  = {Physical Review Letters},
  Year                     = {2009},
  Number                   = {2},
  Pages                    = {020504},
  Volume                   = {102},
  Publisher                = {APS}
}

@Article{Clauser1969,
  Title                    = {Proposed experiment to test local hidden-variable theories},
  Author                   = {Clauser, John F and Horne, Michael A and Shimony, Abner and Holt, Richard A},
  Journal                  = {Physical Review Letters},
  Year                     = {1969},
  Number                   = {15},
  Pages                    = {880},
  Volume                   = {23},
  Publisher                = {APS}
}

@Article{Csiszar1978,
  Title                    = {Broadcast channels with confidential messages},
  Author                   = {Csisz{\'a}r, Imre and K{\"o}rner, Janos},
  Journal                  = {Information Theory, IEEE Transactions on},
  Year                     = {1978},
  Number                   = {3},
  Pages                    = {339--348},
  Volume                   = {24},
  Publisher                = {IEEE}
}

@Article{Devetak2005,
  Title                    = {Distillation of secret key and entanglement from quantum states},
  Author                   = {Devetak, Igor and Winter, Andreas},
  Journal                  = {Proceedings of the Royal Society A: Mathematical, Physical and Engineering Science},
  Year                     = {2005},
  Number                   = {2053},
  Pages                    = {207--235},
  Volume                   = {461},
  Publisher                = {The Royal Society}
}

@Article{Ekert1991,
  Title                    = {Quantum cryptography based on Bell's theorem},
  Author                   = {Ekert, Artur K},
  Journal                  = {Physical Review Letters},
  Year                     = {1991},
  Number                   = {6},
  Pages                    = {661},
  Volume                   = {67},
  Publisher                = {APS}
}

@Article{Filip2008,
  Title                    = {Continuous-variable quantum key distribution with noisy coherent states},
  Author                   = {Filip, Radim},
  Journal                  = {Physical Review A},
  Year                     = {2008},
  Number                   = {2},
  Pages                    = {022310},
  Volume                   = {77},
  Publisher                = {APS}
}

@Article{Fiurasek2001,
  Title                    = {Optical implementation of continuous-variable quantum cloning machines},
  Author                   = {Fiur{\'a}{\v{s}}ek, Jarom{\'\i}r},
  Journal                  = {Physical Review Letters},
  Year                     = {2001},
  Number                   = {21},
  Pages                    = {4942},
  Volume                   = {86},
  Publisher                = {APS}
}

@Article{Fiurasek2012,
  Title                    = {Gaussian postselection and virtual noiseless amplification in continuous-variable quantum key distribution},
  Author                   = {Fiur{\'a}{\v{s}}ek, Jarom{\'\i}r and Cerf, Nicolas J},
  Journal                  = {Physical Review A},
  Year                     = {2012},
  Number                   = {6},
  Pages                    = {060302},
  Volume                   = {86},
  Publisher                = {APS}
}

@Article{Fossier2009a,
  Title                    = {Improvement of continuous-variable quantum key distribution systems by using optical preamplifiers},
  Author                   = {Fossier, Simon and Diamanti, Eleni and Debuisschert, Thierry and Tualle-Brouri, Rosa and Grangier, Philippe},
  Journal                  = {Journal of Physics B},
  Year                     = {2009},
  Number                   = {11},
  Pages                    = {114014},
  Volume                   = {42}
}

@Article{Fossier2009,
  Title                    = {Field test of a continuous-variable quantum key distribution prototype},
  Author                   = {Fossier, Simon and Diamanti, Eleni and Debuisschert, Thierry and Villing, Andr{\'e} and Tualle-Brouri, Rosa and Grangier, Philippe},
  Journal                  = {New Journal of Physics},
  Year                     = {2009},
  Number                   = {4},
  Pages                    = {045023},
  Volume                   = {11}
}

@Article{Furrer2012,
  Title                    = {Continuous variable quantum key distribution: Finite-key analysis of composable security against coherent attacks},
  Author                   = {Furrer, Fabian and Franz, Torsten and Berta, Mario and Leverrier, Anthony and Scholz, Volkher B and Tomamichel, Marco and Werner, Reinhard F},
  Journal                  = {Physical Review Letters},
  Year                     = {2012},
  Number                   = {10},
  Pages                    = {100502},
  Volume                   = {109},
  Publisher                = {APS}
}

@Article{Garcia2009,
  author    = {Garc\'ia-Patr\'on, Ra\'ul and Cerf, Nicolas J},
  title     = {Continuous-variable quantum key distribution protocols over noisy channels},
  journal   = {Physical Review Letters},
  year      = {2009},
  volume    = {102},
  number    = {13},
  pages     = {130501},
  publisher = {APS},
}

@Article{Garcia2006,
  author    = {Garc\'ia-Patr\'on, Ra\'ul and Cerf, Nicolas J},
  title     = {Unconditional optimality of Gaussian attacks against continuous-variable quantum key distribution},
  journal   = {Physical Review Letters},
  year      = {2006},
  volume    = {97},
  number    = {19},
  pages     = {190503},
  publisher = {APS},
}

@Article{Gisin2002,
  Title                    = {Quantum cryptography},
  Author                   = {Gisin, Nicolas and Ribordy, Gr{\'e}goire and Tittel, Wolfgang and Zbinden, Hugo},
  Journal                  = {Reviews of Modern Physics},
  Year                     = {2002},
  Number                   = {1},
  Pages                    = {145},
  Volume                   = {74},
  Publisher                = {APS}
}

@Article{Gottesman2001,
  Title                    = {Secure quantum key distribution using squeezed states},
  Author                   = {Gottesman, Daniel and Preskill, John},
  Journal                  = {Physical Review A},
  Year                     = {2001},

  Month                    = {Jan},
  Number                   = {2},
  Pages                    = {022309},
  Volume                   = {63},

  Doi                      = {10.1103/physreva.63.022309},
  Publisher                = {American Physical Society (APS)}
}

@Article{Grosshans2005,
  Title                    = {Collective attacks and unconditional security in continuous variable quantum key distribution},
  Author                   = {Grosshans, Fr{\'e}d{\'e}ric},
  Journal                  = {Physical Review Letters},
  Year                     = {2005},
  Number                   = {2},
  Pages                    = {020504},
  Volume                   = {94},
  Publisher                = {APS}
}

@Article{Grosshans2003a,
  author     = {Grosshans, Fr{\'e}d{\'e}ric and Cerf, Nicolas J. and Wenger, J{\'e}r\^{o}me and Tualle-Brouri, Rosa and Grangier, Philippe},
  title      = {Virtual entanglement and reconciliation protocols for quantum cryptography with continuous variables},
  journal    = {Quantum Info. Comput.},
  year       = {2003},
  volume     = {3},
  number     = {7},
  pages      = {535--552},
  month      = oct,
  acmid      = {2011570},
  address    = {Paramus, NJ},
  issue_date = {October 2003},
  numpages   = {18},
  publisher  = {Rinton Press, Incorporated},
}

@Article{Grosshans2002,
  Title                    = {Continuous variable quantum cryptography using coherent states},
  Author                   = {Grosshans, Fr{\'e}d{\'e}ric and Grangier, Philippe},
  Journal                  = {Physical Review Letters},
  Year                     = {2002},
  Number                   = {5},
  Pages                    = {057902},
  Volume                   = {88},
  Publisher                = {APS}
}

@Article{Grosshans2003,
  Title                    = {Quantum key distribution using Gaussian-modulated coherent states},
  Author                   = {Grosshans, Fr{\'e}d{\'e}ric and Van Assche, Gilles and Wenger, J{\'e}r{\^o}me and Brouri, Rosa and Cerf, Nicolas J and Grangier, Philippe},
  Journal                  = {Nature},
  Year                     = {2003},
  Number                   = {6920},
  Pages                    = {238--241},
  Volume                   = {421},
  Publisher                = {Nature Publishing Group}
}

@Article{Heid2006,
  Title                    = {Efficiency of coherent-state quantum cryptography in the presence of loss: Influence of realistic error correction},
  Author                   = {Heid, Matthias and L{\"u}tkenhaus, Norbert},
  Journal                  = {Physical Review A},
  Year                     = {2006},
  Number                   = {5},
  Pages                    = {052316},
  Volume                   = {73},
  Publisher                = {APS}
}

@Article{Hillery2000,
  Title                    = {Quantum cryptography with squeezed states},
  Author                   = {Hillery, Mark},
  Journal                  = {Physical Review A},
  Year                     = {2000},
  Number                   = {2},
  Pages                    = {022309},
  Volume                   = {61},
  Publisher                = {APS}
}

@Article{Huang2014,
  Title                    = {Quantum hacking on quantum key distribution using homodyne detection},
  Author                   = {Huang, Jing-Zheng and Kunz-Jacques, S{\'e}bastien and Jouguet, Paul and Weedbrook, Christian and Yin, Zhen-Qiang and Wang, Shuang and Chen, Wei and Guo, Guang-Can and Han, Zheng-Fu},
  Journal                  = {Physical Review A},
  Year                     = {2014},
  Number                   = {3},
  Pages                    = {032304},
  Volume                   = {89},
  Publisher                = {APS}
}

@Article{Huang2013,
  Title                    = {Quantum hacking of a continuous-variable quantum-key-distribution system using a wavelength attack},
  Author                   = {Huang, Jing-Zheng and Weedbrook, Christian and Yin, Zhen-Qiang and Wang, Shuang and Li, Hong-Wei and Chen, Wei and Guo, Guang-Can and Han, Zheng-Fu},
  Journal                  = {Physical Review A},
  Year                     = {2013},
  Number                   = {6},
  Pages                    = {062329},
  Volume                   = {87},
  Publisher                = {APS}
}

@Article{Jouguet2014,
  Title                    = {High performance error correction for quantum key distribution using polar codes},
  Author                   = {Jouguet, Paul and Kunz-Jacques, S{\'e}bastien},
  Journal                  = {Quantum Information \& Computation},
  Year                     = {2014},
  Number                   = {3-4},
  Pages                    = {329--338},
  Volume                   = {14},
  Publisher                = {Rinton Press, Incorporated}
}

@Article{Jouguet2013a,
  Title                    = {Preventing calibration attacks on the local oscillator in continuous-variable quantum key distribution},
  Author                   = {Jouguet, Paul and Kunz-Jacques, S{\'e}bastien and Diamanti, Eleni},
  Journal                  = {Physical Review A},
  Year                     = {2013},
  Number                   = {6},
  Pages                    = {062313},
  Volume                   = {87},
  Publisher                = {APS}
}

@Article{Jouguet2012,
  Title                    = {Analysis of imperfections in practical continuous-variable quantum key distribution},
  Author                   = {Jouguet, Paul and Kunz-Jacques, S{\'e}bastien and Diamanti, Eleni and Leverrier, Anthony},
  Journal                  = {Physical Review A},
  Year                     = {2012},
  Number                   = {3},
  Pages                    = {032309},
  Volume                   = {86},
  Publisher                = {APS}
}

@Article{Jouguet2013,
  Title                    = {Experimental demonstration of long-distance continuous-variable quantum key distribution},
  Author                   = {Jouguet, Paul and Kunz-Jacques, S{\'e}bastien and Leverrier, Anthony and Grangier, Philippe and Diamanti, Eleni},
  Journal                  = {Nature Photonics},
  Year                     = {2013},
  Number                   = {5},
  Pages                    = {378--381},
  Volume                   = {7},
  Publisher                = {Nature Publishing Group}
}

@Article{Kunz2015,
  Title                    = {Robust shot-noise measurement for continuous-variable quantum key distribution},
  Author                   = {Kunz-Jacques, S{\' e}bastien and Jouguet, Paul},
  Journal                  = {Physical Review A},
  Year                     = {2015},

  Month                    = {Feb},
  Number                   = {2},
  Pages                    = {022307},
  Volume                   = {91},

  Doi                      = {10.1103/physreva.91.022307},
  Publisher                = {American Physical Society (APS)}
}

@Article{Leverrier2010,
  Title                    = {Finite-size analysis of a continuous-variable quantum key distribution},
  Author                   = {Leverrier, Anthony and Grosshans, Fr{\'e}d{\'e}ric and Grangier, Philippe},
  Journal                  = {Physical Review A},
  Year                     = {2010},
  Number                   = {6},
  Pages                    = {062343},
  Volume                   = {81},
  Publisher                = {APS}
}

@Article{Lo2012,
  Title                    = {Measurement-device-independent quantum key distribution},
  Author                   = {Lo, Hoi-Kwong and Curty, Marcos and Qi, Bing},
  Journal                  = {Physical Review Letters},
  Year                     = {2012},
  Number                   = {13},
  Pages                    = {130503},
  Volume                   = {108},
  Publisher                = {APS}
}

@Article{Lodewyck2007,
  Title                    = {Quantum key distribution over 25 km with an all-fiber continuous-variable system},
  Author                   = {Lodewyck, J{\'e}r{\^o}me and Bloch, Matthieu and Garc{\'\i}a-Patr{\'o}n, Ra{\'u}l and Fossier, Simon and Karpov, Evgueni and Diamanti, Eleni and Debuisschert, Thierry and Cerf, Nicolas J and Tualle-Brouri, Rosa and McLaughlin, Steven W and others},
  Journal                  = {Physical Review A},
  Year                     = {2007},
  Number                   = {4},
  Pages                    = {042305},
  Volume                   = {76},
  Publisher                = {APS}
}

@article{Luo2024,
	title = {An {Overview} of {Postprocessing} in {Quantum} {Key} {Distribution}},
	volume = {12},
	doi = {10.3390/math12142243},
	language = {en},
	number = {14},

	journal = {Mathematics},
	author = {Luo, Yi and Cheng, Xi and Mao, Hao-Kun and Li, Qiong},
	month = jan,
	year = {2024},
	pages = {2243},
}

@article{wang2021experimental,
  title={Experimental authentication of quantum key distribution with post-quantum cryptography},
  author={Wang, Liu-Jun and Zhang, Kai-Yi and Wang, Jia-Yong and Cheng, Jie and Yang, Yong-Hua and Tang, Shi-Biao and Yan, Di and Tang, Yan-Lin and Liu, Zhen and Yu, Yu and others},
  journal={npj quantum information},
  volume={7},
  number={1},
  pages={67},
  year={2021},
  publisher={Nature Publishing Group UK London}
}

@InBook{fregona2024authentication,
  author    = {Fregona, Giacomo and De Lazzari, Claudia and Giani, Damiano and Chirici, Fernando and Stocco, Francesco and Signorini, Edoardo and Morgari, Guglielmo and Occhipinti, Tommaso and Zavatta, Alessandro and Bacco, Davide},
  publisher = {IOS Press},
  title     = {Authentication Methods for Quantum Key Distribution: Challenges and Perspectives},
  year      = {2024},
  isbn      = {9781643684994},
  month     = feb,
  booktitle = {Toward a Quantum-Safe Communication Infrastructure},
  doi       = {10.3233/nicsp240007},
}

@article{gilbert1974codes,
  title={Codes which detect deception},
  author={Gilbert, Edgar N and MacWilliams, F Jessie and Sloane, Neil JA},
  journal={Bell System Technical Journal},
  volume={53},
  number={3},
  pages={405--424},
  year={1974},
  publisher={Wiley Online Library}
}

@article{wegman1981new,
  title={New hash functions and their use in authentication and set equality},
  author={Wegman, Mark N and Carter, J Lawrence},
  journal={Journal of computer and system sciences},
  volume={22},
  number={3},
  pages={265--279},
  year={1981},
  publisher={Elsevier}
}

@article{kiktenko2020lightweight,
  title={Lightweight authentication for quantum key distribution},
  author={Kiktenko, Evgeniy O and Malyshev, Aleksei O and Gavreev, Maxim A and Bozhedarov, Anton A and Pozhar, Nikolay O and Anufriev, Maxim N and Fedorov, Aleksey K},
  journal={IEEE Transactions on Information Theory},
  volume={66},
  number={10},
  pages={6354--6368},
  year={2020},
  publisher={IEEE}
}

@inproceedings{abidin2012new,
  title={New universal hash functions},
  author={Abidin, Aysajan and Larsson, Jan-{\AA}ke},
  booktitle={Research in Cryptology: 4th Western European Workshop, WEWoRC 2011, Weimar, Germany, July 20-22, 2011, Revised Selected Papers 4},
  pages={99--108},
  year={2012},
  organization={Springer}
}

@inproceedings{rogaway1995bucket,
  title={Bucket hashing and its application to fast message authentication},
  author={Rogaway, Phillip},
  booktitle={Annual International Cryptology Conference},
  pages={29--42},
  year={1995},
  organization={Springer}
}

@article{stinson1996connections,
  title={On the connections between universal hashing, combinatorial designs and error-correcting codes},
  author={Stinson, Douglas R},
  journal={Congressus Numerantium},
  volume={114},
  pages={7},
  year={1996},
  publisher={Citeseer}
}

@article{frohlich2013quantum,
  title={A quantum access network},
  author={Fr{\"o}hlich, Bernd and Dynes, James F and Lucamarini, Marco and Sharpe, Andrew W and Yuan, Zhiliang and Shields, Andrew J},
  journal={Nature},
  volume={501},
  number={7465},
  pages={69--72},
  year={2013},
  publisher={Nature Publishing Group UK London}
}

@article{mani2021multiedge,
  title={Multiedge-type low-density parity-check codes for continuous-variable quantum key distribution},
  author={Mani, Hossein and Gehring, Tobias and Grabenweger, Philipp and {\"O}mer, Bernhard and Pacher, Christoph and Andersen, Ulrik Lund},
  journal={Physical Review A},
  volume={103},
  number={6},
  pages={062419},
  year={2021},
  publisher={APS}
}

@article{van2004reconciliation,
  title={Reconciliation of a quantum-distributed Gaussian key},
  author={Van Assche, Gilles and Cardinal, Jean and Cerf, Nicolas J},
  journal={IEEE Transactions on Information Theory},
  volume={50},
  number={2},
  pages={394--400},
  year={2004},
  publisher={IEEE}
}

@inproceedings{takahashi2016high,
  title={High-speed implementation of privacy amplification in quantum key distribution},
  author={Takahashi, Ririka and Tanizawa, Yoshimichi and Dixon, Alexander R},
  booktitle={6th Int. Conf. Quantum Cryptography},
  year={2016}
}

@article{yan2022efficient,
  title={An efficient hybrid hash based privacy amplification algorithm for quantum key distribution},
  author={Yan, Bingze and Li, Qiong and Mao, Haokun and Chen, Nan},
  journal={Quantum Information Processing},
  volume={21},
  number={4},
  pages={130},
  year={2022},
  publisher={Springer}
}

@article{leverrier2008multidimensional,
  title={Multidimensional reconciliation for a continuous-variable quantum key distribution},
  author={Leverrier, Anthony and All{\'e}aume, Romain and Boutros, Joseph and Z{\'e}mor, Gilles and Grangier, Philippe},
  journal={Physical Review A},
  volume={77},
  number={4},
  pages={042325},
  year={2008},
  publisher={APS}
}

@article{jouguet2014high,
  title={High-bit-rate continuous-variable quantum key distribution},
  author={Jouguet, Paul and Elkouss, David and Kunz-Jacques, S{\'e}bastien},
  journal={Physical Review A},
  volume={90},
  number={4},
  pages={042329},
  year={2014},
  publisher={APS}
}

@article{zhang2024continuous,
  title={Continuous-variable quantum key distribution system: Past, present, and future},
  author={Zhang, Yichen and Bian, Yiming and Li, Zhengyu and Yu, Song and Guo, Hong},
  journal={Applied Physics Reviews},
  volume={11},
  number={1},
  year={2024},
  publisher={AIP Publishing}
}

@article{bennett1995generalized,
  title={Generalized privacy amplification},
  author={Bennett, Charles H and Brassard, Gilles and Cr{\'e}peau, Claude and Maurer, Ueli M},
  journal={IEEE Transactions on Information theory},
  volume={41},
  number={6},
  pages={1915--1923},
  year={1995},
  publisher={IEEE}
}

@article{weerasinghe2024practical,
  title={Practical, High-Speed, Gaussian Modulated Coherent State Continuous Variable Quantum Key Distribution with Real-Time Post Processing},
  author={Weerasinghe, Weerasinghe},
  year={2024},
  journal={Ph.D. thesis (University of Cambridge)},
url={https://www.repository.cam.ac.uk/handle/1810/375669}
}

@article{tang2019high,
  title={High-speed and large-scale privacy amplification scheme for quantum key distribution},
  author={Tang, Bang-Ying and Liu, Bo and Zhai, Yong-Ping and Wu, Chun-Qing and Yu, Wan-Rong},
  journal={Scientific reports},
  volume={9},
  number={1},
  pages={15733},
  year={2019},
  publisher={Nature Publishing Group UK London}
}

@Article{Ma2014,
  Title                    = {Enhancement of the security of a practical continuous-variable quantum-key-distribution system by manipulating the intensity of the local oscillator},
  Author                   = {Ma, Xiang-Chun and Sun, Shi-Hai and Jiang, Mu-Sheng and Gui, Ming and Zhou, Yan-Li and Liang, Lin-Mei},
  Journal                  = {Physical Review A},
  Year                     = {2014},
  Number                   = {3},
  Pages                    = {032310},
  Volume                   = {89},
  Publisher                = {APS}
}

@Article{Ma2013,
  Title                    = {Wavelength attack on practical continuous-variable quantum-key-distribution system with a heterodyne protocol},
  Author                   = {Ma, Xiang-Chun and Sun, Shi-Hai and Jiang, Mu-Sheng and Liang, Lin-Mei},
  Journal                  = {Physical Review A},
  Year                     = {2013},
  Number                   = {5},
  Pages                    = {052309},
  Volume                   = {87},
  Publisher                = {APS}
}

@Article{Ma2013a,
  Title                    = {Local oscillator fluctuation opens a loophole for Eve in practical continuous-variable quantum-key-distribution systems},
  Author                   = {Ma, Xiang-Chun and Sun, Shi-Hai and Jiang, Mu-Sheng and Liang, Lin-Mei},
  Journal                  = {Physical Review A},
  Year                     = {2013},

  Month                    = {Aug},
  Number                   = {2},
  Pages                    = {022339},
  Volume                   = {88},

  Doi                      = {10.1103/physreva.88.022339},
  Publisher                = {American Physical Society (APS)}
}

@Article{Madsen2012,
  Title                    = {Continuous variable quantum key distribution with modulated entangled states},
  Author                   = {Madsen, Lars S and Usenko, Vladyslav C and Lassen, Mikael and Filip, Radim and Andersen, Ulrik L},
  Journal                  = {Nature Communications},
  Year                     = {2012},
  Pages                    = {1083},
  Volume                   = {3},
  Publisher                = {Nature Publishing Group}
}

@InProceedings{Mayers1998,
  Title                    = {Quantum cryptography with imperfect apparatus},
  Author                   = {Mayers, Dominic and Yao, Adnrew},
  Booktitle                = {Foundations of Computer Science, Palo Alto, 1998. Proceedings. 39th Annual Symposium on},
  Year                     = {1998},

  Address                  = {Washington, DC},
  Organization             = {IEEE},
  Pages                    = {503--509},

  Doi                      = {10.1109/SFCS.1998.743501}
}

@Article{Navascues2005,
  Title                    = {Security bounds for continuous variables quantum key distribution},
  Author                   = {Navascu\'{e}s, Miguel and Ac\'{i}n, Antonio},
  Journal                  = {Physical Review Letters},
  Year                     = {2005},
  Number                   = {2},
  Pages                    = {020505},
  Volume                   = {94},
  Publisher                = {APS}
}

@Article{Navascues2006,
  Title                    = {Optimality of Gaussian attacks in continuous-variable quantum cryptography},
  Author                   = {Navascu{\'e}s, Miguel and Grosshans, Fr{\'e}d{\'e}ric and Ac\'{i}n, Antonio},
  Journal                  = {Physical Review Letters},
  Year                     = {2006},
  Number                   = {19},
  Pages                    = {190502},
  Volume                   = {97},
  Publisher                = {APS}
}

@Article{Pirandola2008,
  Title                    = {Characterization of collective Gaussian attacks and security of coherent-state quantum cryptography},
  Author                   = {Pirandola, Stefano and Braunstein, Samuel L and Lloyd, Seth},
  Journal                  = {Physical Review Letters},
  Year                     = {2008},
  Number                   = {20},
  Pages                    = {200504},
  Volume                   = {101},
  Publisher                = {APS}
}

@Article{Ottaviani2016,
  Title                    = { Secret key capacity of the
thermal-loss channel: improving the lower bound},
  Author                   = {Ottaviani, Carlo and Laurenza, Riccardo and Cope, Thomas P. W. and Spedalieri, Gaetana and Braunstein, Samuel L. and Pirandola, Stefano},
  Journal                  = {Proc. SPIE 9996},
  Year                     = {2016},
  Pages                    = {999609}
}

@Article{Mele2025,
  Title                    = {Maximum tolerable excess noise in continuous-variable quantum key distribution and improved lower bound on two-way capacities},
  Author                   = {Mele, F. A. and Lami, L. and Giovannetti, V.},
  Journal                  = {Nature Photonics},
  Year                     = {2025},
  Volume                   = {19},
  Pages                    = {329-334}
}

@article{Lloyd1996,
  title = {Capacity of the noisy quantum channel},
  author = {Lloyd, Seth},
  journal = {Phys. Rev. A},
  volume = {55},
  issue = {3},
  pages = {1613--1622},
  numpages = {0},
  year = {1997},
  month = {Mar},
  publisher = {American Physical Society},
  doi = {10.1103/PhysRevA.55.1613}
}

@article{Schumacher1996,
  title = {Quantum data processing and error correction},
  author = {Schumacher, Benjamin and Nielsen, M. A.},
  journal = {Phys. Rev. A},
  volume = {54},
  issue = {4},
  pages = {2629--2635},
  numpages = {0},
  year = {1996},
  month = {Oct},
  publisher = {American Physical Society},
  doi = {10.1103/PhysRevA.54.2629}
}

@article{Patron2009,
  title = {Reverse Coherent Information},
  author = {Garc\'{\i}a-Patr\'on, Ra\'ul and Pirandola, Stefano and Lloyd, Seth and Shapiro, Jeffrey H.},
  journal = {Phys. Rev. Lett.},
  volume = {102},
  issue = {21},
  pages = {210501},
  numpages = {4},
  year = {2009},
  month = {May},
  publisher = {American Physical Society},
  doi = {10.1103/PhysRevLett.102.210501}
}

@Article{Pirandola2009,
  Title                    = {Direct and reverse secret-key capacities of a quantum channel},
  Author                   = {Pirandola, Stefano and Garc{\'\i}a-Patr{\'o}n, Raul and Braunstein, Samuel L and Lloyd, Seth},
  Journal                  = {Physical Review Letters},
  Year                     = {2009},
  Number                   = {5},
  Pages                    = {050503},
  Volume                   = {102},
  Publisher                = {APS}
}

@Article{Pirandola2006,
  Title                    = {Quantum teleportation with continuous variables: a survey},
  Author                   = {Pirandola, Stefano and Mancini, Stefano},
  Journal                  = {Laser Physics},
  Year                     = {2006},
  Number                   = {10},
  Pages                    = {1418--1438},
  Volume                   = {16},
  Publisher                = {Springer}
}

@article{Pirandola2006a,
  title = {Macroscopic Entanglement by Entanglement Swapping},
  author = {Pirandola, Stefano and Vitali, David and Tombesi, Paolo and Lloyd, Seth},
  journal = {Phys. Rev. Lett.},
  volume = {97},
  issue = {15},
  pages = {150403},
  year = {2006},
  month = {Oct},
  publisher = {American Physical Society},
  doi = {10.1103/PhysRevLett.97.150403}
}

@Article{Pirandola2008a,
  Title                    = {Continuous-variable quantum cryptography using two-way quantum communication},
  Author                   = {Pirandola, Stefano and Mancini, Stefano and Lloyd, Seth and Braunstein, Samuel L},
  Journal                  = {Nature Physics},
  Year                     = {2008},
  Number                   = {9},
  Pages                    = {726--730},
  Volume                   = {4},
  Publisher                = {Nature Publishing Group}
}

@Article{Ralph1999,
  Title                    = {Continuous variable quantum cryptography},
  Author                   = {Ralph, Timothy C},
  Journal                  = {Physical Review A},
  Year                     = {1999},
  Number                   = {1},
  Pages                    = {010303},
  Volume                   = {61},
  Publisher                = {APS}
}

@Article{Reid2000,
  Title                    = {Quantum cryptography with a predetermined key, using continuous-variable Einstein-Podolsky-Rosen correlations},
  Author                   = {Reid, Margaret D},
  Journal                  = {Physical Review A},
  Year                     = {2000},
  Number                   = {6},
  Pages                    = {062308},
  Volume                   = {62},
  Publisher                = {APS}
}

@Article{Renner2009,
  Title                    = {de Finetti representation theorem for infinite-dimensional quantum systems and applications to quantum cryptography},
  Author                   = {Renner, Renato and Cirac, J Ignacio},
  Journal                  = {Physical Review Letters},
  Year                     = {2009},
  Number                   = {11},
  Pages                    = {110504},
  Volume                   = {102},
  Publisher                = {APS}
}

@Article{Ruppert2014,
  Title                    = {Long-distance continuous-variable quantum key distribution with efficient channel estimation},
  Author                   = {Ruppert, L{\'a}szl{\'o} and Usenko, Vladyslav C and Filip, Radim},
  Journal                  = {Physical Review A},
  Year                     = {2014},
  Number                   = {6},
  Pages                    = {062310},
  Volume                   = {90},
  Publisher                = {APS}
}

@Article{Scarani2004,
  Title                    = {Quantum cryptography protocols robust against photon number splitting attacks for weak laser pulse implementations},
  Author                   = {Scarani, Valerio and Ac\'{i}n, Antonio and Ribordy, Gr{\'e}goire and Gisin, Nicolas},
  Journal                  = {Physical Review Letters},
  Year                     = {2004},
  Number                   = {5},
  Pages                    = {057901},
  Volume                   = {92},
  Publisher                = {APS}
}

@Article{Scarani2009,
  Title                    = {The security of practical quantum key distribution},
  Author                   = {Scarani, Valerio and Bechmann-Pasquinucci, Helle and Cerf, Nicolas J and Du{\v{s}}ek, Miloslav and L{\"u}tkenhaus, Norbert and Peev, Momtchil},
  Journal                  = {Reviews of Modern Physics},
  Year                     = {2009},
  Number                   = {3},
  Pages                    = {1301},
  Volume                   = {81},
  Publisher                = {APS}
}

@Article{Serafini2004,
  Title                    = {Symplectic invariants, entropic measures and correlations of Gaussian states},
  Author                   = {Serafini, Alessio and Illuminati, Fabrizio and De Siena, Silvio},
  Journal                  = {Journal of Physics B},
  Year                     = {2004},
  Number                   = {2},
  Pages                    = {L21},
  Volume                   = {37}
}

@Article{Serafini2005,
  Title                    = {Quantifying decoherence in continuous variable systems},
  Author                   = {Serafini, A and Paris, MGA and Illuminati, F and De Siena, S},
  Journal                  = {Journal of Optics B: Quantum and Semiclassical Optics},
  Year                     = {2005},
  Number                   = {4},
  Pages                    = {R19},
  Volume                   = {7}
}

@Article{Silberhorn2002,
  Title                    = {Quantum key distribution with bright entangled beams},
  Author                   = {Silberhorn, Ch and Korolkova, Natalia and Leuchs, Gerd},
  Journal                  = {Physical Review Letters},
  Year                     = {2002},
  Number                   = {16},
  Pages                    = {167902},
  Volume                   = {88},
  Publisher                = {APS}
}

@Article{Silberhorn2002a,
  Title                    = {Continuous variable quantum cryptography: beating the 3 dB loss limit},
  Author                   = {Silberhorn, Ch and Ralph, Timothy C and L{\"u}tkenhaus, Norbert and Leuchs, Gerd},
  Journal                  = {Physical Review Letters},
  Year                     = {2002},
  Number                   = {16},
  Pages                    = {167901},
  Volume                   = {89},
  Publisher                = {APS}
}

@Article{Su2009,
  Title                    = {Continuous variable quantum key distribution based on optical entangled states without signal modulation},
  Author                   = {Su, Xiaolong and Wang, Wenzhe and Wang, Yu and Jia, Xiaojun and Xie, Changde and Peng, Kunchi},
  Journal                  = {Europhysics Letters},
  Year                     = {2009},
  Number                   = {2},
  Pages                    = {20005},
  Volume                   = {87}
}

@Article{Tomamichel2012,
  Title                    = {Tight finite-key analysis for quantum cryptography},
  Author                   = {Tomamichel, Marco and Lim, Charles Ci Wen and Gisin, Nicolas and Renner, Renato},
  Journal                  = {Nature Communications},
  Year                     = {2012},
  Pages                    = {634},
  Volume                   = {3},
  Publisher                = {Nature Publishing Group}
}

@Article{Usenko2011,
  Title                    = {Squeezed-state quantum key distribution upon imperfect reconciliation},
  Author                   = {Usenko, Vladyslav C and Filip, Radim},
  Journal                  = {New Journal of Physics},
  Year                     = {2011},
  Number                   = {11},
  Pages                    = {113007},
  Volume                   = {13}
}

@Article{Usenko2010a,
  Title                    = {Feasibility of continuous-variable quantum key distribution with noisy coherent states},
  Author                   = {Usenko, Vladyslav C and Filip, Radim},
  Journal                  = {Physical Review A},
  Year                     = {2010},
  Number                   = {2},
  Pages                    = {022318},
  Volume                   = {81},
  Publisher                = {APS}
}

@Article{Usenko2012,
  Title                    = {Entanglement of Gaussian states and the applicability to quantum key distribution over fading channels},
  Author                   = {Usenko, Vladyslav C and Heim, Bettina and Peuntinger, Christian and Wittmann, Christoffer and Marquardt, Christoph and Leuchs, Gerd and Filip, Radim},
  Journal                  = {New Journal of Physics},
  Year                     = {2012},
  Number                   = {9},
  Pages                    = {093048},
  Volume                   = {14}
}

@Article{Usenko2014,
  Title                    = {Entanglement-based continuous-variable quantum key distribution with multimode states and detectors},
  Author                   = {Usenko, Vladyslav C and Ruppert, Laszlo and Filip, Radim},
  Journal                  = {Physical Review A},
  Year                     = {2014},
  Number                   = {6},
  Pages                    = {062326},
  Volume                   = {90},
  Publisher                = {APS}
}

@Article{Walk2013,
  Title                    = {Security of continuous-variable quantum cryptography with Gaussian postselection},
  Author                   = {Walk, Nathan and Ralph, Timothy C and Symul, Thomas and Lam, Ping Koy},
  Journal                  = {Physical Review A},
  Year                     = {2013},
  Number                   = {2},
  Pages                    = {020303},
  Volume                   = {87},
  Publisher                = {APS}
}

@Article{Weedbrook2013,
  Title                    = {Continuous-variable quantum key distribution with entanglement in the middle},
  Author                   = {Weedbrook, Christian},
  Journal                  = {Physical Review A},
  Year                     = {2013},
  Number                   = {2},
  Pages                    = {022308},
  Volume                   = {87},
  Publisher                = {APS}
}

@Article{Weedbrook2004,
  Title                    = {Quantum cryptography without switching},
  Author                   = {Weedbrook, Christian and Lance, Andrew M and Bowen, Warwick P and Symul, Thomas and Ralph, Timothy C and Lam, Ping Koy},
  Journal                  = {Physical Review Letters},
  Year                     = {2004},
  Number                   = {17},
  Pages                    = {170504},
  Volume                   = {93},
  Publisher                = {APS}
}

@Article{Weedbrook2012a,
  author    = {Weedbrook, Christian and Pirandola, Stefano and Garc\'ia-Patr\'on, Ra\'ul and Cerf, Nicolas J and Ralph, Timothy C and Shapiro, Jeffrey H and Lloyd, Seth},
  title     = {Gaussian quantum information},
  journal   = {Reviews of Modern Physics},
  year      = {2012},
  volume    = {84},
  number    = {2},
  pages     = {621},
  publisher = {APS},
}

@Article{Weedbrook2010,
  Title                    = {Quantum cryptography approaching the classical limit},
  Author                   = {Weedbrook, Christian and Pirandola, Stefano and Lloyd, Seth and Ralph, Timothy C},
  Journal                  = {Physical Review Letters},
  Year                     = {2010},
  Number                   = {11},
  Pages                    = {110501},
  Volume                   = {105},
  Publisher                = {APS}
}

@Article{Weedbrook2012,
  Title                    = {Continuous-variable quantum key distribution using thermal states},
  Author                   = {Weedbrook, Christian and Pirandola, Stefano and Ralph, Timothy C},
  Journal                  = {Physical Review A},
  Year                     = {2012},
  Number                   = {2},
  Pages                    = {022318},
  Volume                   = {86},
  Publisher                = {APS}
}

@Article{Wolf2006,
  Title                    = {Extremality of Gaussian quantum states},
  Author                   = {Wolf, Michael M and Giedke, Geza and Cirac, J Ignacio},
  Journal                  = {Physical Review Letters},
  Year                     = {2006},
  Number                   = {8},
  Pages                    = {080502},
  Volume                   = {96},
  Publisher                = {APS}
}

@Article{Usenko2015,
  author    = {Usenko, Vladyslav C and Grosshans, Fr{\'e}d{\'e}ric},
  title     = {Unidimensional continuous-variable quantum key distribution},
  journal   = {Physical Review A},
  year      = {2015},
  volume    = {92},
  number    = {6},
  pages     = {062337},
  publisher = {APS},
}

@Article{Usenko2016,
  author    = {Usenko, Vladyslav C and Filip, Radim},
  title     = {Trusted noise in continuous-variable quantum key distribution: A threat and a defense},
  journal   = {Entropy},
  year      = {2016},
  volume    = {18},
  number    = {1},
  pages     = {20},
  publisher = {Multidisciplinary Digital Publishing Institute},
}

@Article{Leverrier2017,
  author    = {Leverrier, Anthony},
  title     = {Security of continuous-variable quantum key distribution via a Gaussian de Finetti reduction},
  journal   = {Physical Review Letters},
  year      = {2017},
  volume    = {118},
  number    = {20},
  pages     = {200501},
  publisher = {APS},
}

@Article{Leverrier2009,
  author    = {Leverrier, Anthony and Karpov, Evgueni and Grangier, Philippe and Cerf, Nicolas J},
  title     = {Security of continuous-variable quantum key distribution: towards a de Finetti theorem for rotation symmetry in phase space},
  journal   = {New Journal of Physics},
  year      = {2009},
  volume    = {11},
  number    = {11},
  pages     = {115009},
  }

@Article{Renner2008,
  author    = {Renner, Renato},
  title     = {Security of quantum key distribution},
  journal   = {International Journal of Quantum Information},
  year      = {2008},
  volume    = {6},
  number    = {01},
  pages     = {1--127},
  publisher = {World Scientific},
}

@Article{Leverrier2011,
  author    = {Leverrier, Anthony and Grangier, Philippe},
  title     = {Continuous-variable quantum-key-distribution protocols with a non-Gaussian modulation},
  journal   = {Physical Review A},
  year      = {2011},
  volume    = {83},
  number    = {4},
  pages     = {042312},
  publisher = {APS},
}

@Article{Zhao2009,
  author    = {Zhao, Yi-Bo and Heid, Matthias and Rigas, Johannes and L{\"u}tkenhaus, Norbert},
  title     = {Asymptotic security of binary modulated continuous-variable quantum key distribution under collective attacks},
  journal   = {Physical Review A},
  year      = {2009},
  volume    = {79},
  number    = {1},
  pages     = {012307},
  publisher = {APS},
}

@Article{Diamanti2015,
  author    = {Diamanti, Eleni and Leverrier, Anthony},
  title     = {Distributing secret keys with quantum continuous variables: principle, security and implementations},
  journal   = {Entropy},
  year      = {2015},
  volume    = {17},
  number    = {9},
  pages     = {6072},
  publisher = {Multidisciplinary Digital Publishing Institute},
}

@Article{Pirandola2015,
  author    = {Pirandola, Stefano and Ottaviani, Carlo and Spedalieri, Gaetana and Weedbrook, Christian and Braunstein, Samuel L and Lloyd, Seth and Gehring, Tobias and Jacobsen, Christian S and Andersen, Ulrik L},
  title     = {High-rate measurement-device-independent quantum cryptography},
  journal   = {Nature Photonics},
  year      = {2015},
  volume    = {9},
  number    = {6},
  pages     = {397--402},
  publisher = {Nature Research},
}

@article{hajomer2022high,
  title={High-rate continuous-variable measurement-device-independent quantum key distribution},
  author={Hajomer, Adnan AE and Nguyen, Huy Q and Gehring, Tobias},
  journal={arXiv:2210.07576},
  year={2022}
}

@Article{Soh2015,
  author    = {Soh, Daniel BS and Brif, Constantin and Coles, Patrick J and L{\"u}tkenhaus, Norbert and Camacho, Ryan M and Urayama, Junji and Sarovar, Mohan},
  title     = {Self-referenced continuous-variable quantum key distribution protocol},
  journal   = {Physical Review X},
  year      = {2015},
  volume    = {5},
  number    = {4},
  pages     = {041010},
  publisher = {APS},
}

@Article{Gehring2015,
  author    = {Gehring, Tobias and H{\"a}ndchen, Vitus and Duhme, J{\"o}rg and Furrer, Fabian and Franz, Torsten and Pacher, Christoph and Werner, Reinhard F and Schnabel, Roman},
  title     = {Implementation of continuous-variable quantum key distribution with composable and one-sided-device-independent security against coherent attacks},
  journal   = {Nature communications},
  year      = {2015},
  volume    = {6},
  pages     = {8795},
  publisher = {Nature Publishing Group},
}

@Article{Qi2015,
  author    = {Qi, Bing and Lougovski, Pavel and Pooser, Raphael and Grice, Warren and Bobrek, Miljko},
  title     = {Generating the local oscillator locally in continuous-variable quantum key distribution based on coherent detection},
  journal   = {Physical Review X},
  year      = {2015},
  volume    = {5},
  number    = {4},
  pages     = {041009},
  publisher = {APS},
}

@Article{Huang2016,
  author    = {Huang, Duan and Huang, Peng and Lin, Dakai and Zeng, Guihua},
  journal   = {Scientific reports},
  title     = {Long-distance continuous-variable quantum key distribution by controlling excess noise},
  year      = {2016},
  volume    = {6},
  pages     = {19201},
  publisher = {Nature Publishing Group},
}

@Article{Huang2015,
  author    = {Huang, Duan and Lin, Dakai and Wang, Chao and Liu, Weiqi and Fang, Shuanghong and Peng, Jinye and Huang, Peng and Zeng, Guihua},
  journal   = {Optics express},
  title     = {Continuous-variable quantum key distribution with 1 Mbps secure key rate},
  year      = {2015},
  number    = {13},
  pages     = {17511--17519},
  volume    = {23},
  publisher = {Optical Society of America},
}

@Article{Pirandola2020,
  author    = {S. Pirandola and U. L. Andersen and L. Banchi and M. Berta and D. Bunandar and R. Colbeck and D. Englund and T. Gehring and C. Lupo and C. Ottaviani and J. L. Pereira and M. Razavi and J. Shamsul Shaari and M. Tomamichel and V. C. Usenko and G. Vallone and P. Villoresi and P. Wallden},
  journal   = {Advances in Optics and Photonics},
  title     = {Advances in quantum cryptography},
  year      = {2020},
  month     = {dec},
  number    = {4},
  pages     = {1012},
  volume    = {12},
  doi       = {10.1364/aop.361502},
  publisher = {The Optical Society},
}

@Article{Xu2020,
  author    = {Feihu Xu and Xiongfeng Ma and Qiang Zhang and Hoi-Kwong Lo and Jian-Wei Pan},
  journal   = {Reviews of Modern Physics},
  title     = {Secure quantum key distribution with realistic devices},
  year      = {2020},
  month     = {may},
  number    = {2},
  pages     = {025002},
  volume    = {92},
  doi       = {10.1103/revmodphys.92.025002},
  publisher = {American Physical Society ({APS})},
}

@Article{Zhang2020,
  author    = {Yichen Zhang and Ziyang Chen and Stefano Pirandola and Xiangyu Wang and Chao Zhou and Binjie Chu and Yijia Zhao and Bingjie Xu and Song Yu and Hong Guo},
  journal   = {Physical Review Letters},
  title     = {Long-Distance Continuous-Variable Quantum Key Distribution over 202.81~km of Fiber},
  year      = {2020},
  month     = {jun},
  number    = {1},
  pages     = {010502},
  volume    = {125},
  doi       = {10.1103/physrevlett.125.010502},
  publisher = {American Physical Society ({APS})},
}

@Article{Eberle2013,
  author    = {Tobias Eberle and Vitus H\"andchen and J?rg Duhme and Torsten Franz and Fabian Furrer and Roman Schnabel and Reinhard F Werner},
  journal   = {New Journal of Physics},
  title     = {Gaussian entanglement for quantum key distribution from a single-mode squeezing source},
  year      = {2013},
  month     = {may},
  number    = {5},
  pages     = {053049},
  volume    = {15},
  doi       = {10.1088/1367-2630/15/5/053049},
  publisher = {{IOP} Publishing},
}

@Article{Derkach2020,
  author    = {Ivan Derkach and Vladyslav C Usenko and Radim Filip},
  journal   = {New Journal of Physics},
  title     = {Squeezing-enhanced quantum key distribution over atmospheric channels},
  year      = {2020},
  month     = {may},
  number    = {5},
  pages     = {053006},
  volume    = {22},
  doi       = {10.1088/1367-2630/ab7f8f},
  publisher = {{IOP} Publishing},
}

@Article{Qi2018,
  author    = {Bing Qi and Philip G. Evans and Warren P. Grice},
  journal   = {Physical Review A},
  title     = {Passive state preparation in the Gaussian-modulated coherent-states quantum key distribution},
  year      = {2018},
  month     = {jan},
  number    = {1},
  pages     = {012317},
  volume    = {97},
  doi       = {10.1103/physreva.97.012317},
  publisher = {American Physical Society ({APS})},
}

@Article{Usenko2018,
  author    = {Vladyslav C. Usenko},
  journal   = {Physical Review A},
  title     = {Unidimensional continuous-variable quantum key distribution using squeezed states},
  year      = {2018},
  month     = {sep},
  number    = {3},
  pages     = {032321},
  volume    = {98},
  doi       = {10.1103/physreva.98.032321},
  publisher = {American Physical Society ({APS})},
}

@Article{Jacobsen2018,
  author    = {Christian S. Jacobsen and Lars S. Madsen and Vladyslav C. Usenko and Radim Filip and Ulrik L. Andersen},
  journal   = {npj Quantum Information},
  title     = {Complete elimination of information leakage in continuous-variable quantum communication channels},
  year      = {2018},
  month     = {jul},
  number    = {1},
  volume    = {4},
pages={32},
  doi       = {10.1038/s41534-018-0084-0},
  publisher = {Springer Science and Business Media {LLC}},
}

@Article{Kovalenko2019,
  author    = {Olena Kovalenko and Kirill Yu. Spasibko and Maria V. Chekhova and Vladyslav C. Usenko and Radim Filip},
  journal   = {Optics Express},
  title     = {Feasibility of quantum key distribution with macroscopically bright coherent light},
  year      = {2019},
  month     = {nov},
  number    = {25},
  pages     = {36154},
  volume    = {27},
  doi       = {10.1364/oe.27.036154},
  publisher = {The Optical Society},
}

@Article{Ruppert2019,
  author    = {L{\'{a}}szl{\'{o}} Ruppert and Christian Peuntinger and Bettina Heim and Kevin G{\"u}nthner and Vladyslav C Usenko and Dominique Elser and Gerd Leuchs and Radim Filip and Christoph Marquardt},
  journal   = {New Journal of Physics},
  title     = {Fading channel estimation for free-space continuous-variable secure quantum communication},
  year      = {2019},
  month     = {dec},
  number    = {12},
  pages     = {123036},
  volume    = {21},
  doi       = {10.1088/1367-2630/ab5dd3},
  publisher = {{IOP} Publishing},
}

@Article{Lin2019,
  author    = {Jie Lin and Twesh Upadhyaya and Norbert L{\"u}tkenhaus},
  journal   = {Physical Review X},
  title     = {Asymptotic Security Analysis of Discrete-Modulated Continuous-Variable Quantum Key Distribution},
  year      = {2019},
  month     = {dec},
  number    = {4},
  pages     = {041064},
  volume    = {9},
  doi       = {10.1103/physrevx.9.041064},
  publisher = {American Physical Society ({APS})},
}

@Article{Denys2021,
  author    = {Aur{\'{e}}lie Denys and Peter Brown and Anthony Leverrier},
  journal   = {Quantum},
  title     = {Explicit asymptotic secret key rate of continuous-variable quantum key distribution with an arbitrary modulation},
  year      = {2021},
  month     = {sep},
  pages     = {540},
  volume    = {5},
  doi       = {10.22331/q-2021-09-13-540},
  publisher = {Verein zur Forderung des Open Access Publizierens in den Quantenwissenschaften},
}

@Article{Kaur2021,
  author    = {Eneet Kaur and Saikat Guha and Mark M. Wilde},
  journal   = {Physical Review A},
  title     = {Asymptotic security of discrete-modulation protocols for continuous-variable quantum key distribution},
  year      = {2021},
  month     = {jan},
  number    = {1},
  pages     = {012412},
  volume    = {103},
  doi       = {10.1103/physreva.103.012412},
  publisher = {American Physical Society ({APS})},
}

@Article{Pirandola2021a,
  author    = {Stefano Pirandola},
  journal   = {Physical Review Research},
  title     = {Composable security for continuous variable quantum key distribution: Trust levels and practical key rates in wired and wireless networks},
  year      = {2021},
  month     = {oct},
  number    = {4},
  pages     = {043014},
  volume    = {3},
  doi       = {10.1103/physrevresearch.3.043014},
  publisher = {American Physical Society ({APS})},
}

@Article{Ghorai2019,
  author    = {Shouvik Ghorai and Eleni Diamanti and Anthony Leverrier},
  journal   = {Physical Review A},
  title     = {Composable security of two-way continuous-variable quantum key distribution without active symmetrization},
  year      = {2019},
  month     = {jan},
  number    = {1},
  pages     = {012311},
  volume    = {99},
  doi       = {10.1103/physreva.99.012311},
  publisher = {American Physical Society ({APS})},
}

@Article{Hosseinidehaj2021,
  author    = {Nedasadat Hosseinidehaj and Nathan Walk and Timothy C. Ralph},
  journal   = {Physical Review A},
  title     = {Composable finite-size effects in free-space continuous-variable quantum-key-distribution systems},
  year      = {2021},
  month     = {jan},
  number    = {1},
  pages     = {012605},
  volume    = {103},
  doi       = {10.1103/physreva.103.012605},
  publisher = {American Physical Society ({APS})},
}

@Article{Matsuura2021,
  author    = {Takaya Matsuura and Kento Maeda and Toshihiko Sasaki and Masato Koashi},
  journal   = {Nature Communications},
  title     = {Finite-size security of continuous-variable quantum key distribution with digital signal processing},
  year      = {2021},
  month     = {jan},
  number    = {1},
  volume    = {12},
pages={252},
  doi       = {10.1038/s41467-020-19916-1},
  publisher = {Springer Science and Business Media {LLC}},
}

@Article{Wang2018,
  author    = {Ning Wang and Shanna Du and Wenyuan Liu and Xuyang Wang and Yongmin Li and Kunchi Peng},
  journal   = {Physical Review Applied},
  title     = {Long-Distance Continuous-Variable Quantum Key Distribution with Entangled States},
  year      = {2018},
  month     = {dec},
  number    = {6},
  pages     = {064028},
  volume    = {10},
  doi       = {10.1103/physrevapplied.10.064028},
  publisher = {American Physical Society ({APS})},
}

@Article{Ghalaii2020,
  author    = {Masoud Ghalaii and Carlo Ottaviani and Rupesh Kumar and Stefano Pirandola and Mohsen Razavi},
  journal   = {{IEEE} Journal of Selected Topics in Quantum Electronics},
  title     = {Long-Distance Continuous-Variable Quantum Key Distribution With Quantum Scissors},
  year      = {2020},
  month     = {may},
  number    = {3},
  pages     = {1--12},
  volume    = {26},
  doi       = {10.1109/jstqe.2020.2964395},
  publisher = {Institute of Electrical and Electronics Engineers ({IEEE})},
}

@Article{Mario2018,
  author    = {Milicevic, M and Feng, C. and Zhang, L. M. and Gulak, P. G.},
  journal   = {npj Quantum Information},
  title     = {Quasi-cyclic multi-edge {LDPC} codes for long-distance quantum cryptography},
  year      = {2018},
  volume    = {4},
  pages     = {21}
}

@Article{Mountogiannakis2022,
  author    = {Alexander G. Mountogiannakis and Panagiotis Papanastasiou and Boris Braverman and Stefano Pirandola},
  journal   = {Physical Review Research},
  title     = {Composably secure data processing for Gaussian-modulated continuous-variable quantum key distribution},
  year      = {2022},
  month     = {feb},
  number    = {1},
  pages     = {013099},
  volume    = {4},
  doi       = {10.1103/physrevresearch.4.013099},
  publisher = {American Physical Society ({APS})},
}

@Article{Leverrier2009a,
  author    = {Anthony Leverrier and Philippe Grangier},
  journal   = {Physical Review Letters},
  title     = {Unconditional Security Proof of Long-Distance Continuous-Variable Quantum Key Distribution with Discrete Modulation},
  year      = {2009},
  month     = {may},
  number    = {18},
  pages     = {180504},
  volume    = {102},
  doi       = {10.1103/physrevlett.102.180504},
  publisher = {American Physical Society ({APS})},
}

@Article{Adesso2014,
  author    = {Gerardo Adesso and Sammy Ragy and Antony R. Lee},
  journal   = {Open Systems \& Information Dynamics},
  title     = {Continuous Variable Quantum Information: Gaussian States and Beyond},
  year      = {2014},
  month     = {mar},
  number    = {01n02},
  pages     = {1440001},
  volume    = {21},
  doi       = {10.1142/s1230161214400010},
  publisher = {World Scientific Pub Co Pte Lt},
}

@Article{Nikolopoulos2017,
  author    = {Nikolopoulos, Georgios M. and  Diamanti, Eleni},
  journal   = {Scientific Reports},
  title     = {Continuous-variable quantum authentication of physical unclonable keys},
  year      = {2017},
  month     = {apr},
  number    = {1},
  volume    = {7},
pages={46047},
  doi       = {10.1038/srep46047},
  publisher = {Springer Science and Business Media {LLC}},
}

@Article{Dias2020,
  author    = {Josephine Dias and Matthew S. Winnel and Nedasadat Hosseinidehaj and Timothy C. Ralph},
  journal   = {Physical Review A},
  title     = {Quantum repeater for continuous-variable entanglement distribution},
  year      = {2020},
  month     = {nov},
  number    = {5},
  pages     = {052425},
  volume    = {102},
  doi       = {10.1103/physreva.102.052425},
  publisher = {American Physical Society ({APS})},
}

@Book{Nielsen2012,
  author    = {Michael A. Nielsen and Isaac L. Chuang},
  publisher = {Cambridge University Press},
  title     = {Quantum Computation and Quantum Information},
  year      = {2012},
  month     = {jun},
  doi       = {10.1017/cbo9780511976667},
}

@Article{Huang2015a,
  author    = {Duan Huang and Peng Huang and Dakai Lin and Chao Wang and Guihua Zeng},
  journal   = {Optics Letters},
  title     = {High-speed continuous-variable quantum key distribution without sending a local oscillator},
  year      = {2015},
  month     = {aug},
  number    = {16},
  pages     = {3695},
  volume    = {40},
  doi       = {10.1364/ol.40.003695},
  publisher = {The Optical Society},
}

@Article{Kleis2017,
  author    = {Sebastian Kleis and Max Rueckmann and Christian G. Schaeffer},
  journal   = {Optics Letters},
  title     = {Continuous variable quantum key distribution with a real local oscillator using simultaneous pilot signals},
  year      = {2017},
  month     = {apr},
  number    = {8},
  pages     = {1588},
  volume    = {42},
  doi       = {10.1364/ol.42.001588},
  publisher = {The Optical Society},
}

@ARTICLE{Kleis2019,
  author={Kleis, Sebastian and Schaeffer, Christian G.},
  journal={Journal of Lightwave Technology}, 
  title={Improving the Secret Key Rate of Coherent Quantum Key Distribution With Bayesian Inference}, 
  year={2019},
  volume={37},
  number={3},
  pages={722-728},
  doi={10.1109/JLT.2018.2877823}
}

@Article{Qin2018,
  author    = {Hao Qin and Rupesh Kumar and Vadim Makarov and Romain All\'{e}aume},
  journal   = {Physical Review A},
  title     = {Homodyne-detector-blinding attack in continuous-variable quantum key distribution},
  year      = {2018},
  month     = {jul},
  number    = {1},
  pages     = {012312},
  volume    = {98},
  doi       = {10.1103/physreva.98.012312},
  publisher = {American Physical Society ({APS})},
}

@Article{Marie2017,
  author    = {Adrien Marie and Romain All\'{e}aume},
  journal   = {Physical Review A},
  title     = {Self-coherent phase reference sharing for continuous-variable quantum key distribution},
  year      = {2017},
  month     = {jan},
  number    = {1},
  pages     = {012316},
  volume    = {95},
  doi       = {10.1103/physreva.95.012316},
  publisher = {American Physical Society ({APS})},
}

@Article{Qin2016,
  author    = {Hao Qin and Rupesh Kumar and Romain All\'{e}aume},
  journal   = {Physical Review A},
  title     = {Quantum hacking: Saturation attack on practical continuous-variable quantum key distribution},
  year      = {2016},
  month     = {jul},
  number    = {1},
  pages     = {012325},
  volume    = {94},
  doi       = {10.1103/physreva.94.012325},
  publisher = {American Physical Society ({APS})},
}

@Article{Kumar2015,
  author    = {Rupesh Kumar and Hao Qin and Romain All\'{e}aume},
  journal   = {New Journal of Physics},
  title     = {Coexistence of continuous variable {QKD} with intense {DWDM} classical channels},
  year      = {2015},
  month     = {apr},
  number    = {4},
  pages     = {043027},
  volume    = {17},
  doi       = {10.1088/1367-2630/17/4/043027},
  publisher = {{IOP} Publishing},
}

@Article{Dequal2021,
  author    = {Daniele Dequal and Luis Trigo Vidarte and Victor Roman Rodriguez and Giuseppe Vallone and Paolo Villoresi and Anthony Leverrier and Eleni Diamanti},
  journal   = {npj Quantum Information},
  title     = {Feasibility of satellite-to-ground continuous-variable quantum key distribution},
  year      = {2021},
  month     = {jan},
  number    = {1},
  volume    = {7},
pages={3},
  doi       = {10.1038/s41534-020-00336-4},
  publisher = {Springer Science and Business Media {LLC}},
}

@Article{Jain2022,
  author    = {Nitin Jain and Hou-Man Chin and Hossein Mani and Cosmo Lupo and Dino Solar Nikolic and Arne Kordts and Stefano Pirandola and Thomas Brochmann Pedersen and Matthias Kolb and Bernhard {\"O}mer and Christoph Pacher and Tobias Gehring and Ulrik L. Andersen},
  journal   = {Nature Communications},
  title     = {Practical continuous-variable quantum key distribution with composable security},
  year      = {2022},
  month     = {aug},
  number    = {1},
  volume    = {13},
pages={4740},
  doi       = {10.1038/s41467-022-32161-y},
  publisher = {Springer Science and Business Media {LLC}},
}

@Article{Kovalenko2021,
  author    = {Olena Kovalenko and Young-Sik Ra and Yin Cai and Vladyslav C. Usenko and Claude Fabre and Nicolas Treps and Radim Filip},
  journal   = {Photonics Research},
  title     = {Frequency-multiplexed entanglement for continuous-variable quantum key distribution},
  year      = {2021},
  month     = {nov},
  number    = {12},
  pages     = {2351},
  volume    = {9},
  doi       = {10.1364/prj.434979},
  publisher = {Optica Publishing Group},
}

@Article{Jain2021,
  author    = {Nitin Jain and Ivan Derkach and Hou-Man Chin and Radim Filip and Ulrik L Andersen and Vladyslav C Usenko and Tobias Gehring},
  journal   = {Quantum Science and Technology},
  title     = {Modulation leakage vulnerability in continuous-variable quantum key distribution},
  year      = {2021},
  month     = {aug},
  number    = {4},
  pages     = {045001},
  volume    = {6},
  doi       = {10.1088/2058-9565/ac0d4c},
  publisher = {{IOP} Publishing},
}

@article{Kanitschar2022,
  title = {Optimizing Continuous-Variable Quantum Key Distribution with Phase-Shift Keying Modulation and Postselection},
  author = {Kanitschar, Florian and Pacher, Christoph},
  journal = {Phys. Rev. Appl.},
  volume = {18},
  issue = {3},
  pages = {034073},
  numpages = {32},
  year = {2022},
  month = {Sep},
  publisher = {American Physical Society},
  doi = {10.1103/PhysRevApplied.18.034073}
}

@article{KanitscharT2021,
author={Florian Peter Kanitschar},
title={{Postselection Strategies for CV-QKD protocols with Phase-Shift Keying Modulation}},
year={2021},
journal={Master thesis (TU Wien)},
url={https://repositum.tuwien.at/handle/20.500.12708/18394}
}

@article{UpadhyayaT2021,
author={Upadhyaya, Twesh},
title={{Tools for the Security Analysis of Quantum Key Distribution in Infinite Dimensions}},
year={2021},
journal={Master thesis (University of Waterloo)},
url={http://hdl.handle.net/10012/17209}
}

@Article{Laudenbach2018,
  author    = {Fabian Laudenbach and Christoph Pacher and Chi-Hang Fred Fung and Andreas Poppe and Momtchil Peev and Bernhard Schrenk and Michael Hentschel and Philip Walther and Hannes H{\"u}bel},
  journal   = {Advanced Quantum Technologies},
  title     = {Continuous-Variable Quantum Key Distribution with Gaussian Modulation-The Theory of Practical Implementations},
  year      = {2018},
  month     = {jun},
  number    = {1},
  pages     = {1800011},
  volume    = {1},
  doi       = {10.1002/qute.201800011},
  publisher = {Wiley},
}

@article{Laudenbach2019,
  doi = {10.22331/q-2019-10-07-193},
   title = {Pilot-assisted intradyne reception for high-speed continuous-variable quantum key distribution with true local oscillator},
  author = {Laudenbach, Fabian and Schrenk, Bernhard and Pacher, Christoph and Hentschel, Michael and Fung, Chi-Hang Fred and Karinou, Fotini and Poppe, Andreas and Peev, Momtchil and H{\"{u}}bel, Hannes},
  journal = {{Quantum}},
  publisher = {{Verein zur F{\"{o}}rderung des Open Access Publizierens in den Quantenwissenschaften}},
  volume = {3},
  pages = {193},
  month = oct,
  year = {2019}
}

@Article{Croal2016,
  author    = {Callum Croal and Christian Peuntinger and Bettina Heim and Imran Khan and Christoph Marquardt and Gerd Leuchs and Petros Wallden and Erika Andersson and Natalia Korolkova},
  journal   = {Physical Review Letters},
  title     = {Free-Space Quantum Signatures Using Heterodyne Measurements},
  year      = {2016},
  month     = {sep},
  number    = {10},
  pages     = {100503},
  volume    = {117},
  doi       = {10.1103/physrevlett.117.100503},
  publisher = {American Physical Society ({APS})},
}

@Article{Pereira2018,
  author    = {Jason Pereira and Stefano Pirandola},
  journal   = {Physical Review A},
  title     = {Hacking Alice's box in continuous-variable quantum key distribution},
  year      = {2018},
  month     = {dec},
  number    = {6},
  pages     = {062319},
  volume    = {98},
  doi       = {10.1103/physreva.98.062319},
  publisher = {American Physical Society ({APS})},
}

@Article{Derkach2016,
  author    = {Ivan Derkach and Vladyslav C. Usenko and Radim Filip},
  journal   = {Physical Review A},
  title     = {Preventing side-channel effects in continuous-variable quantum key distribution},
  year      = {2016},
  month     = {mar},
  number    = {3},
  pages     = {032309},
  volume    = {93},
  doi       = {10.1103/physreva.93.032309},
  publisher = {American Physical Society ({APS})},
}

@Article{Eriksson2019,
  author    = {Tobias A. Eriksson and Takuya Hirano and Benjamin J. Puttnam and Georg Rademacher and Ruben S. Lu{\'{\i}}s and Mikio Fujiwara and Ryo Namiki and Yoshinari Awaji and Masahiro Takeoka and Naoya Wada and Masahide Sasaki},
  journal   = {Communications Physics},
  title     = {Wavelength division multiplexing of continuous variable quantum key distribution and 18.3 Tbit/s data channels},
  year      = {2019},
  month     = {jan},
  number    = {1},
  volume    = {2},
pages={9},
  doi       = {10.1038/s42005-018-0105-5},
  publisher = {Springer Science and Business Media {LLC}},
}

@Article{Eriksson2020,
  author    = {Tobias A. Eriksson and Ruben S. Luis and Benjamin J. Puttnam and Georg Rademacher and Mikio Fujiwara and Yoshinari Awaji and Hideaki Furukawa and Naoya Wada and Masahiro Takeoka and Masahide Sasaki},
  journal   = {Journal of Lightwave Technology},
  title     = {Wavelength Division Multiplexing of 194 Continuous Variable Quantum Key Distribution Channels},
  year      = {2020},
  month     = {apr},
  number    = {8},
  pages     = {2214--2218},
  volume    = {38},
  doi       = {10.1109/jlt.2020.2970179},
  publisher = {Institute of Electrical and Electronics Engineers ({IEEE})},
}

@Article{Zhang2019,
  author    = {G. Zhang and J. Y. Haw and H. Cai and F. Xu and S. M. Assad and J. F. Fitzsimons and X. Zhou and Y. Zhang and S. Yu and J. Wu and W. Ser and L. C. Kwek and A. Q. Liu},
  journal   = {Nature Photonics},
  title     = {An integrated silicon photonic chip platform for continuous-variable quantum key distribution},
  year      = {2019},
  month     = {aug},
  number    = {12},
  pages     = {839--842},
  volume    = {13},
  doi       = {10.1038/s41566-019-0504-5},
  publisher = {Springer Science and Business Media {LLC}},
}

@Article{Eriksson2019a,
  author    = {Tobias A. Eriksson and Benjamin J. Puttnam and Georg Rademacher and Ruben S. Luis and Mikio Fujiwara and Masahiro Takeoka and Yoshinari Awaji and Masahide Sasaki and Naoya Wada},
  journal   = {{IEEE} Photonics Technology Letters},
  title     = {Crosstalk Impact on Continuous Variable Quantum Key Distribution in Multicore Fiber Transmission},
  year      = {2019},
  month     = {mar},
  number    = {6},
  pages     = {467--470},
  volume    = {31},
  doi       = {10.1109/lpt.2019.2898458},
  publisher = {Institute of Electrical and Electronics Engineers ({IEEE})},
}

@Article{Gisin2007,
  author    = {Nicolas Gisin and Rob Thew},
  journal   = {Nature Photonics},
  title     = {Quantum communication},
  year      = {2007},
  month     = {mar},
  number    = {3},
  pages     = {165--171},
  volume    = {1},
  doi       = {10.1038/nphoton.2007.22},
  publisher = {Springer Science and Business Media {LLC}},
}

@Article{Bennett1993,
  author    = {Charles H. Bennett and Gilles Brassard and Claude Cr{\'{e}}peau and Richard Jozsa and Asher Peres and William K. Wootters},
  journal   = {Physical Review Letters},
  title     = {Teleporting an unknown quantum state via dual classical and Einstein-Podolsky-Rosen channels},
  year      = {1993},
  month     = {mar},
  number    = {13},
  pages     = {1895--1899},
  volume    = {70},
  doi       = {10.1103/physrevlett.70.1895},
  publisher = {American Physical Society ({APS})},
}

@Article{Pirandola2015b,
  author    = {S. Pirandola and J. Eisert and C. Weedbrook and A. Furusawa and S. L. Braunstein},
  journal   = {Nature Photonics},
  title     = {Advances in quantum teleportation},
  year      = {2015},
  month     = {sep},
  number    = {10},
  pages     = {641--652},
  volume    = {9},
  doi       = {10.1038/nphoton.2015.154},
  publisher = {Springer Science and Business Media {LLC}},
}

@Article{Bouwmeester1997,
  author    = {Dik Bouwmeester and Jian-Wei Pan and Klaus Mattle and Manfred Eibl and Harald Weinfurter and Anton Zeilinger},
  journal   = {Nature},
  title     = {Experimental quantum teleportation},
  year      = {1997},
  month     = {dec},
  number    = {6660},
  pages     = {575--579},
  volume    = {390},
  doi       = {10.1038/37539},
  publisher = {Springer Science and Business Media {LLC}},
}

@Article{Loock2000,
  author    = {P. van Loock and Samuel L. Braunstein},
  journal   = {Physical Review Letters},
  title     = {Multipartite Entanglement for Continuous Variables: A Quantum Teleportation Network},
  year      = {2000},
  month     = {apr},
  number    = {15},
  pages     = {3482--3485},
  volume    = {84},
  doi       = {10.1103/physrevlett.84.3482},
  publisher = {American Physical Society ({APS})},
}

@Article{Deng2003,
  author    = {Fu-Guo Deng and Gui Lu Long and Xiao-Shu Liu},
  journal   = {Physical Review A},
  title     = {Two-step quantum direct communication protocol using the Einstein-Podolsky-Rosen pair block},
  year      = {2003},
  month     = {oct},
  number    = {4},
  pages     = {042317},
  volume    = {68},
  doi       = {10.1103/physreva.68.042317},
  publisher = {American Physical Society ({APS})},
}

@Article{Deng2004,
  author    = {Fu-Guo Deng and Gui Lu Long},
  journal   = {Physical Review A},
  title     = {Secure direct communication with a quantum one-time pad},
  year      = {2004},
  month     = {may},
  number    = {5},
  pages     = {052319},
  volume    = {69},
  doi       = {10.1103/physreva.69.052319},
  publisher = {American Physical Society ({APS})},
}

@Article{Clarke2012,
  author    = {Patrick J. Clarke and Robert J. Collins and Vedran Dunjko and Erika Andersson and John Jeffers and Gerald S. Buller},
  journal   = {Nature Communications},
  title     = {Experimental demonstration of quantum digital signatures using phase-encoded coherent states of light},
  year      = {2012},
  month     = {nov},
  number    = {1},
pages={1174},
  volume    = {3},
  doi       = {10.1038/ncomms2172},
  publisher = {Springer Science and Business Media {LLC}},
}

@Article{Dunjko2014,
  author    = {Vedran Dunjko and Petros Wallden and Erika Andersson},
  journal   = {Physical Review Letters},
  title     = {Quantum Digital Signatures without Quantum Memory},
  year      = {2014},
  month     = {jan},
  number    = {4},
  pages     = {040502},
  volume    = {112},
  doi       = {10.1103/physrevlett.112.040502},
  publisher = {American Physical Society ({APS})},
}

@Article{Collins2014,
  author    = {Robert J. Collins and Ross J. Donaldson and Vedran Dunjko and Petros Wallden and Patrick J. Clarke and Erika Andersson and John Jeffers and Gerald S. Buller},
  journal   = {Physical Review Letters},
  title     = {Realization of Quantum Digital Signatures without the Requirement of Quantum Memory},
  year      = {2014},
  month     = {jul},
  number    = {4},
  pages     = {040502},
  volume    = {113},
  doi       = {10.1103/physrevlett.113.040502},
  publisher = {American Physical Society ({APS})},
}

@Article{Curty2001,
  author    = {Marcos Curty and David J. Santos},
  journal   = {Physical Review A},
  title     = {Quantum authentication of classical messages},
  year      = {2001},
  month     = {nov},
  number    = {6},
  pages     = {062309},
  volume    = {64},
  doi       = {10.1103/physreva.64.062309},
  publisher = {American Physical Society ({APS})},
}

@Article{Hong2017,
  author    = {Hong, Chang ho and Heo, Jino and  Jang, Jin Gak and Kwon, Daesung},
  journal   = {Quantum Information Processing},
  title     = {Quantum identity authentication with single photon},
  year      = {2017},
  month     = {aug},
  number    = {10},
  volume    = {16},
pages={236},
  doi       = {10.1007/s11128-017-1681-0},
  publisher = {Springer Science and Business Media {LLC}},
}

@Article{Gottesman2001a,
  author    = {Daniel Gottesman and Alexei Kitaev and John Preskill},
  journal   = {Physical Review A},
  title     = {Encoding a qubit in an oscillator},
  year      = {2001},
  month     = {jun},
  number    = {1},
  pages     = {012310},
  volume    = {64},
  doi       = {10.1103/physreva.64.012310},
  publisher = {American Physical Society ({APS})},
}

@Article{Braunstein2000a,
  author    = {Samuel L. Braunstein and H. J. Kimble},
  journal   = {Physical Review A},
  title     = {Dense coding for continuous variables},
  year      = {2000},
  month     = {mar},
  number    = {4},
  pages     = {042302},
  volume    = {61},
  doi       = {10.1103/physreva.61.042302},
  publisher = {American Physical Society ({APS})},
}

@Article{Bennett1992b,
  author    = {Charles H. Bennett and Stephen J. Wiesner},
  journal   = {Physical Review Letters},
  title     = {Communication via one- and two-particle operators on Einstein-Podolsky-Rosen states},
  year      = {1992},
  month     = {nov},
  number    = {20},
  pages     = {2881--2884},
  volume    = {69},
  doi       = {10.1103/physrevlett.69.2881},
  publisher = {American Physical Society ({APS})},
}

@Article{Li2002,
  author    = {Xiaoying Li and Qing Pan and Jietai Jing and Jing Zhang and Changde Xie and Kunchi Peng},
  journal   = {Physical Review Letters},
  title     = {Quantum Dense Coding Exploiting a Bright Einstein-Podolsky-Rosen Beam},
  year      = {2002},
  month     = {jan},
  number    = {4},
  pages     = {047904},
  volume    = {88},
  doi       = {10.1103/physrevlett.88.047904},
  publisher = {American Physical Society ({APS})},
}

@Article{Pirandola2008b,
  author    = {S. Pirandola and S. L. Braunstein and S. Mancini and S. Lloyd},
  journal   = {{EPL} (Europhysics Letters)},
  title     = {Quantum direct communication with continuous variables},
  year      = {2008},
  month     = {oct},
  number    = {2},
  pages     = {20013},
  volume    = {84},
  doi       = {10.1209/0295-5075/84/20013},
  publisher = {{IOP} Publishing},
}

@Article{Hosseinidehaj2020,
  author    = {Nedasadat Hosseinidehaj and Andrew M. Lance and Thomas Symul and Nathan Walk and Timothy C. Ralph},
  journal   = {Physical Review A},
  title     = {Finite-size effects in continuous-variable quantum key distribution with Gaussian postselection},
  year      = {2020},
  month     = {may},
  number    = {5},
  pages     = {052335},
  volume    = {101},
  doi       = {10.1103/physreva.101.052335},
  publisher = {American Physical Society ({APS})},
}

@article{Kanitschar2023,
  title = {Finite-Size Security for Discrete-Modulated Continuous-Variable Quantum Key Distribution Protocols},
  author = {Kanitschar, Florian and George, Ian and Lin, Jie and Upadhyaya, Twesh and L\"utkenhaus, Norbert},
  journal = {PRX Quantum},
  volume = {4},
  issue = {4},
  pages = {040306},
  numpages = {33},
  year = {2023},
  month = {Oct},
  publisher = {American Physical Society},
  doi = {10.1103/PRXQuantum.4.040306}
}

@article{Baeuml2023,
   title={Security of discrete-modulated continuous-variable quantum key distribution},
   volume={8},
   DOI={10.22331/q-2024-07-18-1418},
   journal={Quantum},
   publisher={Verein zur Forderung des Open Access Publizierens in den Quantenwissenschaften},
   author={B\"{a}uml, Stefan and Pascual-Garc\'{i}a, Carlos and Wright, Victoria and Fawzi, Omar and Ac\'{i}n, Antonio},
   year={2024},
   month=jul, pages={1418} }

@article{Kanitschar2024,
      title={Security of Multi-User Quantum Key Distribution with Discrete-Modulated Continuous-Variables}, 
      author={Florian Kanitschar and Christoph Pacher},
      year={2024},
      journal={arXiv:2406.14610}
}

@inproceedings{Metger2022,
   title={Generalised entropy accumulation},
   volume={2},
      DOI={10.1109/focs54457.2022.00085},
   booktitle={2022 IEEE 63rd Annual Symposium on Foundations of Computer Science (FOCS)},
   publisher={IEEE},
   author={Metger, Tony and Fawzi, Omar and Sutter, David and Renner, Renato},
   year={2022},
   month=oct, pages={844-850}
}

@article{Lorente2024,
      title={Quantum key distribution rates from non-symmetric conic optimization}, 
      author={Andrés González Lorente and Pablo V. Parellada and Miguel Castillo-Celeita and Mateus Ara\'{u}jo},
      year={2024},
      journal={arXiv:2407.00152}
    }

@article{Primaatmaja2024,
      title={Discrete-modulated continuous-variable quantum key distribution secure against general attacks}, 
      author={Ignatius William Primaatmaja and Wen Yu Kon and Charles Lim},
      year={2024},
      journal={arXiv:2409.02630}
      }

@article{Araujo2023,
   title={Quantum key distribution rates from semidefinite programming},
   volume={7},
   DOI={10.22331/q-2023-05-24-1019},
   journal={Quantum},
   publisher={Verein zur Forderung des Open Access Publizierens in den Quantenwissenschaften},
   author={Ara\'{u}jo, Mateus and Huber, Marcus and Navascu\'{e}s, Miguel and Pivoluska, Matej and Tavakoli, Armin},
   year={2023},
   month=may, pages={1019} 
}

@article{Metger2023,
   title={Security of quantum key distribution from generalised entropy accumulation},
   volume={14},
   DOI={10.1038/s41467-023-40920-8},
   number={1},
pages={5272},
   journal={Nature Communications},
   publisher={Springer Science and Business Media LLC},
   author={Metger, Tony and Renner, Renato},
   year={2023},
   month=aug 
}

@Article{Lo1997,
  author    = {Hoi-Kwong Lo},
  journal   = {Physical Review A},
  title     = {Insecurity of quantum secure computations},
  year      = {1997},
  month     = {aug},
  number    = {2},
  pages     = {1154--1162},
  volume    = {56},
  doi       = {10.1103/physreva.56.1154},
  publisher = {American Physical Society ({APS})},
}

@Article{Wehner2008,
  author    = {Stephanie Wehner and Christian Schaffner and Barbara M. Terhal},
  journal   = {Physical Review Letters},
  title     = {Cryptography from Noisy Storage},
  year      = {2008},
  month     = {jun},
  number    = {22},
  pages     = {220502},
  volume    = {100},
  doi       = {10.1103/physrevlett.100.220502},
  publisher = {American Physical Society ({APS})},
}

@Article{Furrer2018a,
  author    = {Fabian Furrer and Tobias Gehring and Christian Schaffner and Christoph Pacher and Roman Schnabel and Stephanie Wehner},
  journal   = {Nature Communications},
  title     = {Continuous-variable protocol for oblivious transfer in the noisy-storage model},
  year      = {2018},
  month     = {apr},
  number    = {1},
  volume    = {9},
pages={1450},
  doi       = {10.1038/s41467-018-03729-4},
  publisher = {Springer Science and Business Media {LLC}},
}

@Article{Dowling2003,
  author    = {Dowling, Jonathan P. and Milburn, Gerard J.},
  journal   = {Philosophical Transactions of the Royal Society of London. Series A: Mathematical, Physical and Engineering Sciences},
  title     = {Quantum technology: the second quantum revolution},
  year      = {2003},
  month     = jun,
  number    = {1809},
  pages     = {1655--1674},
  volume    = {361},
  doi       = {10.1098/rsta.2003.1227},
  editor    = {MacFarlane, A. G. J.},
  publisher = {The Royal Society},
}

@Article{Wootters1982,
  author    = {Wootters, W. K. and Zurek, W. H.},
  journal   = {Nature},
  title     = {A single quantum cannot be cloned},
  year      = {1982},
    month     = oct,
  number    = {5886},
  pages     = {802--803},
  volume    = {299},
  doi       = {10.1038/299802a0},
  publisher = {Springer Science and Business Media LLC},
}

@Article{Aspect1982,
  author    = {Aspect, Alain and Grangier, Philippe and Roger, Gérard},
  journal   = {Physical Review Letters},
  title     = {Experimental Realization of Einstein-Podolsky-Rosen-BohmGedankenexperiment: A New Violation of Bells Inequalities},
  year      = {1982},
    month     = jul,
  number    = {2},
  pages     = {91--94},
  volume    = {49},
  doi       = {10.1103/physrevlett.49.91},
  publisher = {American Physical Society (APS)},
}

@Article{Bennett1992c,
  author    = {Bennett, Charles H. and Brassard, Gilles and Mermin, N. David},
  journal   = {Physical Review Letters},
  title     = {Quantum cryptography without Bell's theorem},
  year      = {1992},
    month     = feb,
  number    = {5},
  pages     = {557--559},
  volume    = {68},
  doi       = {10.1103/physrevlett.68.557},
  publisher = {American Physical Society (APS)},
}

@Article{Bennett1992d,
  author    = {Bennett, Charles H. and Bessette, François and Brassard, Gilles and Salvail, Louis and Smolin, John},
  journal   = {Journal of Cryptology},
  title     = {Experimental quantum cryptography},
  year      = {1992},
    month     = jan,
  number    = {1},
  pages     = {3--28},
  volume    = {5},
  doi       = {10.1007/bf00191318},
  publisher = {Springer Science and Business Media LLC},
}

@Article{Brassard2000,
  author    = {Brassard, Gilles and L{\"u}tkenhaus, Norbert and Mor, Tal and Sanders, Barry C.},
  journal   = {Physical Review Letters},
  title     = {Limitations on Practical Quantum Cryptography},
  year      = {2000},
    month     = aug,
  number    = {6},
  pages     = {1330--1333},
  volume    = {85},
  doi       = {10.1103/physrevlett.85.1330},
  publisher = {American Physical Society (APS)},
}

@Article{Hwang2003,
  author    = {Hwang, Won-Young},
  journal   = {Physical Review Letters},
  title     = {Quantum Key Distribution with High Loss: Toward Global Secure Communication},
  year      = {2003},
    month     = aug,
  number    = {5},
  pages     = {057901},
  volume    = {91},
  doi       = {10.1103/physrevlett.91.057901},
  publisher = {American Physical Society (APS)},
}

@Article{Ursin2007,
  author    = {Ursin, R. and Tiefenbacher, F. and Schmitt-Manderbach, T. and Weier, H. and Scheidl, T. and Lindenthal, M. and Blauensteiner, B. and Jennewein, T. and Perdigues, J. and Trojek, P. and {\"O}mer, B. and F{\"u}rst, M. and Meyenburg, M. and Rarity, J. and Sodnik, Z. and Barbieri, C. and Weinfurter, H. and Zeilinger, A.},
  journal   = {Nature Physics},
  title     = {Entanglement-based quantum communication over 144?km},
  year      = {2007},
    month     = jun,
  number    = {7},
  pages     = {481--486},
  volume    = {3},
  doi       = {10.1038/nphys629},
  publisher = {Springer Science and Business Media LLC},
}

@Article{Korzh2015,
  author    = {Korzh, Boris and Lim, Charles Ci Wen and Houlmann, Raphael and Gisin, Nicolas and Li, Ming Jun and Nolan, Daniel and Sanguinetti, Bruno and Thew, Rob and Zbinden, Hugo},
  journal   = {Nature Photonics},
  title     = {Provably secure and practical quantum key distribution over 307?km of optical fibre},
  year      = {2015},
    month     = feb,
  number    = {3},
  pages     = {163--168},
  volume    = {9},
  doi       = {10.1038/nphoton.2014.327},
  publisher = {Springer Science and Business Media LLC},
}

@Article{Liao2018,
  author    = {Liao, Sheng-Kai and Cai, Wen-Qi and Handsteiner, Johannes and Liu, Bo and Yin, Juan and Zhang, Liang and Rauch, Dominik and Fink, Matthias and Ren, Ji-Gang and Liu, Wei-Yue and Li, Yang and Shen, Qi and Cao, Yuan and Li, Feng-Zhi and Wang, Jian-Feng and Huang, Yong-Mei and Deng, Lei and Xi, Tao and Ma, Lu and Hu, Tai and Li, Li and Liu, Nai-Le and Koidl, Franz and Wang, Peiyuan and Chen, Yu-Ao and Wang, Xiang-Bin and Steindorfer, Michael and Kirchner, Georg and Lu, Chao-Yang and Shu, Rong and Ursin, Rupert and Scheidl, Thomas and Peng, Cheng-Zhi and Wang, Jian-Yu and Zeilinger, Anton and Pan, Jian-Wei},
  journal   = {Physical Review Letters},
  title     = {Satellite-Relayed Intercontinental Quantum Network},
  year      = {2018},
    month     = jan,
  number    = {3},
  pages     = {030501},
  volume    = {120},
  doi       = {10.1103/physrevlett.120.030501},
  publisher = {American Physical Society (APS)},
}

@Article{Portmann2022,
  author    = {Portmann, Christopher and Renner, Renato},
  journal   = {Reviews of Modern Physics},
  title     = {Security in quantum cryptography},
  year      = {2022},
    month     = jun,
  number    = {2},
  pages     = {025008},
  volume    = {94},
  doi       = {10.1103/revmodphys.94.025008},
  publisher = {American Physical Society (APS)},
}

@Article{Vazirani2014,
  author    = {Vazirani, Umesh and Vidick, Thomas},
  journal   = {Physical Review Letters},
  title     = {Fully Device-Independent Quantum Key Distribution},
  year      = {2014},
    month     = sep,
  number    = {14},
  pages     = {140501},
  volume    = {113},
  doi       = {10.1103/physrevlett.113.140501},
  publisher = {American Physical Society (APS)},
}

@Article{Lydersen2010,
  author    = {Lydersen, Lars and Wiechers, Carlos and Wittmann, Christoffer and Elser, Dominique and Skaar, Johannes and Makarov, Vadim},
  journal   = {Nature Photonics},
  title     = {Hacking commercial quantum cryptography systems by tailored bright illumination},
  year      = {2010},
    month     = aug,
  number    = {10},
  pages     = {686--689},
  volume    = {4},
  doi       = {10.1038/nphoton.2010.214},
  publisher = {Springer Science and Business Media LLC},
}

@Article{Beige2002,
  author    = {Beige, A. and Englert, B.G. and Kurtsiefer, Ch. and Weinfurter, H.},
  journal   = {Acta Physica Polonica A},
  title     = {Secure Communication with a Publicly Known Key},
  year      = {2002},
    month     = mar,
  number    = {3},
  pages     = {357--368},
  volume    = {101},
  doi       = {10.12693/aphyspola.101.357},
  publisher = {Institute of Physics, Polish Academy of Sciences},
}

@Article{Bostroem2002,
  author    = {Bostr{\"o}m, Kim and Felbinger, Timo},
  journal   = {Physical Review Letters},
  title     = {Deterministic Secure Direct Communication Using Entanglement},
  year      = {2002},
    month     = oct,
  number    = {18},
  pages     = {187902},
  volume    = {89},
  doi       = {10.1103/physrevlett.89.187902},
  publisher = {American Physical Society (APS)},
}

@Article{Zhang2017,
  author    = {Zhang, Wei and Ding, Dong-Sheng and Sheng, Yu-Bo and Zhou, Lan and Shi, Bao-Sen and Guo, Guang-Can},
  journal   = {Physical Review Letters},
  title     = {Quantum Secure Direct Communication with Quantum Memory},
  year      = {2017},
    month     = may,
  number    = {22},
  pages     = {220501},
  volume    = {118},
  doi       = {10.1103/physrevlett.118.220501},
  publisher = {American Physical Society (APS)},
}

@Article{Wang2005,
  author    = {Wang, Chuan and Deng, Fu-Guo and Li, Yan-Song and Liu, Xiao-Shu and Long, Gui Lu},
  journal   = {Physical Review A},
  title     = {Quantum secure direct communication with high-dimension quantum superdense coding},
  year      = {2005},
    month     = apr,
  number    = {4},
  pages     = {044305},
  volume    = {71},
  doi       = {10.1103/physreva.71.044305},
  publisher = {American Physical Society (APS)},
}

@Article{Schaetz2004,
  author    = {Schaetz, T. and Barrett, M. D. and Leibfried, D. and Chiaverini, J. and Britton, J. and Itano, W. M. and Jost, J. D. and Langer, C. and Wineland, D. J.},
  journal   = {Physical Review Letters},
  title     = {Quantum Dense Coding with Atomic Qubits},
  year      = {2004},
    month     = jul,
  number    = {4},
  pages     = {040505},
  volume    = {93},
  doi       = {10.1103/physrevlett.93.040505},
  publisher = {American Physical Society (APS)},
}

@Article{Williams2017,
  author    = {Williams, Brian P. and Sadlier, Ronald J. and Humble, Travis S.},
  journal   = {Physical Review Letters},
  title     = {Superdense Coding over Optical Fiber Links with Complete Bell-State Measurements},
  year      = {2017},
    month     = feb,
  number    = {5},
  pages     = {050501},
  volume    = {118},
  doi       = {10.1103/physrevlett.118.050501},
  publisher = {American Physical Society (APS)},
}

@article{Gottesman2001b,
  author    = {Gottesman, Daniel and Chuang, Isaac},
  title     = {Quantum Digital Signatures},
  year      = {2001},
journal = {arXiv:quant-ph/0105032}
}

@InBook{Crepeau1995,
  author    = {Cr\'epeau, Claude and Salvail, Louis},
  pages     = {133--146},
  publisher = {Springer Berlin Heidelberg},
  title     = {Quantum Oblivious Mutual Identification},
  year      = {1995},
  isbn      = {9783540492641},
  booktitle = {Lecture Notes in Computer Science},
  doi       = {10.1007/3-540-49264-x_11},
  }

@Article{Dusek1999,
  author    = {Du\v{s}ek, Miloslav and Haderka, Ond\v{r}ej and Hendrych, Martin and My\v{s}ka, Robert},
  journal   = {Physical Review A},
  title     = {Quantum identification system},
  year      = {1999},
    month     = jul,
  number    = {1},
  pages     = {149--156},
  volume    = {60},
  doi       = {10.1103/physreva.60.149},
  publisher = {American Physical Society (APS)},
}

@InProceedings{Barnum2002,
  author     = {Barnum, H. and Crepeau, C. and Gottesman, D. and Smith, A. and Tapp, A.},
  booktitle  = {The 43rd Annual IEEE Symposium on Foundations of Computer Science, 2002. Proceedings.},
  title      = {Authentication of quantum messages},
  year       = {2002},
  publisher  = {IEEE Comput. Soc},
  series     = {SFCS-02},
  collection = {SFCS-02},
  doi        = {10.1109/sfcs.2002.1181969},
}

@InProceedings{Kilian1988,
  author     = {Kilian, Joe},
  booktitle  = {Proceedings of the twentieth annual ACM symposium on Theory of computing - STOC '88},
  title      = {Founding crytpography on oblivious transfer},
  year       = {1988},
  publisher  = {ACM Press},
  series     = {STOC '88},
  collection = {STOC '88},
  doi        = {10.1145/62212.62215},
}

@InBook{Bennett1992e,
  author    = {Bennett, Charles H. and Brassard, Gilles and Cr\'{e}peau, Claude and Skubiszewska, Marie-Hél\'ene},
  pages     = {351--366},
  publisher = {Springer Berlin Heidelberg},
  title     = {Practical Quantum Oblivious Transfer},
  year      = {1992},
  isbn      = {9783540551881},
  booktitle = {Lecture Notes in Computer Science},
  doi       = {10.1007/3-540-46766-1_29},
}

@Article{Mayers1997,
  author    = {Mayers, Dominic},
  journal   = {Physical Review Letters},
  title     = {Unconditionally Secure Quantum Bit Commitment is Impossible},
  year      = {1997},
    month     = apr,
  number    = {17},
  pages     = {3414--3417},
  volume    = {78},
  doi       = {10.1103/physrevlett.78.3414},
  publisher = {American Physical Society (APS)},
}

@Article{Lo1997a,
  author    = {Lo, Hoi-Kwong and Chau, H. F.},
  journal   = {Physical Review Letters},
  title     = {Is Quantum Bit Commitment Really Possible?},
  year      = {1997},
    month     = apr,
  number    = {17},
  pages     = {3410--3413},
  volume    = {78},
  doi       = {10.1103/physrevlett.78.3410},
  publisher = {American Physical Society (APS)},
}

@Article{Chailloux2010,
  author    = {Chailloux, André and Kerenidis, Iordanis and Sikora, Jamie},
  title     = {Lower Bounds for Quantum Oblivious Transfer},
  year      = {2010},
  journal={arXiv:quant-ph/1007.1875}
}

@Article{Paris2003,
  author    = {Paris, Matteo G. A. and Illuminati, Fabrizio and Serafini, Alessio and De Siena, Silvio},
  journal   = {Physical Review A},
  title     = {Purity of Gaussian states: Measurement schemes and time evolution in noisy channels},
  year      = {2003},
    month     = jul,
  number    = {1},
  pages     = {012314},
  volume    = {68},
  doi       = {10.1103/physreva.68.012314},
  publisher = {American Physical Society (APS)},
}

@Article{Eisert2002,
  author    = {Eisert, J. and Scheel, S. and Plenio, M. B.},
  journal   = {Physical Review Letters},
  title     = {Distilling Gaussian States with Gaussian Operations is Impossible},
  year      = {2002},
    month     = sep,
  number    = {13},
  pages     = {137903},
  volume    = {89},
  doi       = {10.1103/physrevlett.89.137903},
  publisher = {American Physical Society (APS)},
}

@Article{Fiurasek2002a,
  author    = {Fiur{\'a}{\v{s}}ek, Jarom{\'\i}r },
  journal   = {Physical Review Letters},
  title     = {Gaussian Transformations and Distillation of Entangled Gaussian States},
  year      = {2002},
    month     = sep,
  number    = {13},
  pages     = {137904},
  volume    = {89},
  doi       = {10.1103/physrevlett.89.137904},
  publisher = {American Physical Society (APS)},
}

@article{GarciaPatron2007,
  author = {Garc\'{i}a-Patr\'{o}n, Ra\'{u}l},
    title  = {Quantum Information with Optical Continuous Variables: from Bell Tests to Key Distribution},
  year   = {2007},
journal={Ph.D. thesis (Universit\'e Libre de Bruxelles)}
}

@article{Pirandola2024,
  title = {Improved composable key rates for CV-QKD},
  author = {Pirandola, Stefano and Papanastasiou, Panagiotis},
  journal = {Phys. Rev. Res.},
  volume = {6},
  issue = {2},
  pages = {023321},
  numpages = {13},
  year = {2024},
  month = {Jun},
  publisher = {American Physical Society},
  doi = {10.1103/PhysRevResearch.6.023321}  
}

@Article{Papanastasiou2021,
  author    = {Papanastasiou, Panagiotis and Pirandola, Stefano},
  journal   = {Physical Review Research},
  title     = {Continuous-variable quantum cryptography with discrete alphabets: Composable security under collective Gaussian attacks},
  year      = {2021},
  month     = jan,
  number    = {1},
  pages     = {013047},
  volume    = {3},
  doi       = {10.1103/physrevresearch.3.013047},
  publisher = {American Physical Society (APS)},
}

@Article{Ghorai2019a,
  author    = {Ghorai, Shouvik and Grangier, Philippe and Diamanti, Eleni and Leverrier, Anthony},
  journal   = {Physical Review X},
  title     = {Asymptotic Security of Continuous-Variable Quantum Key Distribution with a Discrete Modulation},
  year      = {2019},
    month     = jun,
  number    = {2},
  pages     = {021059},
  volume    = {9},
  doi       = {10.1103/physrevx.9.021059},
  publisher = {American Physical Society (APS)},
}

@Article{Upadhyaya2021,
  author    = {Upadhyaya, Twesh and van Himbeeck, Thomas and Lin, Jie and L{\"u}tkenhaus, Norbert},
  journal   = {PRX Quantum},
  title     = {Dimension Reduction in Quantum Key Distribution for Continuous- and Discrete-Variable Protocols},
  year      = {2021},
    month     = may,
  number    = {2},
  pages     = {020325},
  volume    = {2},
  doi       = {10.1103/prxquantum.2.020325},
  publisher = {American Physical Society (APS)},
}

@article{Sych2010,
	doi = {10.1088/1367-2630/12/5/053019},
		year = 2010,
	month = {may},
	publisher = {{IOP} Publishing},
	volume = {12},
	number = {5},
	pages = {053019},
	author = {Denis Sych and Gerd Leuchs},
	title = {Coherent state quantum key distribution with multi letter phase-shift keying},
	journal = {New Journal of Physics}
}

@article{Rigas2006,
author    = {Rigas, J.},
 title     = {Detection of prepare \& measurement entanglement in continuous variable quantu mkey distirbution},
journal = {Ph.D. thesis (University of Erlangen-Nuremberg)},
 year      =  {2006}
}

@article{Bradler2018,
   title={{Security proof of continuous-variable quantum key distribution using three coherent states}},
   volume={97},
pages={022310},
   DOI={10.1103/physreva.97.022310},
   number={2},
   journal={Physical Review A},
   publisher={American Physical Society (APS)},
   author={Br{\'a}dler, Kamil and Weedbrook, Christian},
   year={2018},
   month={Feb}
}

@article{Lin2020,
   title={Trusted Detector Noise Analysis for Discrete Modulation Schemes of Continuous-Variable Quantum Key Distribution},
   volume={14},
      DOI={10.1103/physrevapplied.14.064030},
   number={6},
   journal={Physical Review Applied},
   publisher={American Physical Society (APS)},
   author={Lin, Jie and L{\"u}tkenhaus, Norbert},
   year={2020},
   month={Dec},
pages={064030}
}

@article{Coles2016,
  title = {{Numerical approach for unstructured quantum key distribution}},
  author = {Coles, Patrick J. and Metodiev, Eric M. and L\"{u}tkenhaus, Norbert},
  journal = {Nature Communications},
  publisher = {Nature Publishing Group},
  volume = {7},
  issue = {1},
  pages = {11712},
  numpages = {9},
  year = {2016},
  month = {May},
  doi = {10.1038/ncomms11712}
}

@ARTICLE{Winick2018,
   author       = "Winick, Adam and L{\"u}tkenhaus, Norbert and Coles, Patrick J.",
   title        = "Reliable numerical key rates for quantum key distribution",
   journal      = "Quantum",
   volume       = "2", 
   pages        = "77",
   year         = "2018",
   month        = "Jul",
   publisher    = "Verein zur Forderung des Open Access Publizierens in den Quantenwissenschaften",
      DOI          = "10.22331/q-2018-07-26-77"
   }

@ARTICLE{Frank1956,
title = {An algorithm for quadratic programming},
author = {Frank, Marguerite and Wolfe, Philip},
year = {1956},
journal = {Nav. Res. Logist. Q.},
volume = {3},
number = {1-2},
pages = {95-110}
}

@article{Wang2023,
doi = {10.1088/1367-2630/acb964},
year = {2023},
month = {feb},
volume = {25},
number = {2},
pages = {023019},
author = {Pu Wang and Yu Zhang and Zhenguo Lu and Xuyang Wang and Yongmin Li},
title = {Discrete-modulation continuous-variable quantum key distribution with a high key rate},
journal = {New Journal of Physics}
}

@misc{cvx1,
  author       = {Michael Grant and Stephen Boyd},
  title        = {{CVX}: Matlab Software for Disciplined Convex Programming, version 2.1},
  howpublished = {\url{http://cvxr.com/cvx}},
  month        = mar,
  year         = 2014
}

@incollection{cvx2,
  author    = {Michael Grant and Stephen Boyd},
  title     = {Graph implementations for nonsmooth convex programs},
  booktitle = {Recent Advances in Learning and Control},
  series    = {Lecture Notes in Control and Information Sciences},
  editor    = {V. Blondel and S. Boyd and H. Kimura},
  publisher = {Springer-Verlag Limited},
  pages     = {95--110},
  year      = 2008,
}

@Article{SDPT3a,
  author    = {Toh, K. C. and Todd, M. J. and T\"{u}t\"{u}nc\"{u}, R. H.},
  journal   = {Optimization Methods and Software},
  title     = {SDPT3 - A Matlab software package for semidefinite programming, Version 1.3},
  year      = {1999},
  month     = jan,
  number    = {1-4},
  pages     = {545--581},
  volume    = {11},
  doi       = {10.1080/10556789908805762},
  publisher = {Informa UK Limited},
}

@Article{SDPT3b,
  author    = {T\"{u}t\"{u}nc\"{u}, R. H. and Toh, K. C. and Todd, M. J.},
  journal   = {Mathematical Programming},
  title     = {Solving semidefinite-quadratic-linear programs using SDPT3},
  year      = {2003},
  month     = feb,
  number    = {2},
  pages     = {189--217},
  volume    = {95},
  doi       = {10.1007/s10107-002-0347-5},
  publisher = {Springer Science and Business Media LLC},
}

@Article{Renner2006,
      title={Security of Quantum Key Distribution}, 
      author={Renato Renner},
      year={2006},
      journal={Ph.D. thesis (ETH Zurich)}
}

@ARTICLE{Tomamichel2012a,

  author={Tomamichel, Marco and Schaffner, Christian and Smith, Adam and Renner, Renato},
  journal={IEEE Transactions on Information Theory}, 
  title={Leftover Hashing Against Quantum Side Information}, 
  year={2011},
  volume={57},
  number={8},
  pages={5524-5535},
  doi={10.1109/TIT.2011.2158473}
}

@ARTICLE{Tomamichel2009,

  author={Tomamichel, Marco and Colbeck, Roger and Renner, Renato},
  journal={IEEE Transactions on Information Theory}, 
  title={A Fully Quantum Asymptotic Equipartition Property}, 
  year={2009},
  volume={55},
  number={12},
  pages={5840-5847},
  doi={10.1109/TIT.2009.2032797}}

@article{Yamano2024,
	title = {Finite-size security proof of binary-modulation continuous-variable quantum key distribution using only heterodyne measurement},
	volume = {99},
	doi = {10.1088/1402-4896/ad1022},
	language = {en},
	number = {2},
	urldate = {2024-02-07},
	journal = {Physica Scripta},
	author = {Yamano, Shinichiro and Matsuura, Takaya and Kuramochi, Yui and Sasaki, Toshihiko and Koashi, Masato},
	month = jan,
	year = {2024},
		pages = {025115}
}

@article{Winter1999,
   title={Coding theorem and strong converse for quantum channels},
   volume={45},
      DOI={10.1109/18.796385},
   number={7},
   journal={IEEE Transactions on Information Theory},
   publisher={Institute of Electrical and Electronics Engineers (IEEE)},
   author={Winter, A.},
   year={1999},
   pages={2481-2485} }

@article{Shirokov2017,
   title={Tight uniform continuity bounds for the quantum conditional mutual information, for the Holevo quantity, and for capacities of quantum channels},
   volume={58},
      DOI={10.1063/1.4987135},
   number={10},
   journal={Journal of Mathematical Physics},
   publisher={AIP Publishing},
   author={Shirokov, M. E.},
   year={2017},
pages={102202},
   month=oct }

@article{Dupuis2020,
   title={Entropy Accumulation},
   volume={379},
   DOI={10.1007/s00220-020-03839-5},
   number={3},
   journal={Communications in Mathematical Physics},
   publisher={Springer Science and Business Media LLC},
   author={Dupuis, Frédéric and Fawzi, Omar and Renner, Renato},
   year={2020},
   month=sep, pages={867-913} }

@article{Dupuis2019,
   title={Entropy Accumulation With Improved Second-Order Term},
   volume={65},
   DOI={10.1109/tit.2019.2929564},
   number={11},
   journal={IEEE Transactions on Information Theory},
   publisher={Institute of Electrical and Electronics Engineers (IEEE)},
   author={Dupuis, Frederic and Fawzi, Omar},
   year={2019},
   month=nov, pages={7596-7612} }

@article{Leverrier2015,
  title = {Composable Security Proof for Continuous-Variable Quantum Key Distribution with Coherent States},
  author = {Leverrier, Anthony},
  journal = {Phys. Rev. Lett.},
  volume = {114},
  issue = {7},
  pages = {070501},
  numpages = {5},
  year = {2015},
  month = {Feb},
  publisher = {American Physical Society},
  doi = {10.1103/PhysRevLett.114.070501}
}

@article{Shor2000,
   title={Simple Proof of Security of the BB84 Quantum Key Distribution Protocol},
   volume={85},
      DOI={10.1103/physrevlett.85.441},
   number={2},
   journal={Physical Review Letters},
   publisher={American Physical Society (APS)},
   author={Shor, Peter W. and Preskill, John},
   year={2000},
   month=jul, pages={441-444} }

@article{Koenig2007,
  title = {Small Accessible Quantum Information Does Not Imply Security},
  author = {K\"onig, Robert and Renner, Renato and Bariska, Andor and Maurer, Ueli},
  journal = {Phys. Rev. Lett.},
  volume = {98},
  issue = {14},
  pages = {140502},
  numpages = {4},
  year = {2007},
  month = {Apr},
  publisher = {American Physical Society},
  doi = {10.1103/PhysRevLett.98.140502}
}

@article{Canetti2000,
author = {Canetti, Ran},
title = {Security and Composition of Multiparty Cryptographic Protocols},
year = {2000},
issue_date = {January   2000},
publisher = {Springer-Verlag},
address = {Berlin, Heidelberg},
volume = {13},
number = {1},
doi = {10.1007/s001459910006},
journal = {J. Cryptol.},
month = {jan},
pages = {143-202},
numpages = {60}
}

@inproceedings{Canetti2001,
  author = {Canetti, Ran},
  booktitle = {FOCS},
   ee = {http://doi.ieeecomputersociety.org/10.1109/SFCS.2001.959888},
  isbn = {0-7695-1390-5},
  pages = {136-145},
  publisher = {IEEE Computer Society},
  title = {Universally Composable Security: A New Paradigm for Cryptographic Protocols.},
    year = 2001
}

@INPROCEEDINGS{Pfitzmann2001,
  author={Pfitzmann, B. and Waidner, M.},
  booktitle={Proceedings 2001 IEEE Symposium on Security and Privacy. S\&P 2001}, 
  title={A model for asynchronous reactive systems and its application to secure message transmission}, 
  year={2001},
  volume={},
  number={},
  pages={184-200},
  doi={10.1109/SECPRI.2001.924298}
  }

@InProceedings{BenOr2004a,
author="Ben-Or, Michael
and Horodecki, Micha{\l}
and Leung, Debbie W.
and Mayers, Dominic
and Oppenheim, Jonathan",
editor="Kilian, Joe",
title="The Universal Composable Security of Quantum Key Distribution",
booktitle="Theory of Cryptography",
year="2005",
publisher="Springer Berlin Heidelberg",
address="Berlin, Heidelberg",
pages="386--406",
isbn="978-3-540-30576-7"
}

@article{BenOr2004b,
    author = {Ben-Or, Michael and Mayers, Dominic},
  title = {General Security Definition and Composability for Quantum \& Classical Protocols},
  journal = {arXiv:quant-ph/0409062},
  year = {2004}
}

@article{Unruh2004,
  author = {Unruh, Dominique},
  title = {Simulatable security for quantum protocols},
  journal = {arXiv:quant-ph/0409125},
  year = {2004}
}

@book{Barnett2009,
series = {Oxford master series in atomic, optical and laser physics},
publisher = {Oxford Univ. Pr.},
isbn = {0198527624},
year = {2009},
title = {Quantum information},
edition = {1. publ.},
language = {eng},
address = {Oxford [u.a.]},
author = {Barnett, Stephen M}
}

@article{Furrer2014,
    author = {Fabian Furrer},
	title = {Reverse-reconciliation continuous-variable quantum key distribution based on the uncertainty principle},
	journal = {Phys. Rev. A},
	doi = {10.1103/physreva.90.042325},
		year = 2014,
	month = {oct},
	publisher = {American Physical Society ({APS})},
	volume = {90},
	number = {4},
pages = {042325}
	}

@article{Lupo2022,
   title={Quantum Key Distribution with Nonideal Heterodyne Detection: Composable Security of Discrete-Modulation Continuous-Variable Protocols},
   volume={3},
pages={010341},
      DOI={10.1103/prxquantum.3.010341},
   number={1},
   journal={PRX Quantum},
   publisher={American Physical Society (APS)},
   author={Lupo, Cosmo and Ouyang, Yingkai},
   year={2022},
   month=mar }

@ARTICLE{Ottaviani2020,
  author={Ottaviani, Carlo and Woolley, Matthew J. and Erementchouk, Misha and Federici, John F. and Mazumder, Pinaki and Pirandola, Stefano and Weedbrook, Christian},
  journal={IEEE Journal on Selected Areas in Communications}, 
  title={Terahertz Quantum Cryptography}, 
  year={2020},
  volume={38},
  number={3},
  pages={483-495},
  doi={10.1109/JSAC.2020.2968973}}

@article{Papanastasiou2018,
  title = {Gaussian one-way thermal quantum cryptography with finite-size effects},
  author = {Papanastasiou, Panagiotis and Ottaviani, Carlo and Pirandola, Stefano},
  journal = {Phys. Rev. A},
  volume = {98},
  issue = {3},
  pages = {032314},
  numpages = {9},
  year = {2018},
  month = {Sep},
  publisher = {American Physical Society},
  doi = {10.1103/PhysRevA.98.032314}
}

@Article{Weedbrook2014,
  author    = {Weedbrook, Christian and Ottaviani, Carlo and Pirandola, Stefano},
  journal   = {Physical Review A},
  title     = {Two-way quantum cryptography at different wavelengths},
  year      = {2014},
    month     = jan,
  number    = {1},
  pages     = {012309},
  volume    = {89},
  doi       = {10.1103/physreva.89.012309},
  publisher = {American Physical Society (APS)},
}

@Article{Maurer1993,
  author    = {Maurer, U.M.},
  journal   = {IEEE Transactions on Information Theory},
  title     = {Secret key agreement by public discussion from common information},
  year      = {1993},
    month     = may,
  number    = {3},
  pages     = {733--742},
  volume    = {39},
  doi       = {10.1109/18.256484},
  publisher = {Institute of Electrical and Electronics Engineers (IEEE)},
}

@Article{Araki1970,
  author    = {Araki, Huzihiro and Lieb, Elliott H.},
  journal   = {Communications in Mathematical Physics},
  title     = {Entropy inequalities},
  year      = {1970},
    month     = jun,
  number    = {2},
  pages     = {160--170},
  volume    = {18},
  doi       = {10.1007/bf01646092},
  publisher = {Springer Science and Business Media LLC},
}

@Article{Pirandola2017,
  author    = {Pirandola, Stefano and Laurenza, Riccardo and Ottaviani, Carlo and Banchi, Leonardo},
  journal   = {Nature Communications},
  title     = {Fundamental limits of repeaterless quantum communications},
  year      = {2017},
    month     = apr,
  number    = {1},
  volume    = {8},
pages={15043},
  doi       = {10.1038/ncomms15043},
  publisher = {Springer Science and Business Media LLC},
}

@Article{Lasota2023,
  author    = {Lasota, MikoЕ‚aj and Kovalenko, Olena and Usenko, Vladyslav C},
  journal   = {New Journal of Physics},
  title     = {Robustness of entanglement-based discrete- and continuous-variable quantum key distribution against channel noise},
  year      = {2023},
    month     = dec,
  number    = {12},
  pages     = {123003},
  volume    = {25},
  doi       = {10.1088/1367-2630/ad0e8c},
  }

@Article{Lasota2017,
  author    = {Lasota, MikoЕ‚aj and Filip, Radim and Usenko, Vladyslav C.},
  journal   = {Physical Review A},
  title     = {Robustness of quantum key distribution with discrete and continuous variables to channel noise},
  year      = {2017},
    month     = jun,
  number    = {6},
  pages     = {062312},
  volume    = {95},
  doi       = {10.1103/physreva.95.062312},
  publisher = {American Physical Society (APS)},
}

@Article{Scarani2008,
  author    = {Scarani, Valerio and Renner, Renato},
  journal   = {Physical Review Letters},
  title     = {Quantum Cryptography with Finite Resources: Unconditional Security Bound for Discrete-Variable Protocols with One-Way Postprocessing},
  year      = {2008},
    month     = may,
  number    = {20},
  pages     = {200501},
  volume    = {100},
  doi       = {10.1103/physrevlett.100.200501},
  publisher = {American Physical Society (APS)},
}

@Article{Lupo2018,
  author    = {Lupo, Cosmo and Ottaviani, Carlo and Papanastasiou, Panagiotis and Pirandola, Stefano},
  journal   = {Physical Review A},
  title     = {Continuous-variable measurement-device-independent quantum key distribution: Composable security against coherent attacks},
  year      = {2018},
    month     = may,
  number    = {5},
  pages     = {052327},
  volume    = {97},
  doi       = {10.1103/physreva.97.052327},
  publisher = {American Physical Society (APS)},
}

@Article{Papanastasiou2017,
  author    = {Papanastasiou, Panagiotis and Ottaviani, Carlo and Pirandola, Stefano},
  journal   = {Physical Review A},
  title     = {Finite-size analysis of measurement-device-independent quantum cryptography with continuous variables},
  year      = {2017},
    month     = oct,
  number    = {4},
  pages     = {042332},
  volume    = {96},
  doi       = {10.1103/physreva.96.042332},
  publisher = {American Physical Society (APS)},
}

@Article{Papanastasiou2023,
  author    = {Papanastasiou, Panagiotis and Mountogiannakis, Alexandros and Pirandola, Stefano},
  journal   = {Scientific Reports},
  title     = {Composable security of CV-MDI-QKD with secret key rate and data processing},
  year      = {2023},
  pages     = {11636},
  volume    = {13},
  doi       = {10.1038/s41598-023-37699-5},
}

@article{Ghalaii2023,
  title = {Continuous-variable measurement-device-independent quantum key distribution in free-space channels},
  author = {Ghalaii, Masoud and Pirandola, Stefano},
  journal = {Phys. Rev. A},
  volume = {108},
  issue = {4},
  pages = {042621},
  numpages = {10},
  year = {2023},
  month = {Oct},
  publisher = {American Physical Society},
  doi = {10.1103/PhysRevA.108.042621}
}

@article{Ghalaii2022,
  title = {Composable end-to-end security of Gaussian quantum networks with untrusted relays},
  author = {Ghalaii, Masoud and Papanastasiou, Panagiotis and Pirandola, Stefano},
  journal = {npj Quantum Inf},
  volume = {8},
  pages = {105},
  year = {2022},
  doi = {10.1038/s41534-022-00620-5}
}

@Article{Derkach2017,
  author    = {Derkach, Ivan and Usenko, Vladyslav C. and Filip, Radim},
  journal   = {Physical Review A},
  title     = {Continuous-variable quantum key distribution with a leakage from state preparation},
  year      = {2017},
    month     = dec,
  number    = {6},
  pages     = {062309},
  volume    = {96},
  doi       = {10.1103/physreva.96.062309},
  publisher = {American Physical Society (APS)},
}

@Article{Gisin2006,
  author    = {Gisin, N. and Fasel, S. and Kraus, B. and Zbinden, H. and Ribordy, G.},
  journal   = {Physical Review A},
  title     = {Trojan-horse attacks on quantum-key-distribution systems},
  year      = {2006},
    month     = feb,
  number    = {2},
  pages     = {022320},
  volume    = {73},
  doi       = {10.1103/physreva.73.022320},
  publisher = {American Physical Society (APS)},
}

@Article{Kumar2019,
  author    = {Kumar, Rupesh and Tang, Xinke and Wonfor, Adrian and Penty, Richard and White, Ian},
  journal   = {Journal of the Optical Society of America B},
  title     = {Continuous variable quantum key distribution with multi-mode signals for noisy detectors},
  year      = {2019},
    month     = feb,
  number    = {3},
  pages     = {B109},
  volume    = {36},
  doi       = {10.1364/josab.36.00b109},
  publisher = {The Optical Society},
}

@Article{Zheng2019a,
  author    = {Zheng, Yi and Huang, Peng and Huang, Anqi and Peng, Jinye and Zeng, Guihua},
  journal   = {Optics Express},
  title     = {Security analysis of practical continuous-variable quantum key distribution systems under laser seeding attack},
  year      = {2019},
    month     = sep,
  number    = {19},
  pages     = {27369},
  volume    = {27},
  doi       = {10.1364/oe.27.027369},
  publisher = {The Optical Society},
}

@Article{Zheng2019,
  author    = {Zheng, Yi and Huang, Peng and Huang, Anqi and Peng, Jinye and Zeng, Guihua},
  journal   = {Physical Review A},
  title     = {Practical security of continuous-variable quantum key distribution with reduced optical attenuation},
  year      = {2019},
    month     = jul,
  number    = {1},
  pages     = {012313},
  volume    = {100},
  doi       = {10.1103/physreva.100.012313},
  publisher = {American Physical Society (APS)},
}

@Article{Haeseler2008,
  author    = {H\"aseler, Hauke and Moroder, Tobias and L\"utkenhaus, Norbert},
  journal   = {Physical Review A},
  title     = {Testing quantum devices: Practical entanglement verification in bipartite optical systems},
  year      = {2008},
    month     = mar,
  number    = {3},
  pages     = {032303},
  volume    = {77},
  doi       = {10.1103/physreva.77.032303},
  publisher = {American Physical Society (APS)},
}

@Article{Zhao2018,
  author    = {Zhao, Yijia and Zhang, Yichen and Huang, Yundi and Xu, Bingjie and Yu, Song and Guo, Hong},
  journal   = {Journal of Physics B},
  title     = {Polarization attack on continuous-variable quantum key distribution},
  year      = {2018},
    month     = nov,
  number    = {1},
  pages     = {015501},
  volume    = {52},
  doi       = {10.1088/1361-6455/aaf0b7},
  }

@Article{Ren2019,
  author    = {Ren, Shengjun and Kumar, Rupesh and Wonfor, Adrian and Tang, Xinke and Penty, Richard and White, Ian},
  journal   = {Journal of the Optical Society of America B},
  title     = {Reference pulse attack on continuous variable quantum key distribution with local local oscillator under trusted phase noise},
  year      = {2019},
    month     = jan,
  number    = {3},
  pages     = {B7},
  volume    = {36},
  doi       = {10.1364/josab.36.0000b7},
  publisher = {The Optical Society},
}

@Article{Shao2022,
  author    = {Shao, Yun and Pan, Yan and Wang, Heng and Pi, Yaodi and Li, Yang and Ma, Li and Zhang, Yichen and Huang, Wei and Xu, Bingjie},
  journal   = {Entropy},
  title     = {Polarization Attack on Continuous-Variable Quantum Key Distribution with a Local Local Oscillator},
  year      = {2022},
    month     = jul,
  number    = {7},
  pages     = {992},
  volume    = {24},
  doi       = {10.3390/e24070992},
  publisher = {MDPI AG},
}

@Article{Fan2023,
  author    = {Fan, Lu and Bian, Yiming and Wu, Mingze and Zhang, Yichen and Yu, Song},
  journal   = {Physical Review Applied},
  title     = {Quantum Hacking Against Discrete-Modulated Continuous-Variable Quantum Key Distribution Using Modified Local Oscillator Intensity Attack with Random Fluctuations},
  year      = {2023},
    month     = aug,
  number    = {2},
  pages     = {024073},
  volume    = {20},
  doi       = {10.1103/physrevapplied.20.024073},
  publisher = {American Physical Society (APS)},
}

@Article{Hajomer2023,
      title={Continuous-Variable Quantum Key Distribution at 10 GBaud using an Integrated Photonic-Electronic Receiver}, 
      author={Adnan A. E. Hajomer and Cedric Bruynsteen and Ivan Derkach and Nitin Jain and Axl Bomhals and Sarah Bastiaens and Ulrik L. Andersen and Xin Yin and Tobias Gehring},
      year={2024},
      journal={Optica},
      volume={11},
      number={9},
      pages={1197--1204}
}

@Article{Hajomer2024,
  author    = {Hajomer, Adnan A. E. and Derkach, Ivan and Jain, Nitin and Chin, Hou-Man and Andersen, Ulrik L. and Gehring, Tobias},
  journal   = {Science Advances},
  title     = {Long-distance continuous-variable quantum key distribution over 100-km fiber with local local oscillator},
  year      = {2024},
    month     = jan,
  number    = {1},
  volume    = {10},
pages={eadi9474},
  doi       = {10.1126/sciadv.adi9474},
  publisher = {American Association for the Advancement of Science (AAAS)},
}

@Article{Tian2022,
  author    = {Tian, Yan and Wang, Pu and Liu, Jianqiang and Du, Shanna and Liu, Wenyuan and Lu, Zhenguo and Wang, Xuyang and Li, Yongmin},
  journal   = {Optica},
  title     = {Experimental demonstration of continuous-variable measurement-device-independent quantum key distribution over optical fiber},
  year      = {2022},
  month     = apr,
  number    = {5},
  pages     = {492},
  volume    = {9},
  doi       = {10.1364/optica.450573},
  publisher = {Optica Publishing Group},
}

@Article{Tian2023,
  author    = {Tian, Yan and Zhang, Yu and Liu, Shuaishuai and Wang, Pu and Lu, Zhenguo and Wang, Xuyang and Li, Yongmin},
  journal   = {Optics Letters},
  title     = {High-performance long-distance discrete-modulation continuous-variable quantum key distribution},
  year      = {2023},
  month     = may,
  number    = {11},
  pages     = {2953},
  volume    = {48},
  doi       = {10.1364/ol.492082},
  publisher = {Optica Publishing Group},
}

@Article{Xiang2010,
  author    = {Xiang, G. Y. and Ralph, T. C. and Lund, A. P. and Walk, N. and Pryde, G. J.},
  journal   = {Nature Photonics},
  title     = {Heralded noiseless linear amplification and distillation of entanglement},
  year      = {2010},
  month     = mar,
  number    = {5},
  pages     = {316--319},
  volume    = {4},
  doi       = {10.1038/nphoton.2010.35},
  publisher = {Springer Science and Business Media LLC},
}

@Article{Notarnicola2024,
  author    = {Notarnicola, M N and Cieciuch, F and Jarzyna, M},
  journal   = {New Journal of Physics},
  title     = {Continuous-variable quantum key distribution over multispan links employing phase-insensitive and phase-sensitive amplifiers},
  year      = {2024},
  month     = apr,
  number    = {4},
  pages     = {043015},
  volume    = {26},
  doi       = {10.1088/1367-2630/ad3774},
  }

@Article{Zhao2017,
  author    = {Zhao, Jie and Haw, Jing Yan and Symul, Thomas and Lam, Ping Koy and Assad, Syed M.},
  journal   = {Physical Review A},
  title     = {Characterization of a measurement-based noiseless linear amplifier and its applications},
  year      = {2017},
  month     = jul,
  number    = {1},
  pages     = {012319},
  volume    = {96},
  doi       = {10.1103/physreva.96.012319},
  publisher = {American Physical Society (APS)},
}

@Article{Dias2017,
  author    = {Dias, Josephine and Ralph, T. C.},
  journal   = {Physical Review A},
  title     = {Quantum repeaters using continuous-variable teleportation},
  year      = {2017},
  month     = feb,
  number    = {2},
  pages     = {022312},
  volume    = {95},
  doi       = {10.1103/physreva.95.022312},
  publisher = {American Physical Society (APS)},
}

@Article{Furrer2018,
  author    = {Furrer, Fabian and Munro, William J.},
  journal   = {Physical Review A},
  title     = {Repeaters for continuous-variable quantum communication},
  year      = {2018},
  month     = sep,
  number    = {3},
  pages     = {032335},
  volume    = {98},
  doi       = {10.1103/physreva.98.032335},
  publisher = {American Physical Society (APS)},
}

@Article{Ghalaii2020a,
  author    = {Ghalaii, Masoud and Ottaviani, Carlo and Kumar, Rupesh and Pirandola, Stefano and Razavi, Mohsen},
  journal   = {IEEE Journal on Selected Areas in Communications},
  title     = {Discrete-Modulation Continuous-Variable Quantum Key Distribution Enhanced by Quantum Scissors},
  year      = {2020},
  month     = mar,
  number    = {3},
  pages     = {506--516},
  volume    = {38},
  doi       = {10.1109/jsac.2020.2969058},
  publisher = {Institute of Electrical and Electronics Engineers (IEEE)},
}

@Article{Pegg1998,
  author    = {Pegg, David T. and Phillips, Lee S. and Barnett, Stephen M.},
  journal   = {Physical Review Letters},
  title     = {Optical State Truncation by Projection Synthesis},
  year      = {1998},
  month     = aug,
  number    = {8},
  pages     = {1604--1606},
  volume    = {81},
  doi       = {10.1103/physrevlett.81.1604},
  publisher = {American Physical Society (APS)},
}

@Article{Lund2009,
  author    = {Lund, A. P. and Ralph, T. C.},
  journal   = {Physical Review A},
  title     = {Continuous-variable entanglement distillation over a general lossy channel},
  year      = {2009},
  month     = sep,
  number    = {3},
  pages     = {032309},
  volume    = {80},
  doi       = {10.1103/physreva.80.032309},
  publisher = {American Physical Society (APS)},
}

@Article{Fiurasek2010,
  author    = {Fiur{\'a}{\v{s}}ek, Jarom{\'\i}r },
  journal   = {Physical Review A},
  title     = {Distillation and purification of symmetric entangled Gaussian states},
  year      = {2010},
  month     = oct,
  number    = {4},
  pages     = {042331},
  volume    = {82},
  doi       = {10.1103/physreva.82.042331},
  publisher = {American Physical Society (APS)},
}

@Article{Campbell2012,
  author    = {Campbell, Earl T. and Eisert, Jens},
  journal   = {Physical Review Letters},
  title     = {Gaussification and Entanglement Distillation of Continuous-Variable Systems: A Unifying Picture},
  year      = {2012},
  month     = jan,
  number    = {2},
  pages     = {020501},
  volume    = {108},
  doi       = {10.1103/physrevlett.108.020501},
  publisher = {American Physical Society (APS)},
}

@Article{Qu2017,
  author    = {Qu, Zhen and Djordjevic, Ivan B.},
  journal   = {Optics Express},
  title     = {High-speed free-space optical continuous-variable quantum key distribution enabled by three-dimensional multiplexing},
  year      = {2017},
  month     = mar,
  number    = {7},
  pages     = {7919},
  volume    = {25},
  doi       = {10.1364/oe.25.007919},
  publisher = {The Optical Society},
}

\end{document}